\documentclass[twocolumn,showpacs,preprintnumbers,nofootinbib,prd,superscriptaddress,10pt]{revtex4-2}

\usepackage{bm,graphics,graphicx,epsfig,soul,latexsym,mathrsfs,multirow,url,array}
\usepackage[dvipsnames]{xcolor}
\usepackage{amsmath,amssymb,gensymb}
\usepackage{setspace}
\usepackage{booktabs}
\usepackage{subcaption}
\usepackage{lipsum}
\usepackage[flushleft]{threeparttable}
\usepackage{natbib}
\usepackage{hyperref}
\usepackage{placeins}
\usepackage[english]{babel}
\usepackage[normalem]{ulem}
\usepackage{orcidlink}

\graphicspath{{Draft_figures/}{./}}
\newcommand{\red}{\color{red}}

\newcommand{\violet}{\color{black}}

\newcommand{\enigma}{\texttt{ENIGMA}}

\newcommand{\esigma}{\texttt{ESIGMA}}
\newcommand{\esigmahm}{\texttt{ESIGMAHM}}
\newcommand{\inspiralesigma}{\texttt{InspiralESIGMA}}
\newcommand{\inspiralesigmahm}{\texttt{InspiralESIGMAHM}}
\newcommand{\imresigma}{\texttt{IMRESIGMA}}
\newcommand{\imresigmahm}{\texttt{IMRESIGMAHM}}

\newcommand{\seobnrvfour}{\texttt{SEOBNRv4}}

\newcommand{\imrphenom}{\texttt{IMRPhenomXAS}}
\newcommand{\seobnrvfive}{\texttt{SEOBNRv5\_ROM}}
\newcommand{\nrsurdqfour}{\texttt{NRSur7dq4}}

\newcommand{\ftrans}{$f_{\rm{trans}}$}
\newcommand{\tintrans}{$t_{\rm{trans}}^{\rm{Insp}}$}
\newcommand{\tmrtrans}{$t_{\rm{trans}}^{\rm{PMR}}$}
\newcommand{\tintransstart}{$t_{\rm{trans,0}}^{\rm{Insp}}$}
\newcommand{\tmrtransstart}{$t_{\rm{trans,0}}^{\rm{PMR}}$}
\newcommand{\deltangw}{$\Delta \rm{N}_{\rm{GW}}$}
\definecolor{darkgreen}{rgb}{0,0.5,0}

\hypersetup{
    unicode=false,        
    pdftoolbar=true,      
    pdfmenubar=true,      
    pdfstartview={FitH},  
    pdftitle={IMRESIGMA}, 
    colorlinks=true,      
    linkcolor=red,        
    citecolor=blue,       
    filecolor=magenta,    
    urlcolor=darkgreen,   
    linktocpage=true
}

\newcommand{\CSGC}{\affiliation{Centre for Strings, Gravitation and Cosmology, Department of Physics, Indian Institute of Technology Madras, Chennai 600036, India}} %
\newcommand{\AEIP}{\affiliation{Max Planck Institute for Gravitational Physics
    (Albert Einstein Institute), Am M{\"u}hlenberg 1, 14476 Potsdam, Germany}}
\newcommand{\AEIH}{\affiliation{Max Planck Institute for Gravitational Physics
    (Albert Einstein Institute), Callinstra{\ss}e 38, D-30167 Hannover, Germany}}
\newcommand{\IITM}{\affiliation{Department of Physics, Indian Institute of Technology Madras, Chennai 600036, India}} %
\newcommand{\ICTS}{\affiliation{
    International Centre for Theoretical Sciences,
    Tata Institute of Fundamental Research, Bangalore 560089, India
}}
\newcommand{\BITSG}{\affiliation{
    Department of Physics, BITS-Pilani, K K Birla Goa Campus, Zuarinagar, Goa 403726, India
}}
\newcommand{\UFL}{\affiliation{
    Department of Physics, University of Florida, Gainesville, FL 32611, USA
}}

\allowdisplaybreaks
\begin{document}

\title{ESIGMAHM: An Eccentric, Spinning inspiral-merger-ringdown waveform model with Higher Modes for the detection and characterization of binary black holes}

\author{Kaushik Paul\orcidlink{0000-0002-8406-6503}}\email{kpaul.gw@gmail.com}\IITM\CSGC
\author{Akash Maurya\orcidlink{0009-0006-9399-9168}}\email{akash.maurya@icts.res.in}\ICTS
\author{Quentin Henry\orcidlink{0000-0003-4071-2873}}\email{quentin.henry@aei.mpg.de}\AEIP
\author{Kartikey Sharma\orcidlink{0000-0002-9665-5241}}\email{kartikey.sharma@aei.mpg.de}\AEIH\ICTS\BITSG
\author{Pranav Satheesh\orcidlink{0000-0003-0963-4686}}\email{pranavsatheesh@ufl.edu}\UFL\IITM\CSGC
\author{Divyajyoti\orcidlink{0000-0002-2787-1012}}\email{divyajyoti.physics@gmail.com}\IITM\CSGC
\author{Prayush Kumar \orcidlink{0000-0001-5523-4603}}\email{prayush@icts.res.in}\ICTS
\author{Chandra Kant Mishra \orcidlink{0000-0002-8115-8728}}\email{ckm@iitm.ac.in}\IITM\CSGC

\begin{abstract}
With the updated sensitivity of terrestrial gravitational-wave (GW) detectors in their fourth observing run, we expect a high rate of detection of binary black hole mergers.
With this comes the hope that we will detect rarer classes of merger events. 
Compact binaries formed dynamically in dense stellar environments are likely to be detected with a residual eccentricity as they enter the sensitivity band of ground-based GW detectors.
In this paper, we present a time-domain inspiral-merger-ringdowm (IMR) waveform model \esigmahm{} constructed within a framework we named \esigma{} for coalescing binaries of spinning black holes on moderately eccentric orbits~\citep{Huerta:2017kez}. We now include the effect of black hole spins on the dynamics of eccentric binaries, as well as model sub-dominant waveform harmonics emitted by them. The inspiral evolution is described by a consistent combination of latest results from post-Newtonian theory, self-force, and black hole perturbation theory. We assume that these moderately eccentric binaries radiate away most of their orbital eccentricity before merger, and seamlessly connect the eccentric inspiral with a numerical relativity based surrogate waveform model {\violet \nrsurdqfour{}} for mergers of spinning binaries on quasi-circular orbits. We present two variants of \esigmahm{}: the inspiral-only version is named \inspiralesigmahm{}, while the full IMR one is termed \imresigmahm{}; {or \inspiralesigma{} and \imresigma{} when only dominant modes are used}.
We validate \esigmahm{} against eccentric Numerical Relativity simulations, and also against contemporary effective-one-body and phenomenological models in the quasi-circular limit. We find that \esigmahm{} achieves match values greater than 99\% for quasi-circular spin-aligned binaries with mass ratios up to 8, and above 97\% for non-spinning and spinning eccentric systems with small or positively aligned spins. 
Using \imresigma{}, we quantify the impact of orbital eccentricity on GW signals, showing that next-generation detectors can detect eccentric sources up to 10\% louder than quasi-circular ones. We also show that current templated LIGO-Virgo-KAGRA searches will lose more than 10\% of optimal SNR for about 20\% of all eccentric sources by using only quasi-circular waveform templates. The same will result in a 25\% loss in detection rate for eccentric sources with mass ratios $m_1/m_2\geq 4$. Our results highlight the need for including eccentricity and higher-order modes in GW source models and searches for asymmetric eccentric BBH signals. 

\end{abstract}
\date{\today}
\maketitle
\section{Introduction} 
The detection of gravitational waves (GWs) by LIGO and Virgo detectors~\citep{LIGOScientific:2016aoc,LIGOScientific:2017vwq} has opened a new window  to probe the astrophysics of binary compact objects. By the end of the third observing run (O3), the ground-based network of detectors comprising of LIGO~\citep{LIGOScientific:2014pky} and Virgo~\citep{VIRGO:2014yos} detectors had detected {\violet more than 90} compact binary coalescence (CBC) events~\citep{LIGOScientific:2018mvr, LIGOScientific:2020ibl, LIGOScientific:2021usb, LIGOScientific:2021djp}. Compact binaries of stellar-mass black holes (BHs) and neutron stars (NSs) are prime targets for the existing and upcoming ground-based detectors such as LIGO A+~\cite{Shoemaker:2019bqt, McClelland:T1500290-v3}, Voyager~\cite{McClelland:T1500290-v3}, Cosmic Explorer (CE)~\cite{Dwyer:2014fpa}, and Einstein Telescope (ET)~\cite{Punturo:2010zz}. A majority of these binaries are expected to evolve from binary stars in galactic fields~\citep{Postnov:2006hka}. Their orbits are expected to circularize before reaching the frequency band of ground-based detectors by radiating eccentricity away during the long inspiral~\citep{Peters:1963ux,Peters:1964zz}. However, for a sub-population of compact binaries formed via dynamical interaction in dense stellar environments~\cite{Clausen:2012zu,Samsing:2013kua,Samsing:2017xmd,Rodriguez:2017pec,Zevin:2018kzq} or through Kozai-Lidov processes~\citep{Kozai:1962zz, LIDOV1962719}, this will not be the case. Such sources can have measurable residual eccentricities as they enter the sensitivity bands of ground-based detectors~\citep{LIGOScientific:2018mvr, LIGOScientific:2020ufj, Antonini:2019ulv,LIGOScientific:2020ibl,LIGOScientific:2021usb,KAGRA:2021vkt,LIGOScientific:2018jsj,LIGOScientific:2020kqk,KAGRA:2021duu,Gondan:2020svr,Tiwari:2015gal}. {\violet Indeed, recent follow-up analyses of GW190521~\citep{Abbott:2020tfl, CalderonBustillo:2020xms,Gamba:2021gap} claim this event to be consistent with a binary black hole (BBH) system with eccentricity ranging from $\sim 0.1-0.7$~\citep{Romero-Shaw:2020thy,Gayathri:2020coq,OShea:2021faf}.} Carefully measuring the imprints of eccentricity in GW signals from CBCs will allow detailed studies of binary formation and evolution channels, as well as supporting astrophysical processes in globular clusters and galactic nuclei, which otherwise could remain inaccessible~\cite{Samsing:2013kua,Rodriguez:2016kxx,Antonini:2015zsa,Chomiuk:2013qya,Strader:2012wj,Osburn:2015duj,VanLandingham:2016ccd,Hoang:2017fvh,Gondan:2017hbp,Gondan:2020svr,Kumamoto:2018gdg,Fragione:2019hqt,OLeary:2005vqo}. 
As the sensitivities of the current network of detectors~\citep{KAGRA:2013rdx} improve at lower frequencies, we expect to observe an ever-increasing number of GW events, improving the possibility of detecting an eccentric CBC population~\citep{LIGOScientific:2019dag,Saini:2023wdk}.

Detection of GW signals from CBCs in noise-dominated interferometric data involves matched filtering~\citep{Cutler:1994ys, Poisson:1995ef, Krolak:1995md}, where pre-computed waveforms (called templates) for the expected sources are cross-correlated with noisy detector output~\citep{LIGOScientific:2016vbw}. This matched-filtering technique is very sensitive to filter templates. Extracting information optimally from GW signals therefore requires highly accurate templates~\citep{Lindblom:2008cm,Lindblom:2009ux,Lindblom:2010mh,Kumar:2015tha,Kumar:2016dhh,Purrer:2019jcp} that include all physical phenomenology we expect our sources to exhibit~\citep{OShea:2021faf}. 
The presence of eccentricity alters the length of the signal when compared to a circular counterpart with exact same parameters (see Fig. 9 of Ref.~\citep{Chen:2020lzc}).  It also introduces amplitude and phase modulations at orbital timescale~\citep{Ramos-Buades:2019uvh}. Spins of binary constituents also modify the size and shape of the waveform. For example, the length of the gravitational waveform for an (anti-) aligned spin BBH can be significantly (shorter) longer compared to a non-spinning one~\citep{Campanelli:2006uy} (see Fig. 3-6 of Ref.~\citep{Taracchini:2012ig}). 
Thus the combined effect of component spins and orbital eccentricity could be moderately degenerate~\citep{OShea:2021faf}. Further, in order to describe binaries with asymmetric mass components, subdominant higher-order modes (HMs) play a crucial role. The inclusion of HMs introduces additional dependencies between binary parameters, which are useful in resolving degeneracies between different binary parameters such as luminosity distance and orbital inclination~\citep{Brown:2012nn,Kumar:2018hml,Usman:2018imj,Wadekar:2024zdq}. HMs are also key to measure recoils due to BBH kicks, in particular its direction~\citep{Campanelli:2007ew, Herrmann:2007ac,CalderonBustillo:2018zuq, CalderonBustillo:2022ldv}, as well as for robust tests of General Relativity~\citep{Pang:2018hjb,Gupta:2024gun}.

Orbital eccentricity has been mostly ignored in traditional GW data analyses (owing in part to the expected circularization, but also due to unavailability of reliable waveform models), and that could lead to a significant reduction in detection rates of eccentric binaries by current GW detectors~\citep{Huerta:2013qb}.
Furthermore, ignoring eccentricity in our models can lead to systematic biases if the actual signal corresponds to an eccentric system~\citep{Ramos-Buades:2019uvh,OShea:2021faf,Divyajyoti:2023rht}. Such biases can lead to eccentric sources mimicking more massive binaries~\citep{OShea:2021faf,Divyajyoti:2023rht}, and manifesting false violations of General Relativity (GR)~\citep{Narayan:2023vhm,Shaikh:2024wyn,Gupta:2024gun}.
Ignoring eccentricity could potentially also lead to misleading evidence of spin-orbit precession as well~\cite{CalderonBustillo:2020xms}, although this is still debated in literature~\cite{Divyajyoti:2023rht}.
Naturally, it is crucial to simultaneously account for orbital eccentricity, component spins, and higher-order GW modes in filter templates in order to optimize detection efficiency and extract realistic source information from eccentric GW events~\citep{Brown:2009ng,Brown:2012nn,Huerta:2013qb,Coughlin:2014swa}. And, even if we only observe binaries that are quasi-circular, accurate eccentric models will be required to place bounds on their eccentricity.

Numerical relativity (NR) simulations provide the most accurate gravitational waveforms for BBHs, but they are prohibitively computationally expensive~\citep{SXS:catalog} to be used directly, except in restricted applications~\citep{Kumar:2013gwa, Lange:2017wki, LIGOScientific:2016kms}. Data-driven surrogate modeling strategies have also been developed that can produce waveforms that are arbitrarily indistinguishable from NR, with relatively \textit{trivial computational cost}~\citep{Field:2013cfa, Blackman:2015pia, Blackman:2017dfb, Blackman:2017pcm, Varma:2018mmi, Rifat:2019ltp, Varma:2019csw}. Surrogate models are usually restricted in parameter space however, and are still in early development when it comes to the space of eccentric compact binaries~\citep{Islam:2021mha}.

Contemporary GW data analyses therefore predominantly rely on waveform models that arise out of an amalgamation of multiple analytical approaches to the coalescing two-body problem in GR, calibrated to NR simulations of the late stages of binary mergers~\citep{Abbott:2016wiq}. {\violet Most of these models in literature have been traditionally focused on quasi-circular spinning compact binaries~\citep{Buonanno:1998gg, Buonanno:2006ui, Pound:2012nt,Hannam:2013oca,Damour:2014sva,Bohe:2016gbl, Nagar:2018zoe,Cotesta:2018fcv, Pound:2019lzj,Nagar:2020pcj,Estelles:2020osj,Estelles:2020twz, Ossokine:2020kjp, Akcay:2020qrj, Gamba:2021ydi, Estelles:2021gvs,Riemenschneider:2021ppj, Wardell:2021fyy,Mathews:2021rod,Nagar:2022icd, Thomas:2022rmc, Gadre:2022sed,Khalil:2023kep, Pompili:2023tna, Ramos-Buades:2023ehm, Garcia-Quiros:2020qpx, Pratten:2020ceb,Yu:2023lml, Thompson:2023ase}. 
In the last few years, especially since the discovery of GW150914, progress in constructing waveform models including eccentricity effects has also been made~\citep{Konigsdorffer:2006zt,Yunes:2009yz,Klein:2010ti,Mishra:2015bqa,Moore:2016qxz, Tanay:2016zog, Klein:2018ybm,Boetzel:2019nfw, Ebersold:2019kdc, Moore:2019xkm, Klein:2021jtd, Khalil:2021txt, Paul:2022xfy,Nagar:2022fep, Henry:2023tka, Albanesi:2023bgi,Albertini:2023aol, Nagar:2024dzj, Nagar:2024oyk, Klein:2013qda,Huerta:2014eca,Moore:2018kvz,Klein:2018ybm,Tanay:2019knc,Liu:2019jpg,Tiwari:2020hsu,Klein:2021jtd,Huerta:2016rwp,Hinderer:2017jcs,Huerta:2017kez,Hinder:2017sxy,Huerta:2017kez,Cao:2017ndf,Taracchini:2012ig,Chen:2020lzc,Chiaramello:2020ehz,Nagar:2018zoe, Nagar:2020pcj,Nagar:2021gss,Ramos-Buades:2021adz,Chattaraj:2022tay, Manna:2024ycx, Carullo:2023kvj, Carullo:2024smg,Islam:2021mha,Yun:2021jnh,Becker:2024xdi,Nagar:2024oyk}.} {\violet A series of eccentric inspiral-only~\citep{Klein:2013qda,Huerta:2014eca,Moore:2018kvz,Klein:2018ybm,Tanay:2019knc,Liu:2019jpg,Tiwari:2020hsu,Klein:2021jtd} and inspiral-merger-ringdown (IMR) models~\citep{Huerta:2016rwp,Hinderer:2017jcs,Huerta:2017kez,Hinder:2017sxy,Huerta:2017kez,Cao:2017ndf,Taracchini:2012ig,Chen:2020lzc,Chiaramello:2020ehz,Nagar:2018zoe, Nagar:2020pcj,Nagar:2021gss,Ramos-Buades:2021adz,Chattaraj:2022tay, Manna:2024ycx, Carullo:2023kvj, Carullo:2024smg,Islam:2021mha,Yun:2021jnh,Becker:2024xdi,Nagar:2024oyk,Nagar:2024dzj,Albertini:2023aol,Albanesi:2023bgi} have become available.} We highlight some recent ones. 
The first IMR model, \enigma{}~\citep{Huerta:2016rwp,Huerta:2017kez}, was aimed at moderate initial eccentricities (up to $\sim 0.4$), and combined an eccentric inspiral description with a \textit{quasi-circular} merger model~\citep{Huerta:2016rwp}. This paradigm was subsequently {\violet extended} by many modeling efforts. 
\texttt{SEOBNRE} modified an aligned-spin quasi-circular EOB waveform model (\texttt{SEOBNRv1}) to include effects of eccentricity in the radiative dynamics and waveform multipoles~\citep{Cao:2017ndf,Taracchini:2012ig}. 
Similarly, Ref.~\citep{Chiaramello:2020ehz} modified a different aligned-spin EOB waveform model (\texttt{TEOBResumS\_SM})~\citep{Nagar:2018zoe, Nagar:2020pcj} to include effects of eccentricity. This model was further generalized by replacing the base quasi-circular model with one capable of describing generic-orbits~\citep{Nagar:2021gss}.  {\violet All of these models extend upon the paradigm of \enigma{} by using an eccentric description all the way up to the binary merger.}
There are other progresses in eccentric waveform modeling~\citep{Chattaraj:2022tay, Manna:2024ycx, Carullo:2023kvj, Carullo:2024smg} {\violet in surrogate models~\citep{Islam:2021mha, Yun:2021jnh}} and in the EOB framework~\citep{Ramos-Buades:2021adz}. {\violet In addition to these models, Ref.~\citep{Setyawati:2021gom, Islam:2024bza,Islam:2024rhm,Islam:2024zqo}, recently developed a method to add eccentricity induced modulations to existing quasi-circular BBH waveforms. 
The space of the eccentric spinning IMR models is sparsely populated in the literature though, especially since {\it long} eccentric simulations are the ones required for meaningful comparison with PN/BHP/self-force solutions~\citep{Gold:2012tk,Healy:2022wdn}. And their accuracy has not been as rigorously tested as it has been for quasi-circular waveform models over the past decades~\citep{Kumar:2016dhh,Kumar:2015tha,MacUilliam:2024oif}.}

In this paper, we present a full IMR gravitational waveform model \esigmahm{} (\textbf{E}ccentric \textbf{S}pinning \textbf{I}nspiral with \textbf{G}eneralized \textbf{M}erger \textbf{A}pproximant and \textbf{H}igher-order \textbf{M}odes). It is a time-domain aligned-spin eccentric waveform model that includes higher gravitational wave harmonics, and is a significant extension of the \enigma{} framework~\citep{Huerta:2017kez}. \esigmahm{} carefully combines results from PN theory and self-force program to describe the low-frequency, weak field inspiral dynamics and GW emission. The high-frequency, strong field {\violet plunge-merger-ringdown (PMR)} uses a NR based surrogate model. The fundamental assumption we invoke remains that by the time the evolution of the binary is transitioning from inspiral to plunge and merger, it has lost most of its eccentricity via GW emission. This allows us to attach quasi-circular {\violet PMR} to an eccentric inspiral. 
The inspiral segment of \esigmahm{} has the latest spinning eccentric PN corrections reported in literature, and we use \nrsurdqfour{}~\citep{Varma:2019csw} as our {\violet PMR} model, which is nearly as accurate as NR simulations themselves in describing quasi-circular BBH mergers. We present two versions of \esigmahm{}: the inspiral-only version is named \inspiralesigmahm{}, while the full IMR one is termed \imresigmahm{}; { or \inspiralesigma{} and \imresigma{} when only dominant $\ell = |m| = 2$ modes are chosen. Note that, we use \imresigma{} and \imresigmahm{} interchangeably in the rest of the paper to refer to our IMR model and clarify the inclusion of HMs wherever they're used.}

The \esigmahm{} model was validated by comparing it with existing quasi-circular spinning waveform models and NR simulations.
The first comparisons aimed to quantify the model’s accuracy in reproducing the true inspiral dynamics of spinning systems in the quasi-circular limit. Matches against models like \seobnrvfour{}, \imrphenom{}, and \seobnrvfive{} showed (see Fig.~\ref{fig:mismatch_qc}) that \inspiralesigma{} successfully reproduced the dynamics of spinning systems with match values greater than 99\% for spin-aligned binaries with mass-ratios up to 8.
Further validation using {\it{both quasi-circular and eccentric}} NR waveforms from the public SXS catalog~\citep{Boyle:2019kee} demonstrated (see Figs.~\ref{fig:NR_qc_mismatches},~\ref{fig:NR_comp_p1},~\ref{fig:NR_comp},~\ref{fig:mismatch_NR_mass} \&~\ref{fig:mismatch_anti_aligned}) that \imresigmahm{} closely reproduced NR simulations for both non-spinning and spinning systems, with matches well above 97\% for all mass-ratios and small or positively aligned-spin cases over a range of binary masses from $20M_\odot$ to $100M_\odot$.
For both quasi-circular and eccentric binaries with large anti-aligned component spins (with respect to the orbital angular momentum) and large mass-ratios, however, matches smaller than 97\% were observed between \imresigma{} and NR simulations. This is likely due to combination of (a) weakening of the assumption that binaries circularize beyond the inspiral-merger transition frequency, and (b) late-inspiral requiring higher-order spin information than currently incorporated in the model. We defer a detailed investigation to future work, and recommend the use of \esigmahm{} to model binaries of moderately spinning black holes merging on moderately eccentric orbits.

Using the \imresigma{} model, we quantify the impact of orbital eccentricity (see Figs.~\ref{fig:Ngw_M_q_e} \&~\ref{fig:SNR_e0}) on GW signals and the contribution of subdominant waveform harmonics (see Fig.~\ref{fig:SNR_ET_CE}) for eccentric binaries, focusing on current and third-generation ground-based GW observatories. The increased inspiral rate in eccentric binaries results in a shorter time to merger, quantified by the difference in the number of GW cycles, \deltangw{}, between eccentric and quasi-circular cases. The study shows that \deltangw{} increases with mass ratio for small total masses and exhibits a sharp non-linear growth with increasing initial eccentricity and mass ratio. Additionally, we analyze the change in optimal signal-to-noise ratio (SNR) for eccentric binaries compared to quasi-circular ones, revealing that next-generation detectors like the Einstein Telescope and Cosmic Explorer can detect eccentric sources up to 10\% louder than quasi-circular ones, thereby increasing the likelihood of discovering a significant population of dynamically captured binaries.

We also investigate the effectiveness of current GW searches on LIGO-Virgo-KAGRA data in detecting eccentric binaries of spinning black holes (see Figs.~\ref{fig:FF_dist} \& \ref{fig:effectualness}). As is used in contemporary matched-filtering searches with quasi-circular aligned-spin binary templates~\citep{LIGOScientific:2016vbw}, we create a standard search template bank for aligned-spin binaries~\citep{Brown:2012nn,Brown:2012qf} with component masses between $5-50M_\odot$ and dimensionless spins in the range $[-0.9,0.9]$. Validated with $50,000$ quasi-circular signals, the bank recovers nearly 99\% of signals with fitting factors above 97\%. However, for eccentric signals with eccentricities up to 0.4, the bank recovers only 40\% of sources with the intended 97\% optimal SNR, with 10\% of signals losing more than 10\% of optimal SNR. The effect of sub-dominant modes is smaller but still significant, with 20\% of signals losing more than 10\% of SNR when the eccentric signals included sub-dominant modes with $(\ell, |m|)=(2,1), (3, 3), (3,2), (4,4), (4,3)$ in addition to the dominant $\ell=|m|=2$ modes. For sources with mass ratios $q\geq 4$, current searches can miss 10\% of optimal SNR for even small eccentricities, leading to a 25\% reduction in detection rates for high mass-ratio eccentric sources. This highlights the importance of including orbital eccentricity effects and higher-order modes in GW searches for asymmetric eccentric BBH signals.

This article is organized as follows: Section~\ref{sec:exec_summary} gives a summary of the current work. Section~\ref{sec:wf_model} provides a description of our model and its validity in quasi-circular limits with {{NR simulations}} and existing models. We also discuss the modifications we made in \esigmahm{} over its previous version \enigma{}. In Section~\ref{sec:validation_NR}, a comparison of \imresigma{} with the publicly available spinning eccentric NR simulations is shown. Section~\ref{sec:HM} estimates the importance of including {{eccentricity and}} higher-order modes in the waveform model. Finally, in Section~\ref{sec:conclusion}, we summarize our findings and propose our future plans with \esigmahm{}.

\section{Executive Summary}
\label{sec:exec_summary}

The inaugural work~\citep{Huerta:2016rwp} in the \enigma{} series introduced a non-spinning, moderately eccentric, dominant mode, time-domain, IMR waveform model. {\violet While the inspiral segment was constructed using a consistent combination of post-Newtonian, self-force and black hole perturbation theory results, the PMR part was based on an implicit-rotating-source (IRS) model based on Ref.~\citep{Baker:2008mj,Kelly:2011bp}.}  It reproduced the accurate dynamics of moderately eccentric BBH mergers with mass-ratios $q \in \{1, 2\}$ from inspiral through merger. In the non-spinning limit, this waveform agreed well with an IMR effective-one-body (EOB) model ~\citep{Taracchini:2012ig, Taracchini:2013rva}.

The second paper~\citep{Huerta:2017kez} introduced a Gaussian-process interpolated IMR model named as \textbf{E}ccentric \textbf{N}on-spinning \textbf{I}nspiral \textbf{G}aussian \textbf{M}erger \textbf{A}pproximant (hence \enigma{}). The inspiral segment was improved by including the quasi-circular corrections in energy flux from black hole perturbation theory (BHPT) up to {\violet 4PN} order. \enigma{} was validated using a set of Einstein Toolkit eccentric NR waveforms, with mass-ratios between $1 \leq q \leq 5.5$, and eccentricities $e_0 \leq 0.2$ ten orbits before the merger. \enigma{} reproduced the EOB model, \seobnrvfour{}~\citep{Bohe:2016gbl} in the non-spinning, quasi-circular limit with a match greater than $99\%$.

Further, Ref.~\citep{Chen:2020lzc} upgraded the model by improving how the inspiral and PMR portions are stitched together. This improved version reproduced the physics of quasi-circular mergers with very high accuracy over the applicable parameter space of \enigma{}. 

This paper extends these works~\citep{Huerta:2016rwp, Huerta:2017kez, Chen:2020lzc} and introduces our new waveform model \esigmahm{}:  \textbf{E}ccentric \textbf{S}pinning \textbf{I}nspiral \textbf{G}eneralized \textbf{M}erger \textbf{A}pproximant with \textbf{H}igher-order \textbf{M}odes, which incorporates component spin effects and includes sub-dominant waveform multipoles.
\esigmahm{} includes recently computed high-order PN corrections to radiation reaction and gravitational waveform for compact binaries on eccentric orbits with spinning components~\citep{Henry:2023tka} in the inspiral piece. It employs a recent NR surrogate model \nrsurdqfour{}~\citep{Varma:2019csw} for quasi-circular spinning binaries as the PMR approximant. This construction is valid for eccentric CBC sources that radiate away eccentricity by the end of inspiral to negligible levels. 
In rest of this paper we will refer to two variants of our model:  \inspiralesigmahm{} will refer to an inspiral-only version with waveform multipoles included up to $\ell = 8$; and \imresigmahm{}, which is the full IMR version  that will include modes $(2, \pm 1), (3, \pm 3), (3, \pm 2), (4, \pm 4), (4, \pm 3)$ apart from dominant $(2, \pm 2)$ modes. 
\imresigma{} can be used for the detection and follow-up of binaries with moderate eccentricities $e_0 \le 0.4$~\citep{Hinder:2008kv, Henry:2023tka}, mass-ratios up to $q \le 6$ and positive-aligned or small anti-aligned component spins.

We investigated the impact of orbital eccentricity on (GW) signals using the \imresigma{} model. Our focus is on both current and third-generation ground-based GW observatories. We show that eccentric binaries exhibit an increased inspiral rate, leading to a significantly shorter time to merger compared to quasi-circular cases. Specifically, we show that the difference in the number of GW cycles increases very rapidly with mass ratio and initial eccentricity. Additionally, we analyze the optimal signal-to-noise ratio (SNR) for eccentric binaries, revealing that next-generation detectors like the Einstein Telescope and Cosmic Explorer can detect eccentric sources up to 10\% louder than quasi-circular ones. We also show that current matched-filtering searches may miss 10\% or more of optimal SNR for eccentric sources, and lead to a $25\%$ reduction in the detection rates of those with mass ratios above $4$. Note that the SNR losses we find due to the omission of HMs for eccentric systems are qualitatively consistent with the impact such omission has in non-eccentric cases~\citep{Capano:2013raa, CalderonBustillo:2015lrt,CalderonBustillo:2016rlt, Varma:2014jxa}. These results emphasize the importance of including orbital eccentricity effects and higher-order modes in GW waveform models as well as using such models in GW searches for BBHs. {\violet Further, a long term goal of the \esigma{} framework would be to iteratively develop better waveform models for the eccentric inspiral/plunge and ultimately use eccentric NR surrogates to represent the {\violet plunge-merger-ringdown.}}

\section{Waveform Model Construction}
\label{sec:wf_model}

The framework \esigma{} is built with two primary components. The first one deals with the inspiral evolution that combines input from post-Newtonian theory~\citep{Blanchet:2013haa} and from the self-force approach~\citep{Barausse:2011dq,Pound:2021qin} describing the dynamics of spinning, eccentric compact binary systems. While the second is the PMR segment that has been described by an NR based surrogate model. The following subsections elaborate on these individual pieces.

\subsection{Inspiral Model}
\label{sec:insp_evol}

The inspiral segment of \esigma{} framework is modeled using inputs from the PN theory and self-force program. These corrections are incorporated in the framework in two forms. First is in the evolution of orbital elements such as eccentricity $(e)$, PN parameter $(x)$, mean anomaly $(l)$, and orbital phase $(\phi)$ that describe the dynamical evolution of the system.
The evolution equations for these orbital variables, including spins and eccentricity, were computed up to 3PN order in Ref.~\citep{Klein:2018ybm,Henry:2023tka}. The generalized quasi-Keplerian (QK) representation for non-spinning~\citep{damour-1985, Memmesheimer:2004cv, Hinder:2008kv} and spinning systems~\citep{Klein:2018ybm, Henry:2023tka} are used to describe the inspiral evolution. Besides, {\violet 4PN} corrections from the gravitational self-force approach to the binding energy of compact binaries and energy flux of quasi-circular binaries are also included. 
When modeling the inspiral we work with the adiabatic approximation in which radiation reaction time-scale is much longer compared to orbital and precessional time scales associated with eccentric orbits~\citep{Blanchet:1989cu}. Under this assumption, we can use an orbit-averaged description of the variables, describing the conservative and radiative dynamics of the system. 
 
Second is the spherical harmonic modes of the gravitational waveform. 
There have been several efforts in the past to compute the effects of spins in spherical harmonic mode amplitudes~\citep{Kidder:1995zr, Kidder:2007gz, Kidder:2007rt,Buonanno:2012rv, Mishra:2015bqa,  Blanchet:2008je, Paul:2022xfy, Henry:2021cek, Henry:2022dzx, Henry:2023tka}. The 3PN mode results for non-spinning generic orbits were computed in Ref.~\citep{Mishra:2015bqa}. The spinning generic orbit mode expressions were computed to 2PN order in Ref.~\citep{Khalil:2021txt, Paul:2022xfy} \& pushed to 3PN in Ref.~\citep{Henry:2023tka}.

In the quasi-Keplerian approach~\citep{damour-1985}, the dynamics of a compact binary is described using the orbital variables such as, eccentricity ($e$), mean anomaly ($l$), eccentric anomaly ($u$), and a PN parameter $x=\left(M\Omega\right)^{2/3}$, where $M$ and $\Omega$ represent the total mass and the GW half frequency of the binary respectively.\footnote{This choice of the PN parameter $(x)$ is motivated by exact recovery of circular results in the limit $e\rightarrow0$~\citep{Arun:2007sg}.}\footnote{The GW half frequency $(\Omega)$ coincides with the orbital frequency $(\omega)$ up to $3.5$PN order, and the relation between them is given in Eq.~(8) of Ref.~\citep{Blanchet:2023bwj}.} In a structural form, the PN relations connecting the relative separation ($r$), $l$ to $e$, and $u$ are given by,
\begin{align}
\label{eq:rel_sep}
\frac{r}{M}&= \frac{1-e\,\cos u}{x} + \sum_{i=1}^{3}a_{i}x^{i-1}\,, \\
\label{eq:mean_ano}
l &= u -e\, \sin u + \sum_{i=2}^{3} b_{i}x^{i}\,,
\end{align}
where, $a_{i}$, $b_{i}$ represent the PN corrections up to 3PN~\citep{Huerta:2017kez, Chen:2020lzc, Henry:2023tka}, and index $i$ can take half-integer values. Note that the eccentricity $(e)$ used here is time-eccentricity $(e_t)$ defined in Ref.~\citep{damour-1985,Memmesheimer:2004cv, Arun:2007sg}. Beyond the leading order in quasi-Keplerian representation there exist two other eccentricities related to $r$ \& $\phi$ coordinates $(e_r, e_{\phi})$ which in turn can be written in terms of $e_t$. Besides, we use modified harmonic (MH) gauge (see Ref.~\citep{Memmesheimer:2004cv, Arun:2007sg} for a discussion related to different gauges), i.e., all the derived quantities are in MH coordinates.

The binary's conservative dynamics is given by the time evolution equations for instantaneous angular position $\phi$, and mean anomaly $l$. In a structural form, they're given by,
\begin{align}
\label{eq:con_dyn}
M \dot{\phi}&= x^{3/2}\bigg[\sum_{i=0}^{3}c_{i}x^{i}+\mathcal{O}\Big(x^{4 }\Big)\bigg]\,, \\
M\dot{l}&=x^{3/2}\bigg[1+\sum_{i=1}^{3}d_{i}x^{i}+\mathcal{O}\Big(x^{4}\Big)\bigg]\,.
\end{align} 
Explicit expressions of the PN coefficients $c_{i}$ and $d_{i}$, which contain the eccentric corrections to 3PN, are provided in Ref.~\citep{Hinder:2008kv, Henry:2023tka}.

Over radiation reaction time scales, the binary loses energy and angular momentum through the emission of GWs. This loss in energy and angular momentum can be described by a set of orbit-averaged time evolution equations for $e$ and $x$, and can be written compactly as,
\begin{align}
   \label{eq:rad_dyn}
    M\dot{x} &= x^5\,\bigg[\sum_{i=0}^{4} y_{i}x^{i} +\mathcal{O}\Big(x^{9/2}\Big)\bigg]\,, \\
    M\dot{e}&=x^{4} \bigg[\sum_{i=0}^{3}z_{i}x^{i} + \mathcal{O}\Big(x^4\Big)\bigg] \,.
\end{align} 
The PN coefficients $y_{i}$ and $ z_{i}$ to 3PN order for non-spinning and spinning case for eccentric orbits are derived in Ref.~\citep{Hinder:2008kv, Huerta:2016rwp, Henry:2023tka}. Additionally, they also list spin-orbit (SO) and spin-spin(SS) terms to 4PN order for binaries on quasi-circular orbits.
The 4PN non-spinning (NS) and 3.5PN cubic-in-spin (SSS)  quasi-circular contributions to $\dot{x}$ are derived here using the expressions for the energy and the flux~\citep{Cho:2022syn, Blanchet:2023bwj} in the energy balance equation, and we provide them in Appendix~\ref{sec:appendix}.

The GW strain is represented as a linear superposition of spherical harmonic modes of the waveform using a spin-weighted spherical harmonic basis of weight -2 $(Y^{\ell m}_{-2})$ as follows,
\begin{align}
    h\left(t; \Theta, \Phi\right)  = \sum_{\ell=2}^{\infty}\sum_{m=-\ell}^{+\ell}h^{\ell m}\left(t\right) Y^{\ell m}_{-2}\Big(\Theta,\Phi\Big)\,,
    \label{eq:GW_strain}
\end{align}
where the angles $\left(\Theta, \Phi\right)$ specify the binary's orientation with respect to the observer at a reference. Both instantaneous (the part of GW radiation that depends on the state of the source at a given retarded time) and hereditary effects (the part of GW radiation which depends on the dynamical history of the source) included in the inspiral version of our model. The instantaneous terms in waveform modes include non-spinning and spinning corrections for general orbits, while the hereditary part only includes corrections in the quasi-circular orbit.\footnote{Hereditary contributions include tails, tails-of-tails, tail-of-memory and the non-linear memory contributions in the gravitational waveform~\citep{Blanchet:2013haa}. {\violet Note that $(\ell, 0)$ modes are not included in the inspiral part of the model. Consequently, the non-linear Christodoulou memory~\citep{Thorne:1992sdb} is not taken into account. The tail-of-memory and non-linear memory (except Christodoulou memory) contributions are present in modes other than $(\ell, 0)$.}}
\begin{table*}[t]
\begin{center}
\begin{threeparttable}
\caption{New post-Newtonian effects included in \esigma{} framework as part of upgrades to the previous version \enigma{}~\citep{Chen:2020lzc}.\\}
\begin{tabular}{c|c|c|c}
\hline\hline
& & &\\
Effect & Nature & Order & References\\
& & &\\
\hline
\multirow{4}{*}{$h^{\ell m}$} & quasi-circular (QC), non-spinning (NS) & 3PN, 3.5PN, 4PN ($h^{22}$ only)  & \citep{Kidder:2007rt,Blanchet:2008je,Favata:2008yd, Faye:2012we,Faye:2014fra, Henry:2021cek, Henry:2022ccf} \\
& QC, spinning (S) & up to 3.5PN  & \citep{Buonanno:2012rv,Henry:2022dzx} \\
& eccentric\tnote{\red{*}}
(E), NS & up to 3.5PN  & \citep{Gopakumar:2001dy,Mishra:2015bqa}, App.~\ref{sec:appendix_hlm}\\
& E\tnote{\red{*}}, S & up to 3.5PN  & \citep{Majar:2008zz,Khalil:2021txt,Paul:2022xfy, Henry:2023tka}, App.~\ref{sec:appendix_hlm}\\
\hline
\multirow{4}{*}{$\dot{x}$} & QC, NS & 4PN  & \citep{Marchand:2016vox, Blanchet:2023sbv, Blanchet:2023bwj}, App.~\ref{sec:appendix} \\
& QC, S & up to 4PN  & \citep{Marsat:2012fn, Bohe:2013cla, Marsat:2013caa, Cho:2022syn}, App.~\ref{sec:appendix} \\
& E, NS & ---  & \citep{Damour:2004bz,Hinder:2008kv,Arun:2009mc, Ebersold:2019kdc}\\
& E, S & up to 3PN  & \citep{Klein:2018ybm, Henry:2023tka}\\
\hline
\multirow{2}{*}{$\dot{e}$,\,$\dot{l}$,\,$\dot{\phi}$} & E, NS & --- & \citep{Damour:2004bz, Hinder:2008kv, Arun:2009mc}\\
& E, S & up to 3PN & \citep{Klein:2018ybm, Henry:2023tka} \\
\hline
\hline
\end{tabular}
\label{table:PN_order}
\begin{tablenotes}
{\item[\red{*}]\small{Only instantaneous contributions to GW modes are included. 
(See Sec.~\ref{sec:wf_model} for a discussion)
}}.
\end{tablenotes}
\end{threeparttable}
\end{center}
\end{table*}

Table~\ref{table:PN_order} summarizes all the effects and their corresponding PN orders that are included in the conservative, radiative dynamics, and spherical harmonic modes of the \esigma{} framework. The radiative and the conservative sector is updated with the recently reported spinning, eccentric corrections up to 3PN~\citep{Henry:2023tka}. 
In this paper, we performed additional calculations to include as many terms as possible. We computed the instantaneous 3.5PN pieces of the $(2,2)$, $(2,1)$, $(3,3)$, $(3,2)$, $(4,4)$, $(4,3)$ modes in terms of dynamical variables such as relative separation $(r)$, radial velocity $(\dot{r})$, and orbital angular velocity $(\dot{\phi})$ valid for generic orbits for both non-spinning and spinning cases. 
The $3.5$PN term of $(2,2)$ mode's instantaneous part is given in Appendix~\ref{sec:appendix_hlm}. {{Additionally, latest quasi-circular non-spinning 4PN correction in $(2,2)$ mode~\citep{Blanchet:2023sbv,Blanchet:2023bwj} is also included in our framework.}} 
\subsection{Merger and Ringdown}
\label{sec:mr}
\begin{figure*}[t]
    \centering
    \includegraphics[width=0.49\textwidth]{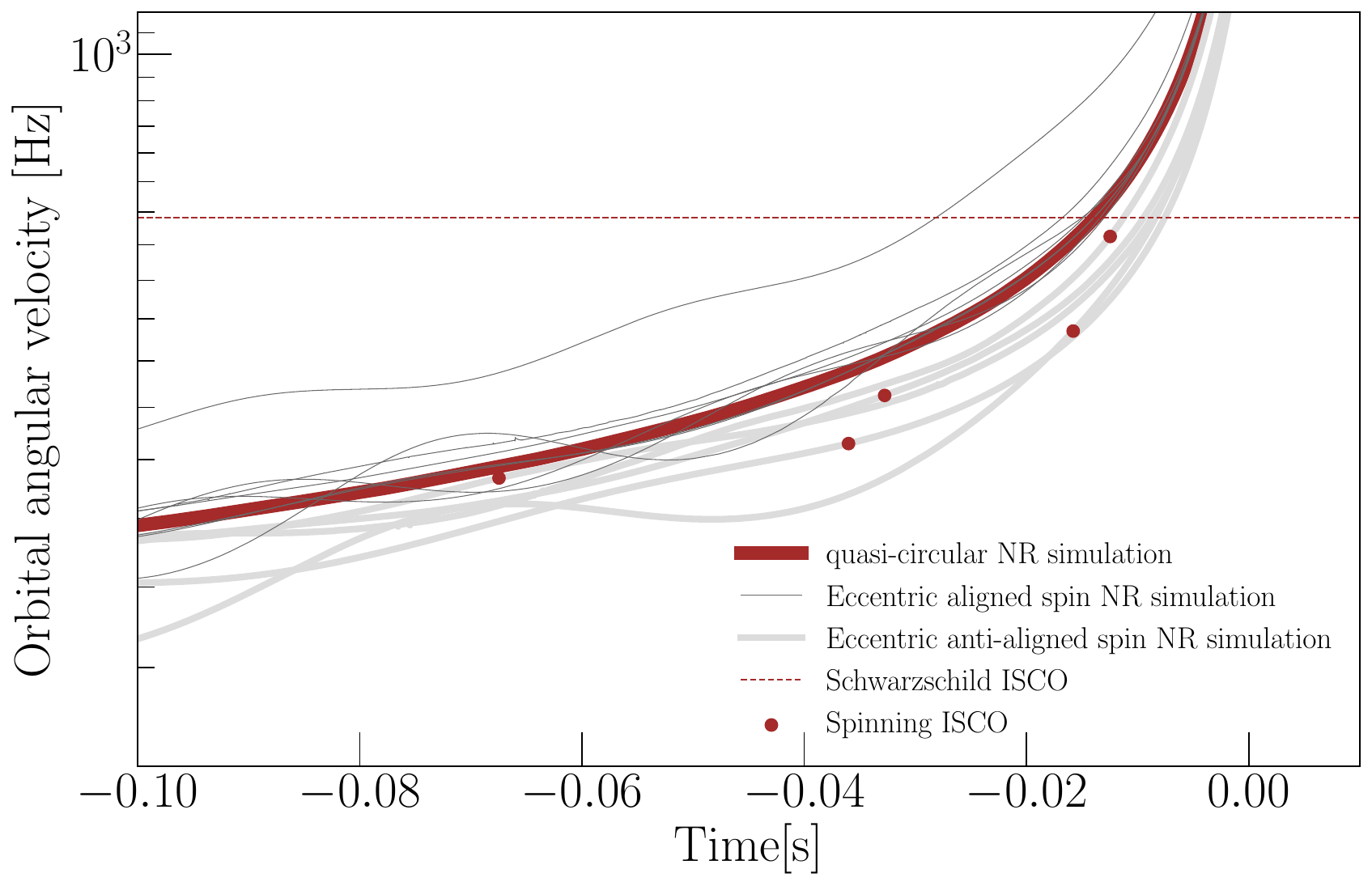}
    \includegraphics[width=0.49\textwidth]{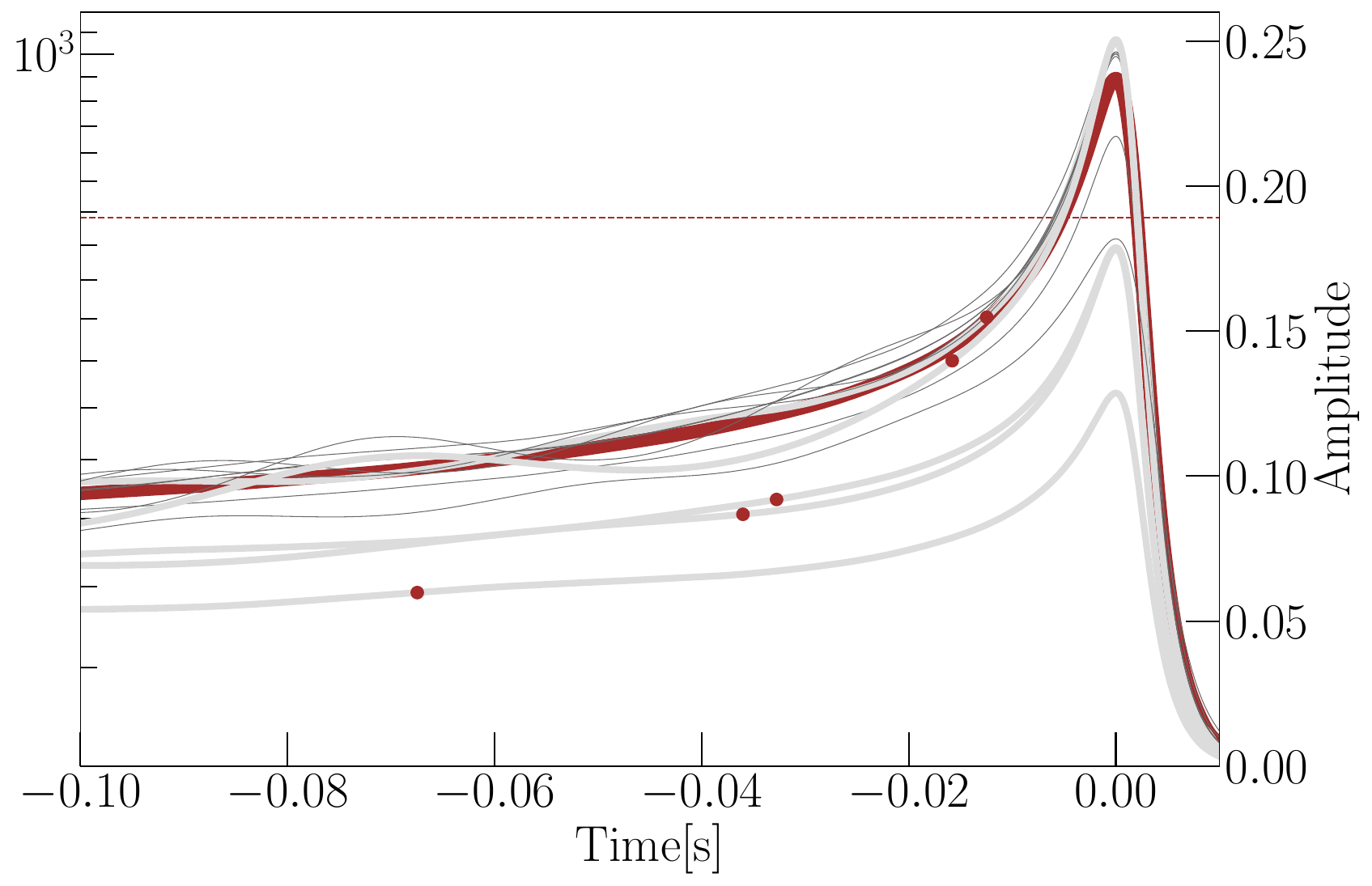}
    \caption{As expected, post-inspiral circularisation of gravitational waveforms for \textit{nearly} all spinning eccentric NR simulations listed in Table~\ref{table:sxs_id} is observed (grey curves); allowing for quasi-circular PMR attachment at a frequency \ftrans{}. For comparison, a quasi-circular simulation is also plotted (brown curve) and since Kerr ISCO frequency is always larger than the Schwarzschild ISCO frequency for positive-aligned spins, solid circles (brown) are shown only for anti-aligned cases.\\ 
    }
    \label{fig:orbital_evol}
\end{figure*}
While post-Newtonian approximation accurately describes binary's early inspiral dynamics, it breaks down toward the late inspiral stage as the system evolves into highly relativistic, non-perturbative regime. This later stage is most accurately described by solutions to Einstein's equations within NR. Significant efforts have been made to develop NR waveforms including the effects of spins and eccentricity with equal-mass to high mass-ratio cases~\citep{Szilagyi:2014fna, Cook:2000vr, Grandclement:2007sb}. As in the \enigma{} framework, we assume that most of the orbital eccentricity has decayed by the end of the inspiral, and post-inspiral is indistinguishable from coalescences on quasi-circular orbits. 
We use \nrsurdqfour{}~\citep{Varma:2019csw}, a quasi-circular spinning model for merger and ringdown. 

To construct the full IMR model, we extract individual spherical harmonic modes of the waveforms and combine the inspiral portion with the PMR by creating an interpolant that stitches the amplitude and frequency of the modes in a smooth manner. The interpolated amplitude and frequency are then compiled to create the IMR modes, which are then summed over to generate the GW strain of Eq.~(\ref{eq:GW_strain}).

This algorithm has three key steps: matching the transition time-window  between inspiral and PMR, aligning their phase in time domain, and applying the stitch. We start with specifying an attachment frequency \ftrans{}, and locate the times \tintrans{} and \tmrtrans{} in the inspiral and PMR waveforms where their orbital (or half of the dominant-mode) frequency crosses \ftrans{}. For the eccentric inspiral, we consider the orbit-averaged (and not the instantaneous) orbital frequency. These times are identified with each other, and denote the \textit{end} of the transition time-window. 
We then compute the integral of the orbit-averaged frequency of the inspiral model backward in time starting from \tintrans{}, and locate the time where the integral crosses a specified value $\delta_{\rm{trans}}$. This is the time $t_\delta$ with:
\begin{equation}
    \phi(t_\delta) := \phi(t_{\rm{trans}}^{\rm{Insp}}) - 2 \pi \delta_{\rm{trans}},
\end{equation}
where
\begin{equation}
    \phi(t) = \int_{t_{\rm{trans}}^{\rm{Insp}}}^t x^{3/2}(t') {\rm{d}}t'.
\end{equation}
The start of the transition windows then correspond to the times \tintransstart{} $=$ \tintrans{} $-t_\delta$ and \tmrtransstart{} $=$ \tmrtrans{} $-t_\delta$ in the inspiral and PMR waveforms. We find that setting \ftrans{} to the minimum of Schwarzschild innermost stable circular orbit (ISCO) frequency and spinning Kerr ISCO frequency~\citep{Husa:2015iqa}\footnote{\violet In order to obtain the Kerr ISCO frequency, we first obtain the dimensionless spin $a^{\rm{eff}}_f \left(\nu, S\right)$ of the Kerr remnant for the binary using Eq.~(3.6) of Ref.~\citep{Husa:2015iqa},
where $\nu$ is the dimensionless mass-ratio, and $S$ is the total spin of the binary. The ISCO frequency is calculated using the prescription in Eq.~(2.23) (and appendix C) of Ref.~\citep{Favata:2021vhw}.} works well in the quasi-circular limit~\footnote{\violet Note that these choices of transition frequency were arrived at by comparison with NR/EOB models in the quasicircular limit. For anti-aligned spin binaries our prescription for transition from the eccentric inspiral to quasicircular NR surrogate implies that it is done at a relatively earlier frequency than for aligned-spin configurations. This naturally results in \esigmahm{} limiting itself to relatively lower values of initial eccentricity for anti-aligned spin binaries. We plan to improve the inspiral model in the late-inspiral in a follow-up work, which will allow the transition to be done closer to merger.}. {\violet Similarly, we set $\delta_{\rm{trans}} = \pi/2$ to set the transition window to be about 1.5 orbital cycles in length.}
{\violet The transition windows are shown as grey bands in all panels of Figs.~\ref{fig:NR_comp_p1} and~\ref{fig:NR_comp}.}

Next, we align the inspiral and PMR portions by minimizing their mismatch within this transition time-window. We add a constant phase offset $\phi_a$ to the phase of the PMR modes $h_{\rm PMR}^{\ell m}$ and
align them with inspiral by minimizing the following mismatch measure $\Delta(\phi_a)$ over $\phi_a$:
\begin{equation}\label{eq:phase_diff}
    \Delta(\phi_a) = \frac{||h^{\ell m}_{\rm{Insp}} - h^{\ell m}_{\rm{PMR}}e^{\mathrm{i}m\phi_a}||^2}{||h^{\ell m}_{\rm{PMR}}||^2}
\end{equation}
We use the dominant mode with $\ell = m = 2$ to find $\phi_a$ to perform this phase alignment, and phase-shift other modes $h^{\ell' m'}$ by $(m'/2)\,\phi_a$ radians. This is motivated by identifying $\phi_a$ as twice the offset needed to align the orbital frequency evolutions between inspiral and PMR. {%
We compared the accuracy of the phase alignment between the following two approaches: 
minimizing the mismatch in Eq.~\eqref{eq:phase_diff} between just the dominant mode of inspiral and PMR waveforms, and between the full inspiral and PMR polarizations constructed by summing over several sub-dominant modes in addition to the dominant ones.
The two approaches showed no significant difference in phase alignment, justifying our choice of using $h^{22}$ mode only.

To apply the stitch, we take individual modes and decompose them into amplitude and frequency. We then create the interpolant using a \textit{blending function ($F_b$)} \citep{Varma:2018mmi} as follows,
\begin{equation}
    X_{\rm{Interp}}(t) = (1-F_b)X_{\rm{Insp}}(t) + (F_b)X_{\rm{PMR}}(t)\,,
    \label{interpolation}
\end{equation}
where $X(t)$ is the time-series being blended, and
\begin{equation}
    F_b = \sin^2 \Big(\frac{\pi}{2}\frac{t-t_i}{t_f-t_i}\Big)\,,
\end{equation}
$t_i$ and $t_f$ being the start and end time of the attachment window. 
We substitute $X(t)$ for mode amplitude $A^{\ell m}$ in Eq.~\eqref{interpolation} to compute the interpolated mode amplitude $A^{\ell m}_{\rm{Interp}}$. The same process is followed to compute the interpolated mode frequency. We integrate the interpolated frequency to procure the interpolated mode phase.  

With these ingredients in hand, the IMR amplitude $A_{\rm{IMR}}^{\ell m}$ and phase $\phi_{\rm{IMR}}^{\ell m}$ take the form  
\[
A_{\rm{IMR}}^{\ell m}(t) = 
\begin{cases}
    A_{\rm{Insp}}^{\ell m}(t) & t < t_i \\
    A_{\rm{Interp}}^{\ell m}(t) & t_i \leq t \leq t_f \\
    A_{\rm{PMR}}^{\ell m} & t_f < t
    
\end{cases},
\]
 
\[
\phi_{\rm{IMR}}^{\ell m}(t) = 
\begin{cases}
    \phi_{\rm{Insp}}^{\ell m}(t) & t < t_i \\
    \phi_{\rm{Interp}}^{\ell m}(t)  & t_i \leq t \leq t_f \\
    \phi_{\rm{PMR}}^{\ell m} + \delta_f^{\ell m} & t_f < t
    
\end{cases} .
\]
The IMR modes are computed using
\begin{equation}
    h_{\rm{IMR}}^{\ell m}(t) = A_{\rm{IMR}}^{\ell m}(t)\, e^{-\mathrm{i} \phi_{\rm{IMR}}^{\ell m}(t)}\,.
\end{equation}
Upon adding all the IMR modes according to Eq.~\eqref{eq:GW_strain}, we get the IMR waveform $h_{\text{IMR}}(t; \Theta, \Phi)$.

\imresigmahm{} includes subdominant modes $(2, \pm 1)$, $(3, \pm 3)$, $(3, \pm 2)$, $(4, \pm 4)$, $(4, \pm 3)$ apart from the dominant $(2, \pm 2)$ mode. These modes are expected to contain most of relevant signal information discernible by current detectors~\citep{Chattaraj:2022tay}.

Finally, to further our understanding of the validity of our model requirement that orbital eccentricity should be small enough initially that it is (nearly) radiated away by the end of inspiral, we show in Fig.~\ref{fig:orbital_evol} the time evolution of orbital angular velocity (\textit{left}) and amplitude (\textit{right}) for eccentric (thin and thick grey lines) and quasi-circular (thick brown line) systems. We show 13 spinning, eccentric (see Table~\ref{table:sxs_id} of Appendix~\ref{sec:appendix_NR}) and one quasi-circular non-spinning (\texttt{SXS:BBH:0007}) NR simulations from the SXS catalog~\citep{SXS:catalog}. The solid brown circles in both figure panels represent \ftrans{} frequencies for the corresponding binary with total mass $40M_{\odot}$. The evolution of spin-aligned eccentric and quasi-circular systems are qualitatively similar beyond this mark, suggesting that the system has circularized by radiating nearly all its eccentricity. Note however, that some of the simulations with anti-aligned spin vectors exhibit mild oscillations beyond the chosen transition frequency \ftrans{}.
As can be seen in Table~\ref{table:sxs_id} of Appendix~\ref{sec:appendix_NR}, the mismatches corresponding to these cases are also relatively larger indicating that one may have to relax the modeling assumption of small eccentricity when dealing with anti-aligned cases with large mass ratios.
 \begin{figure*}
    \centering
    \includegraphics[width=0.32\textwidth]{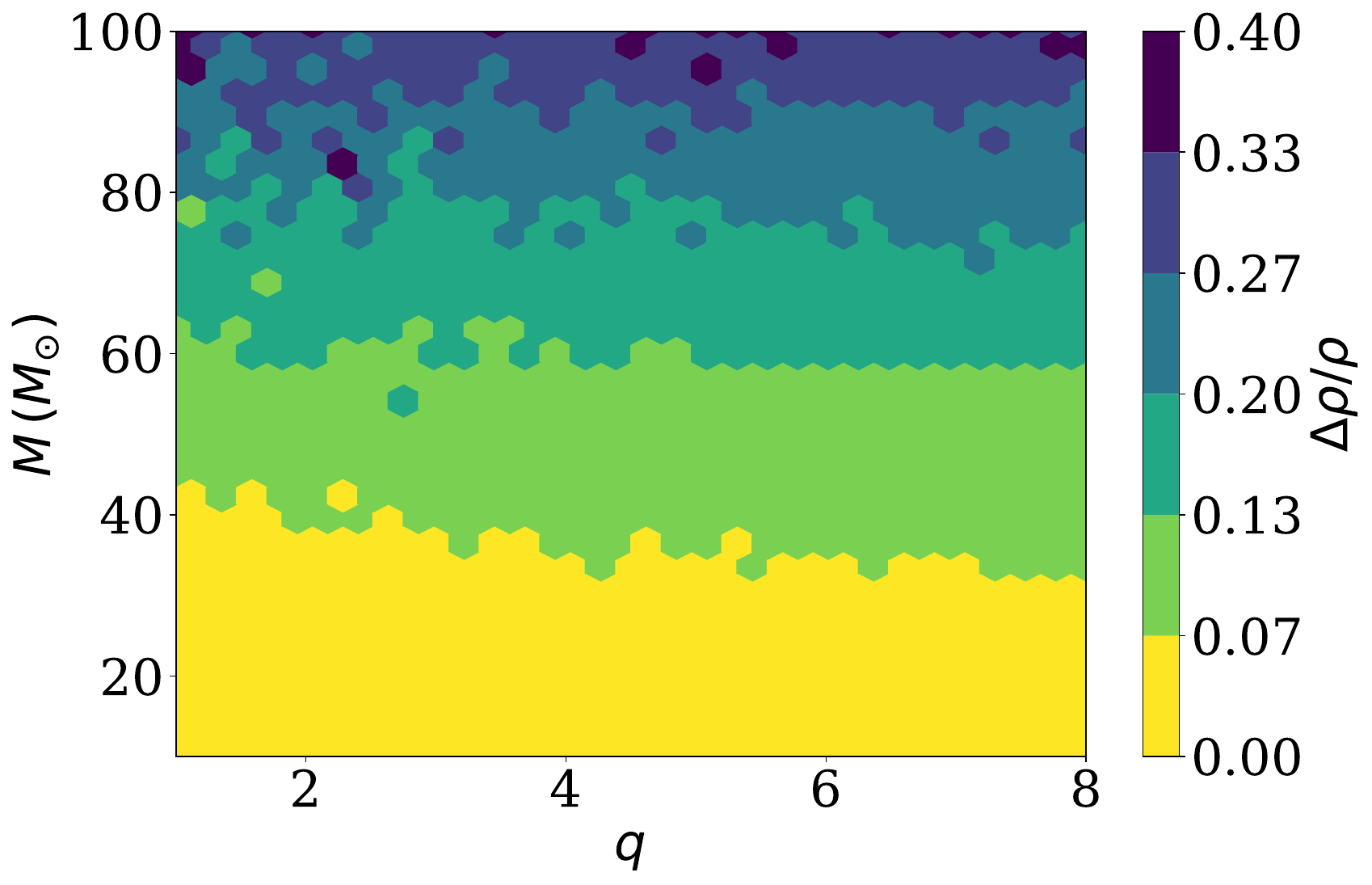}
    \includegraphics[width=0.32\textwidth]{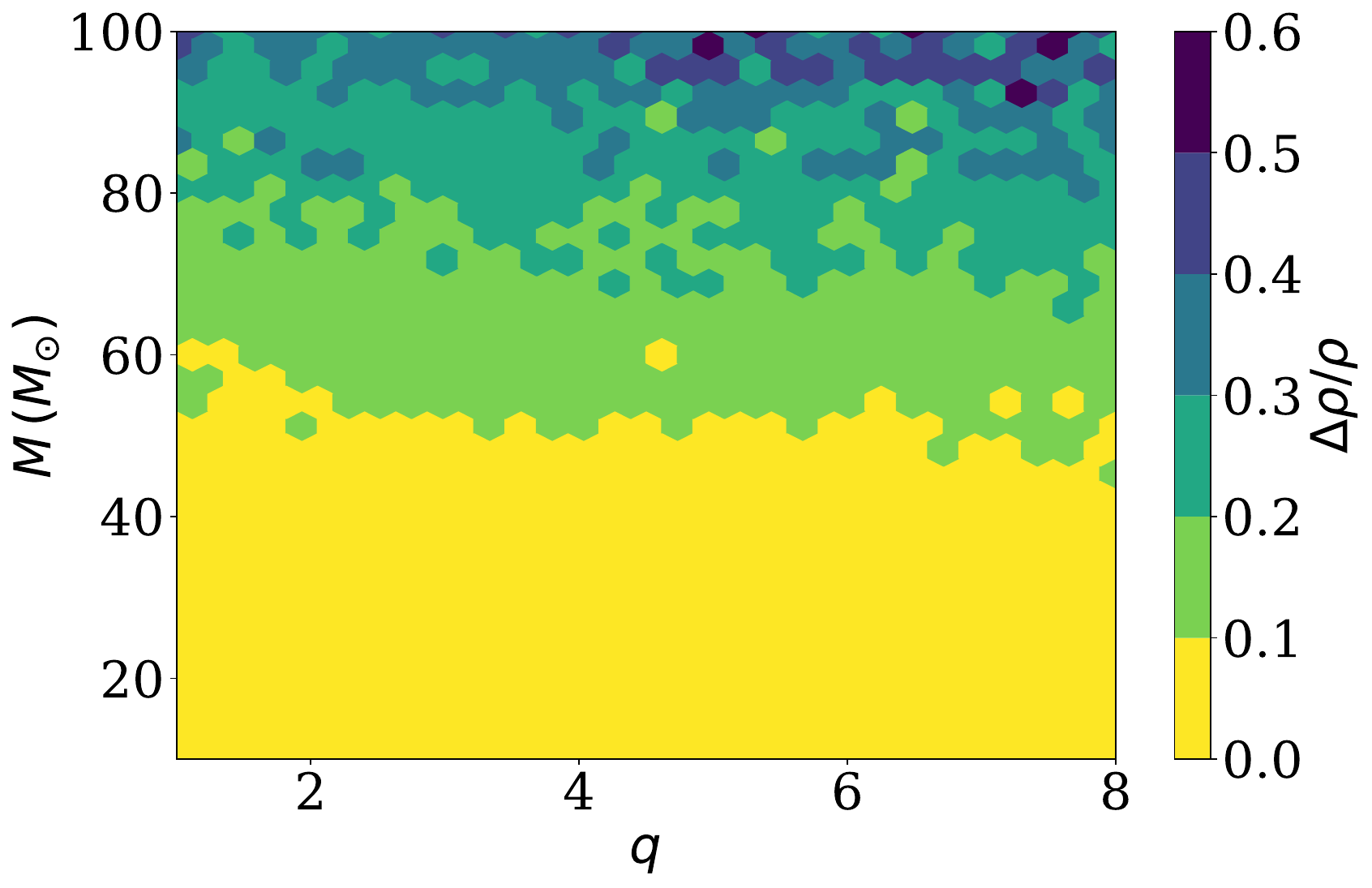}
    \includegraphics[width=0.32\textwidth]{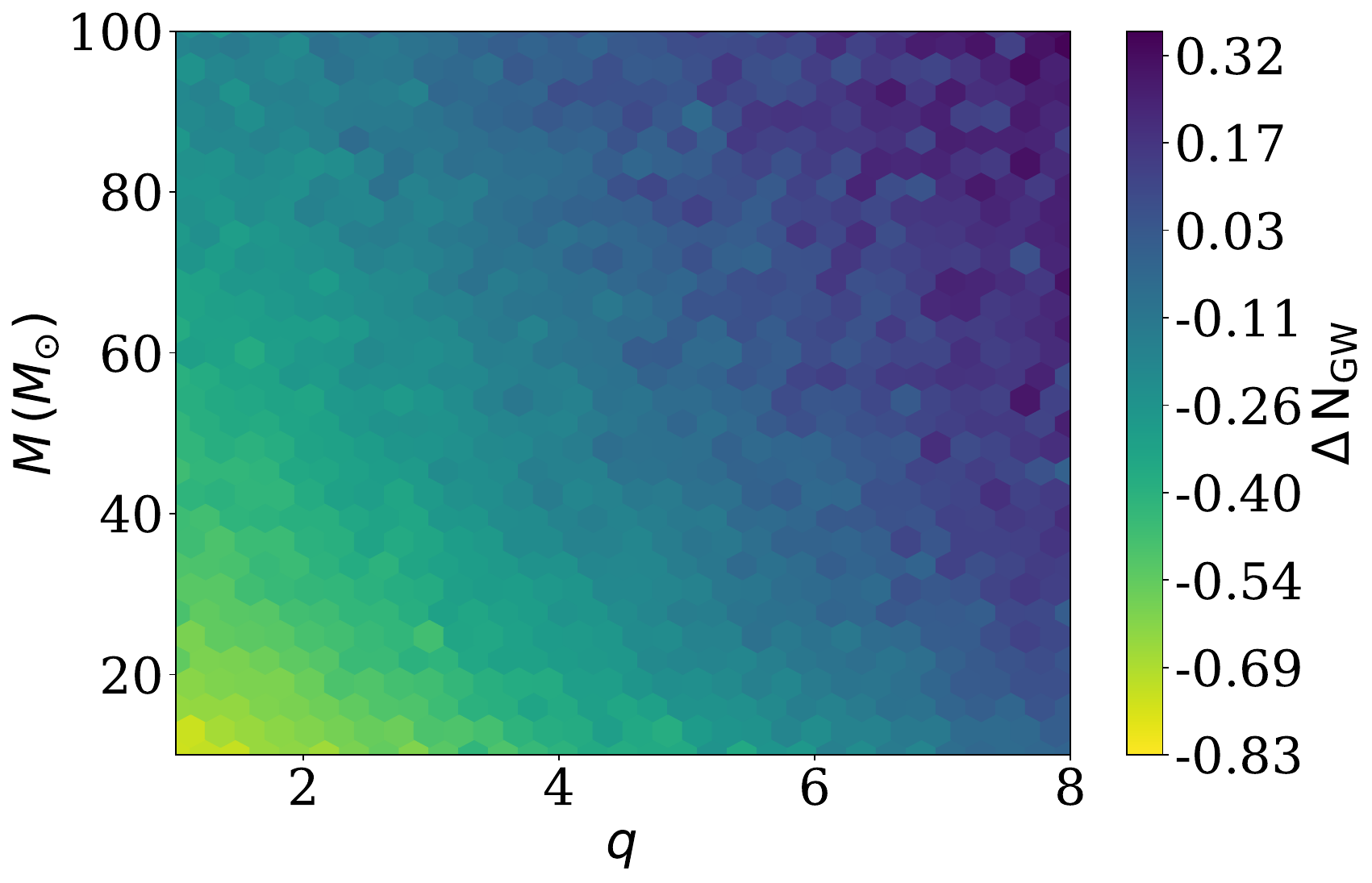}
    \caption{{{Effect of newly added terms in \inspiralesigmahm{} is shown by performing a comparison with \enigma{}. The fractional change in optimal SNR $(\Delta \rho/\rho)$ for sources at $500$Mpc with inclination angle of $0$ (\textit{left}) and $\pi/3$ (\textit{middle}), respectively. Here, the fractional SNR $(\Delta \rho/\rho)$ is defined as the difference in optimal SNR between \inspiralesigmahm{} and \enigma{} divided by optimal SNR obtained by \enigma{}. The \textit{right} figure  shows the ($\Delta \rm{N_{GW}}$)  between \enigma{} and \inspiralesigmahm{} as a function of total mass $(M)$ and mass ratio $(q)$. For all three figures we took eccentricity $(e)$ of $0.3$ and lower cutoff frequency of $10$Hz respectively. The SNR computation is done with the advanced LIGO (aLIGO) zero-detuned high-power noise PSD.}}}
    \label{fig:SNR_Ngw}
\end{figure*}
\subsection{Changes from previous versions}
\label{sec:ENIGMA2ESIGMA}
\enigma{} included general orbit expression for the dominant mode at Newtonian order, and quasi-circular corrections up to 2.5PN following~\citep{Arun:2004ff,Blanchet:2008je}. It also had non-spinning eccentric corrections in evolution of orbital elements up to 3PN order~\citep{Hinder:2008kv, Arun:2009mc}. However, the current inspiral version has up to 3.5PN non-spinning and spinning instantaneous general orbit~\citep{Mishra:2015bqa, Paul:2022xfy, Henry:2023tka}, and 3.5PN non-spinning~\citep{Blanchet:2008je} and spinning~\citep{Henry:2022dzx} hereditary corrections valid for quasi-circular orbits in spherical harmonic modes. Besides, the $(2,2)$ mode is updated up to 4PN order for non-spinning systems evolving in quasi-circular orbits~\citep{Blanchet:2023bwj}. The EOMs are updated with the recently computed 3PN spinning eccentric~\citep{Henry:2023tka}, and 4PN non-spinning quasi-circular terms reported in~\citep{Blanchet:2023bwj}.

We estimate the effects of the upgrades that the inspiral part of \esigmahm{} named \inspiralesigmahm{} has over the previous version \enigma{}~\citep{Chen:2020lzc}. To investigate the effects of these newly added terms in the framework, we perform two studies. First, we estimate the fractional change in optimal SNR between \enigma{} and \inspiralesigmahm{}, keeping all the modes up to $\ell=7$.\footnote{\enigma{} had quasi-circular non-spinning polarizations up to 2.5PN order at which up to $\ell=7$  modes contribute. Hence, we restrict to $\ell=7$ modes for the comparison.} One can define the ``optimal" SNR $(\rho)$ as,
\begin{align}
\rho^{2} &= \Big( h|h\Big),
\label{eq:SNR_optimal}
\end{align}
where $h$ is the GW strain appearing in Eq.~\eqref{eq:GW_strain}, and $\Big(a|b\Big)$ denotes the noise-weighted inner product given by
\begin{equation}
    \Big(a|b\Big) \equiv 4\, \Re \left[ \int_{f_{\rm{low}}}^{f_{\rm{high}}} \dfrac{\Tilde{a}(f) \Tilde{b}^*(f)}{S_n(f)} \rm{d}f \right]\,,
\label{eq:overlap}
\end{equation}
where $\Tilde{a}$ denotes Fourier transform of $a$, $S_n(f)$ is the one-sided power spectral density (PSD) of the GW detector noise, {{and $f_{\rm{low}}$, $f_{\rm{high}}$ are lower and upper cut off frequencies, respectively}}. This inner product can be maximized over an overall phase and time shift between the arguments $a$ and $b$ to arrive at a normalized measure of agreement between the two, colloquially referred to as match $\mathcal{M}$~\citep{Purrer:2014fza}:
\begin{equation}
    \mathcal{M}\left(a, b\right) \equiv \, \underset{\phi_c,  t_c}{\rm{max}}    \dfrac{\left(e^{\rm{i} \left(\phi_c + 2\pi f t_c\right)} a|b\right)}{\sqrt{\left(a|a\right) \left(b|b\right)}}\,,
\label{eq:match}
\end{equation}
which we will use extensively in the following sections. 
Throughout this paper we use the \textit{zero-detuning high-power} noise PSD for the advanced LIGO detectors~\citep{Shoemaker:T0900288-v3}, choose $f_{\rm{low}} = 10$~Hz,  and use a sampling rate of $4096$~Hz, which sets $f_{\rm{high}} = 2048$~Hz in Eq.~\eqref{eq:overlap}.
Second, we compute the change in the number of GW cycles ($\Delta \rm{N_{GW}}$) between \enigma{} and \inspiralesigmahm{}, following the relation~\citep{Blanchet:2013haa},
\begin{align}\label{eq:GW_cycles}
     \rm{N_{GW}} &= \frac{\phi_{\rm{ISCO}} -\phi_{0}}{2\pi}.
\end{align}
Here, $\phi_{\rm{ISCO}}$ is the phase of GW when its frequency equals that of the innermost stable circular orbit (ISCO) for a Schwarzschild black hole, and $\phi_{0}$ is the phase of the wave when it enters detector's sensitivity band at some frequency $f_{0}$. In Fig.~\ref{fig:SNR_Ngw}, the fractional change in SNR ($\Delta \rho/\rho$) (\textit{left} and \textit{middle}) and the difference in number of GW cycles ($\Delta \rm{N}_{\rm{GW}}$) (\textit{right}) between \enigma{} and \inspiralesigmahm{} is shown as a function of total mass ($M$) and mass ratio ($q$) for sources at a distance of $500$~Mpc oriented at an inclination angle of $\pi/3$ and a fixed eccentricity of $0.3$.
We notice that the change in SNR for large total mass and high mass ratio cases easily exceeds $\sim 20\%$ for the chosen highly eccentric binary. A comparison of left and middle panels indicates that some of this can be attributed to the inclusion of several higher-order modes in \inspiralesigmahm{}, and some to intrinsically more information now included in the dominant mode itself. We will revisit the importance of subdominant modes in Sec.~\ref{sec:HM}. The right figure panel shows that the new terms included in the radiative sector of the \esigma{} framework can change the waveforms by nearly a cycle.

\begin{figure}[htbp!]
\centering
\includegraphics[width=\linewidth]{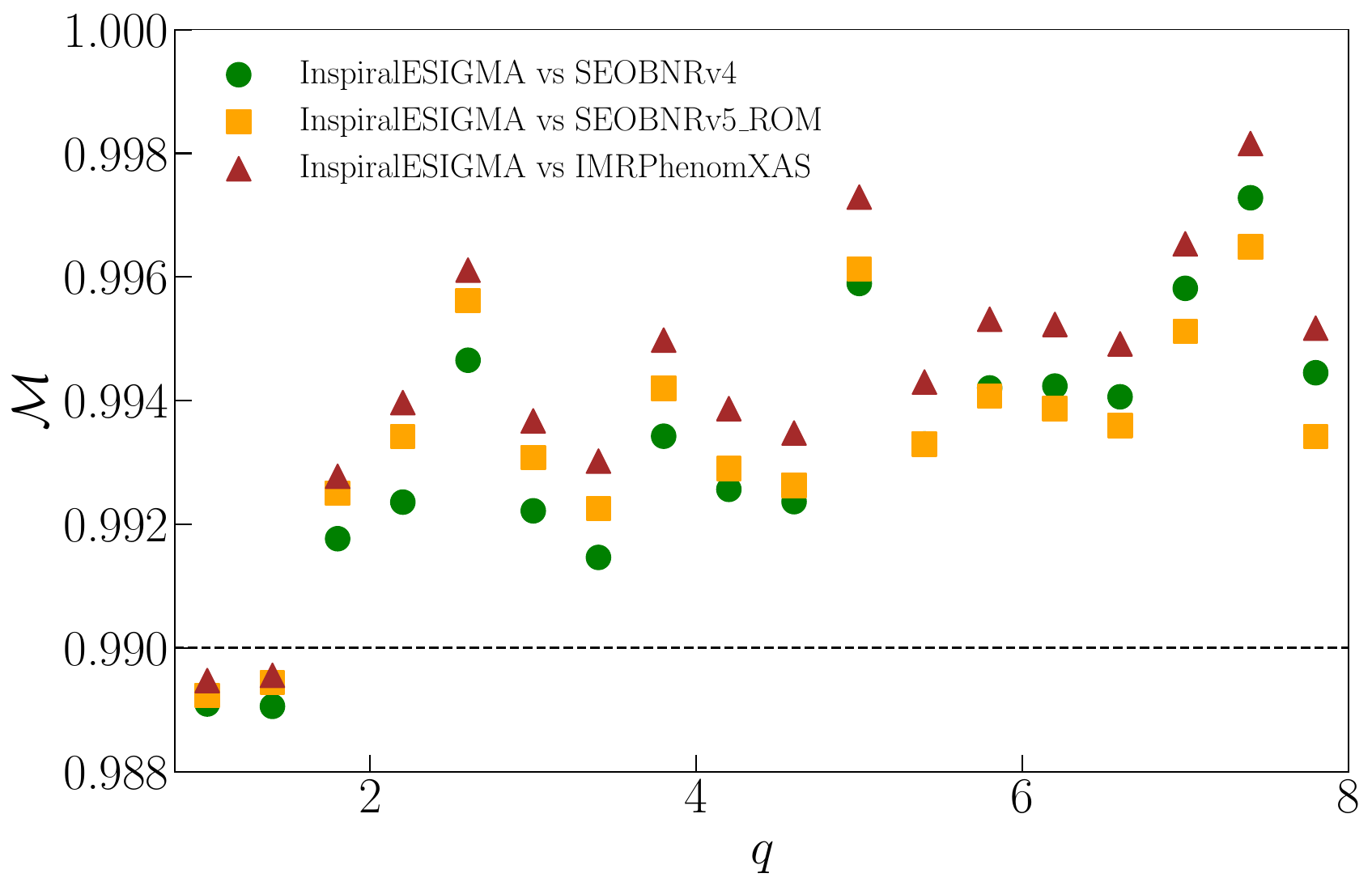}
\caption{{{Agreement between \inspiralesigma{} with existing \texttt{EOB} and \texttt{Phenom} waveforms are shown in quasi-circular limit.}} Match between quasi-circular dominant mode \inspiralesigma{} waveforms with \seobnrvfour{}, \seobnrvfive{} and \imrphenom{} are shown for fixed total mass $40\,M_{\odot}$ and dimensionless spins  $(0.5,0.5)$, respectively. The match computation is performed with a high-frequency cutoff of a minimum of ISCO frequencies between Schwarzschild and spinning cases~\citep{Husa:2015iqa, Favata:2021vhw} and using aLIGO zero-detuned high-power PSD assuming initial gravitational wave frequency of $10$Hz. The horizontal dashed black line denote the 99\% match between the two waveforms.}
\label{fig:mismatch_qc}
\end{figure}
\begin{figure}
    \includegraphics[width=\linewidth]{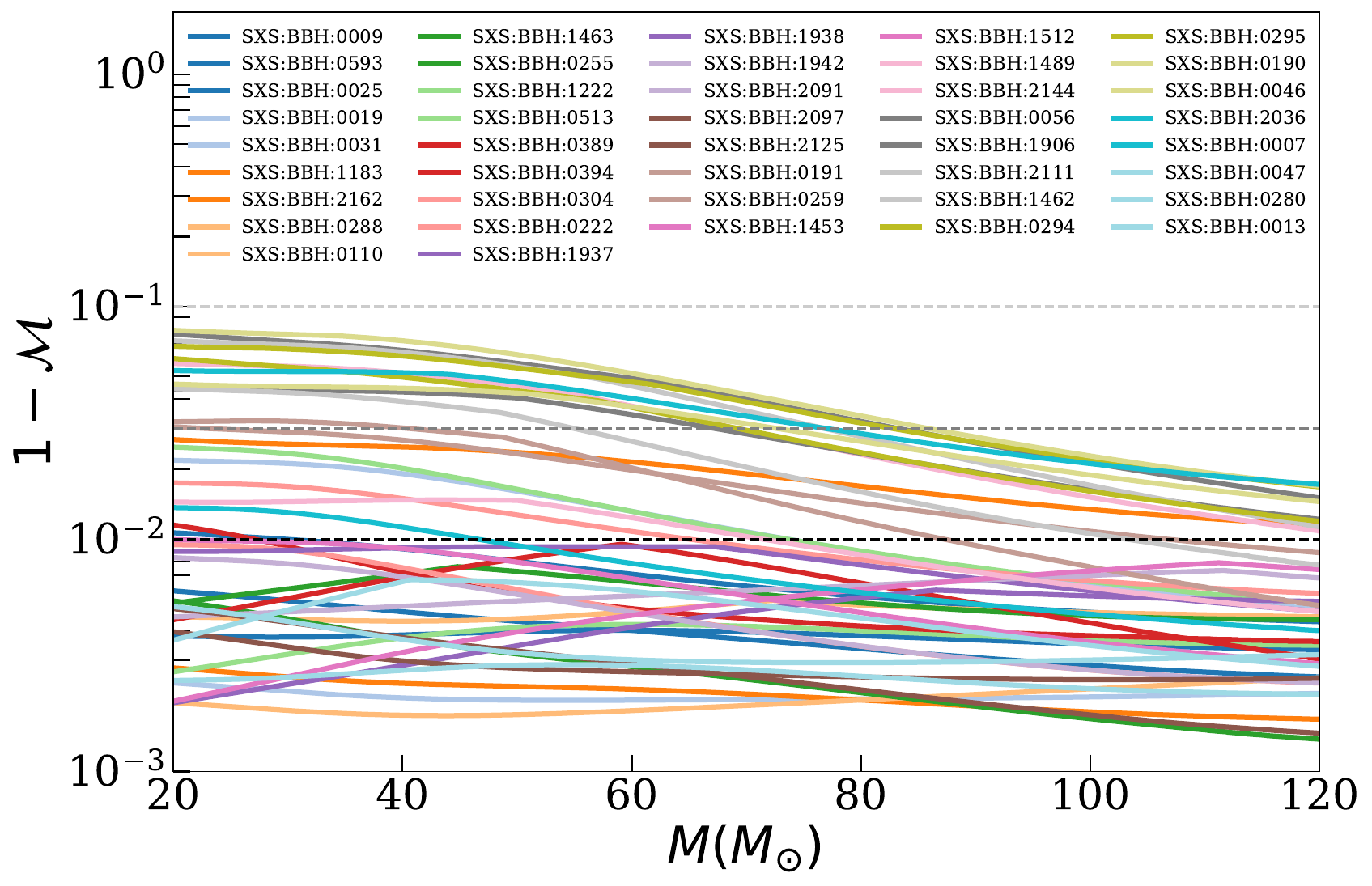}
        \caption{
        Mismatch between quasi-circular NR simulations and \imresigma{} with the dominant $\ell=|m|=2$ modes is shown as a function of binary total mass. Mismatch computations are performed using aLIGO zero-detuned high-power PSD with a lower cutoff frequency 10Hz. The horizontal dashed lines denote the 10\% (light grey), 3\% (grey), and 1\% (black) mismatch marks, respectively. Each curve corresponds to a distinct NR simulation taken from the SXS catalog~\citep{Boyle:2019kee}, whose mass-ratio and component spin values are given in Table~\ref{table:sxs_id_qc_nr_4PN} of Appendix~\ref{sec:appendix_NR}.}
    \label{fig:NR_qc_mismatches}
\end{figure}

\begin{figure*} 
    \centering
     \begin{subfigure}[b]{0.24\textwidth}
        \includegraphics[width=\linewidth]{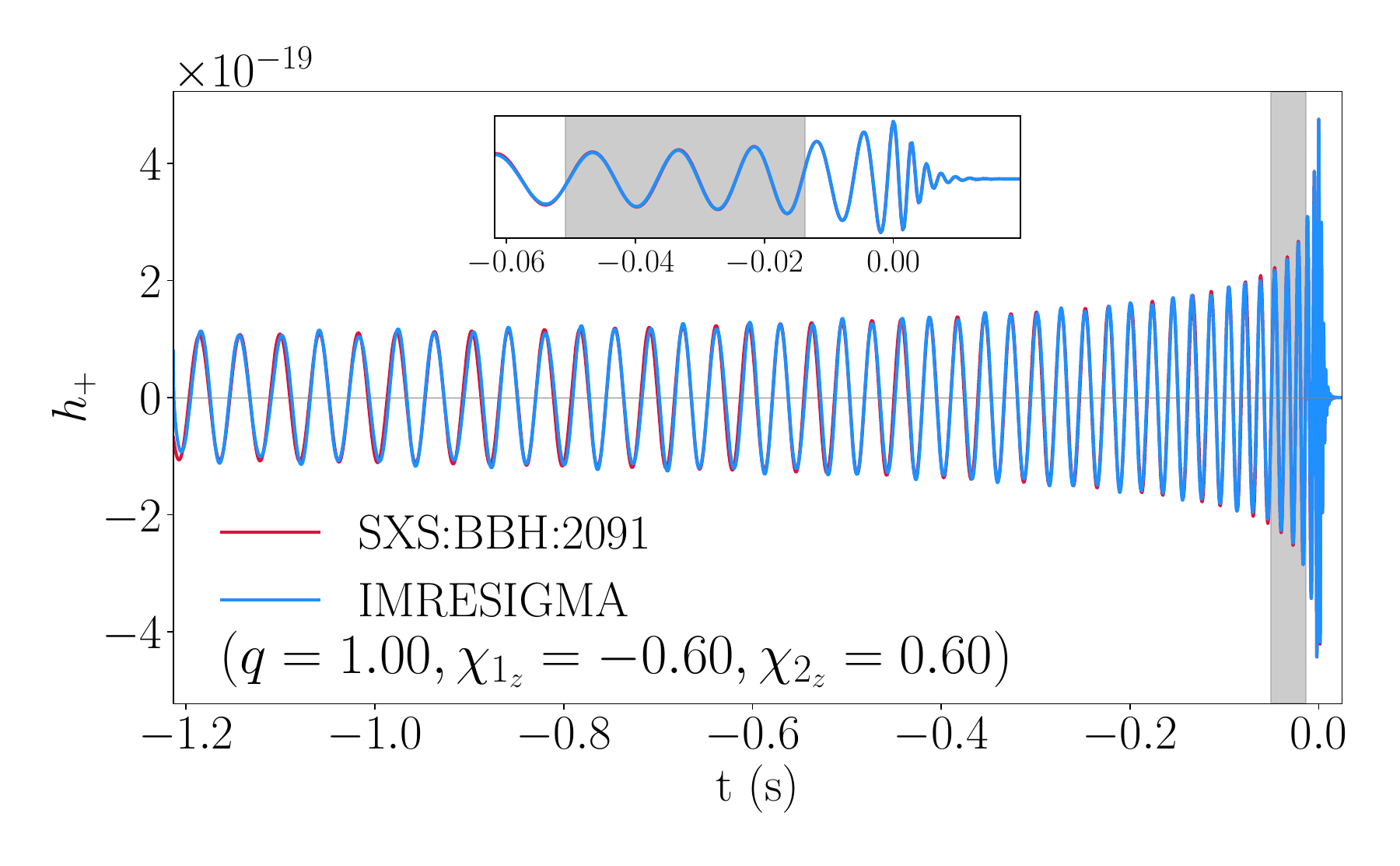}
    \end{subfigure}
     \begin{subfigure}[b]{0.24\textwidth}
        \includegraphics[width=\linewidth]{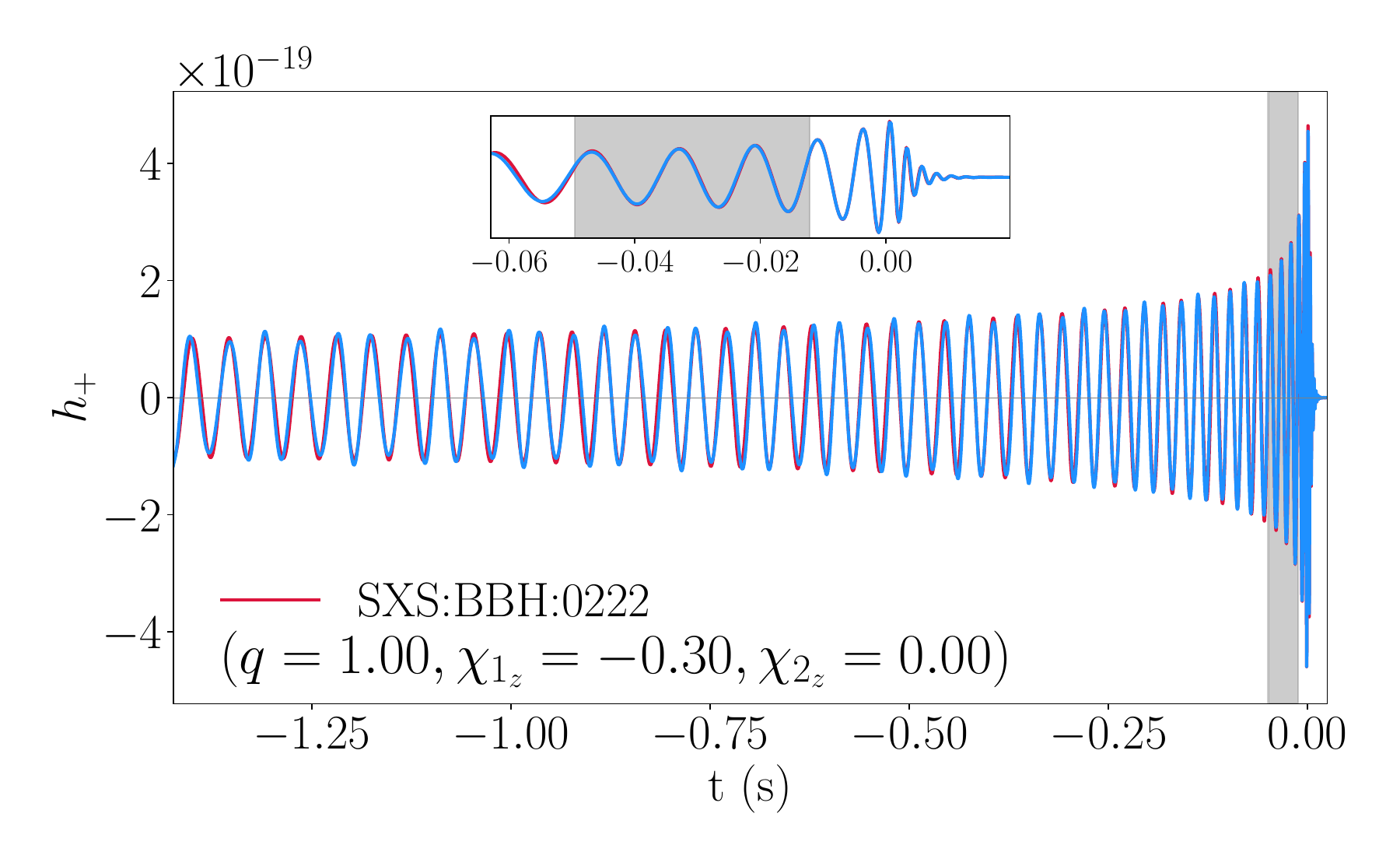}
    \end{subfigure}
     \begin{subfigure}[b]{0.24\textwidth}
            \includegraphics[width=\linewidth]{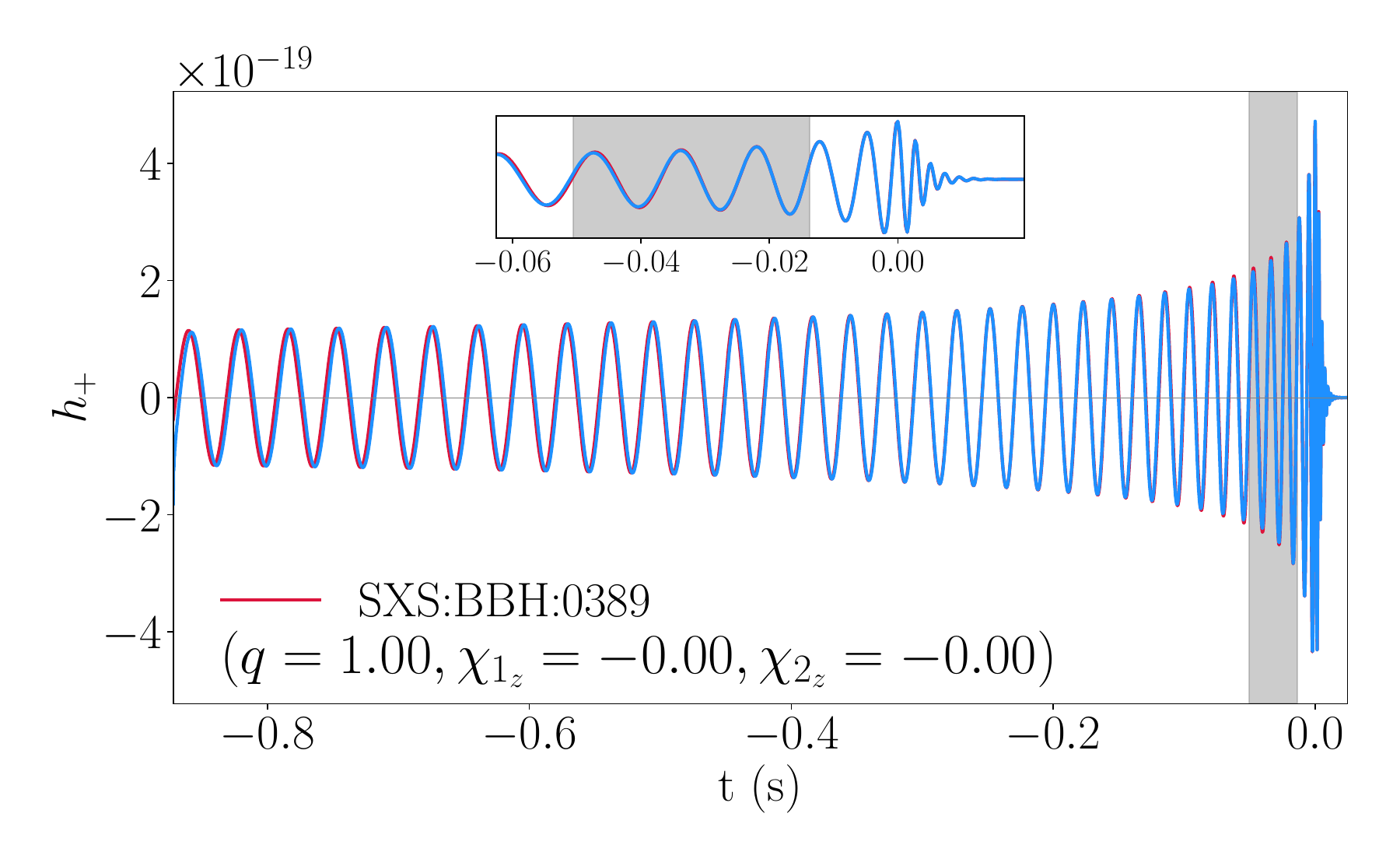}
    \end{subfigure}
    \begin{subfigure}[b]{0.24\textwidth}
        \includegraphics[width=\linewidth]{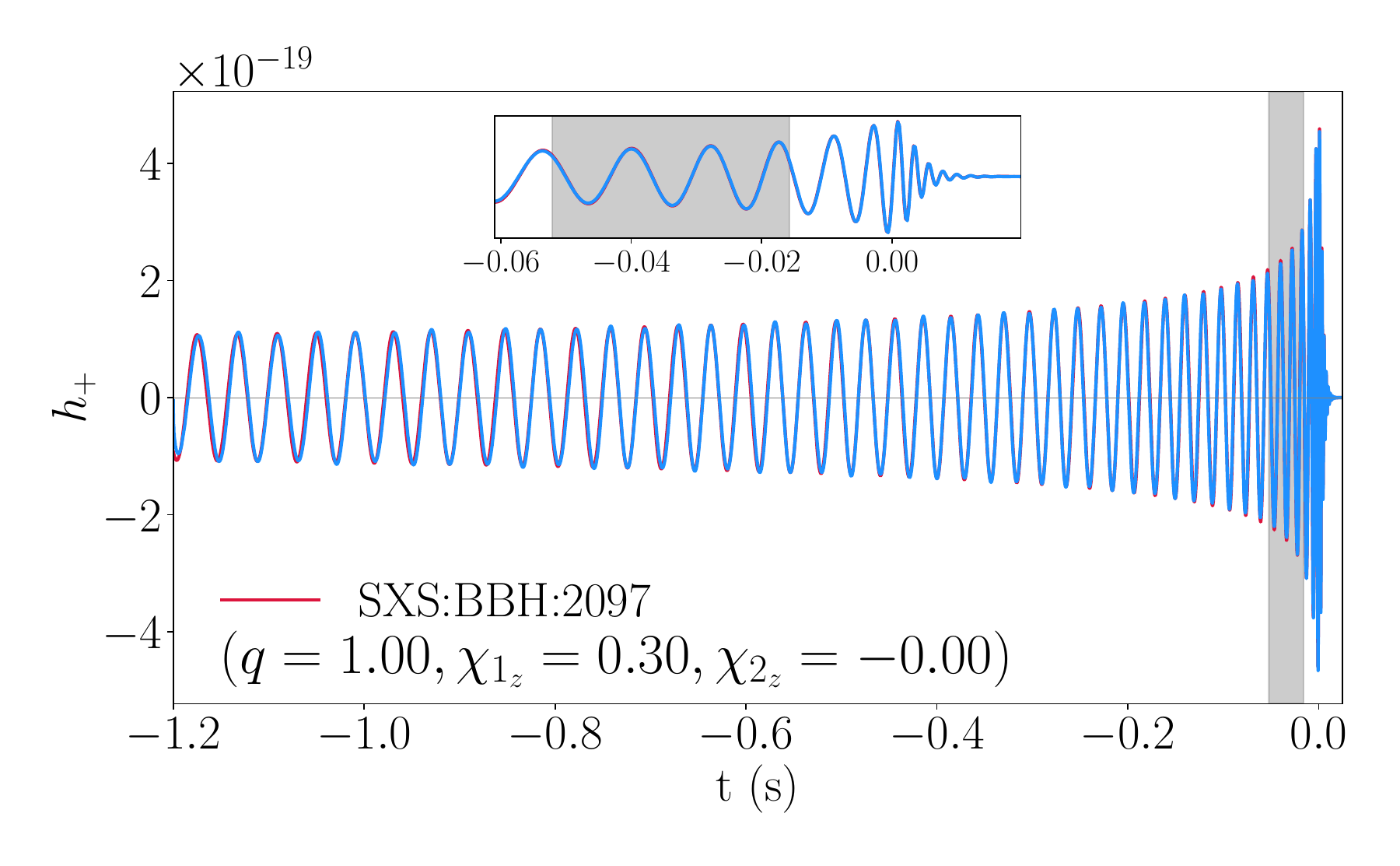}
    \end{subfigure}
     \begin{subfigure}[b]{0.24\textwidth}
        \includegraphics[width=\linewidth]{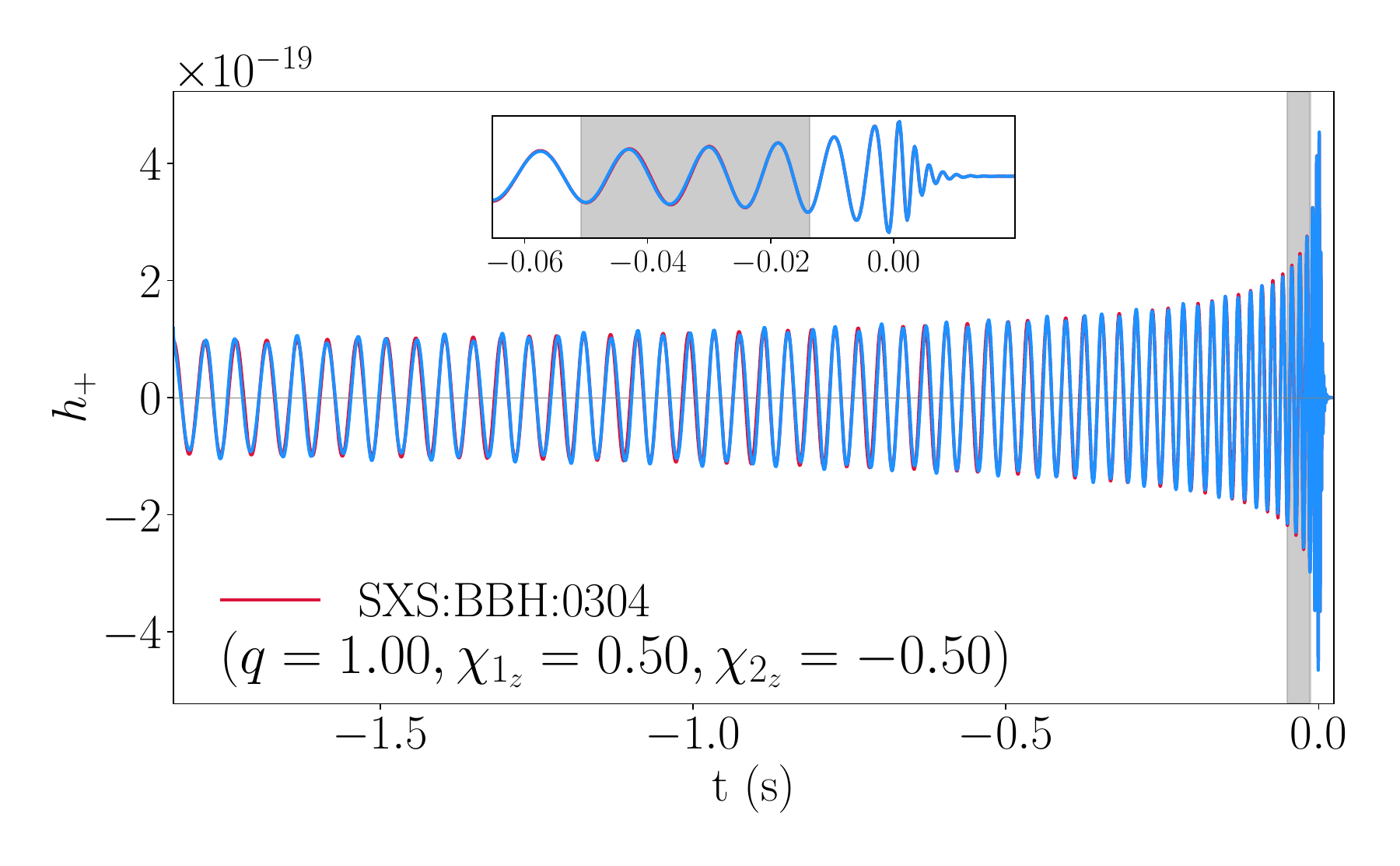}
    \end{subfigure}
      \begin{subfigure}[b]{0.24\textwidth}
        \includegraphics[width=\linewidth]{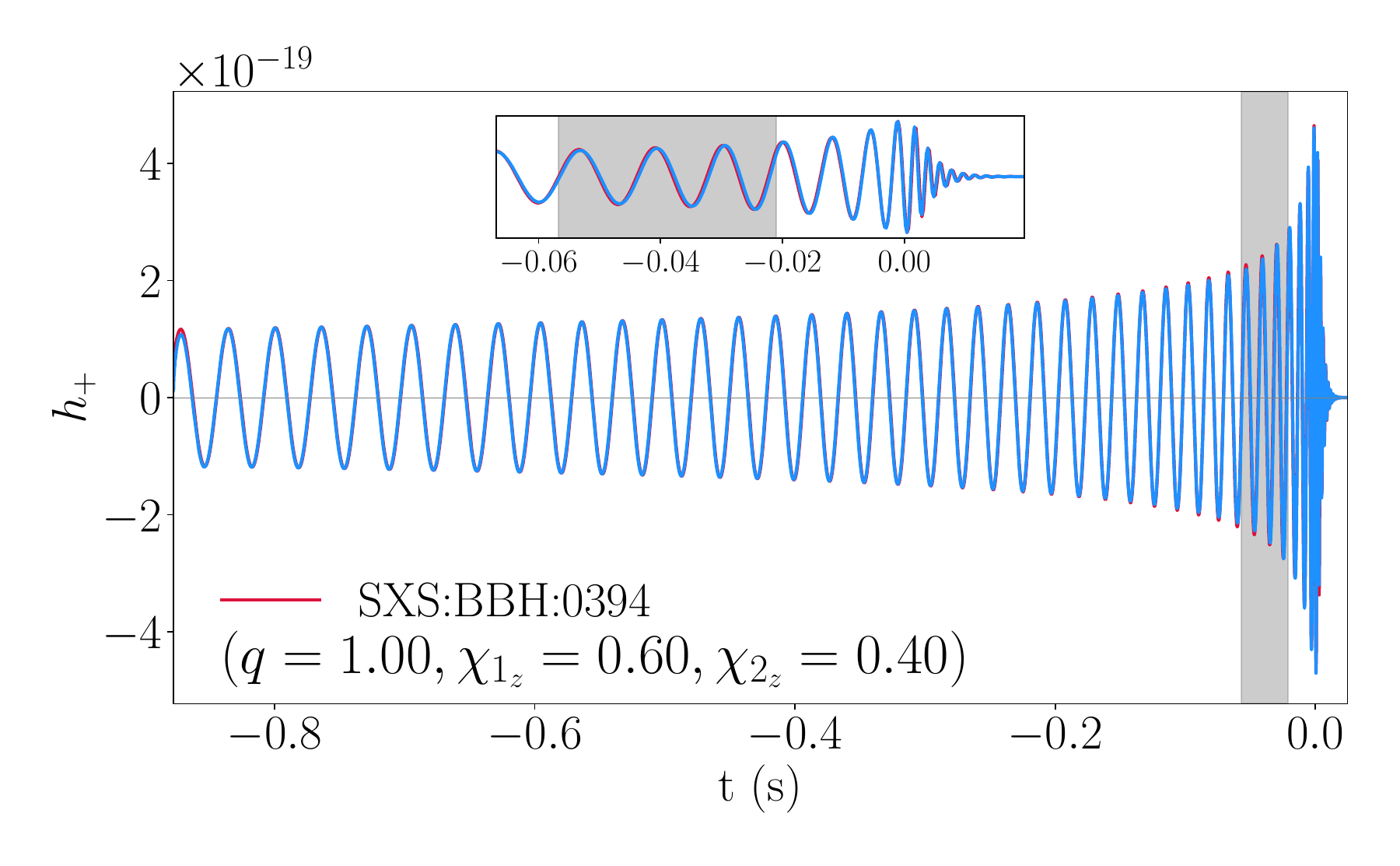}
    \end{subfigure}
     \begin{subfigure}[b]{0.24\textwidth}
        \includegraphics[width=\linewidth]{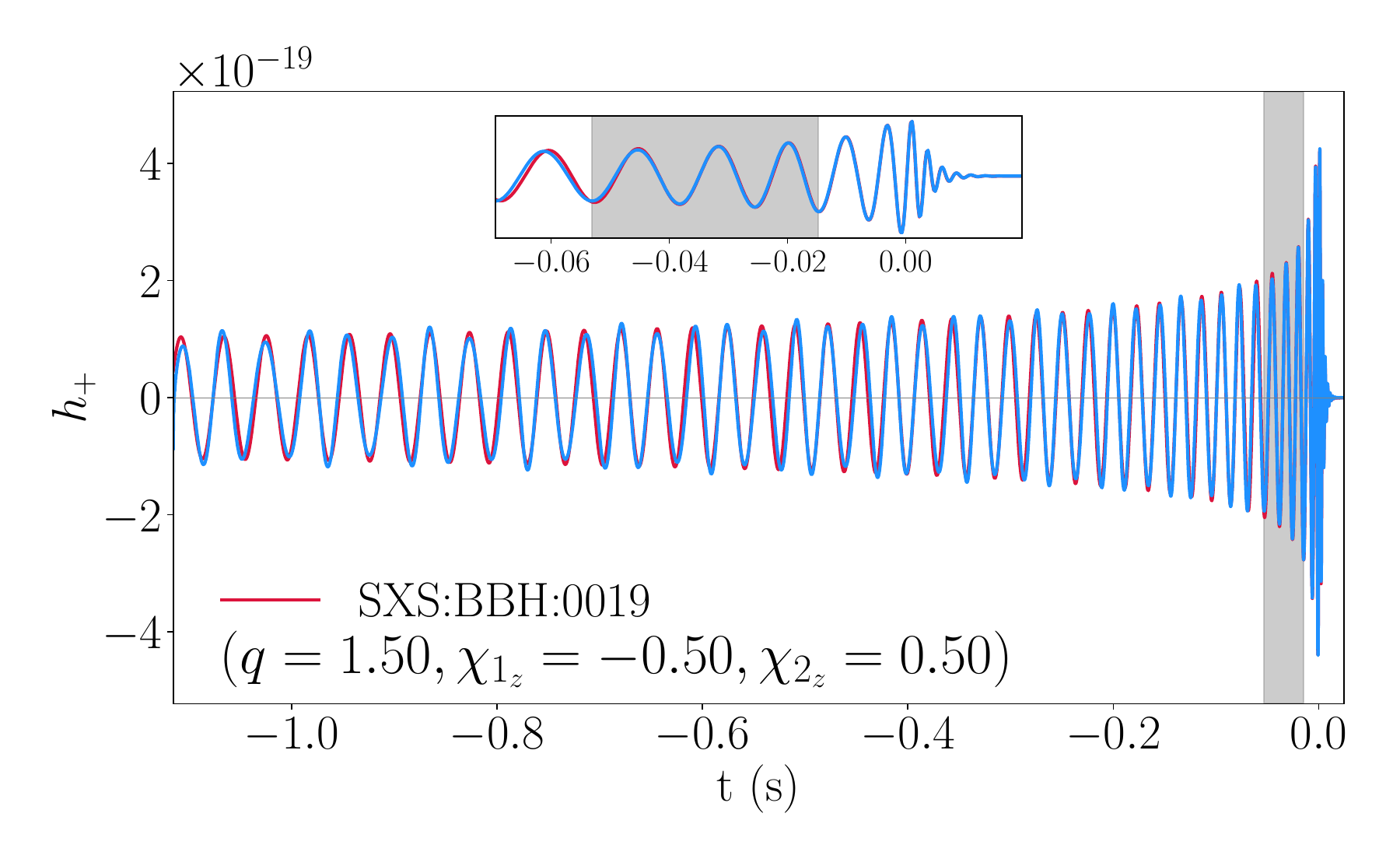}
    \end{subfigure}
      \begin{subfigure}[b]{0.24\textwidth}
        \includegraphics[width=\linewidth]{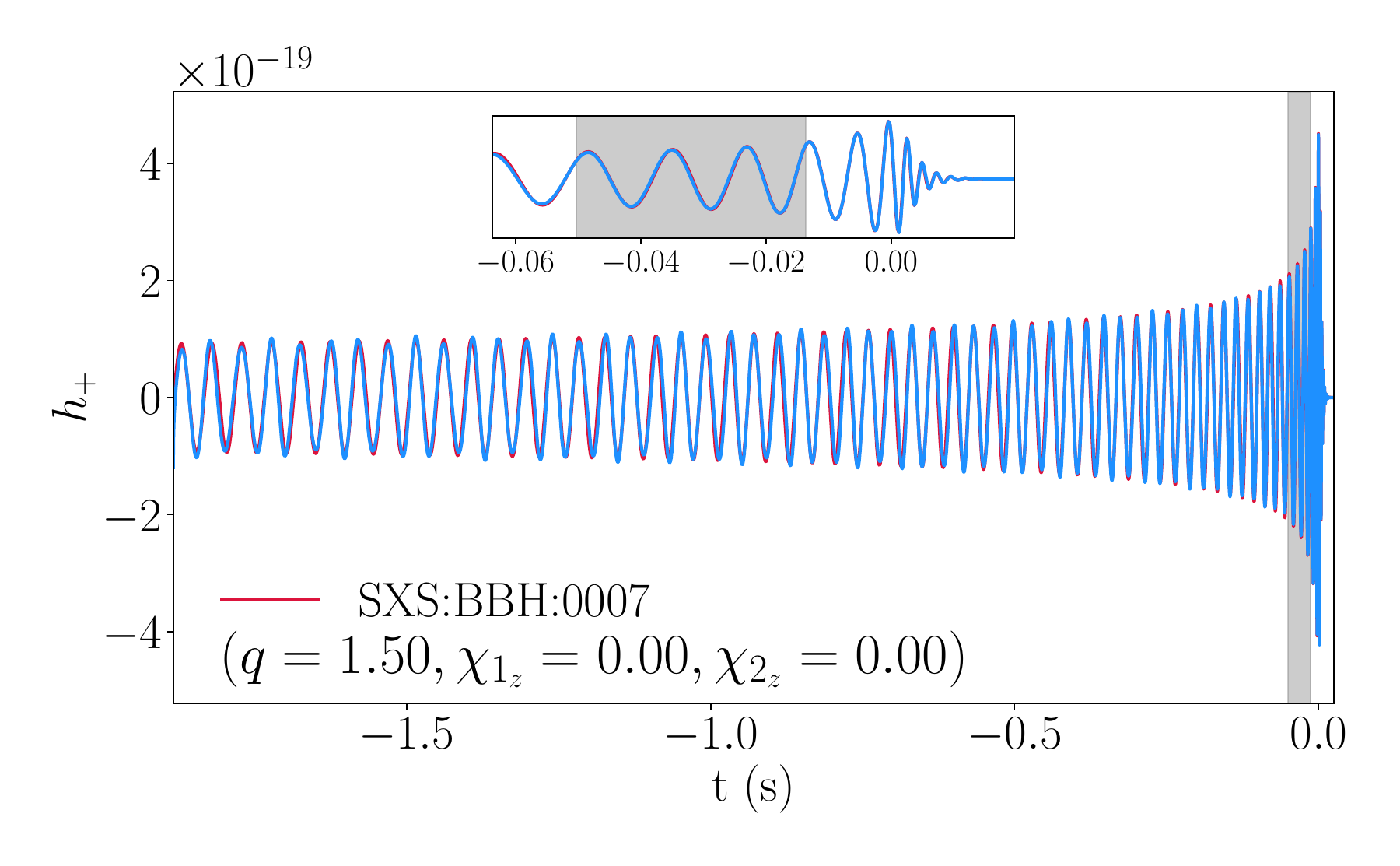}
    \end{subfigure}
     \begin{subfigure}[b]{0.24\textwidth}
        \includegraphics[width=\linewidth]{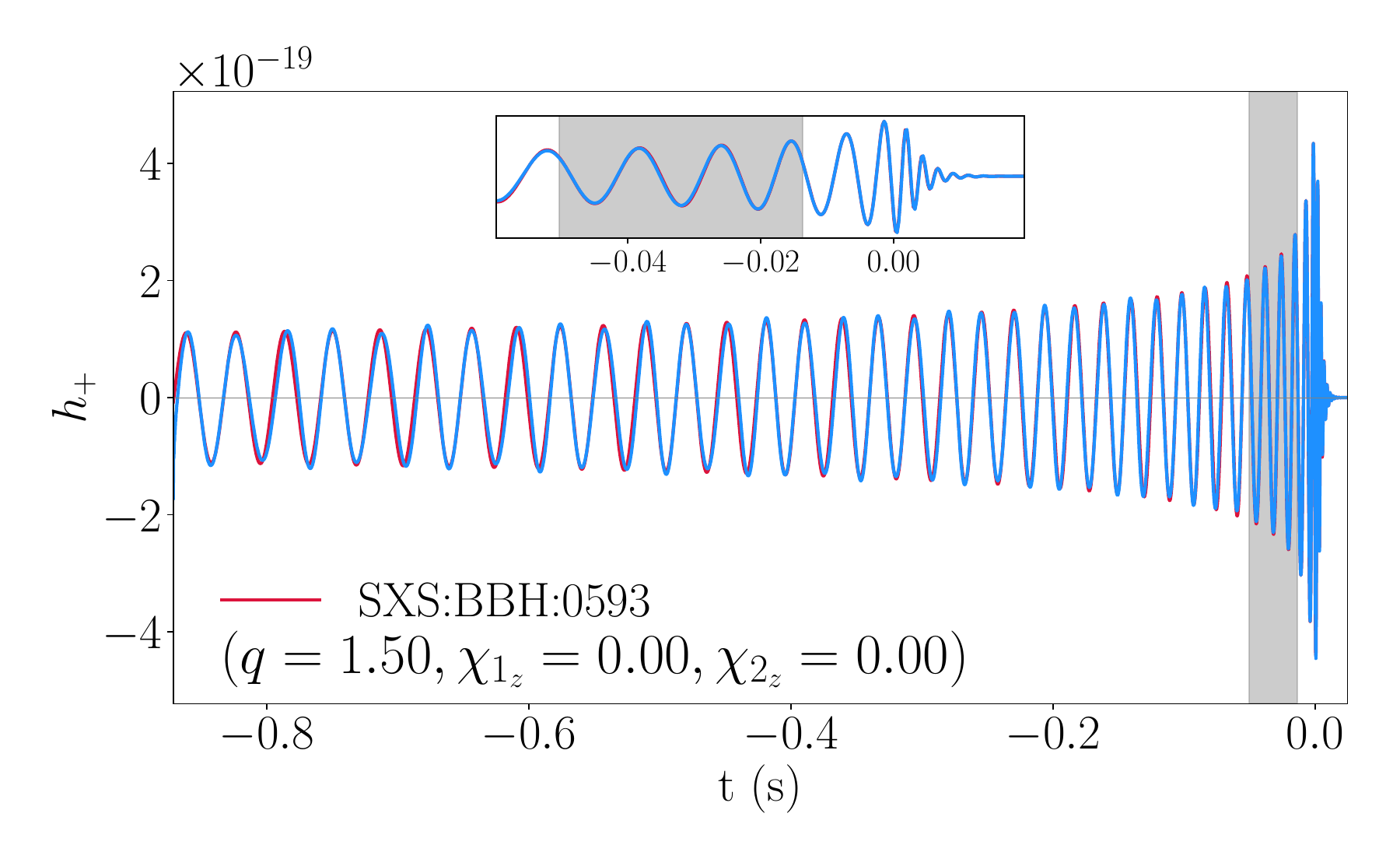}
    \end{subfigure}
      \begin{subfigure}[b]{0.24\textwidth}
        \includegraphics[width=\linewidth]{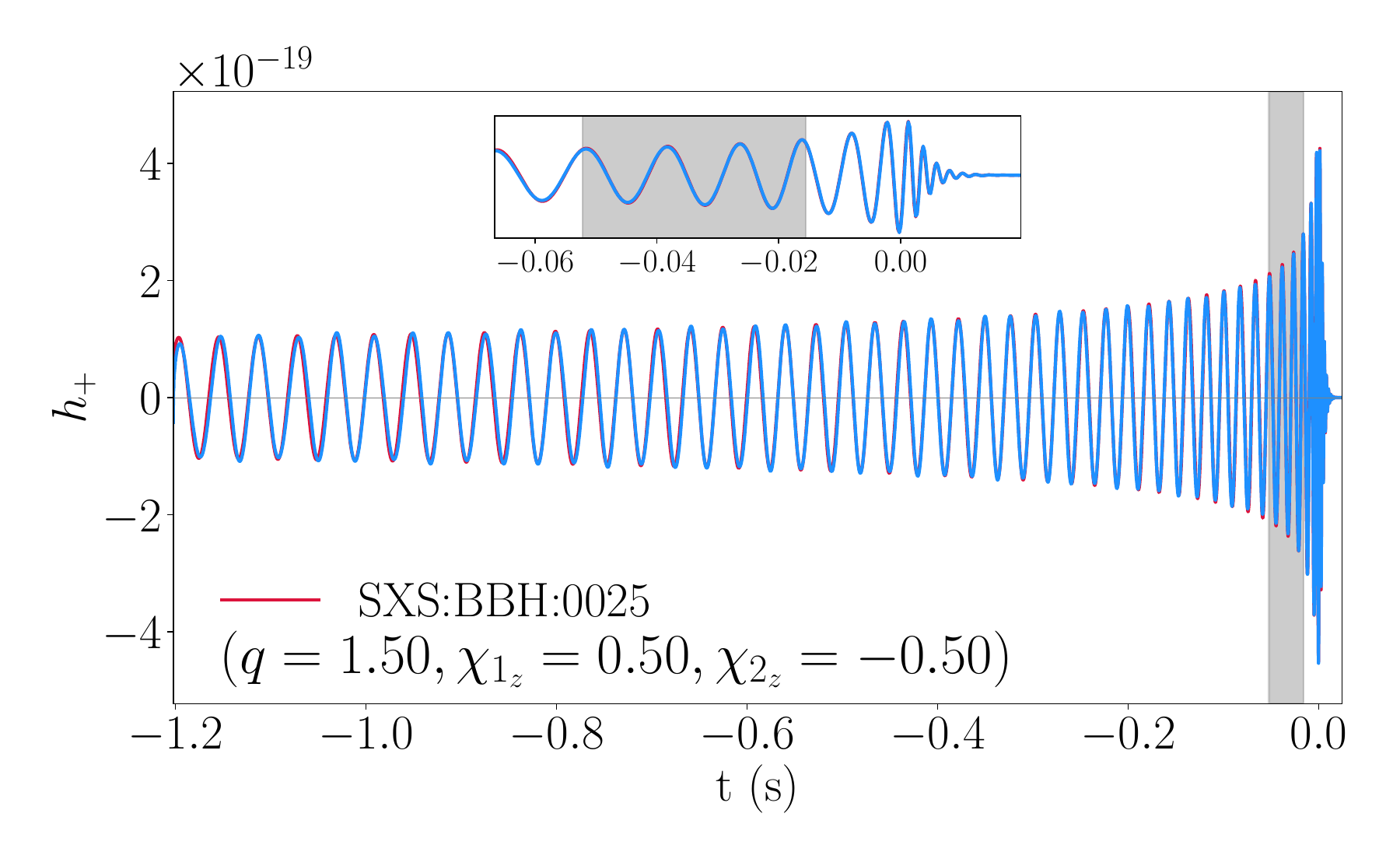}
    \end{subfigure}
       \begin{subfigure}[b]{0.24\textwidth}
            \includegraphics[width=\linewidth]{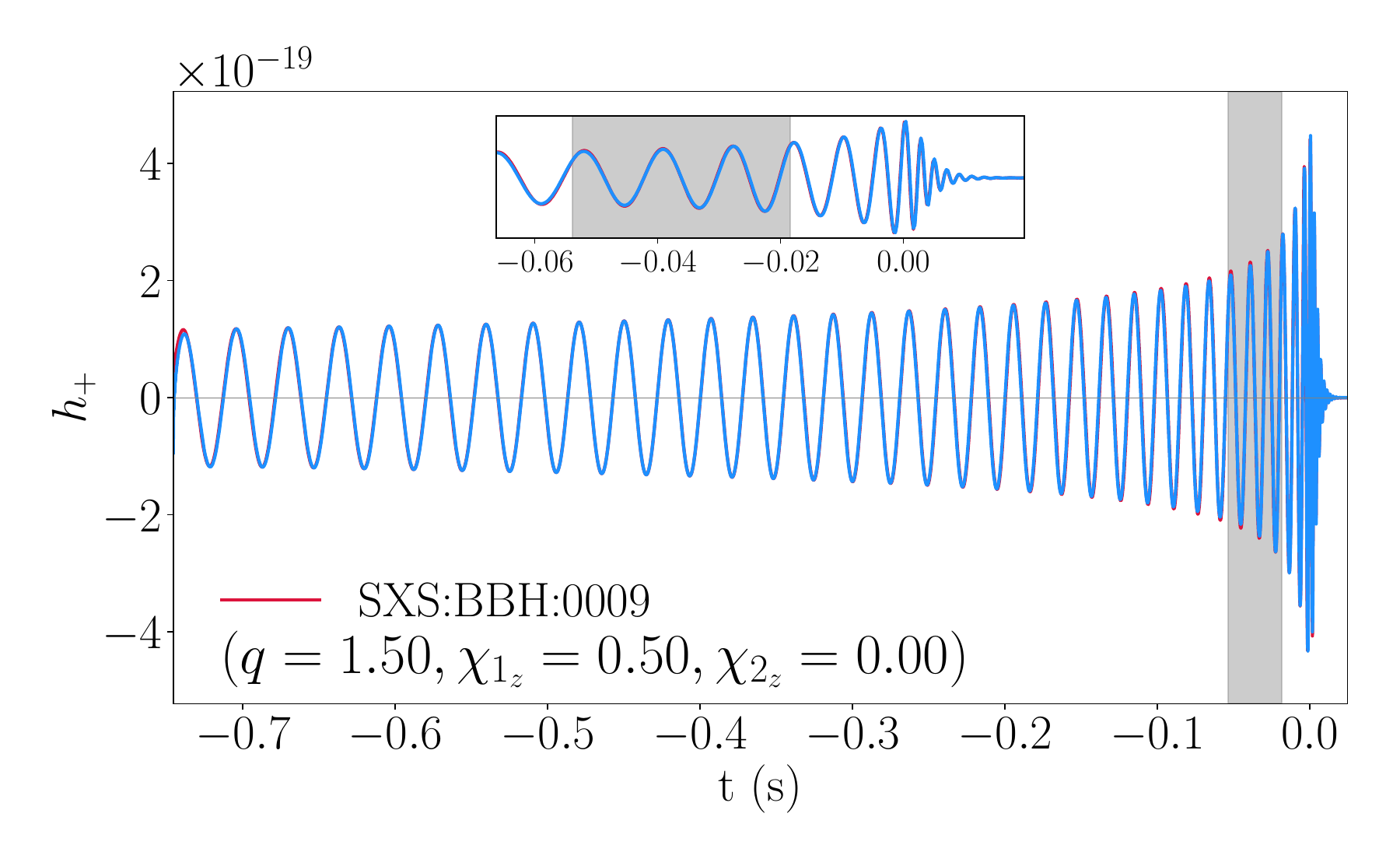}
    \end{subfigure}
    \begin{subfigure}[b]{0.24\textwidth}
        \includegraphics[width=\linewidth]{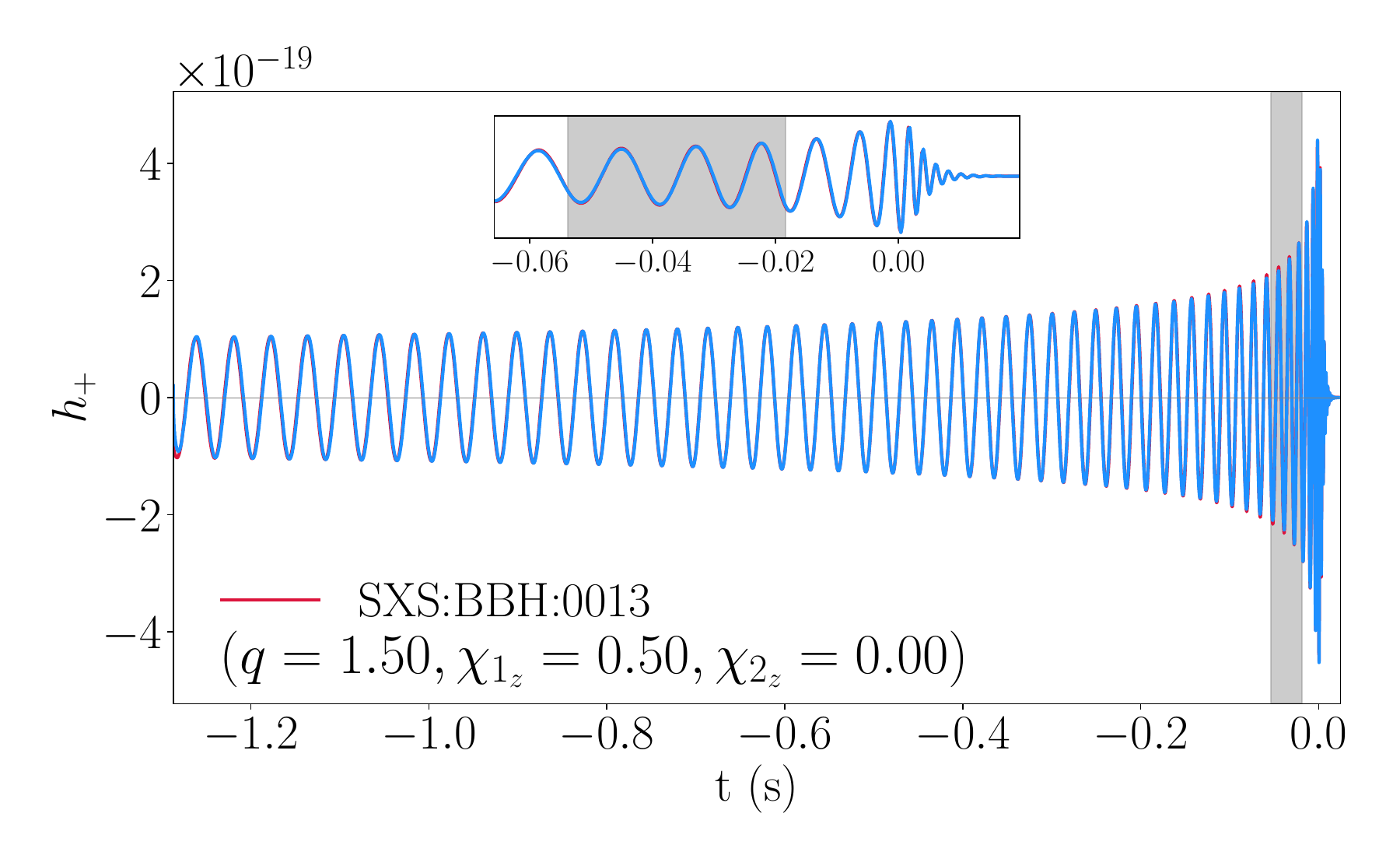}
    \end{subfigure}
      \begin{subfigure}[b]{0.24\textwidth}
        \includegraphics[width=\linewidth]{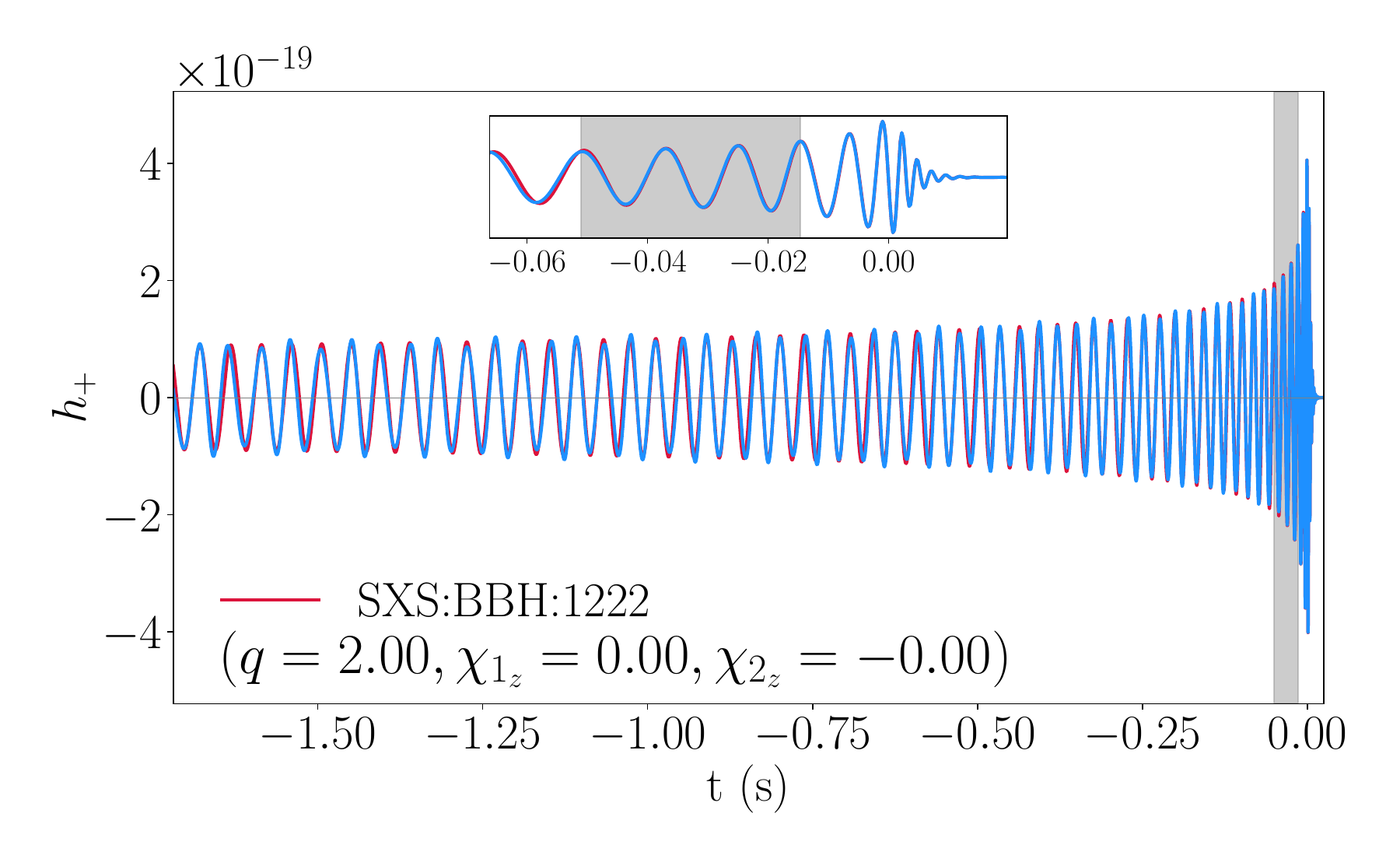}
    \end{subfigure}
        \begin{subfigure}[b]{0.24\textwidth}
        \includegraphics[width=\linewidth]{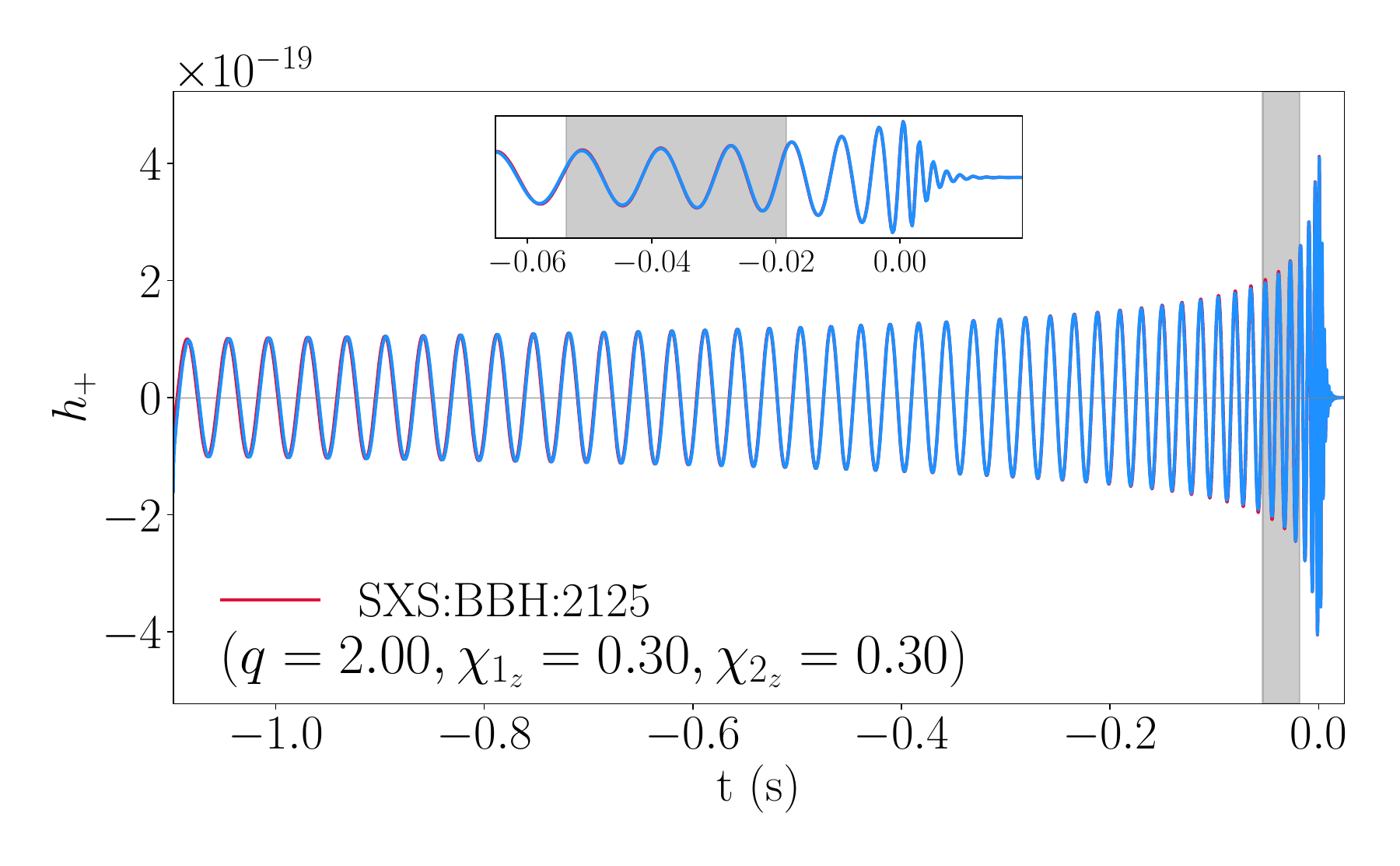}
    \end{subfigure}
     \begin{subfigure}[b]{0.24\textwidth}
        \includegraphics[width=\linewidth]{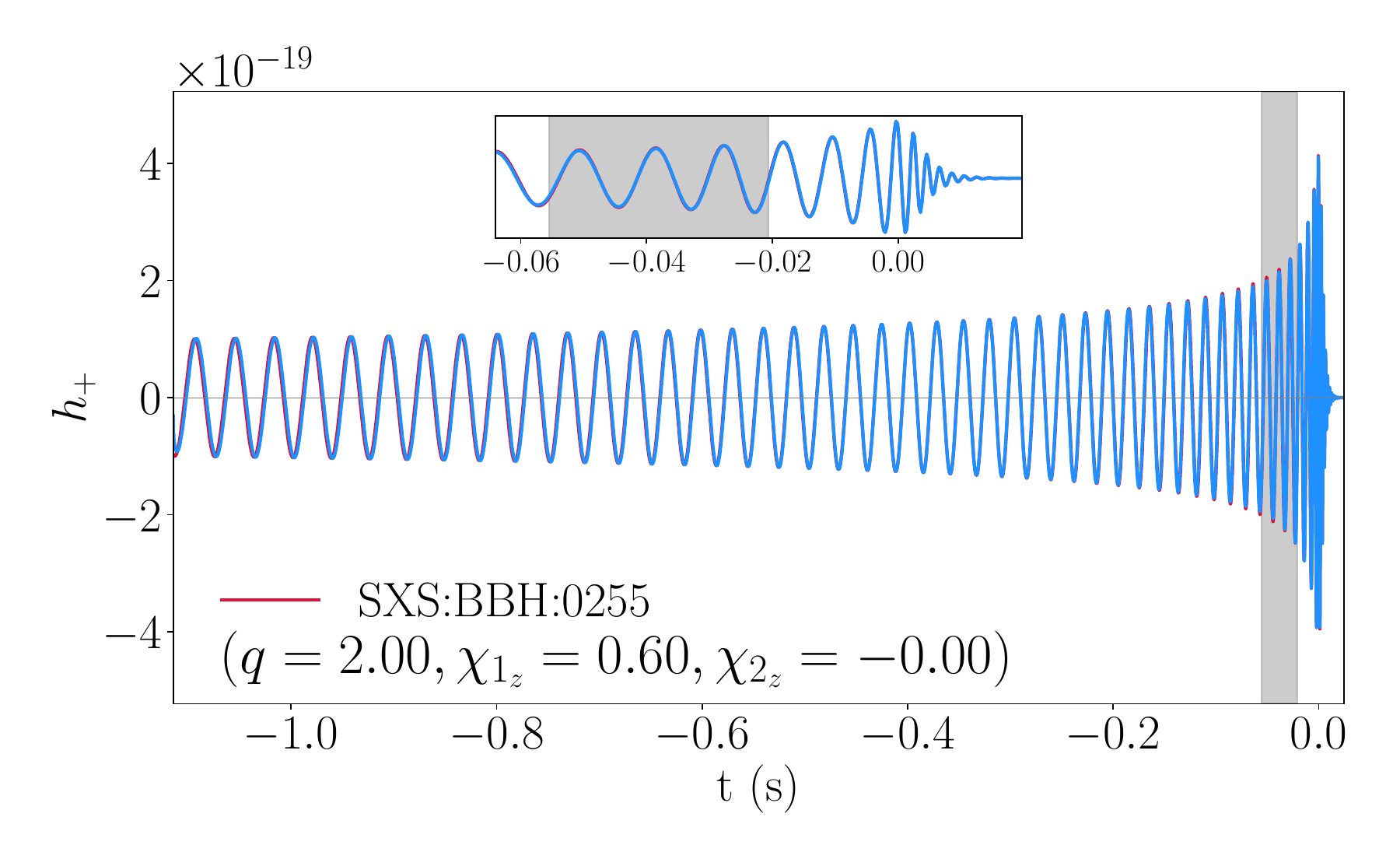}
    \end{subfigure}
       \begin{subfigure}[b]{0.24\textwidth}
        \includegraphics[width=\linewidth]{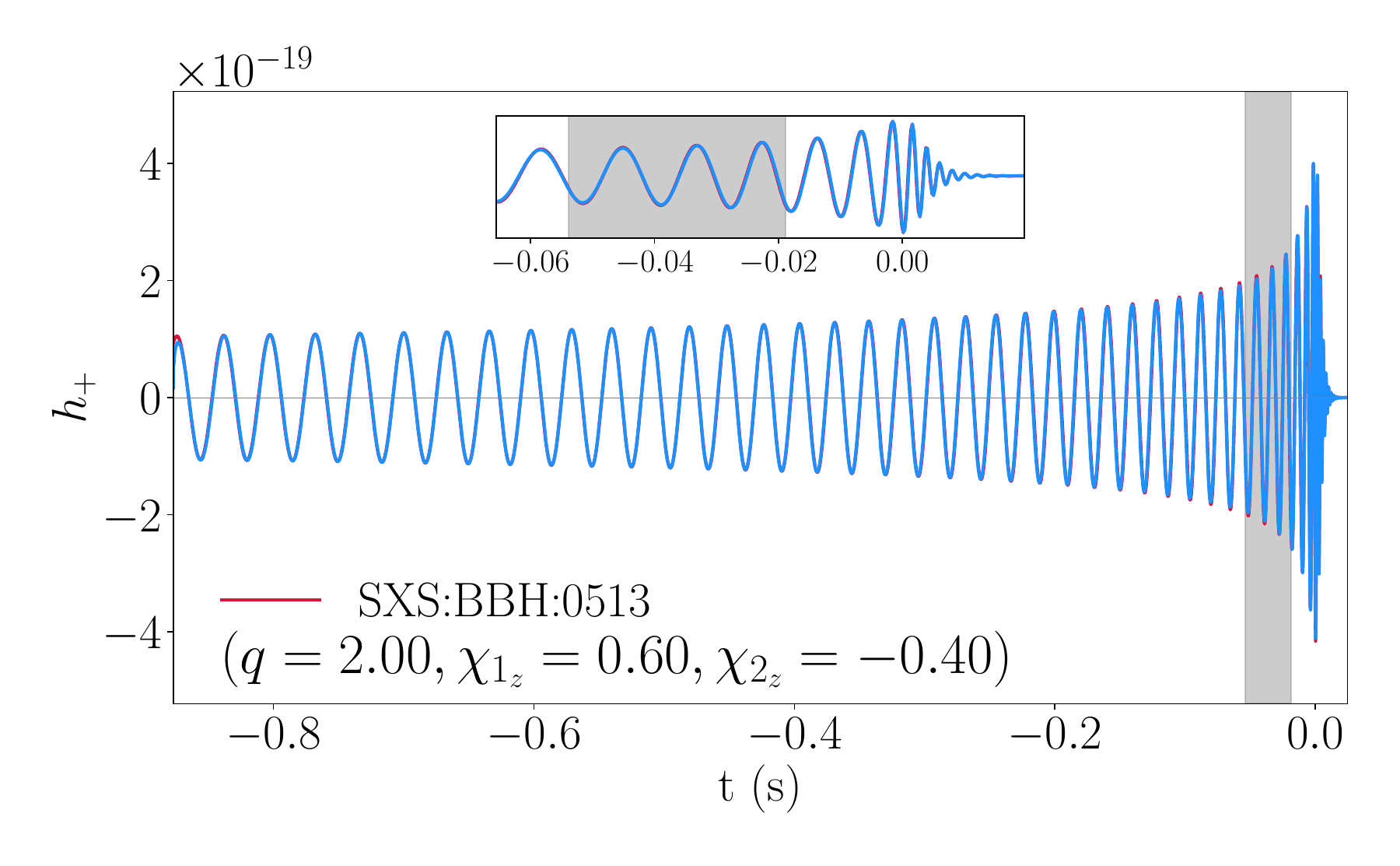}
    \end{subfigure}
    \begin{subfigure}[b]{0.24\textwidth}
        \includegraphics[width=\linewidth]{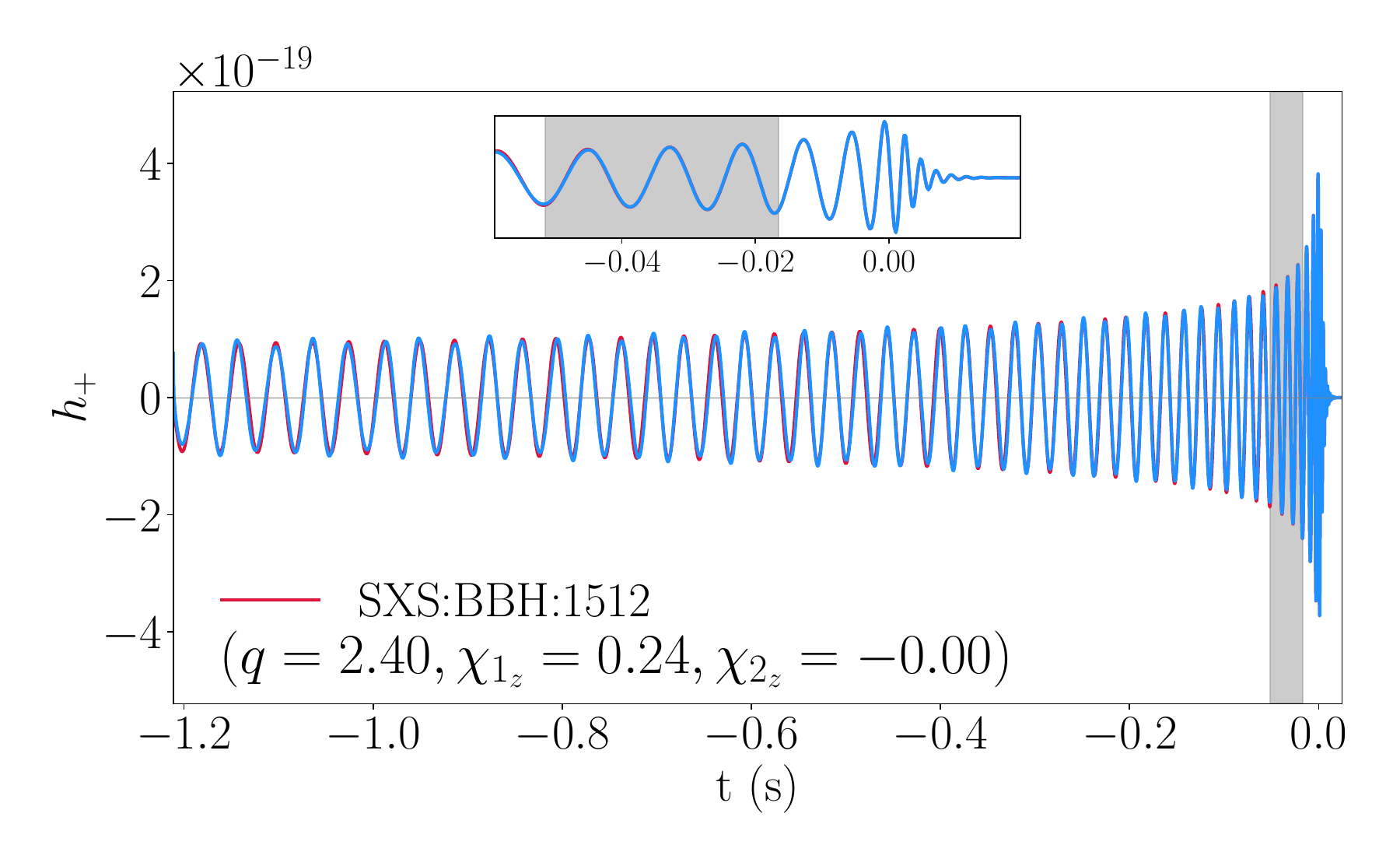}
    \end{subfigure}
     \begin{subfigure}[b]{0.24\textwidth}
        \includegraphics[width=\linewidth]{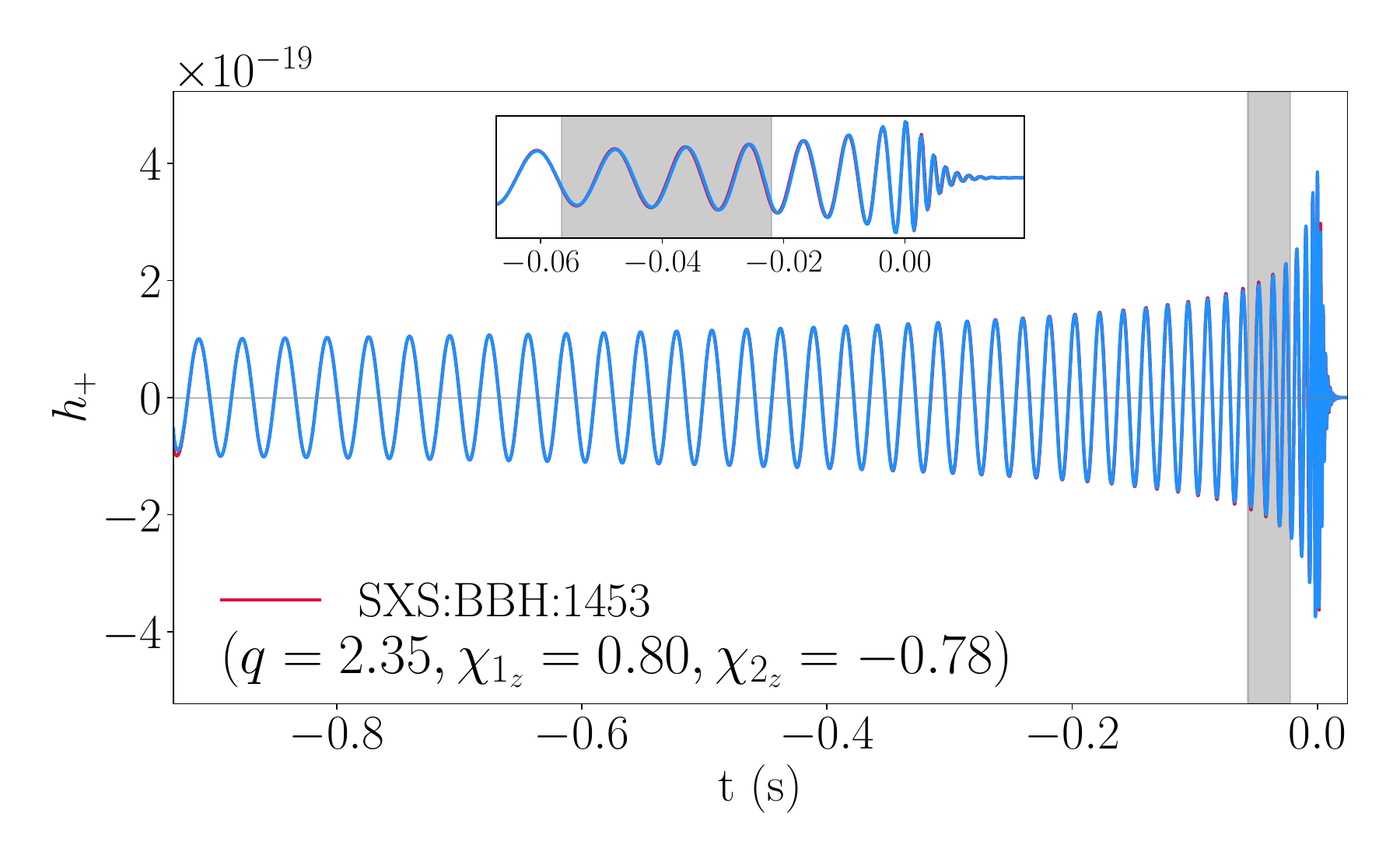}
    \end{subfigure}
      \begin{subfigure}[b]{0.24\textwidth}
        \includegraphics[width=\linewidth]{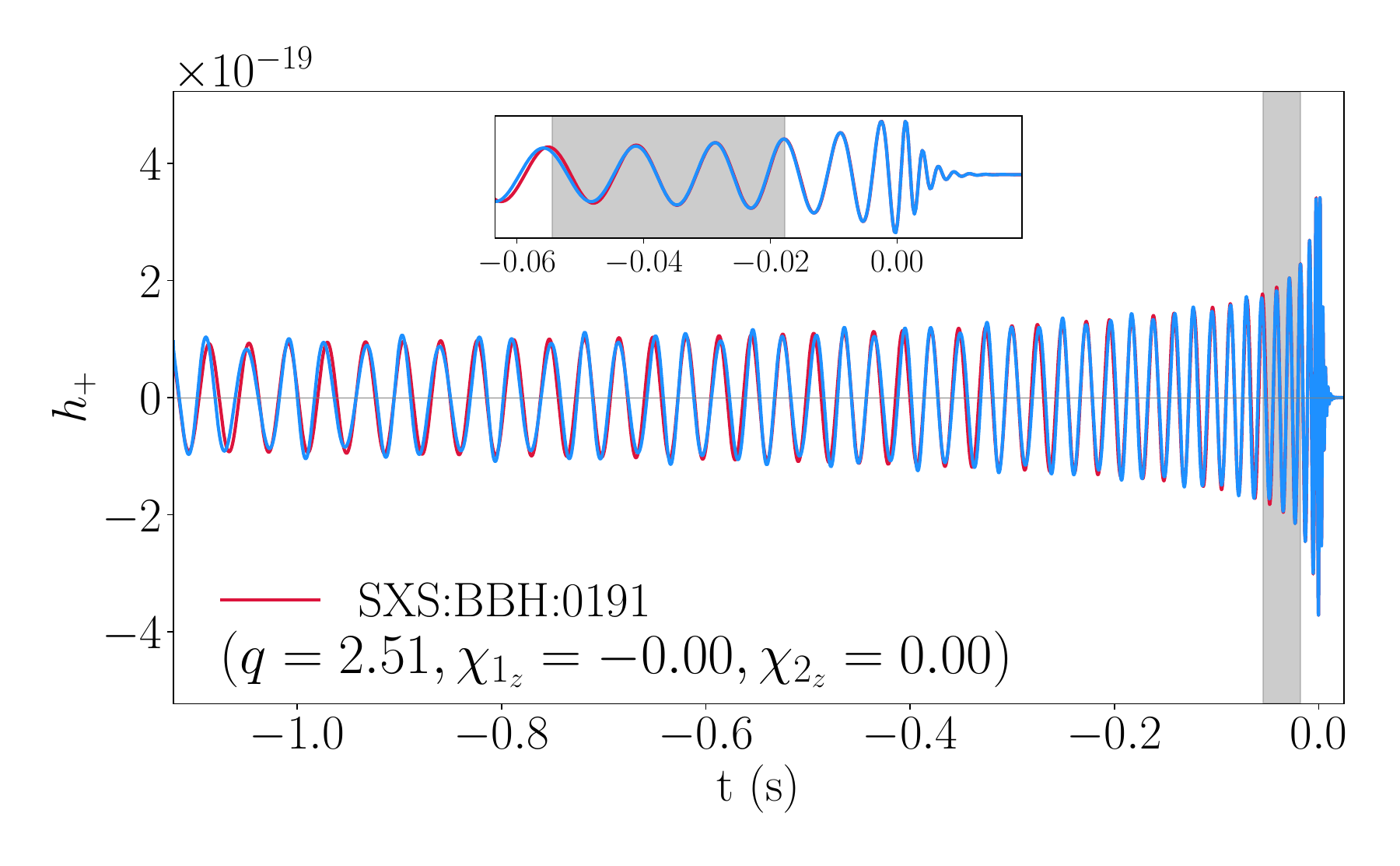}
    \end{subfigure}
      \begin{subfigure}[b]{0.24\textwidth}
        \includegraphics[width=\linewidth]{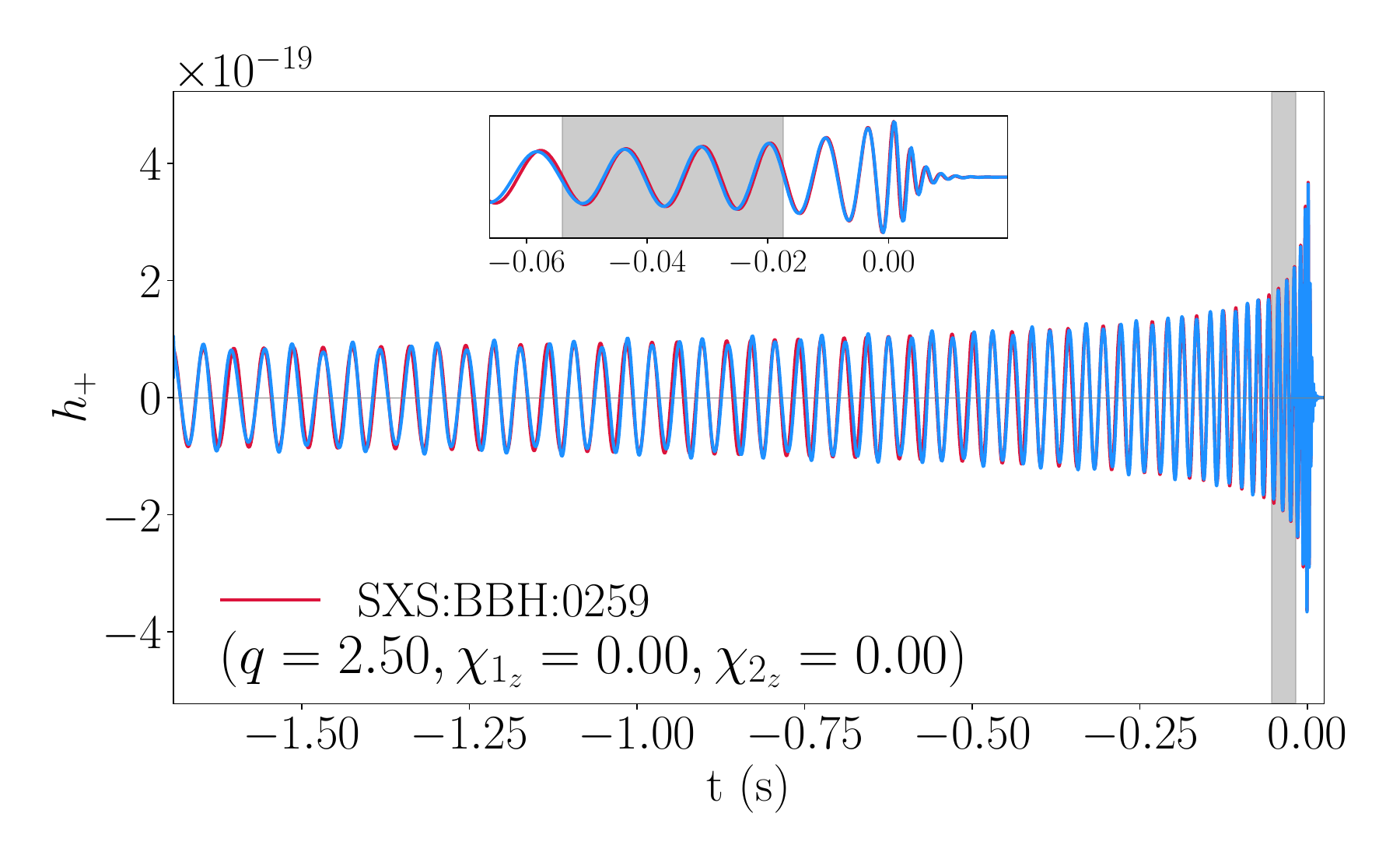}
    \end{subfigure}
     \begin{subfigure}[b]{0.24\textwidth}
        \includegraphics[width=\linewidth]{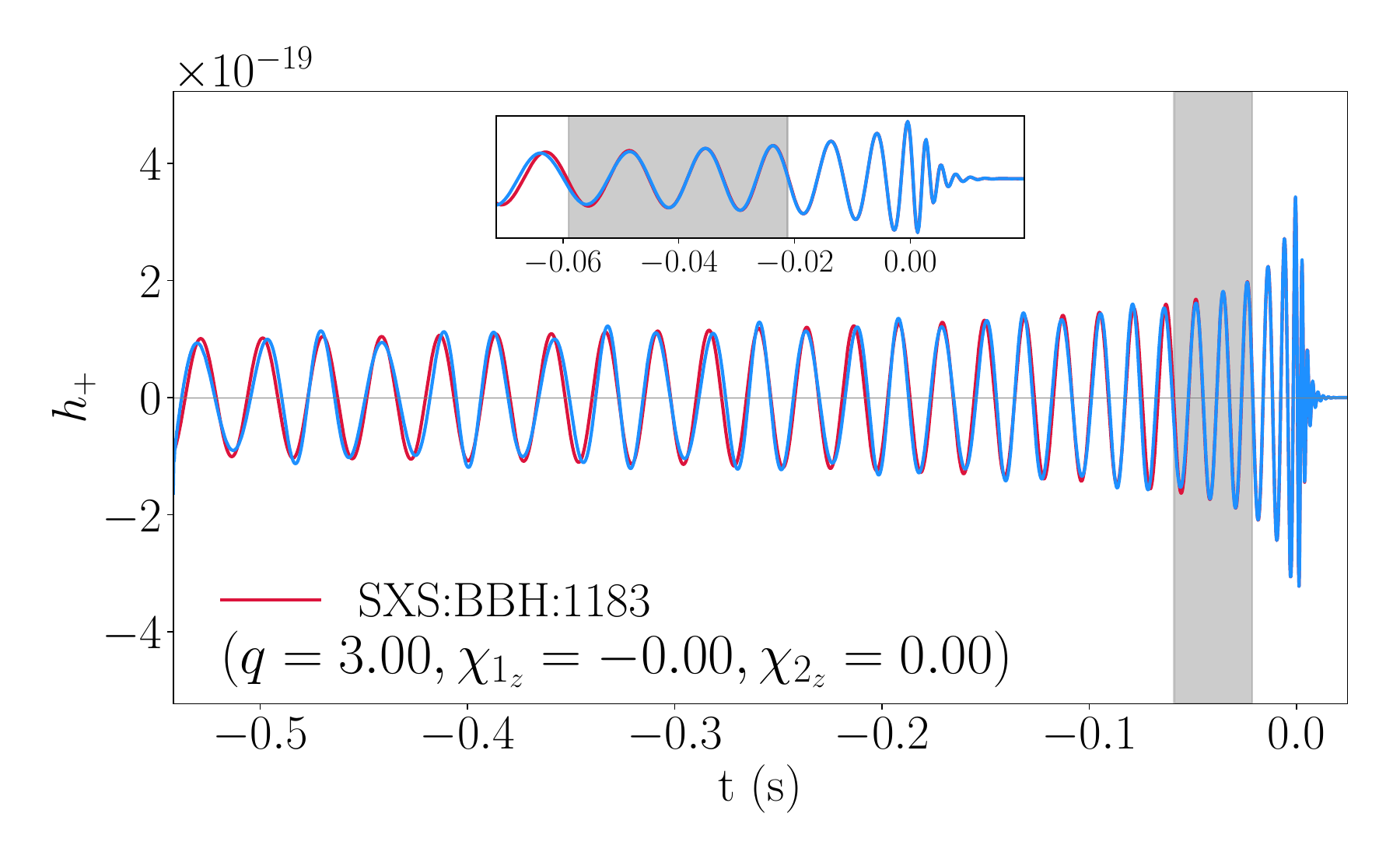}
    \end{subfigure}
       \begin{subfigure}[b]{0.24\textwidth}
        \includegraphics[width=\linewidth]{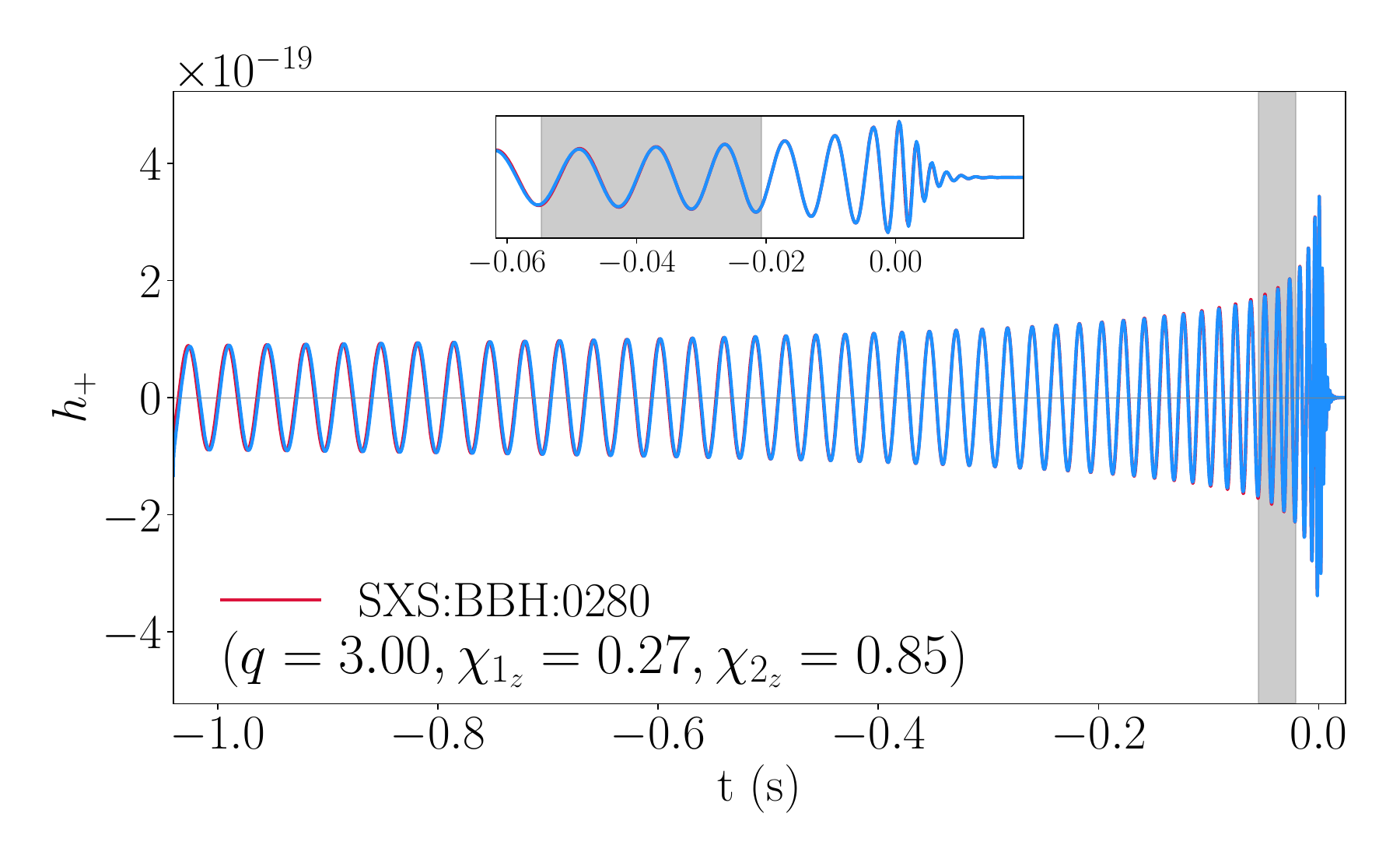}
    \end{subfigure}
      \begin{subfigure}[b]{0.24\textwidth}
        \includegraphics[width=\linewidth]{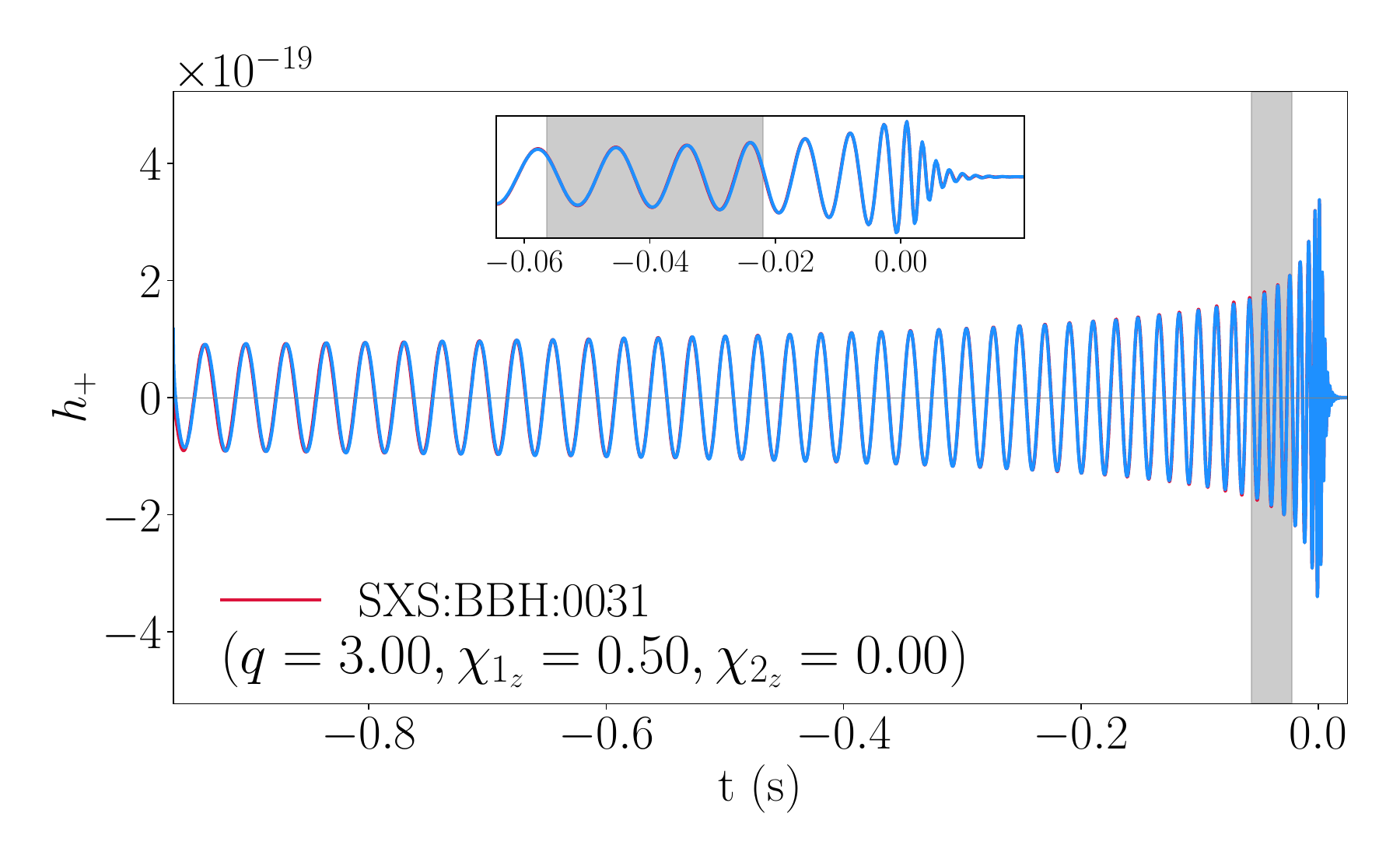}
    \end{subfigure}
     \begin{subfigure}[b]{0.24\textwidth}
        \includegraphics[width=\linewidth]{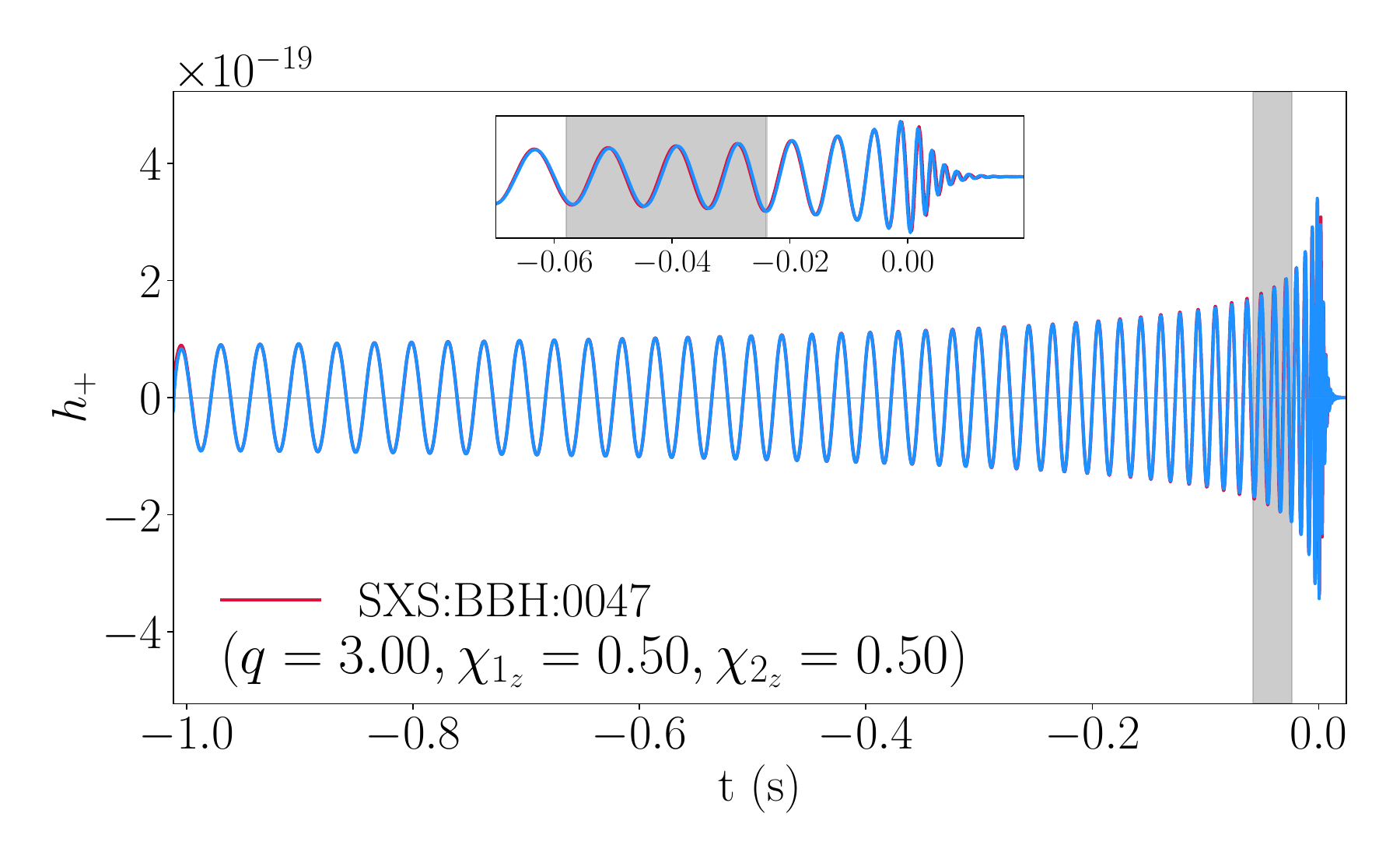}
    \end{subfigure}
    \begin{subfigure}[b]{0.24\textwidth}
        \includegraphics[width=\linewidth]{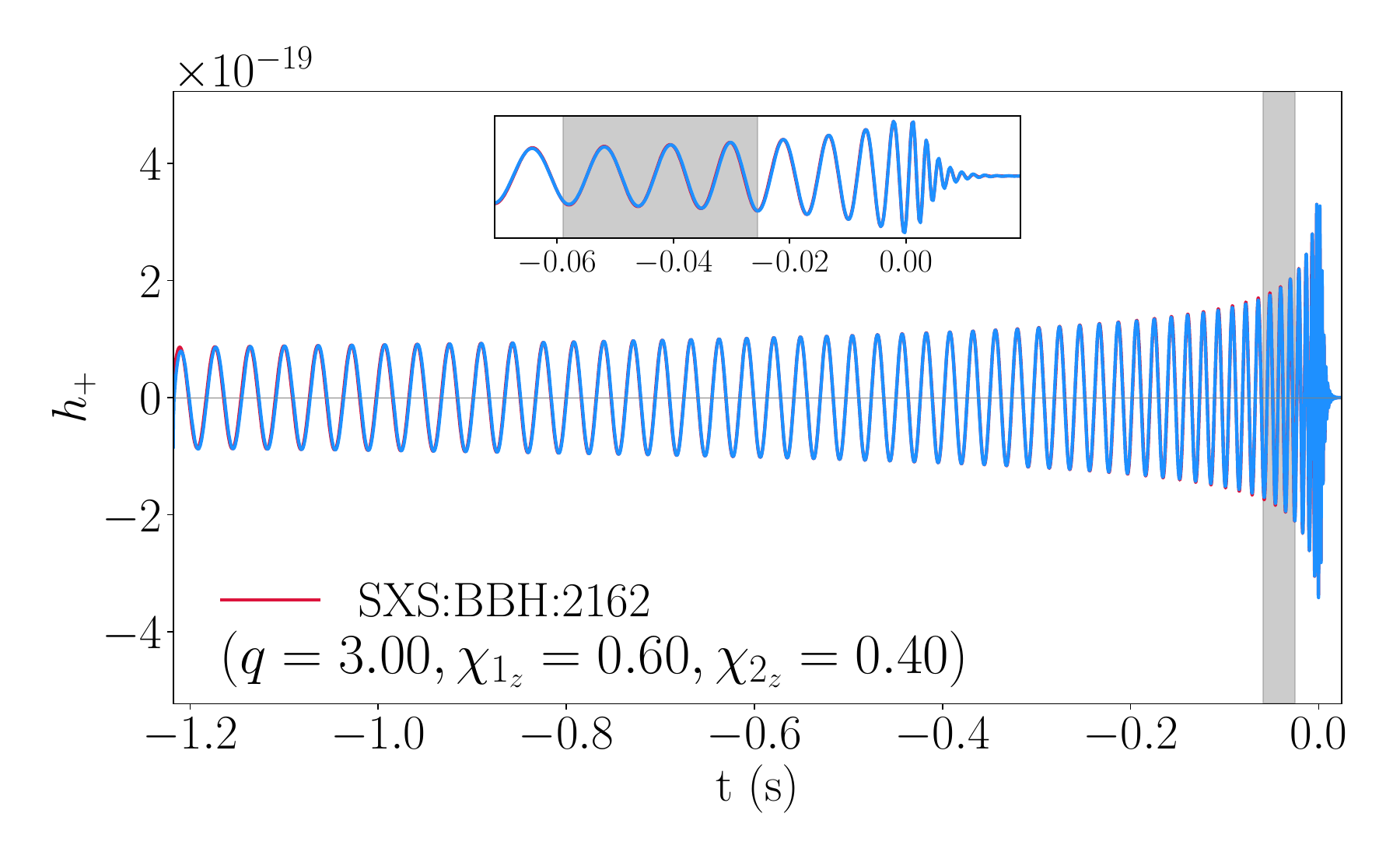}
    \end{subfigure}
    \begin{subfigure}[b]{0.24\textwidth}
        \includegraphics[width=\linewidth]{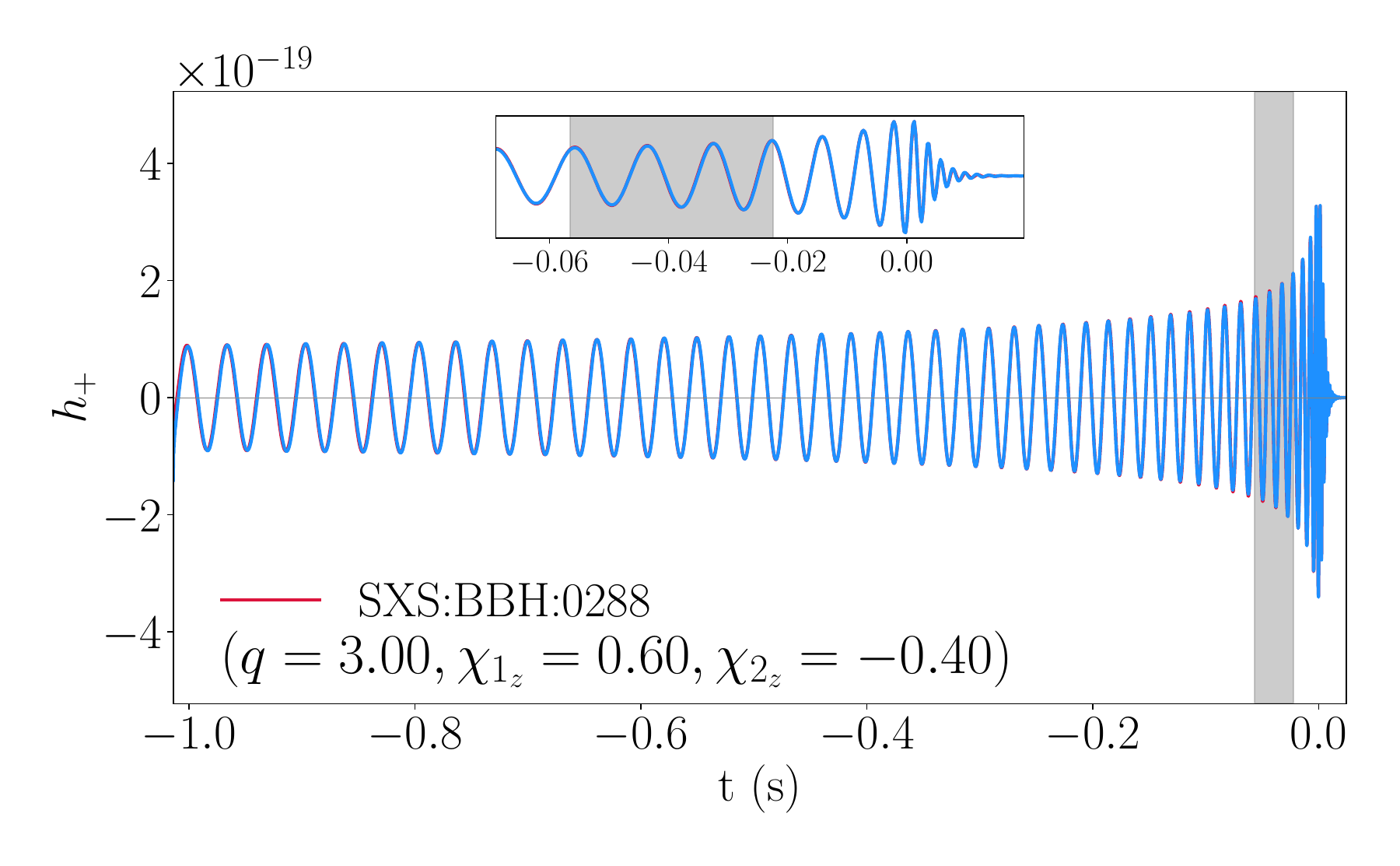}
    \end{subfigure}
    \begin{subfigure}[b]{0.24\textwidth}
        \includegraphics[width=\linewidth]{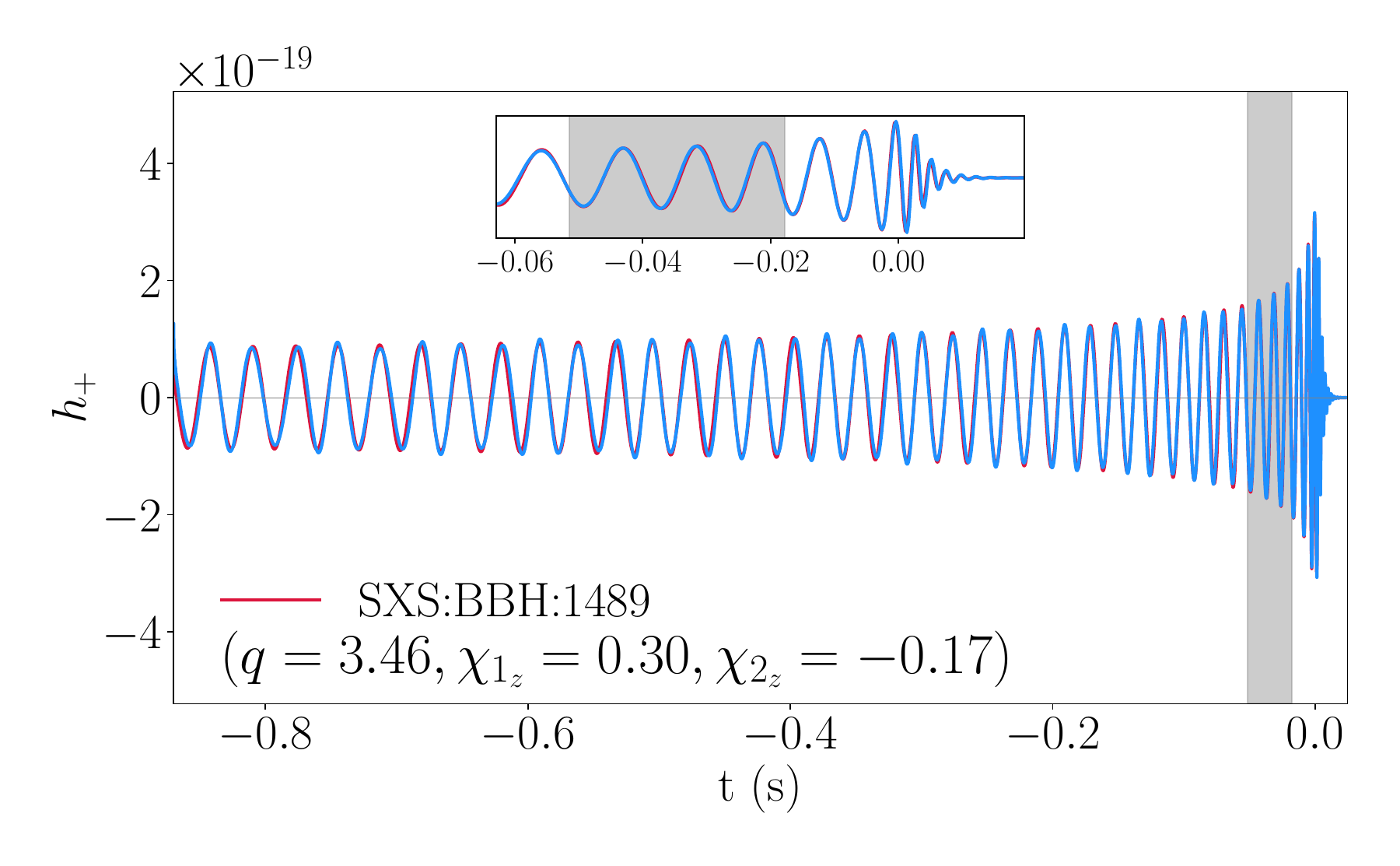}
    \end{subfigure}
    \begin{subfigure}[b]{0.24\textwidth}
        \includegraphics[width=\linewidth]{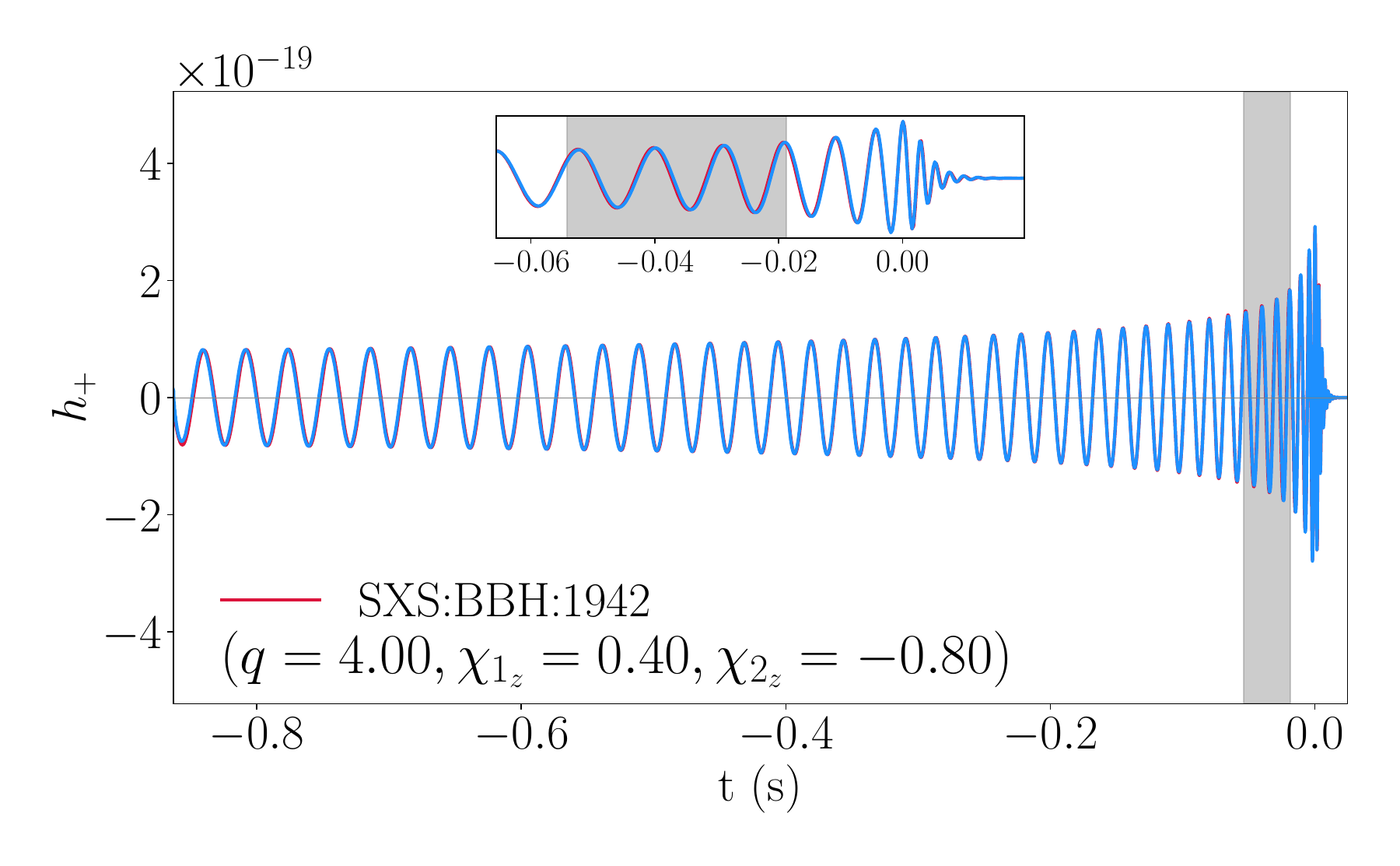}
    \end{subfigure}
    \begin{subfigure}[b]{0.24\textwidth}
        \includegraphics[width=\linewidth]{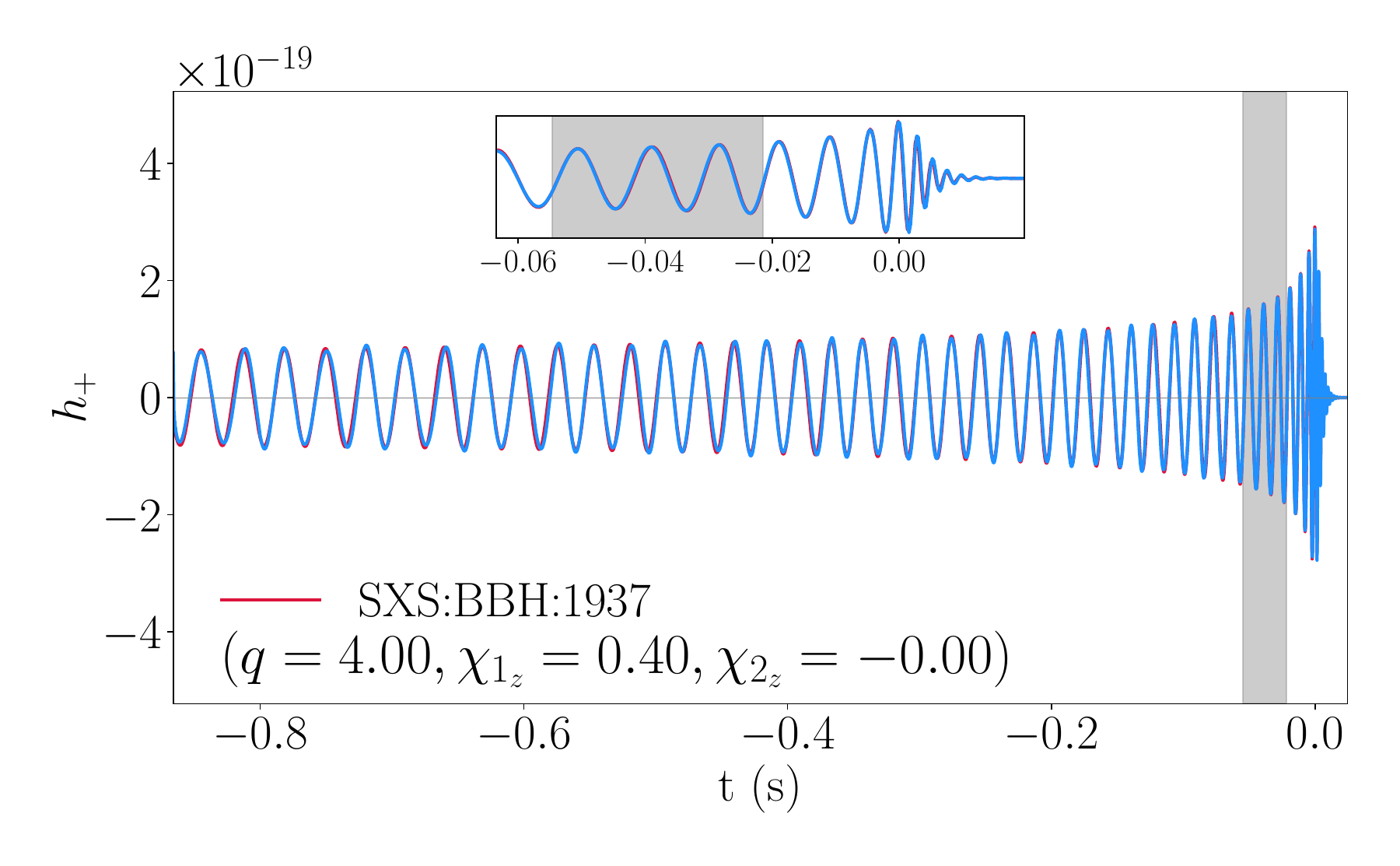}
    \end{subfigure}
    \begin{subfigure}[b]{0.24\textwidth}
        \includegraphics[width=\linewidth]{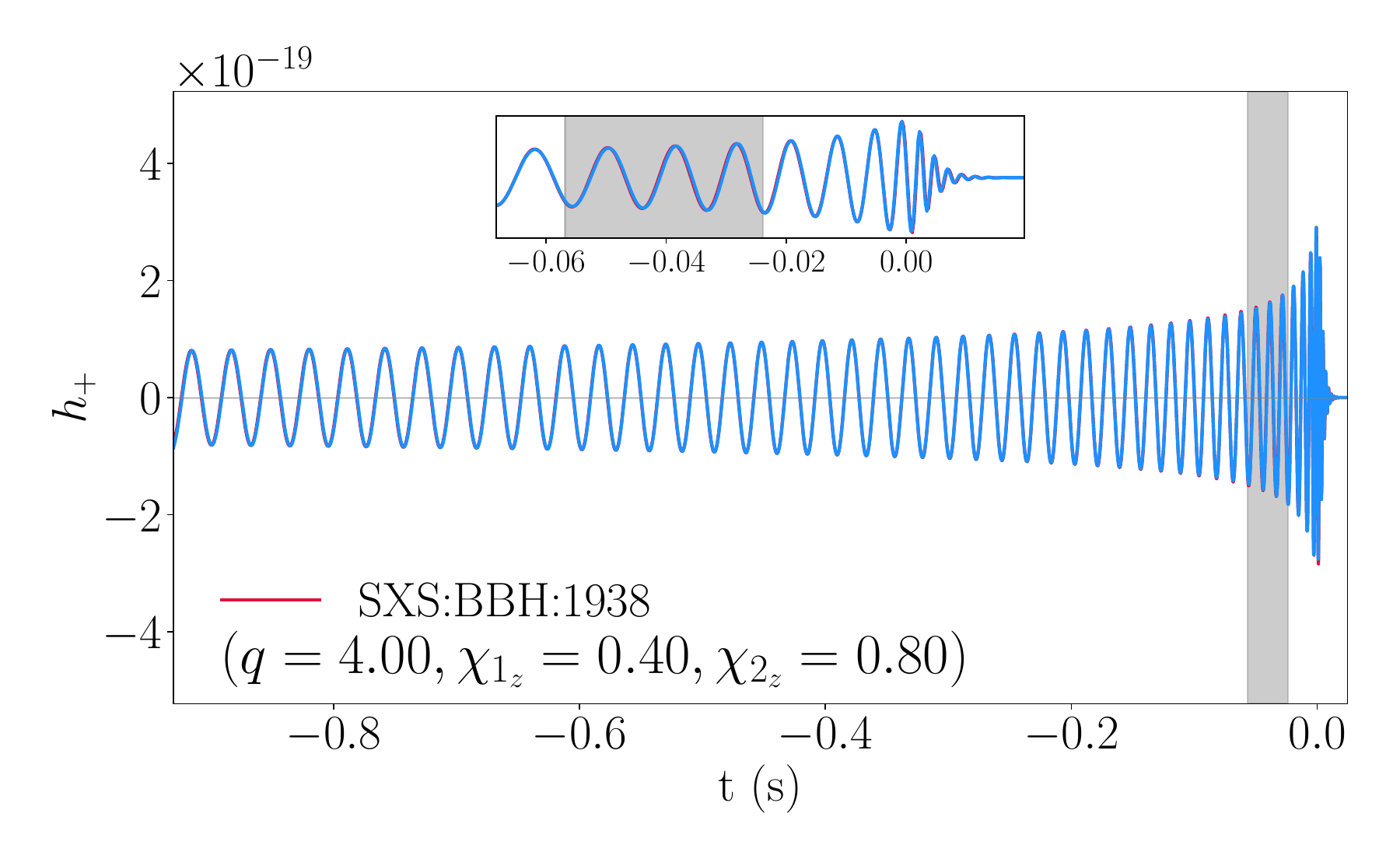}
    \end{subfigure}
    \begin{subfigure}[b]{0.24\textwidth}
        \includegraphics[width=\linewidth]{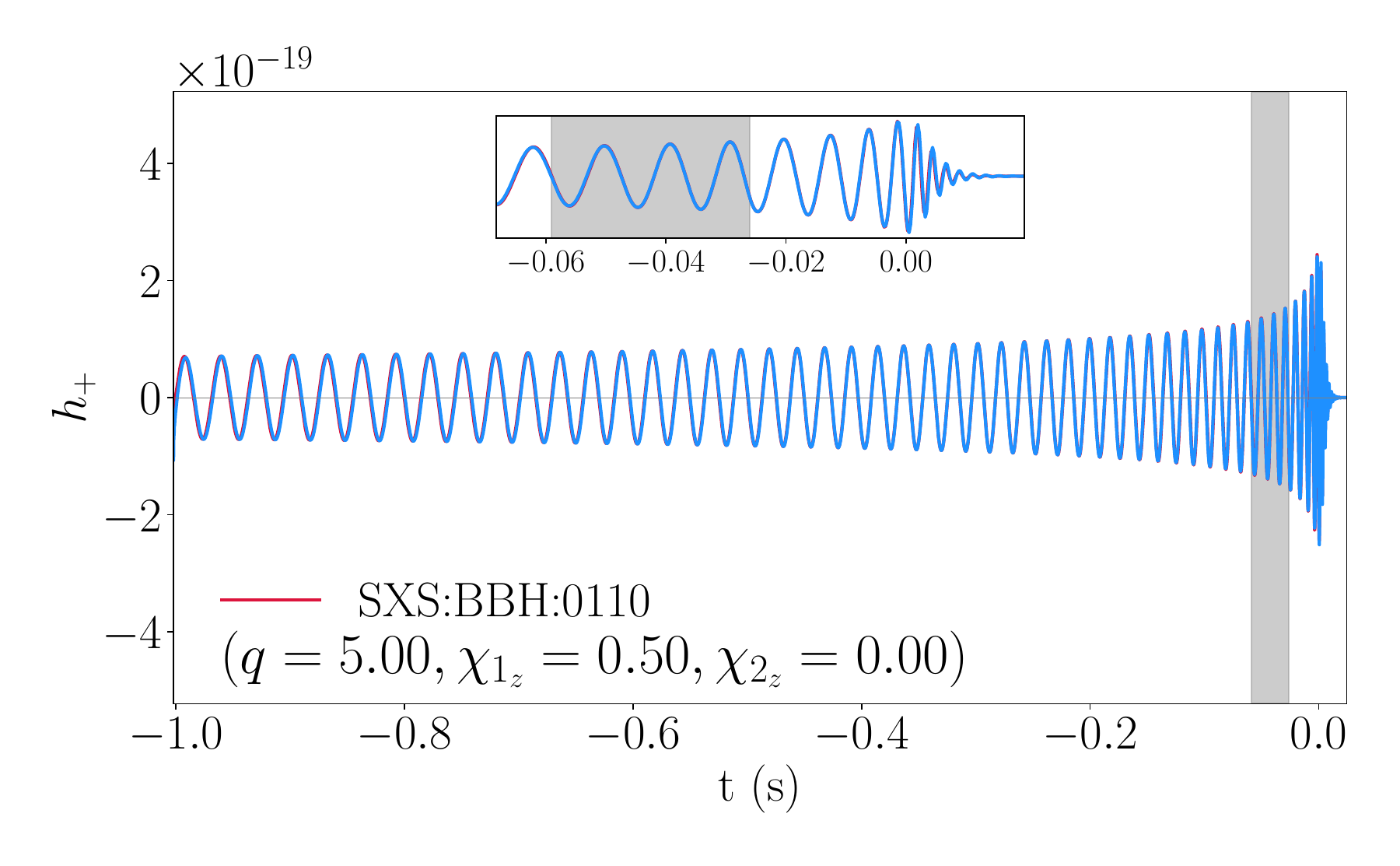}
    \end{subfigure}
    \begin{subfigure}[b]{0.24\textwidth}
        \includegraphics[width=\linewidth]{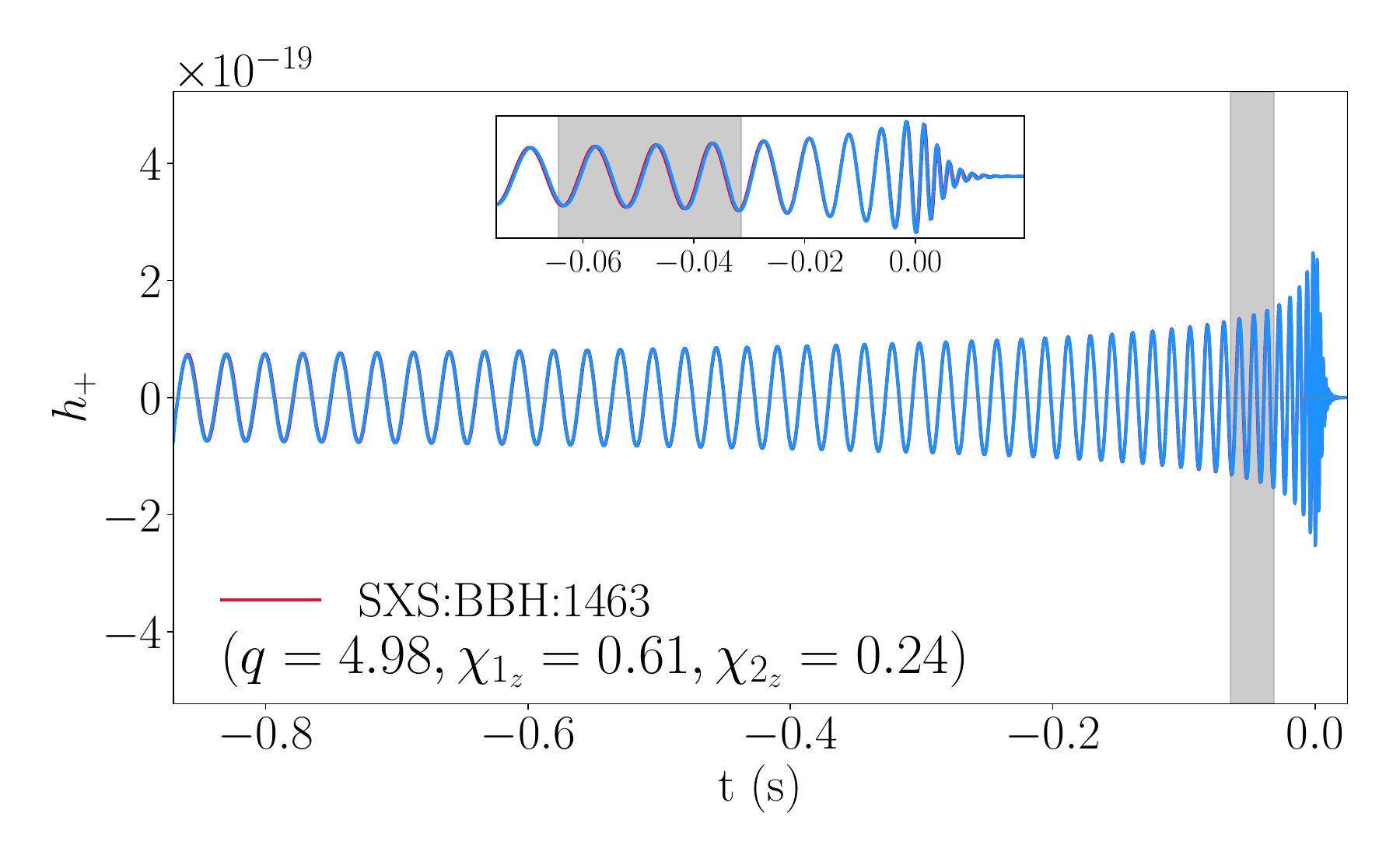}
    \end{subfigure}
    \caption{Comparison of strain data $(h_{+})$ of \imresigma{} with 32 out of 42 (see Table~\ref{table:sxs_id_qc_nr_4PN}) quasi-circular NR simulations taken from SXS catalog to validate \imresigma{} in the zero-eccentricity limit.
    The comparison is done by including the dominant $\ell=|m|=2$ modes and using a total mass of $40M_{\odot}$. The inset shows a zoomed in view of the late inspiral and PMR portion. {\violet The gray shaded region denotes the time interval over which the \imresigma{} transitions from its inspiral prescription to the plunge-merger-ringdown prescription.} The value of reference mass-ratio $(q)$, and the $z$-component of the reference dimensionless spins $(\chi_{1_z},\chi_{2_z})$ of each simulations are given in each figure.}
    \label{fig:NR_comp_p1}
\end{figure*}
\begin{figure*} 
    \centering
     \begin{subfigure}[b]{0.24\textwidth}
        \includegraphics[width=\linewidth]{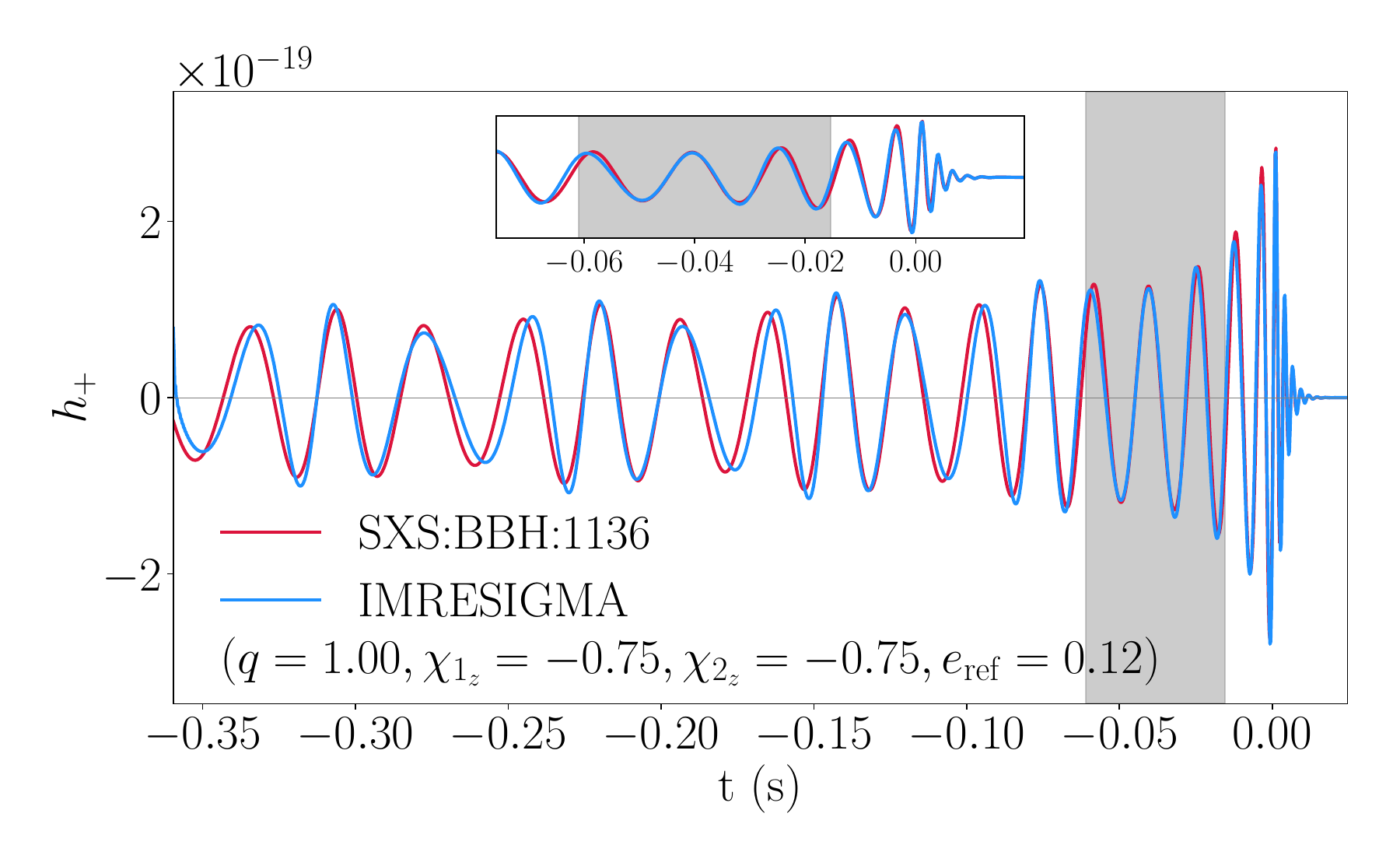}
    \end{subfigure}
    \begin{subfigure}[b]{0.24\textwidth}
        \includegraphics[width=\linewidth]{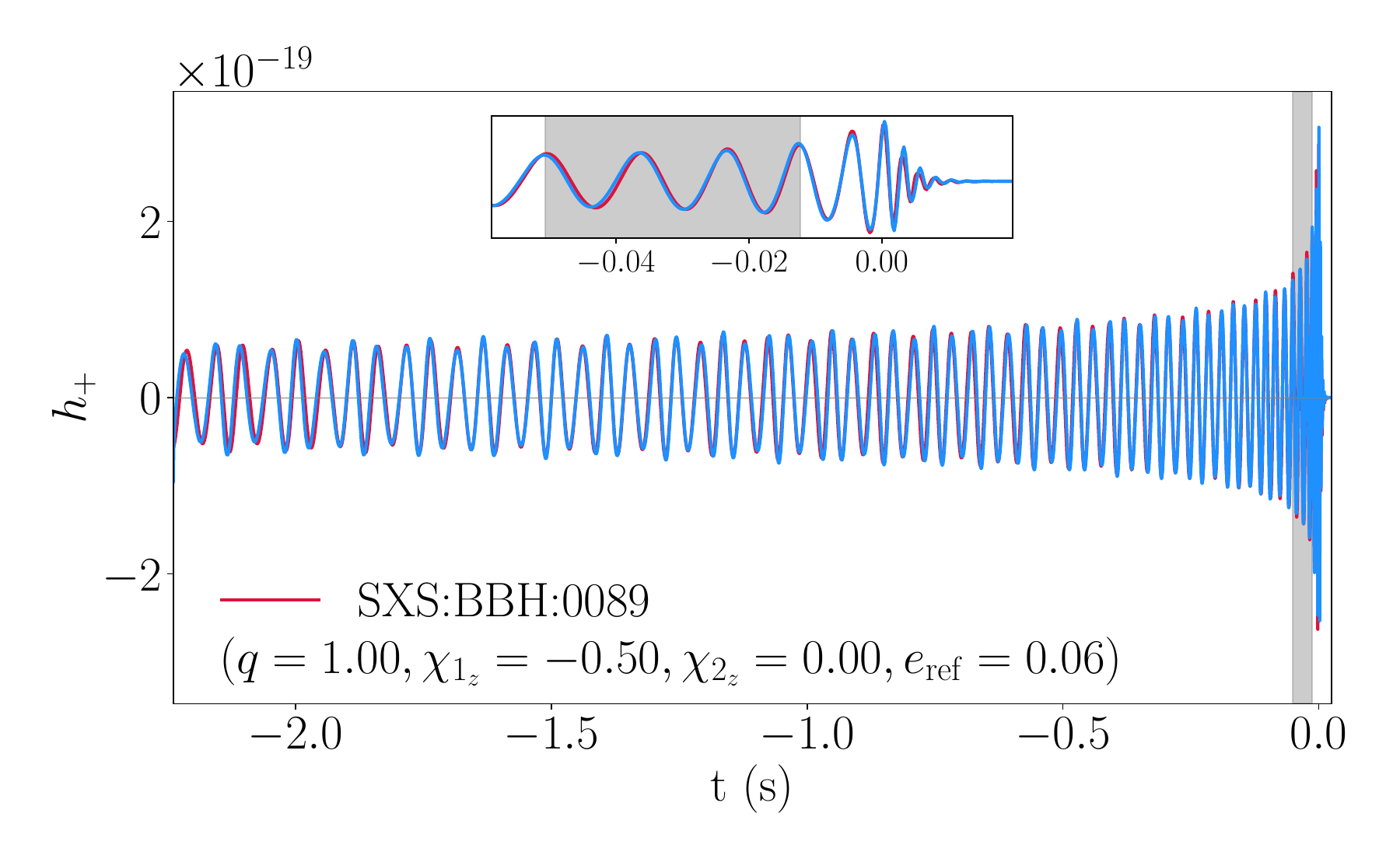}
    \end{subfigure}
    \begin{subfigure}[b]{0.24\textwidth}
        \includegraphics[width=\linewidth]{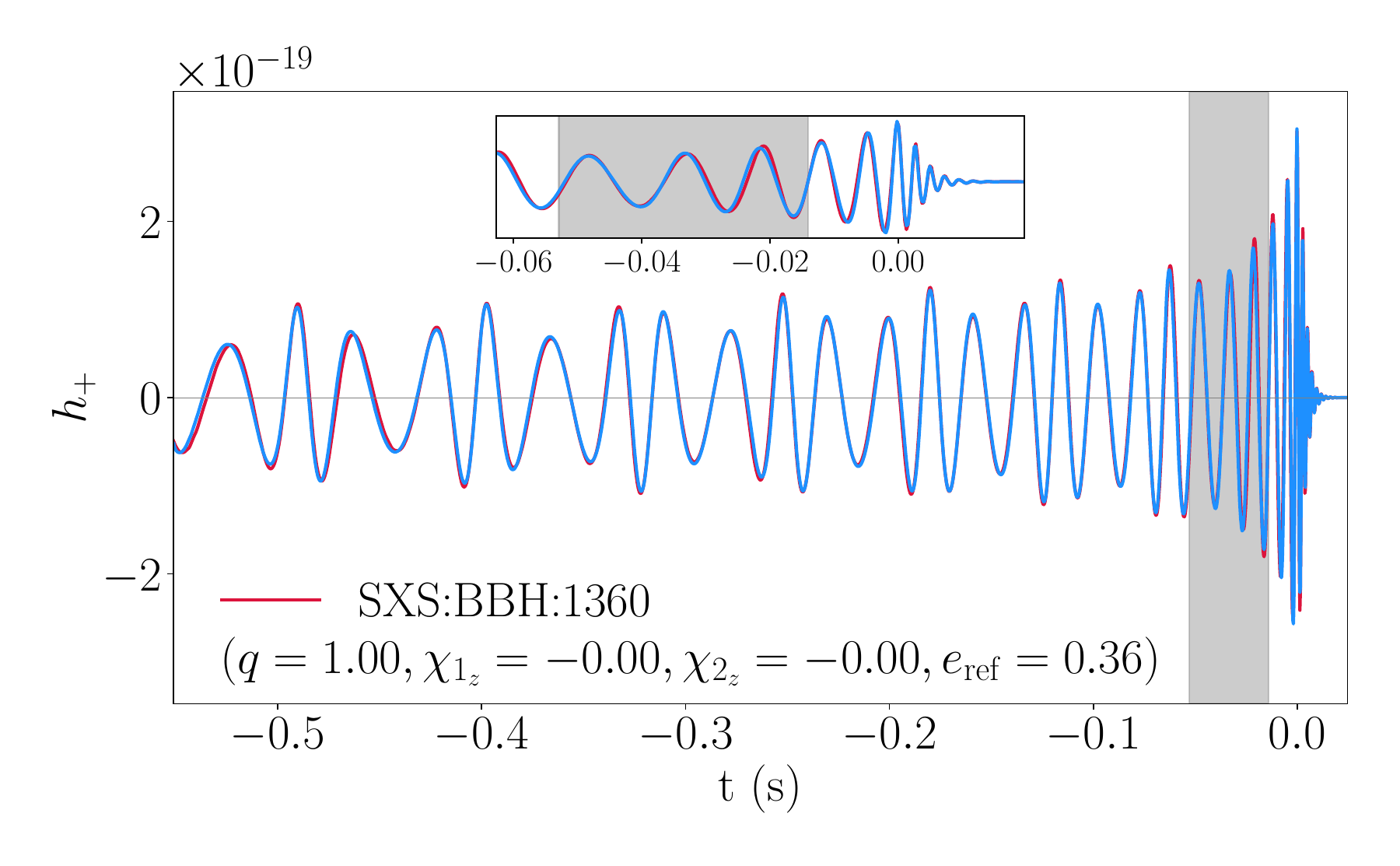}
    \end{subfigure}
     \begin{subfigure}[b]{0.24\textwidth}
        \includegraphics[width=\linewidth]{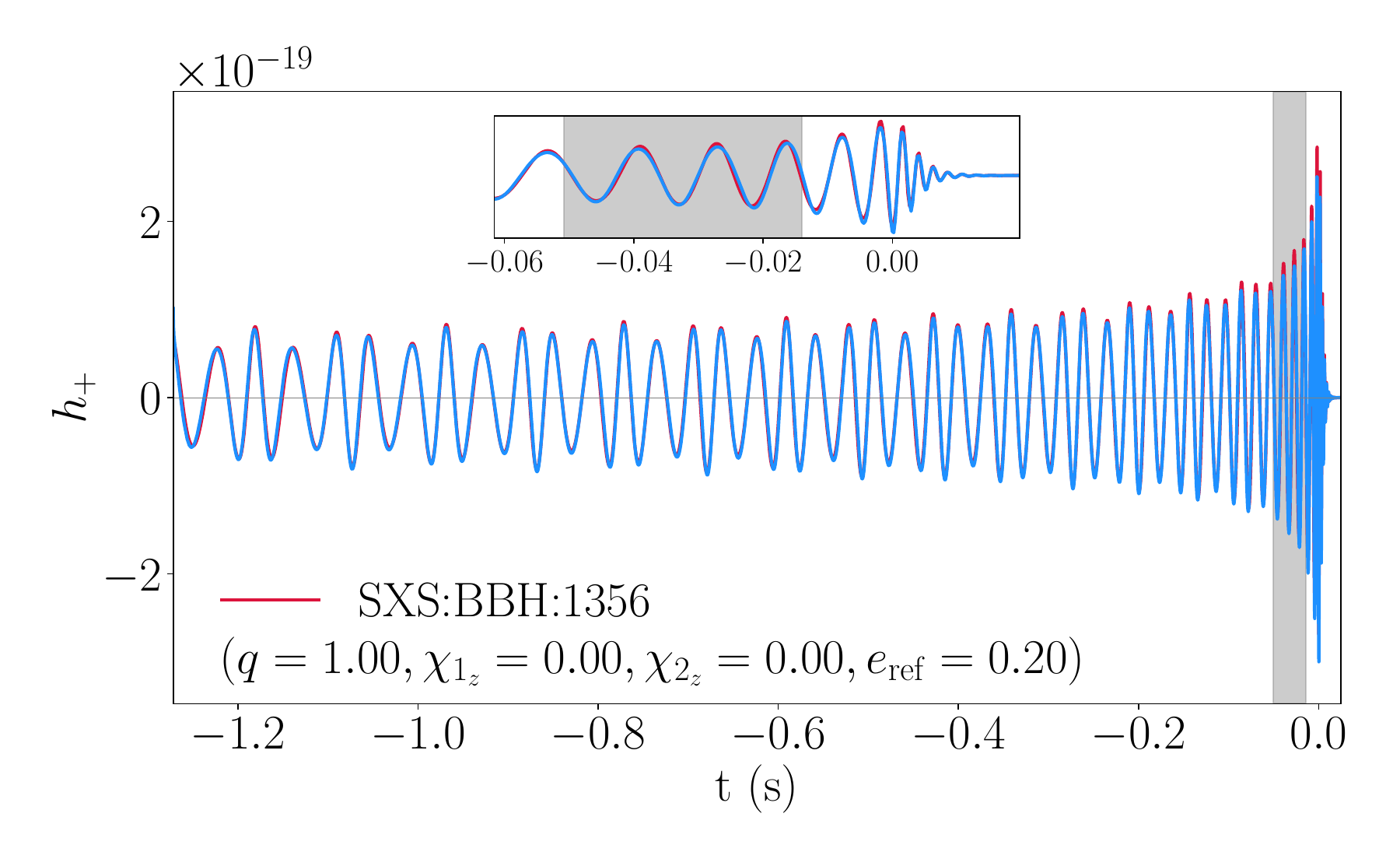}
    \end{subfigure}
    \begin{subfigure}[b]{0.24\textwidth}
        \includegraphics[width=\linewidth]{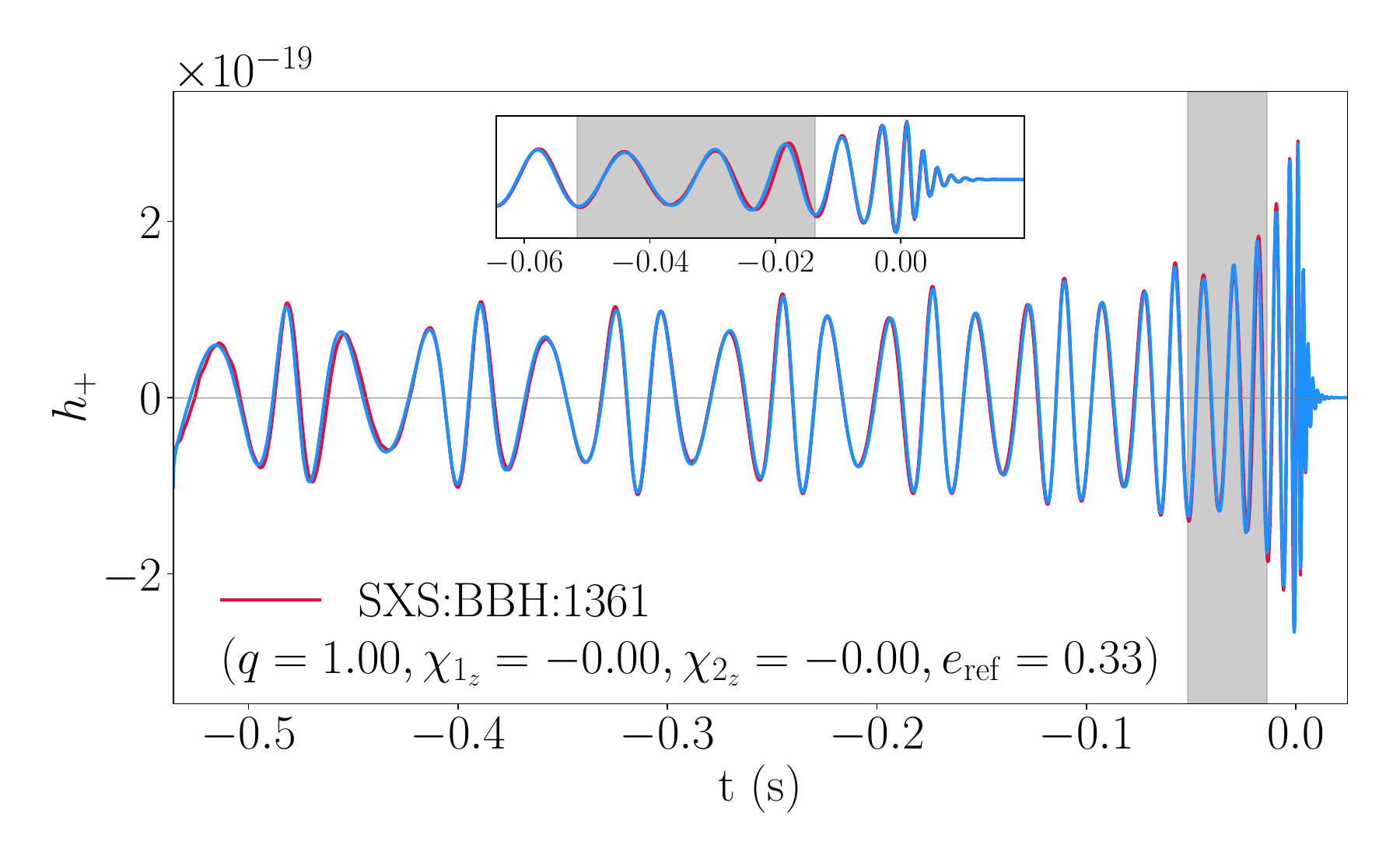}
    \end{subfigure}
    \begin{subfigure}[b]{0.24\textwidth}
        \includegraphics[width=\linewidth]{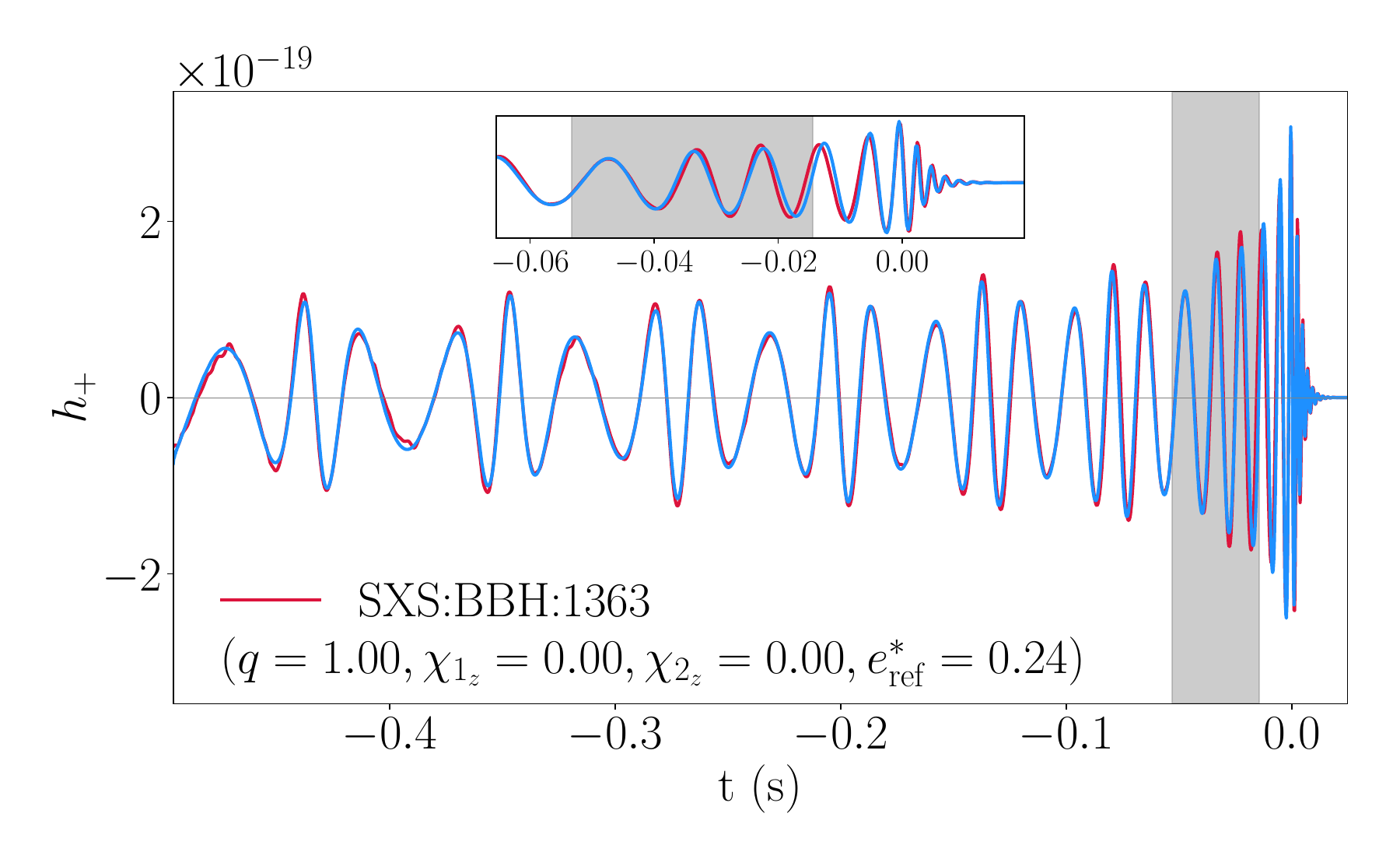}
    \end{subfigure}
       \begin{subfigure}[b]{0.24\textwidth}
        \includegraphics[width=\linewidth]{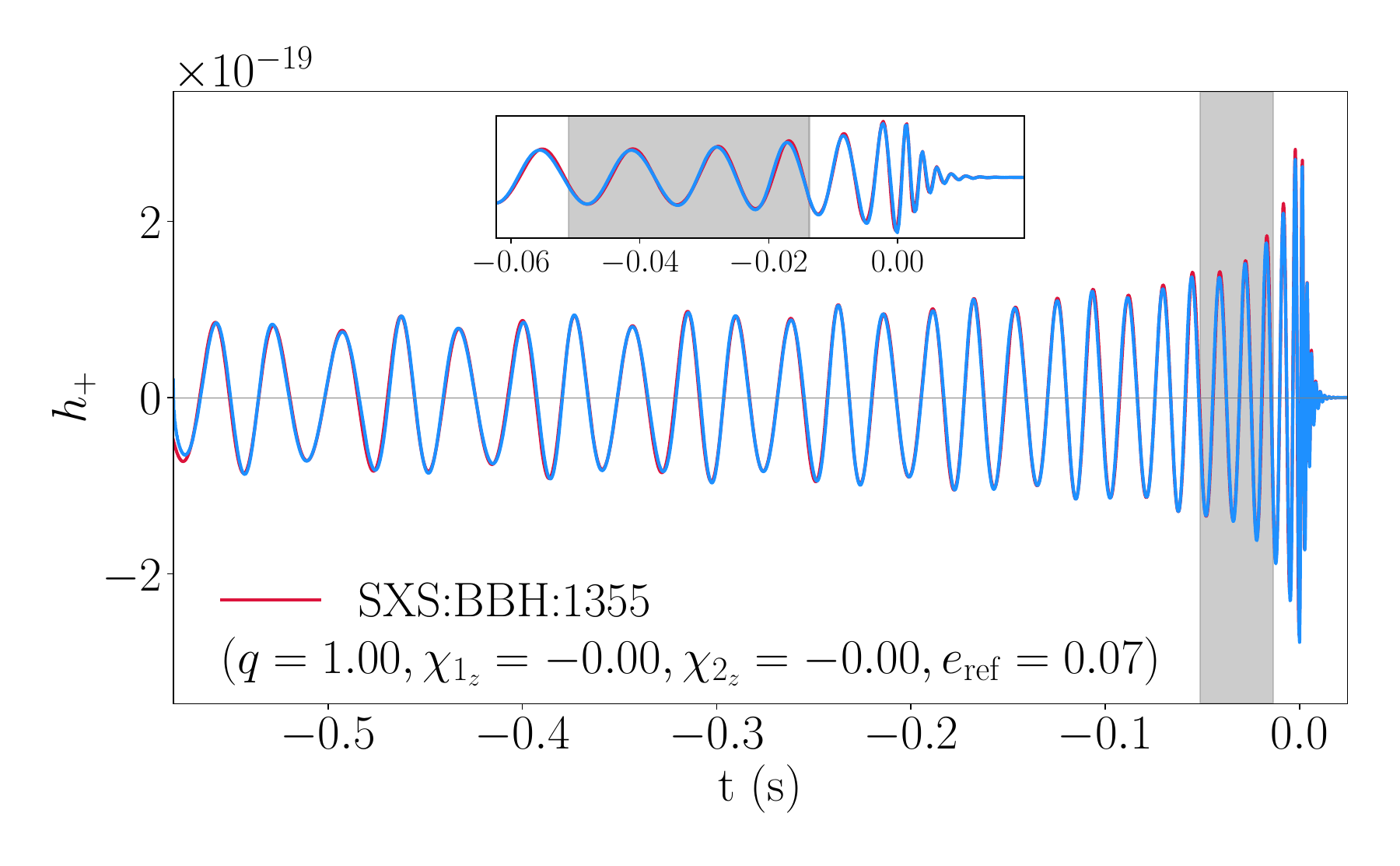}
    \end{subfigure}
    \begin{subfigure}[b]{0.24\textwidth}
        \includegraphics[width=\linewidth]{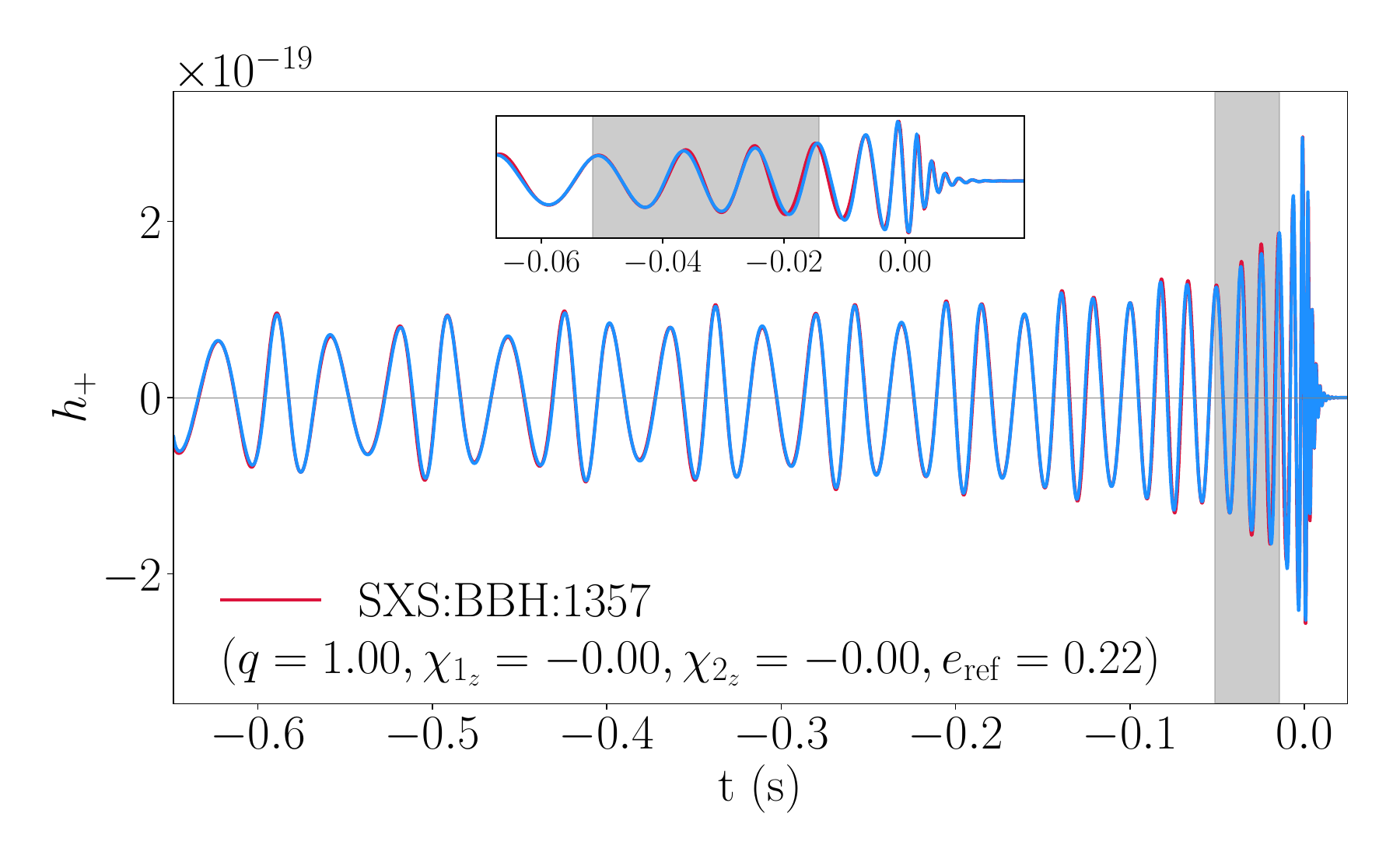}
    \end{subfigure}
      \begin{subfigure}[b]{0.24\textwidth}
        \includegraphics[width=\linewidth]{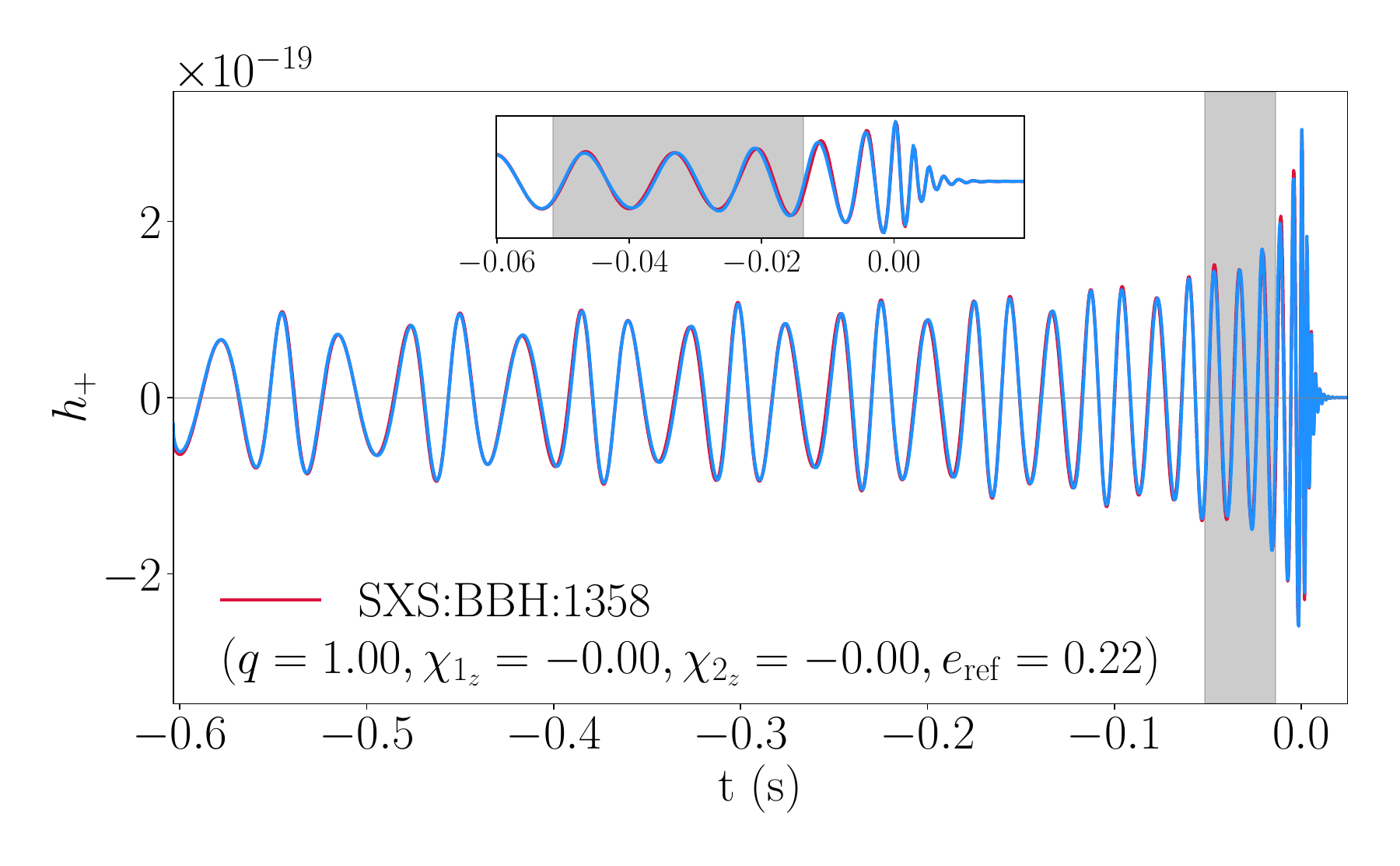}
    \end{subfigure}
    \begin{subfigure}[b]{0.24\textwidth}
        \includegraphics[width=\linewidth]{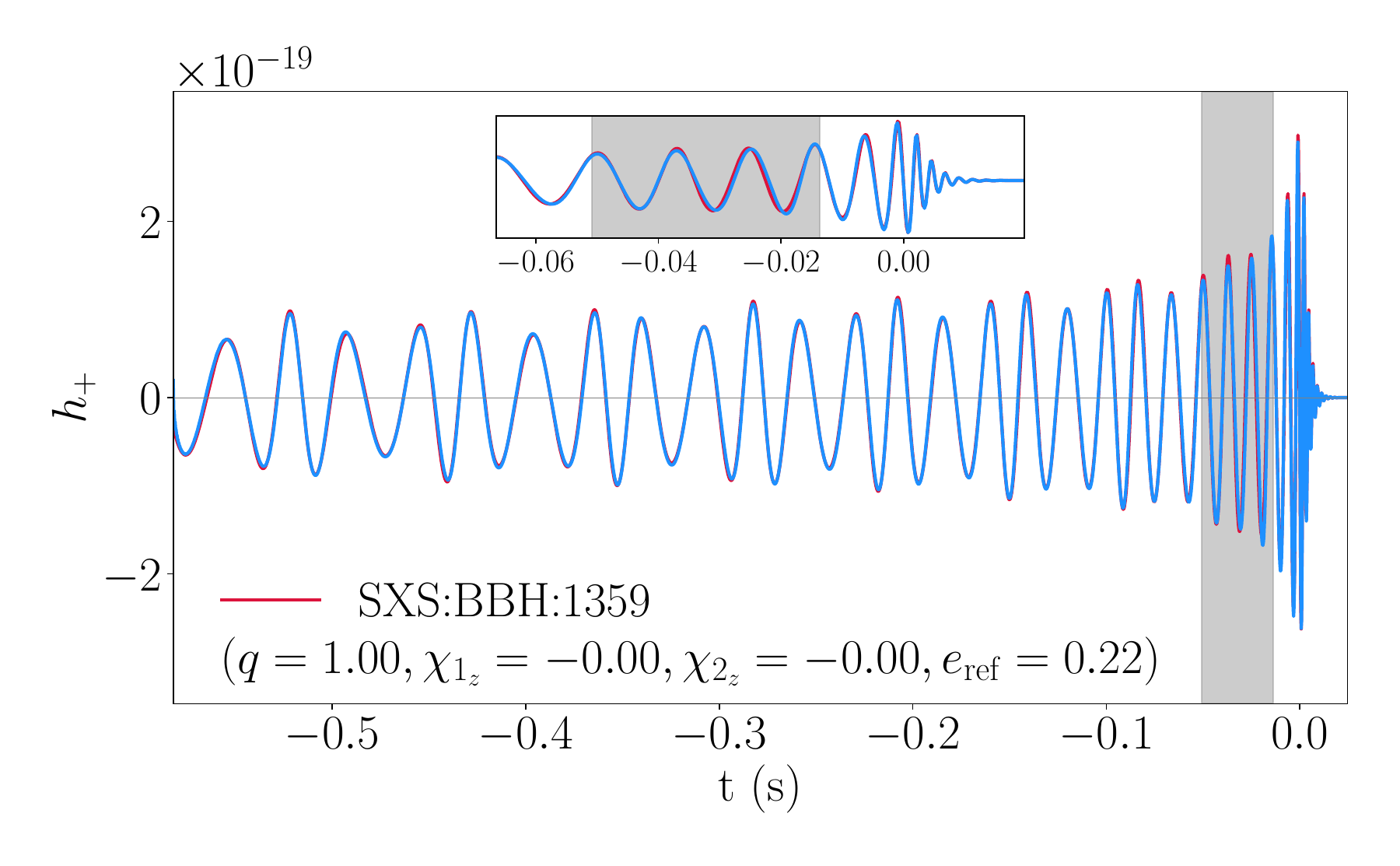}
    \end{subfigure}
    \begin{subfigure}[b]{0.24\textwidth}
        \includegraphics[width=\linewidth]{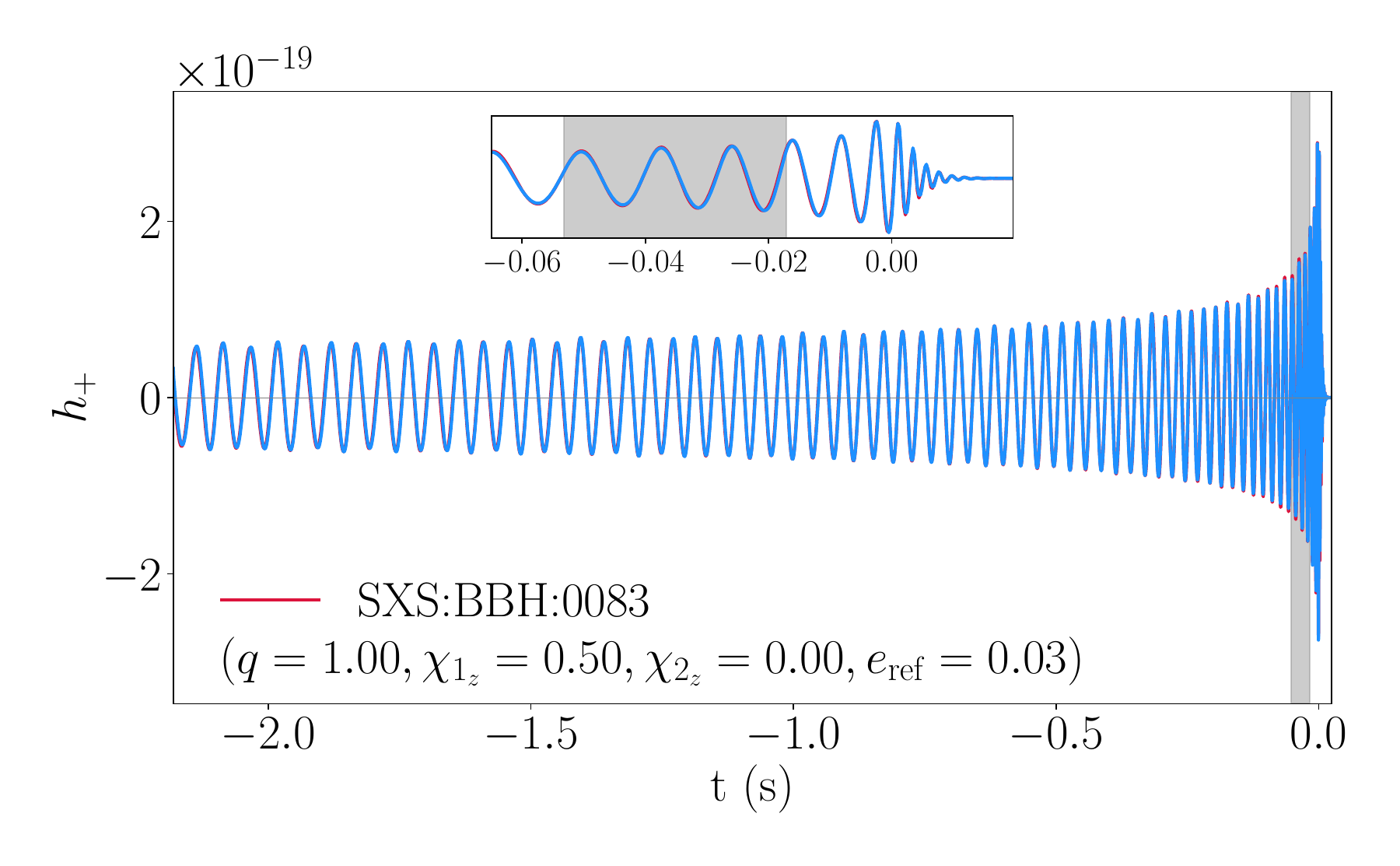}
    \end{subfigure}
      \begin{subfigure}[b]{0.24\textwidth}
        \includegraphics[width=\linewidth]{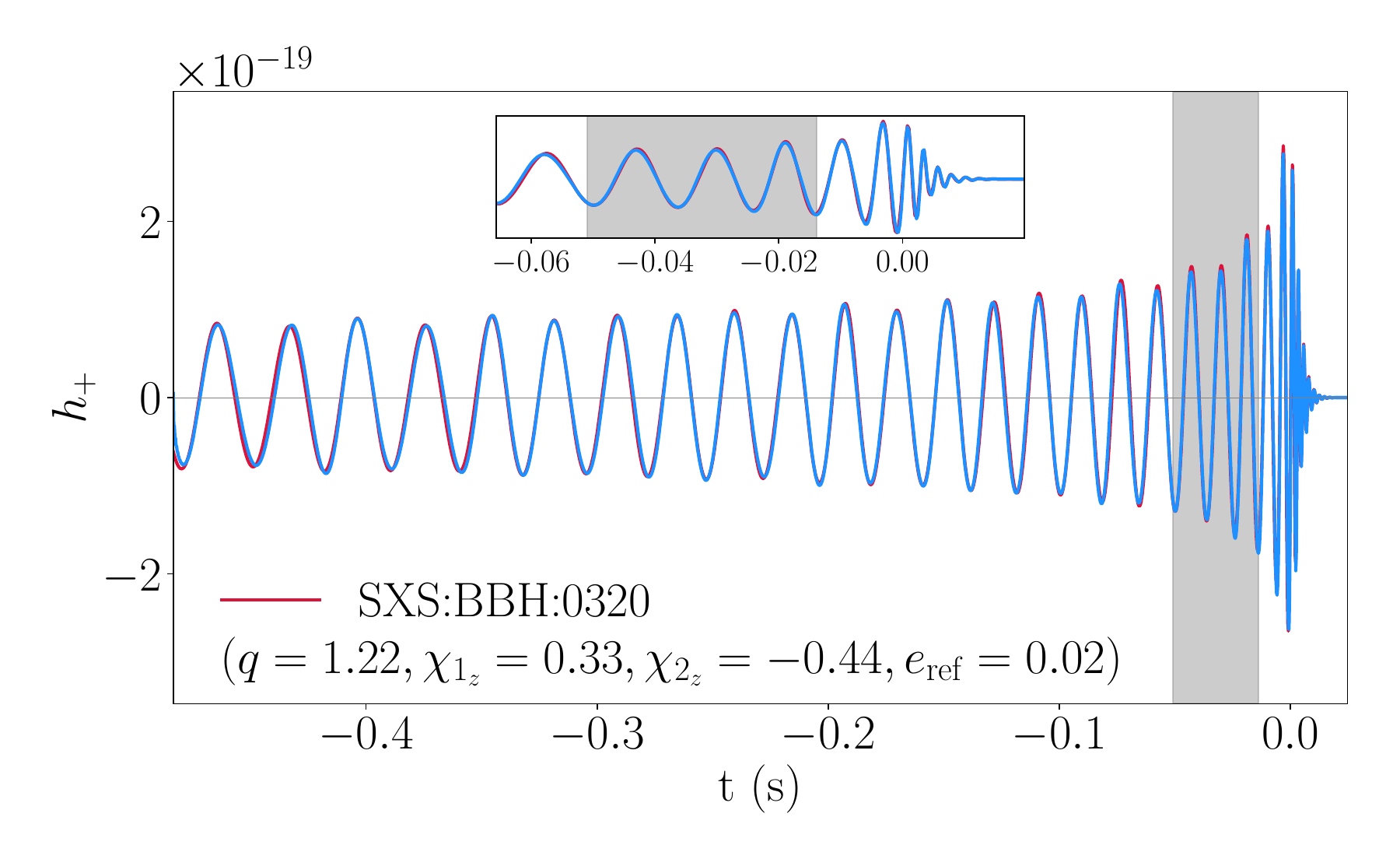}
    \end{subfigure}
    \begin{subfigure}[b]{0.24\textwidth}
        \includegraphics[width=\linewidth]{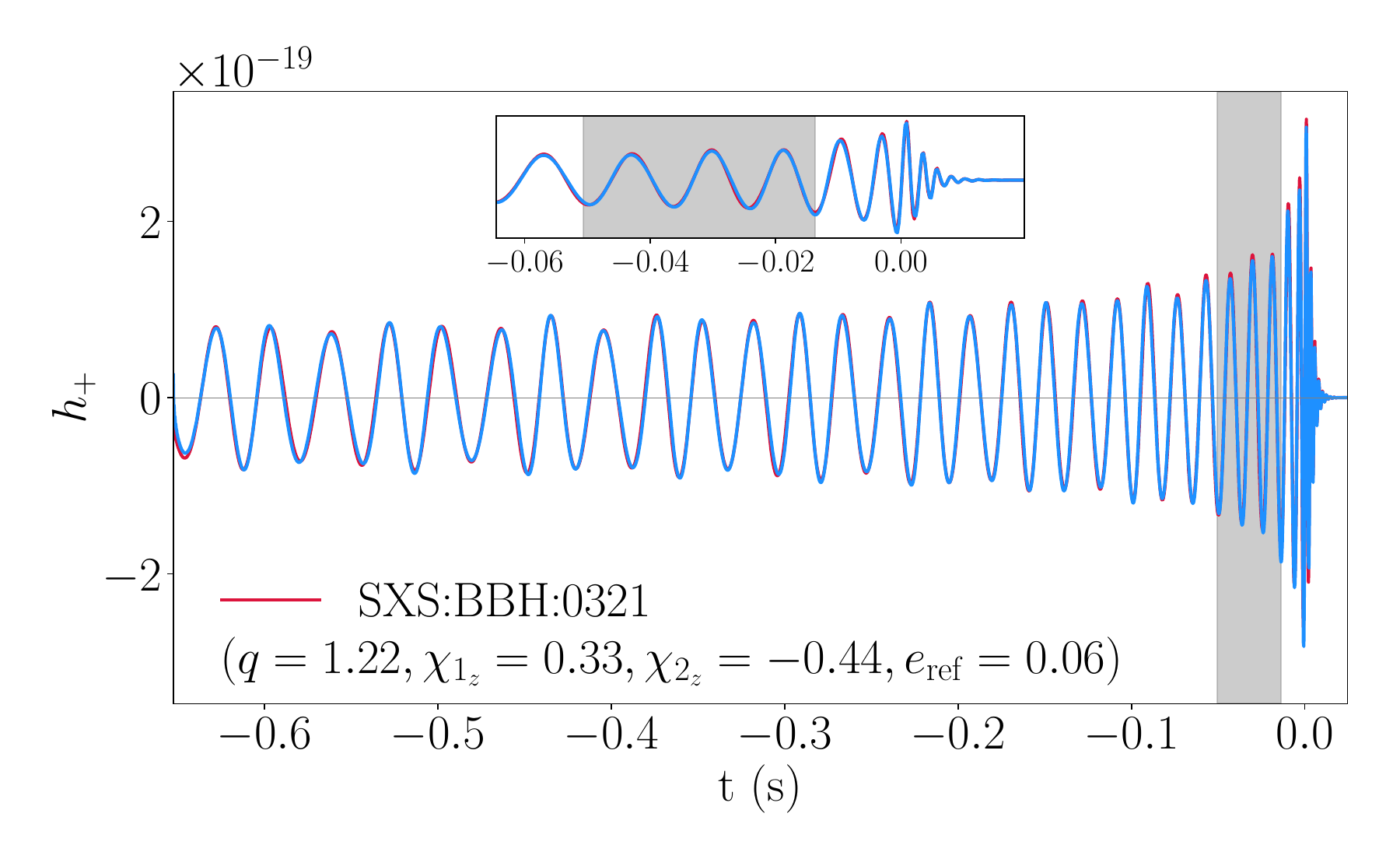}
    \end{subfigure}
    \begin{subfigure}[b]{0.24\textwidth}
        \includegraphics[width=\linewidth]{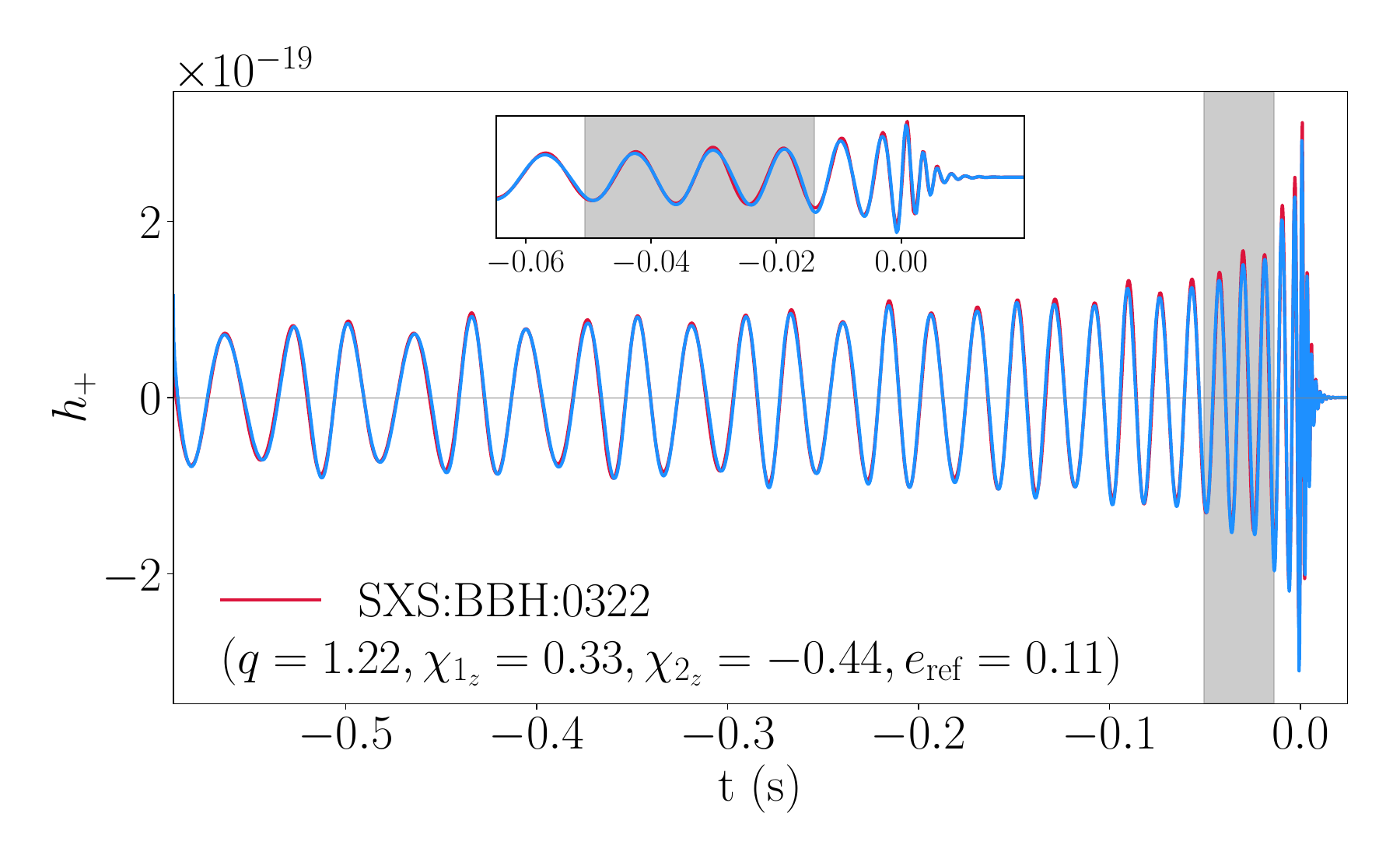}
    \end{subfigure}
    \begin{subfigure}[b]{0.24\textwidth}
        \includegraphics[width=\linewidth]{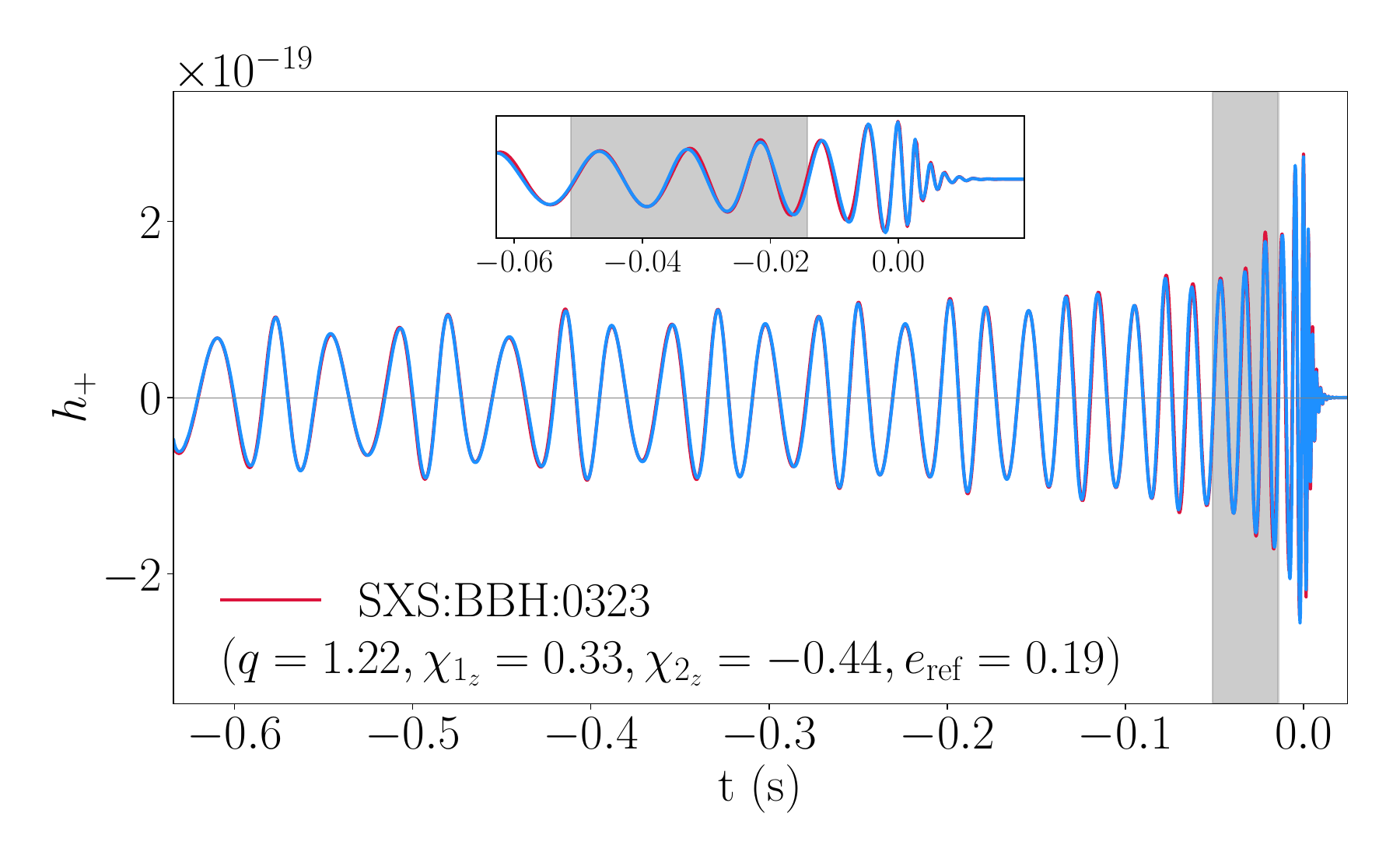}
    \end{subfigure}
      \begin{subfigure}[b]{0.24\textwidth}
        \includegraphics[width=\linewidth]{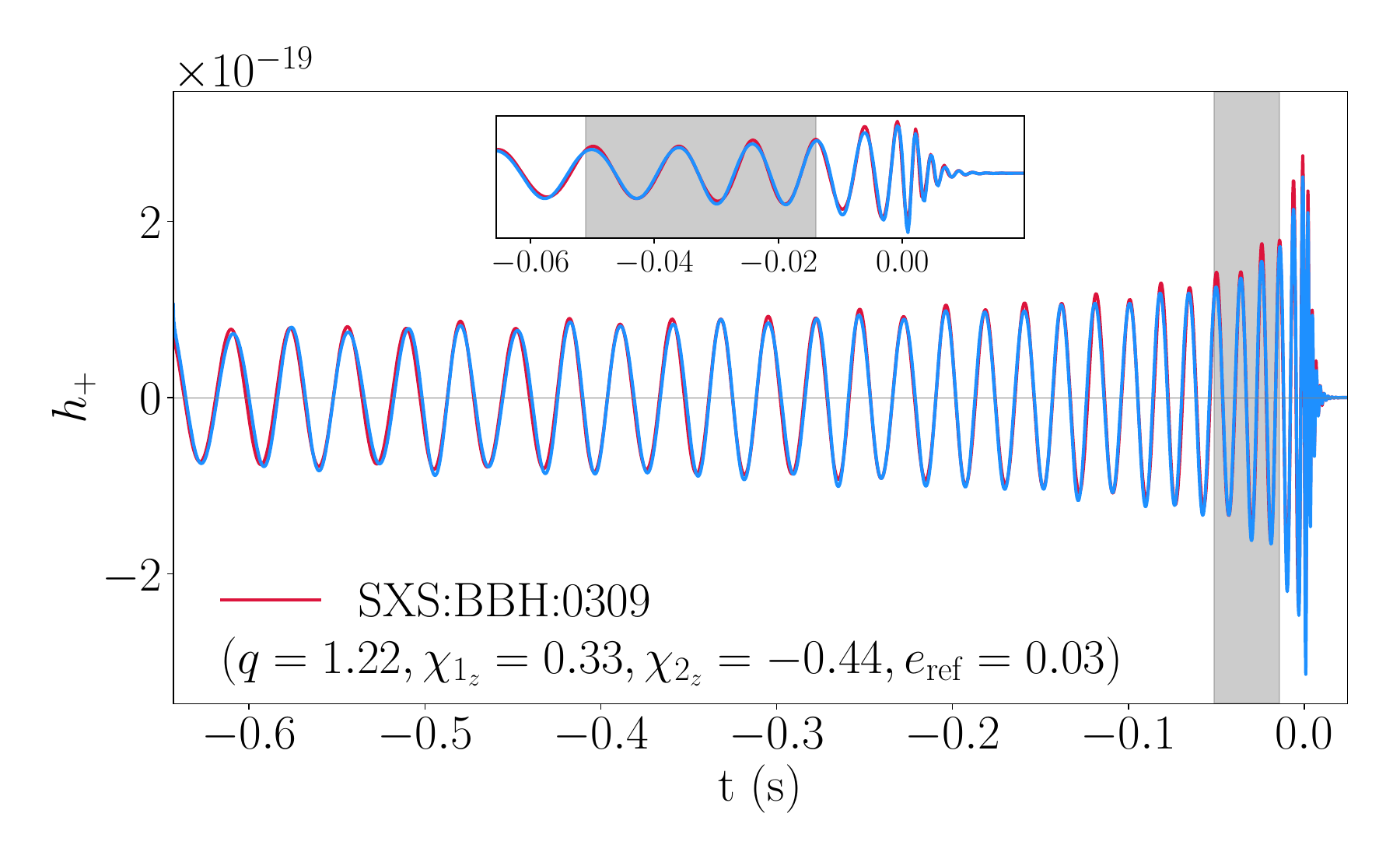}
    \end{subfigure}
      \begin{subfigure}[b]{0.24\textwidth}
        \includegraphics[width=\linewidth]{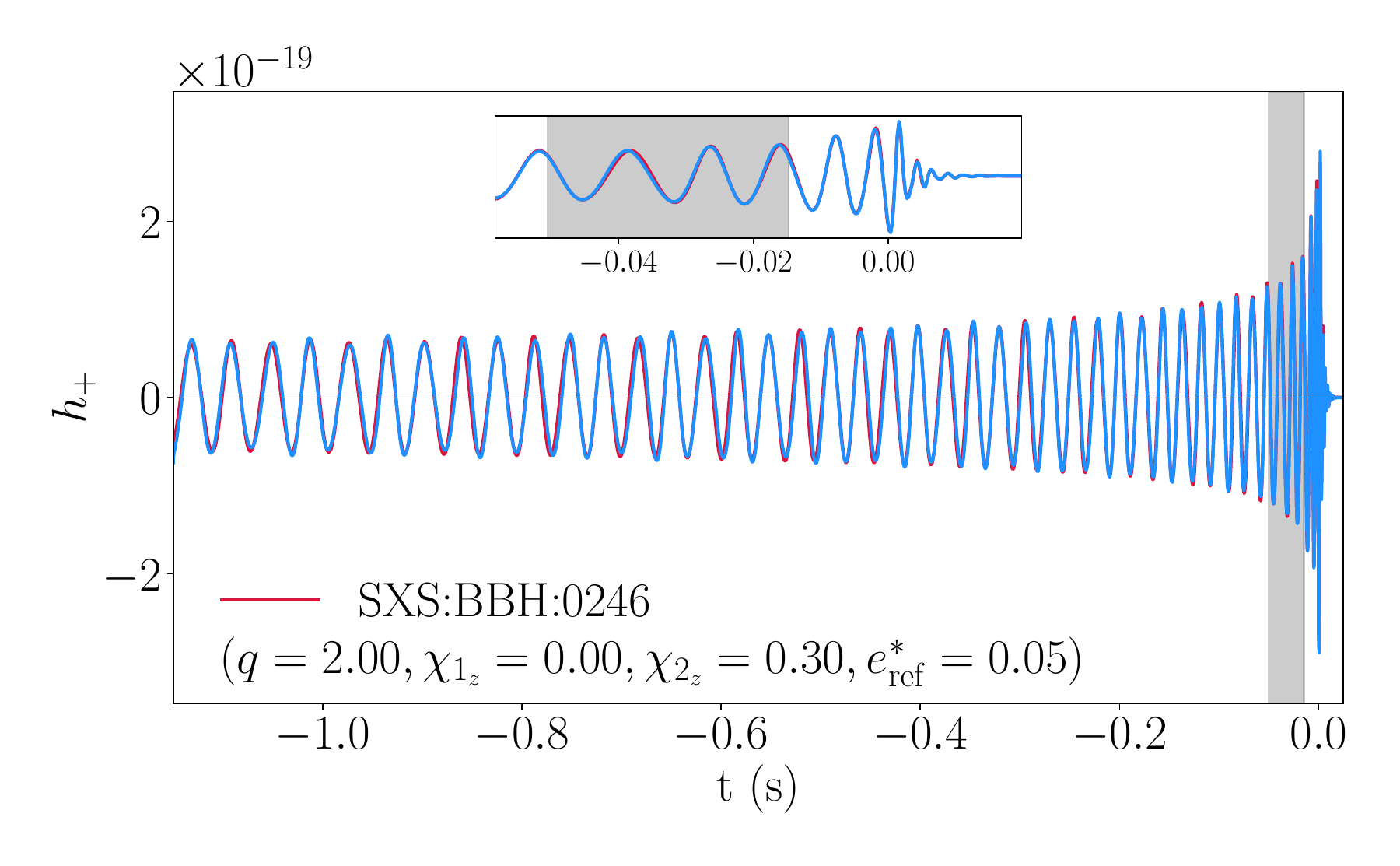}
    \end{subfigure}
    \begin{subfigure}[b]{0.24\textwidth}
        \includegraphics[width=\linewidth]{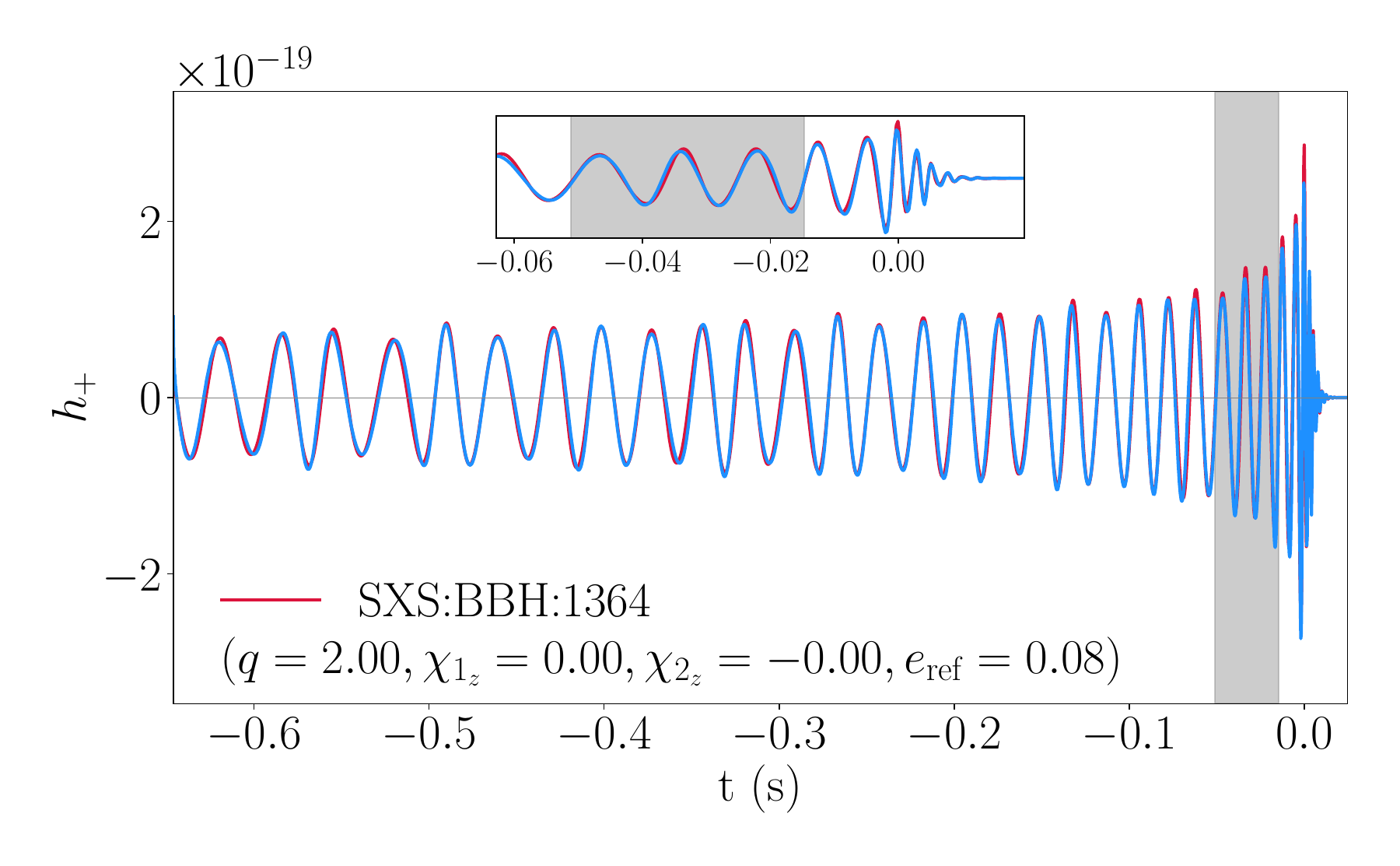}
    \end{subfigure}
    \begin{subfigure}[b]{0.24\textwidth}
        \includegraphics[width=\linewidth]{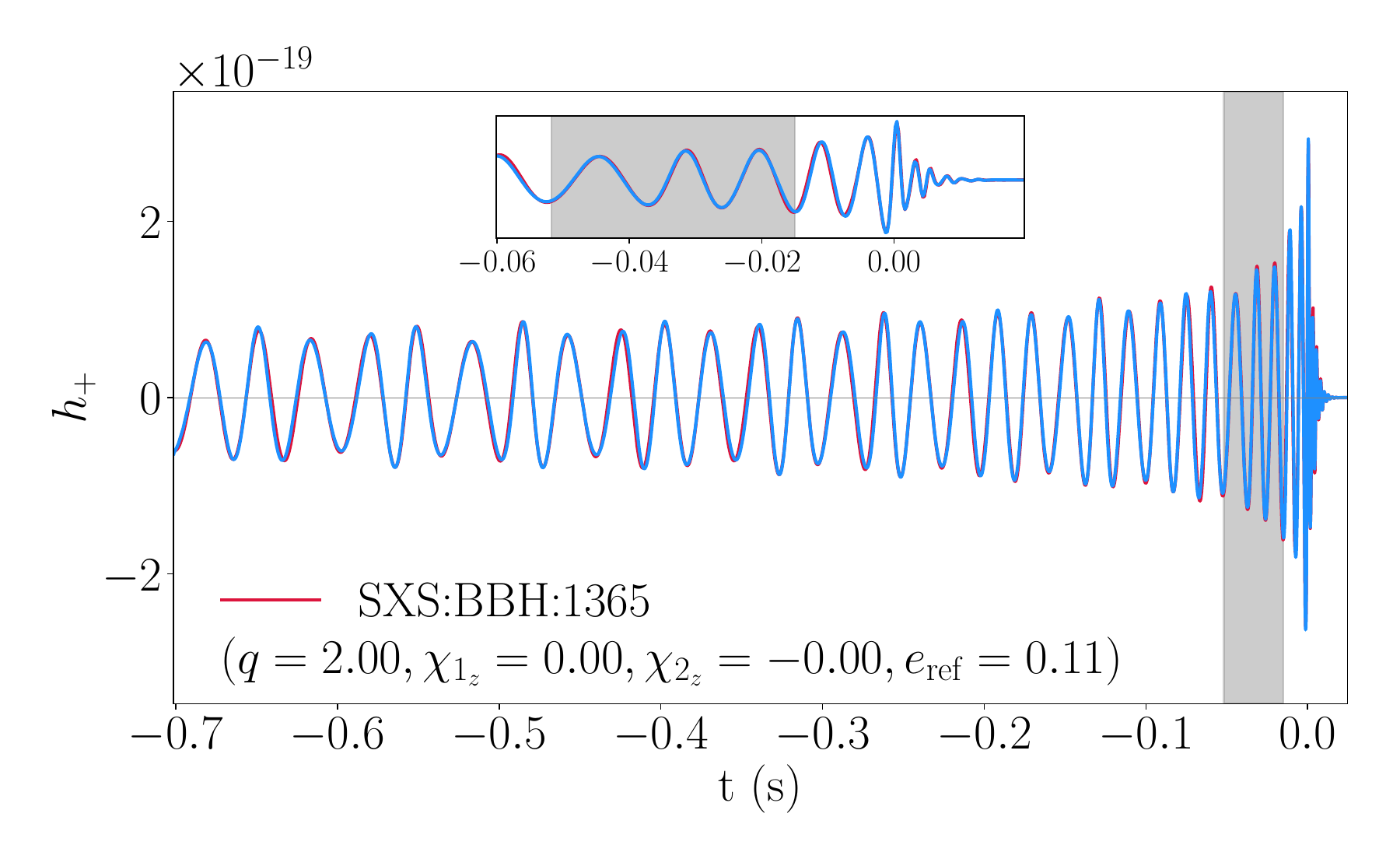}
    \end{subfigure}
    \begin{subfigure}[b]{0.24\textwidth}
        \includegraphics[width=\linewidth]{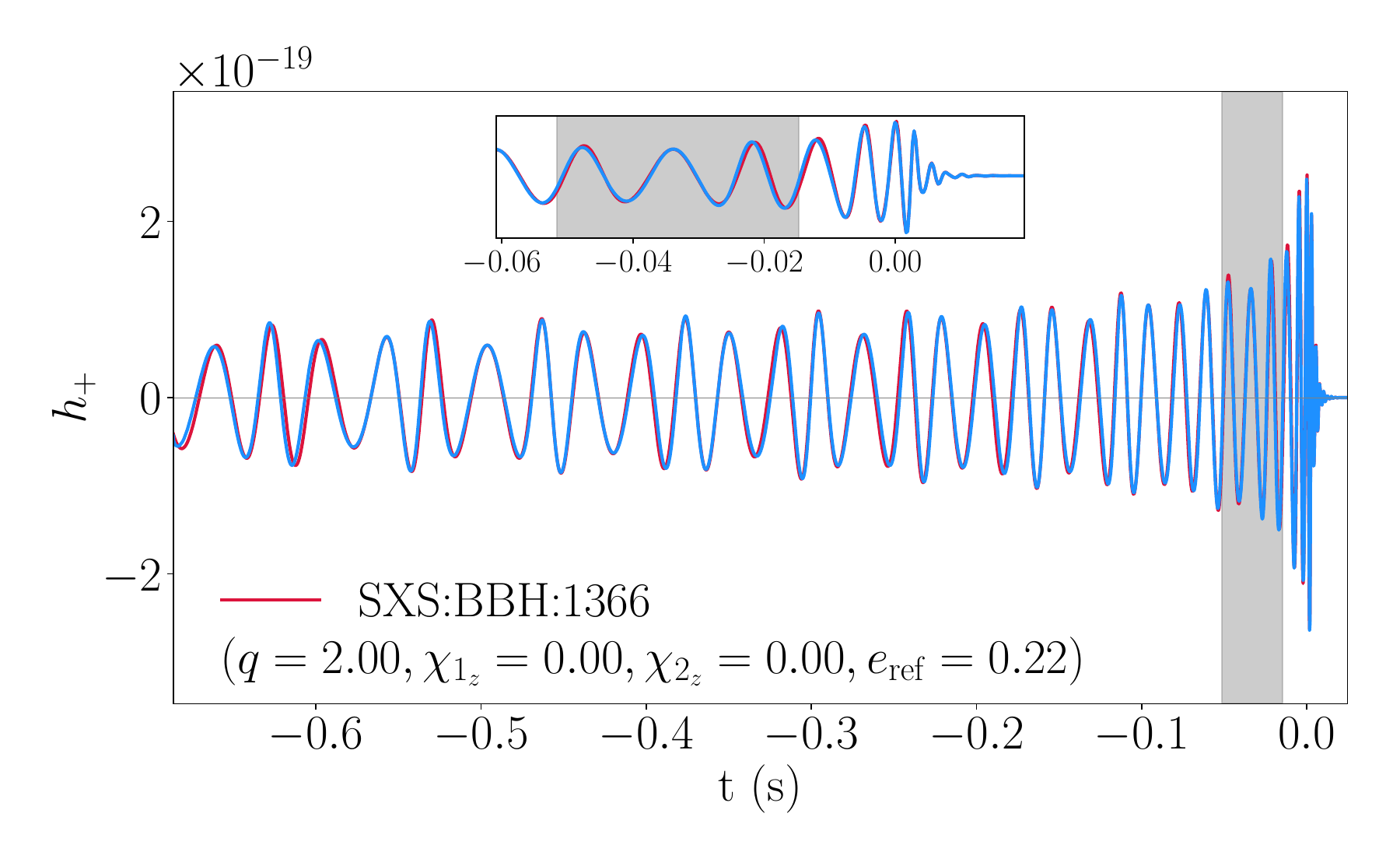}
    \end{subfigure}
    \begin{subfigure}[b]{0.24\textwidth}
        \includegraphics[width=\linewidth]{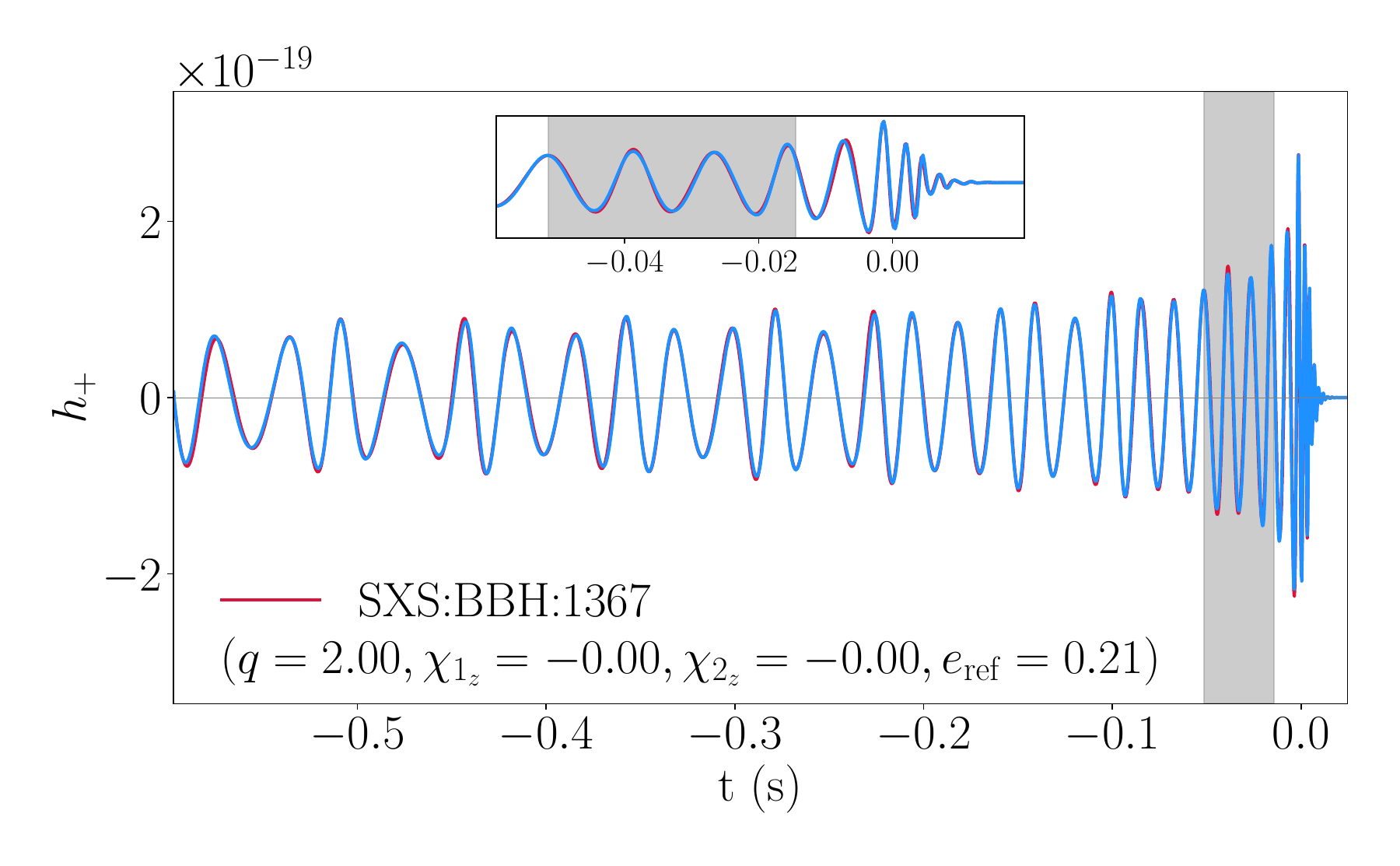}
    \end{subfigure}
     \begin{subfigure}[b]{0.24\textwidth}
        \includegraphics[width=\linewidth]{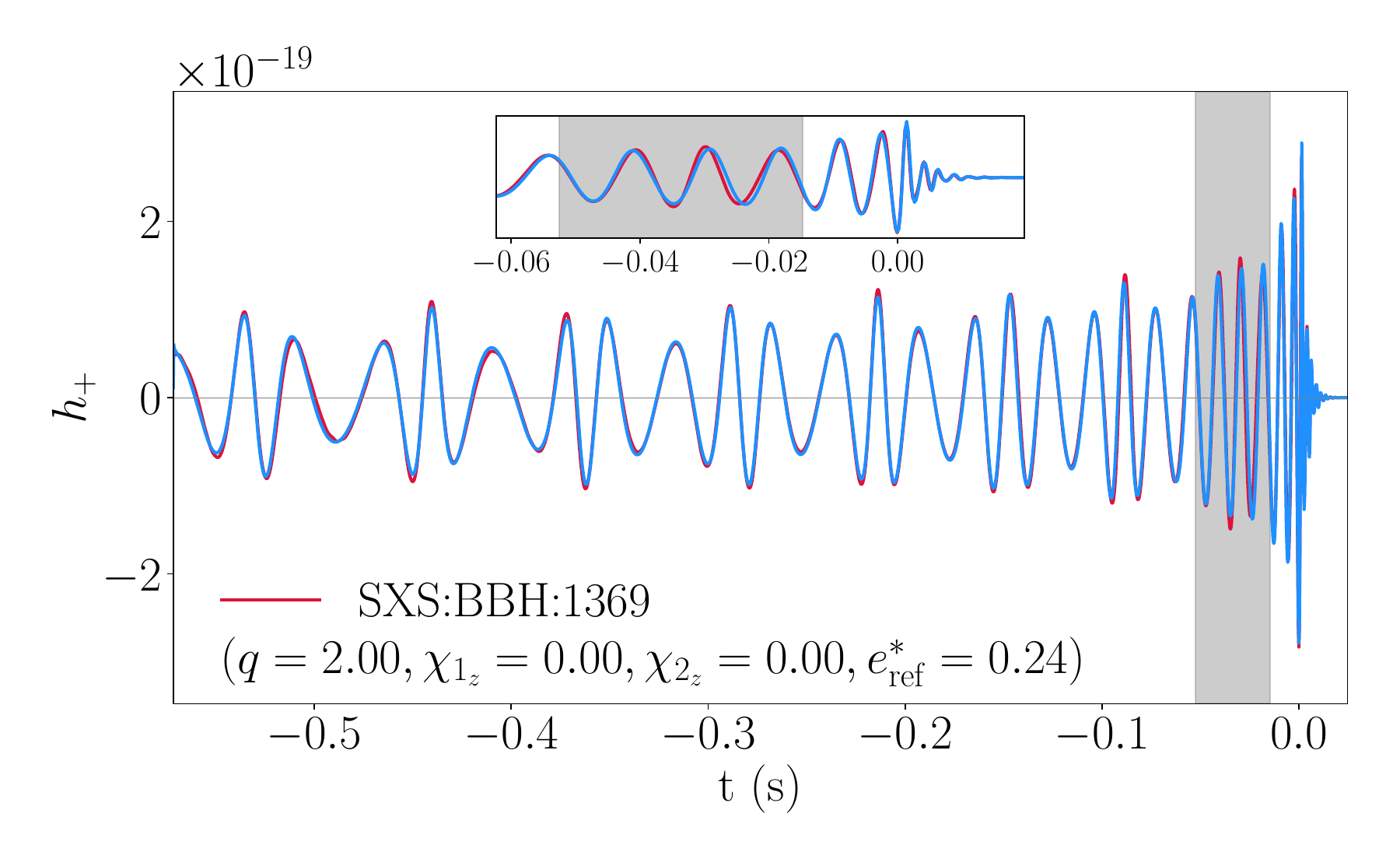}
    \end{subfigure}
     \begin{subfigure}[b]{0.24\textwidth}
        \includegraphics[width=\linewidth]{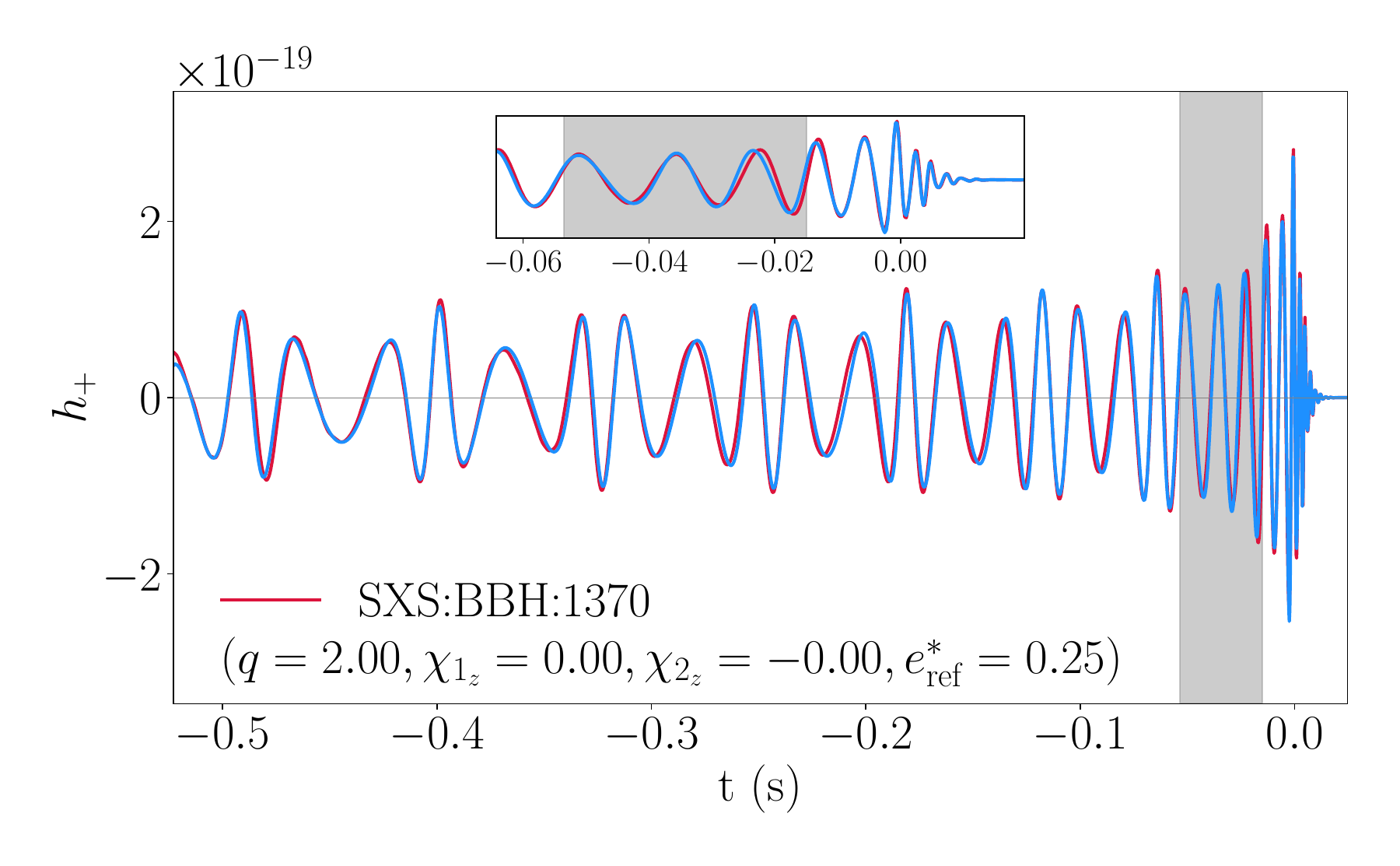}
    \end{subfigure}
    \begin{subfigure}[b]{0.24\textwidth}
        \includegraphics[width=\linewidth]{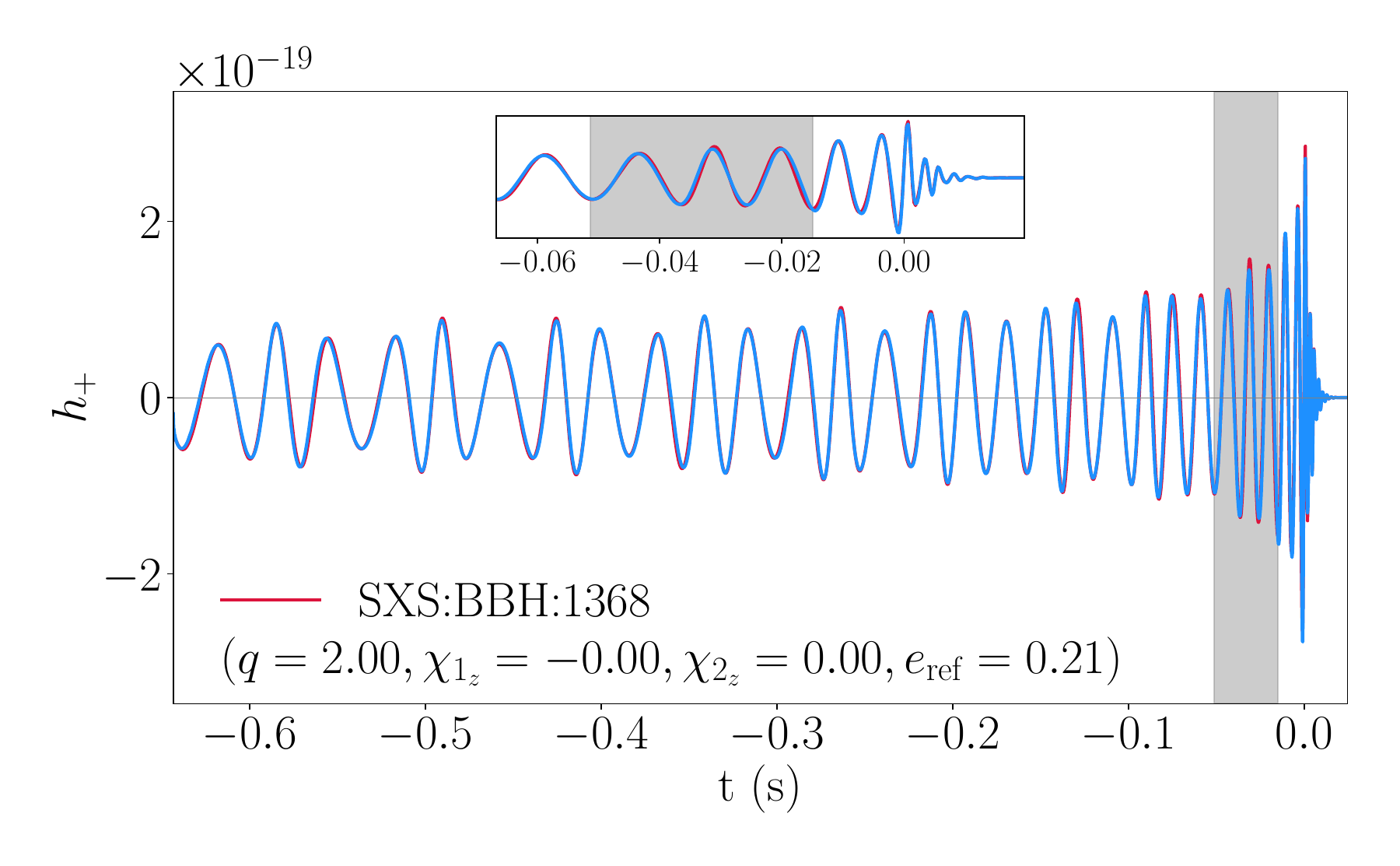}
    \end{subfigure}
      \begin{subfigure}[b]{0.24\textwidth}
        \includegraphics[width=\linewidth]{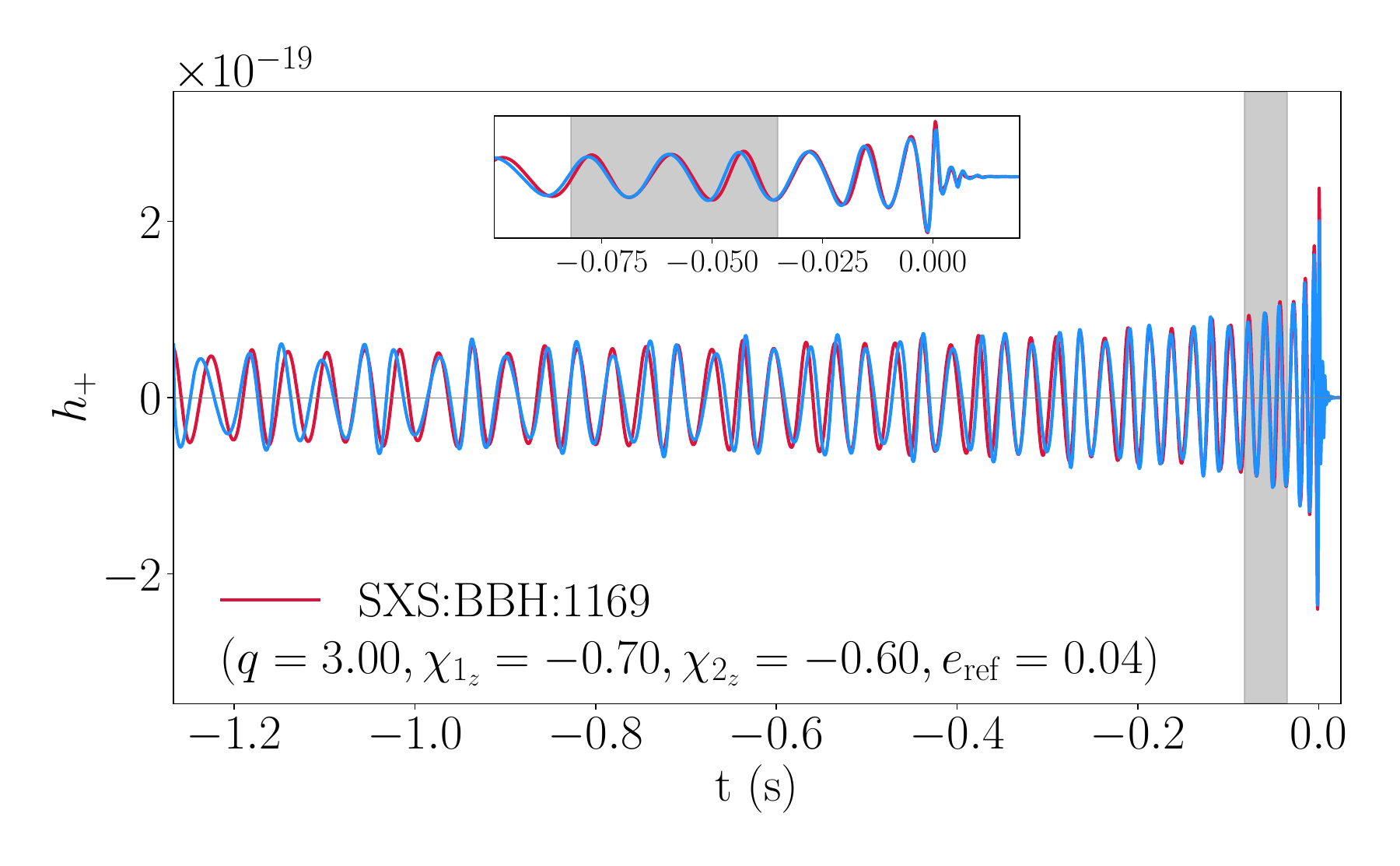}
    \end{subfigure}
    \begin{subfigure}[b]{0.24\textwidth}
        \includegraphics[width=\linewidth]{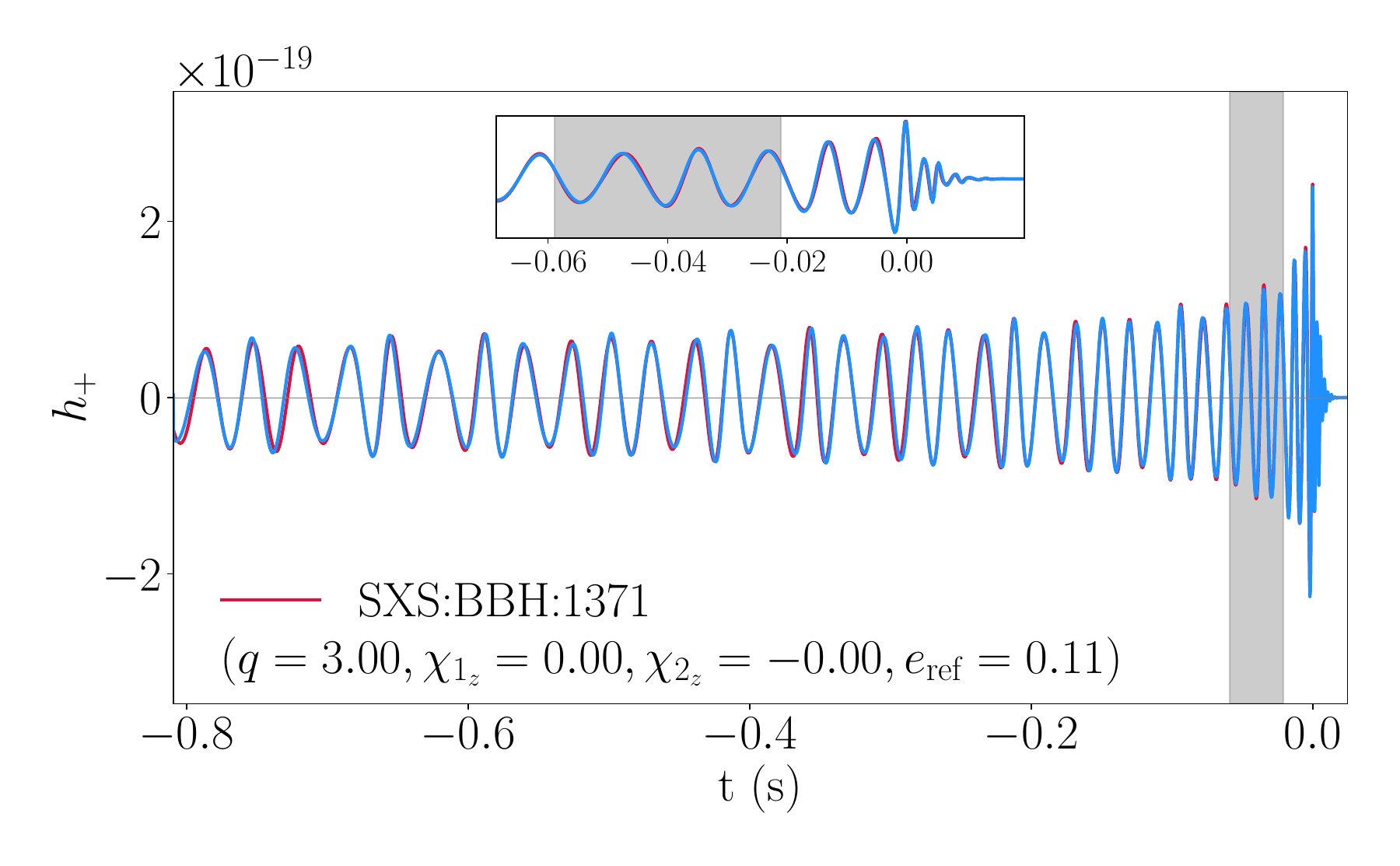}
    \end{subfigure}
    \begin{subfigure}[b]{0.24\textwidth}
        \includegraphics[width=\linewidth]{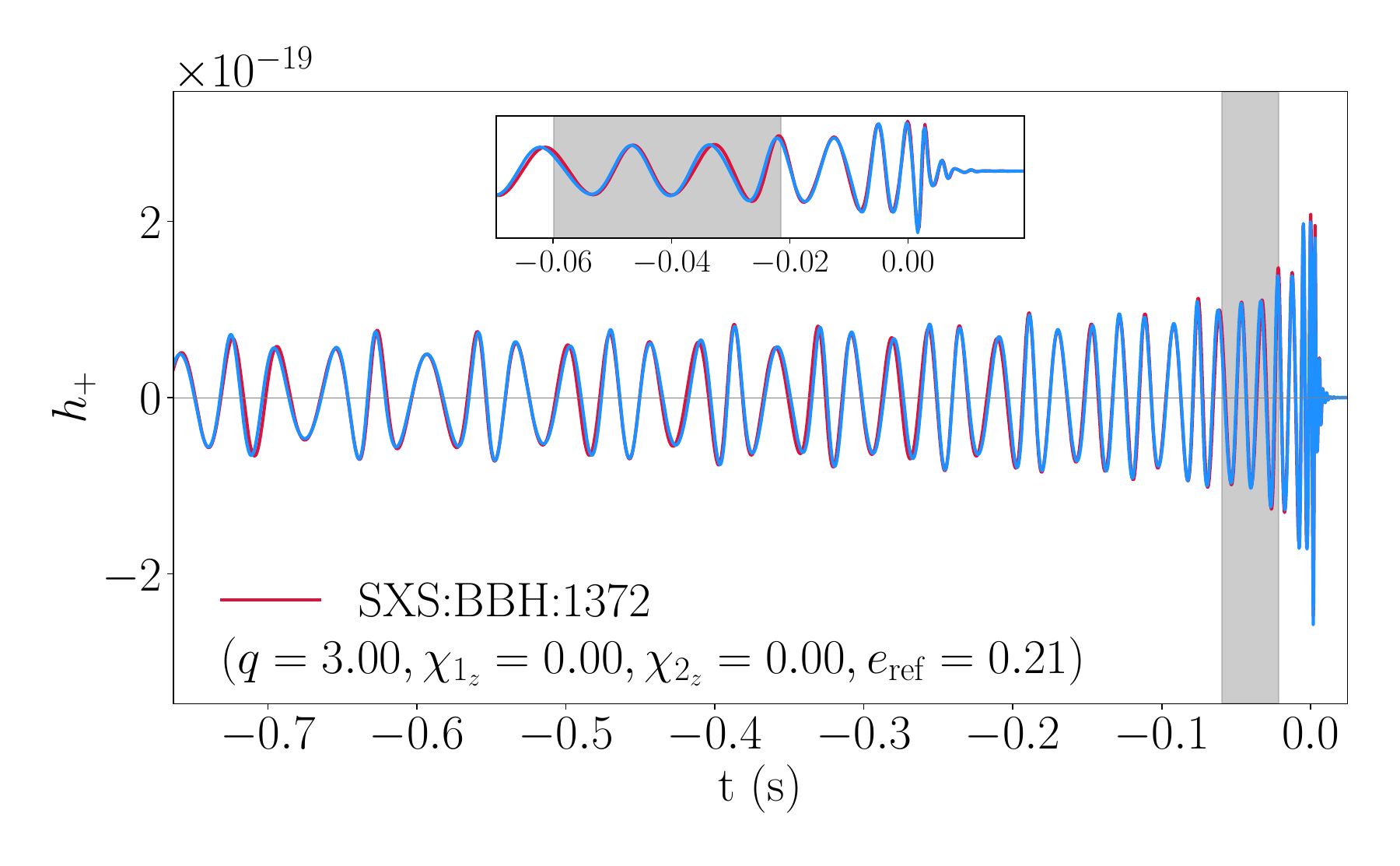}
    \end{subfigure}
    \begin{subfigure}[b]{0.24\textwidth}
        \includegraphics[width=\linewidth]{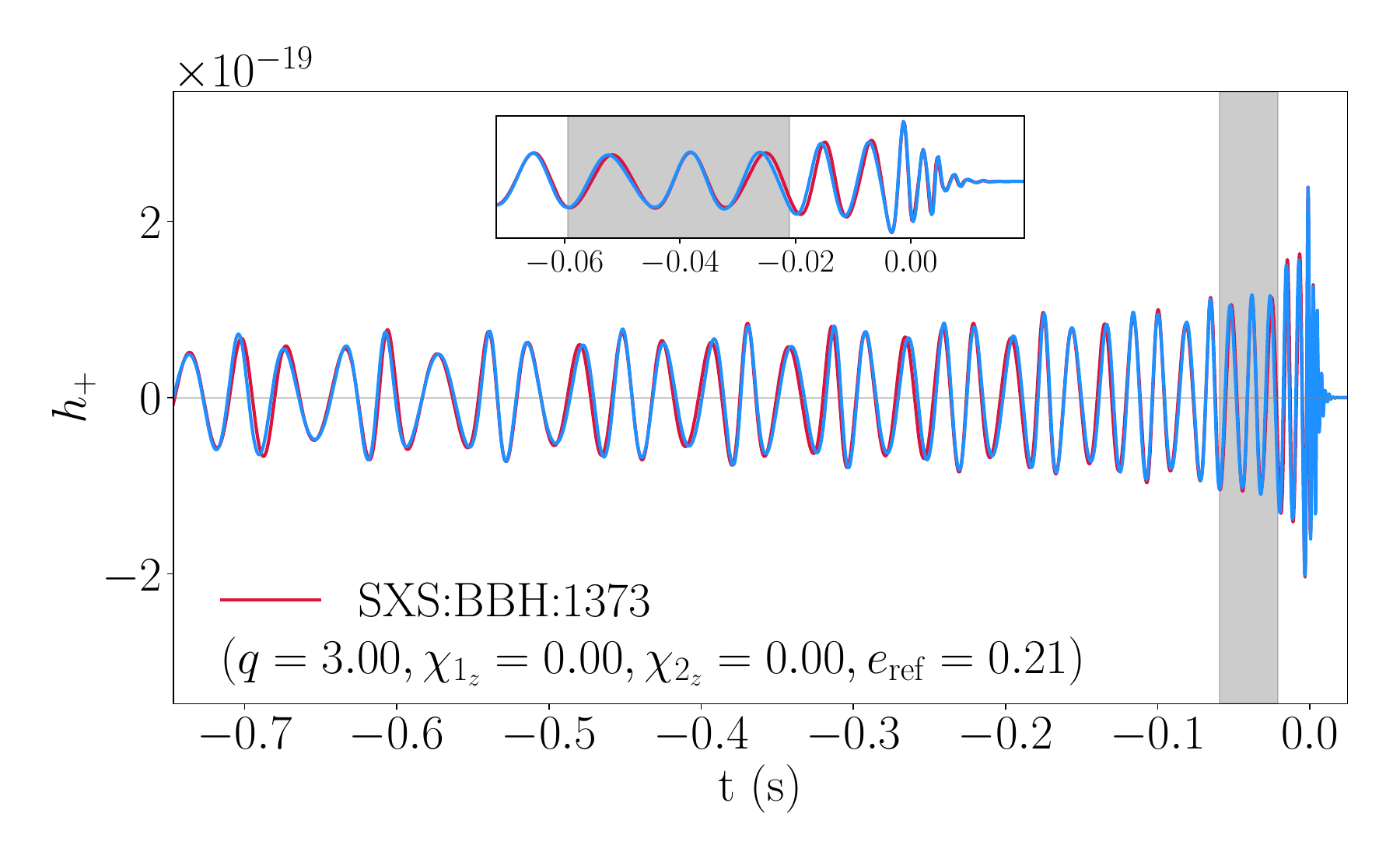}
    \end{subfigure}
    \begin{subfigure}[b]{0.24\textwidth}
        \includegraphics[width=\linewidth]{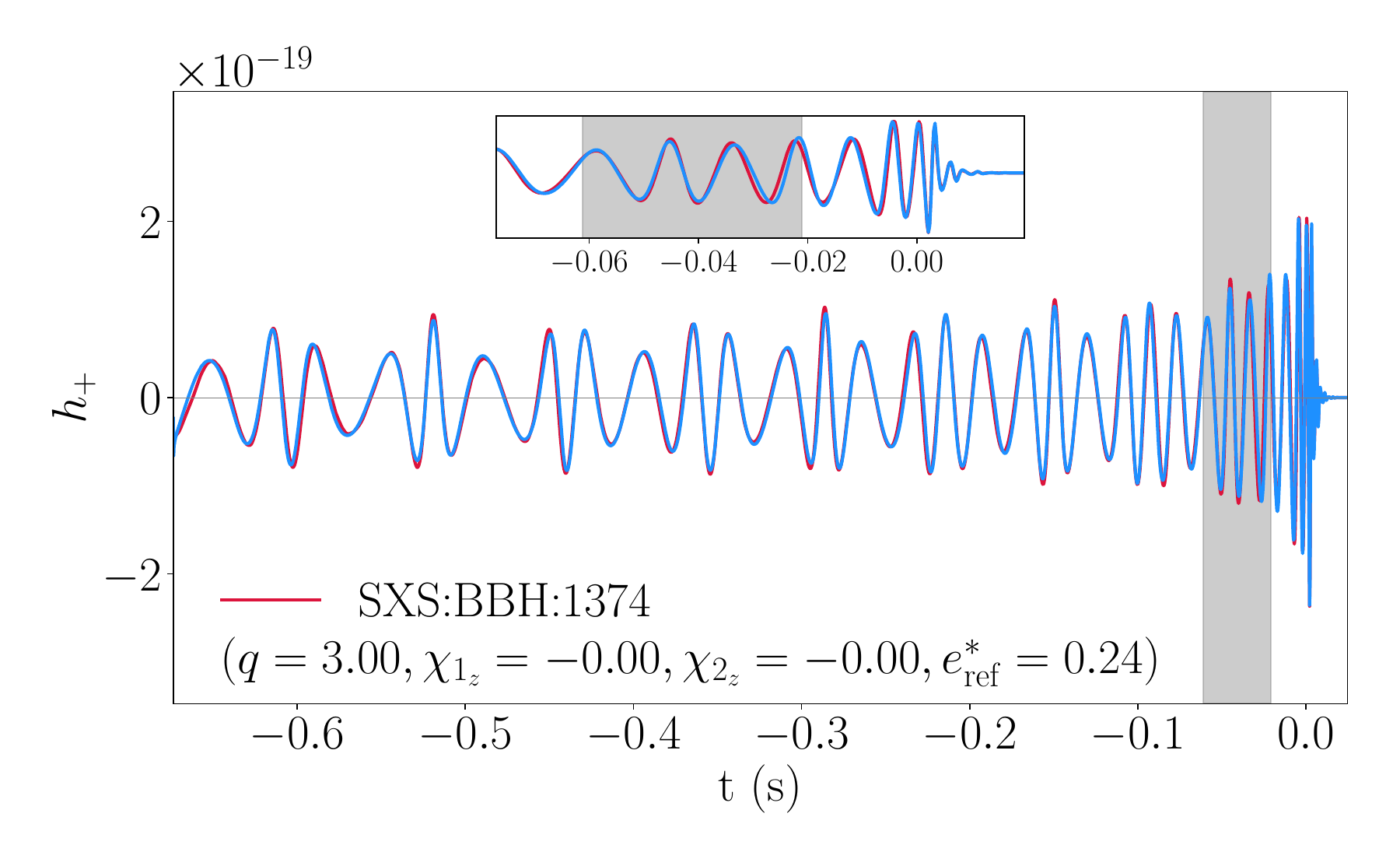}
    \end{subfigure}
      \begin{subfigure}[b]{0.24\textwidth}
        \includegraphics[width=\linewidth]{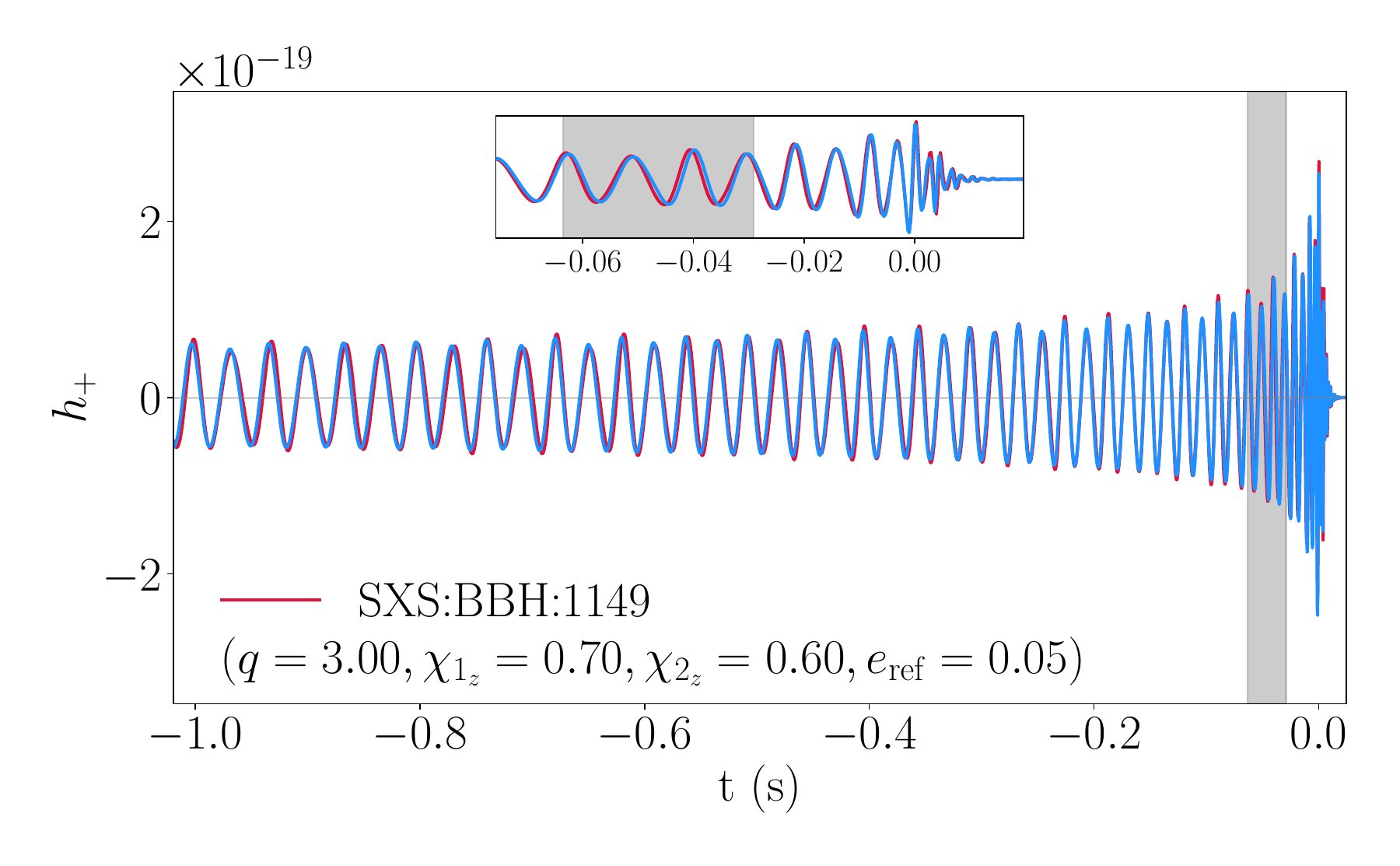}
    \end{subfigure}
    \begin{subfigure}[b]{0.24\textwidth}
        \includegraphics[width=\linewidth]{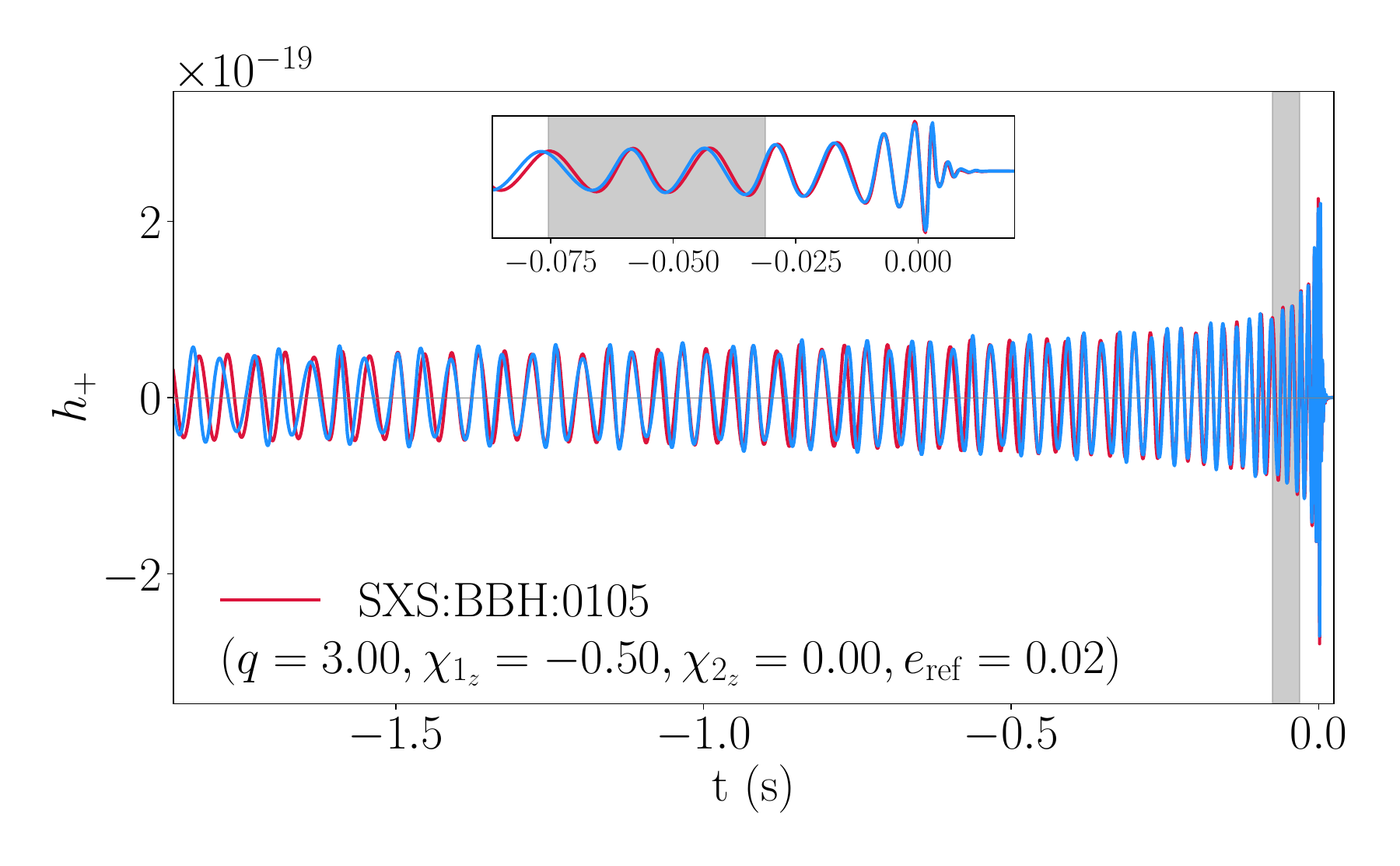}
    \end{subfigure}
    \begin{subfigure}[b]{0.24\textwidth}
        \includegraphics[width=\linewidth]{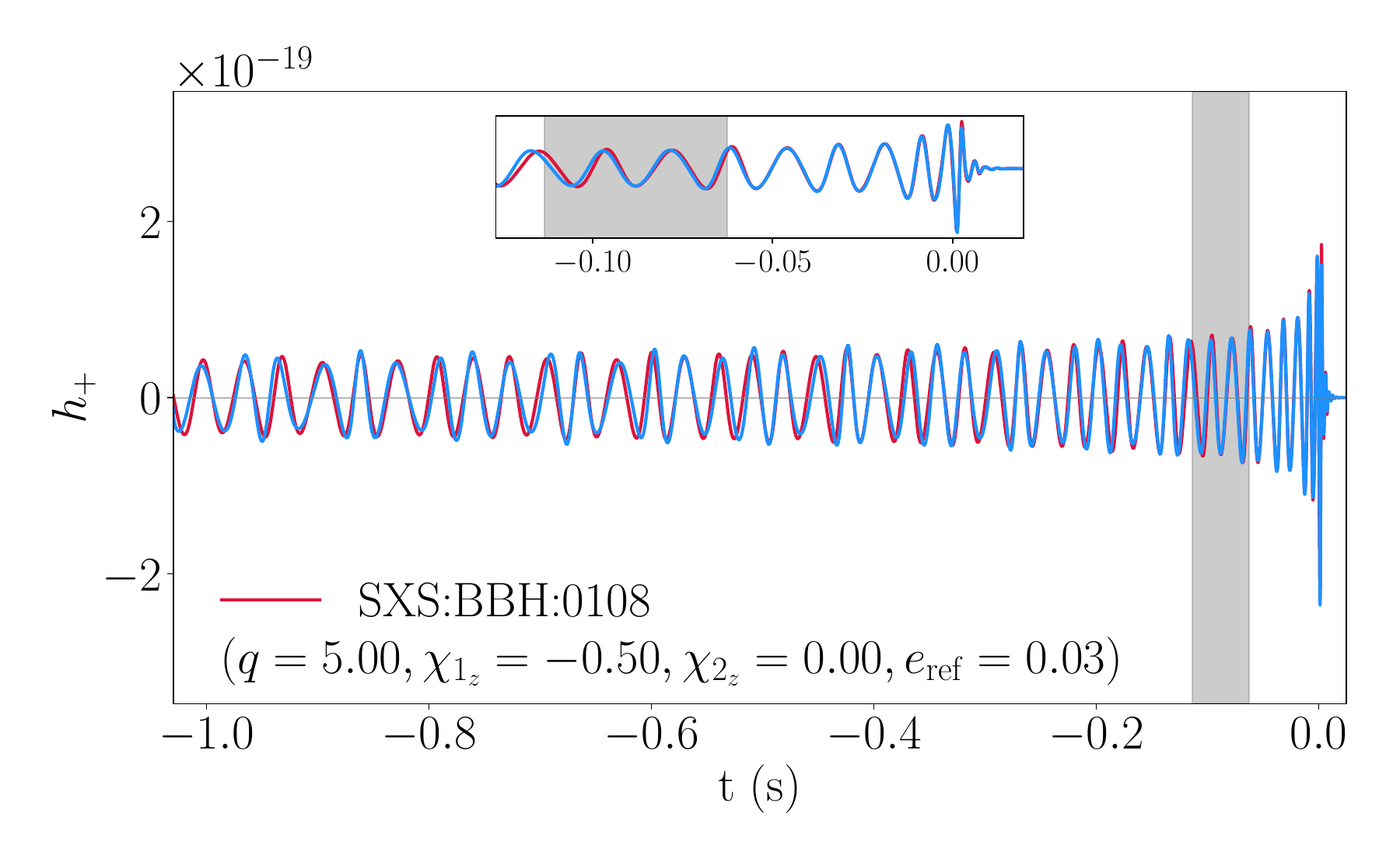}
    \end{subfigure}
    \caption{Comparison of strain data $(h_{+})$ of \imresigma{} with spinning, eccentric NR simulations taken from SXS catalog and listed in Table~\ref{table:sxs_id}. The comparison is done by including $(\ell, |m|)=(2,1), (3,3), (3,2), (4,4), (4,3)$ in addition to $(2,2)$ mode and using a total mass of $40M_{\odot}$. The inset shows a zoomed in view of the late inspiral and PMR portion. {\violet The gray shaded region denotes the time interval over which the \imresigma{} transitions from its inspiral prescription to the plunge-merger-ringdown prescription.} The mass ratio $(q)$, $z$-components of dimensionless spins $(\chi_{1_z}, \chi_{2_z})$, and reference eccentricity $(e_{\rm{ref}})$ are given on the figure corresponding to each simulation. Simulations for which the eccentricity is not well measured, we quote the optimized eccentricity for those cases (denoted as $e^{\ast}_{\rm{ref}})$.}
    \label{fig:NR_comp}
\end{figure*}

\begin{figure*}
    \centering
    \includegraphics[width=0.49\linewidth]{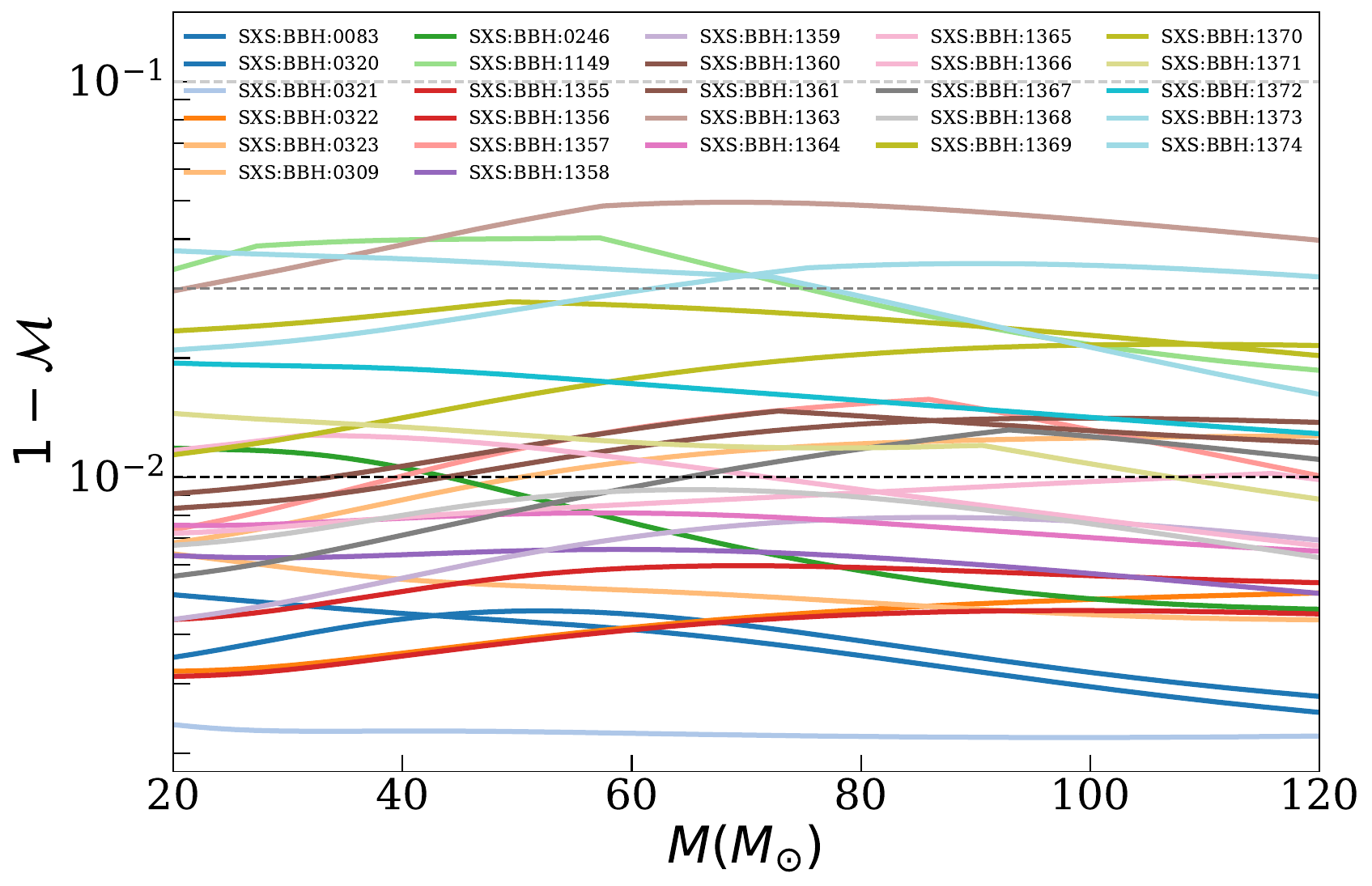}
        \includegraphics[width=0.49\linewidth]{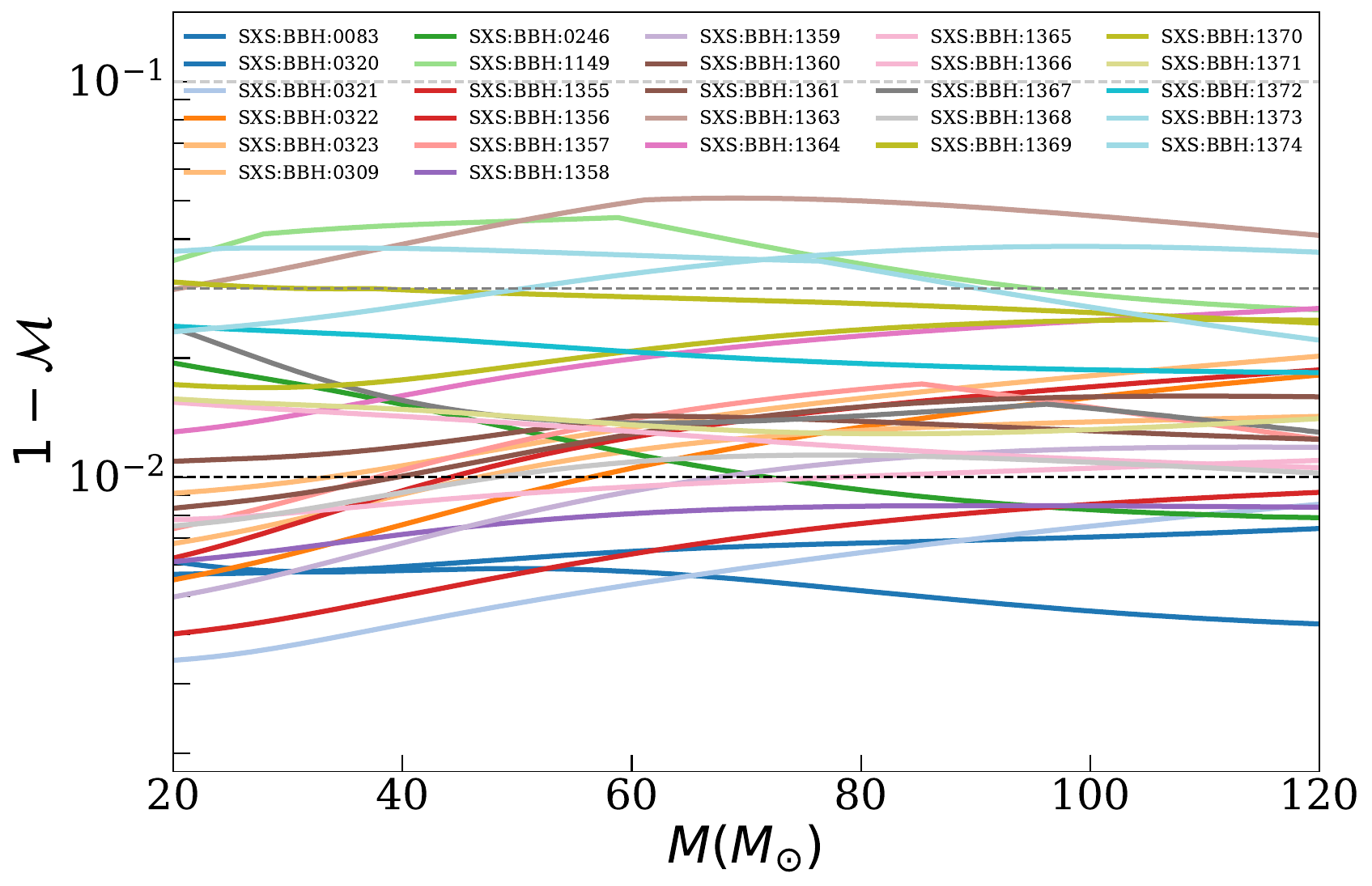}
        \caption{{{Mismatch between \imresigma{} and non-spinning, positive aligned-spin eccentric NR simulations (see Table~\ref{table:sxs_id} for details) are shown as a function of total mass with only $(2, \pm 2)$ (\textit{left}) and including $(2, \pm 1)$, $(3, \pm 3)$, $(3, \pm 2)$, $(4, \pm 4)$, $(4, \pm 3)$ modes in addition to the dominant $\ell = |m| = 2$ modes (\textit{right}). These simulations are taken from the SXS catalog~\citep{Boyle:2019kee}. Mismatch computation is performed using the zero-detuning high-power design noise curve for the LIGO detectors, with a lower cutoff frequency of $10$Hz. The horizontal dashed lines denote the 10\% (light grey), 3\% (grey), and 1\% (black) mismatch marks, respectively.}}}
        \label{fig:mismatch_NR_mass}
\end{figure*}

\begin{figure*}
\includegraphics[width=0.49\linewidth]{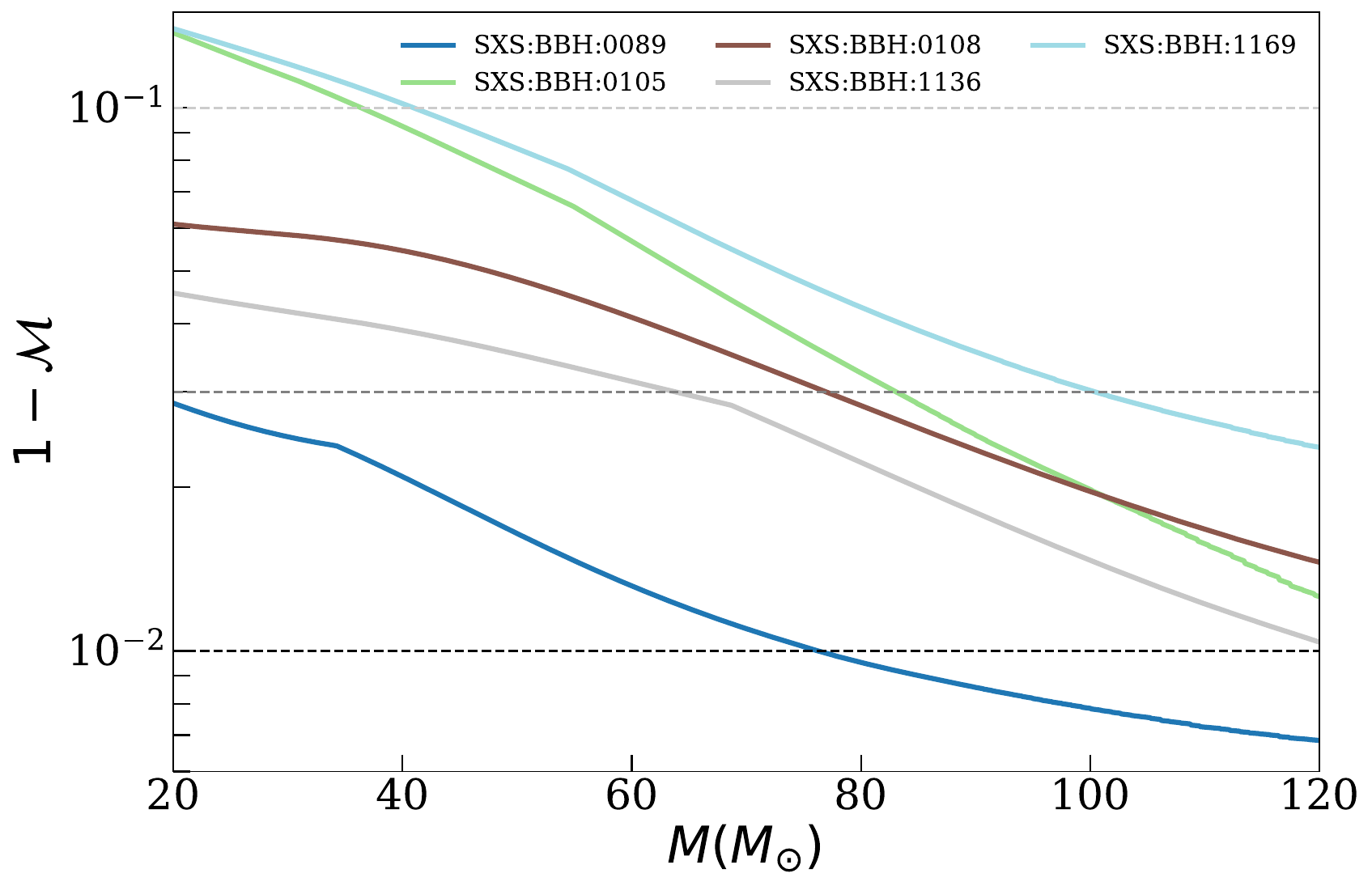}
    \includegraphics[width=0.50\linewidth]{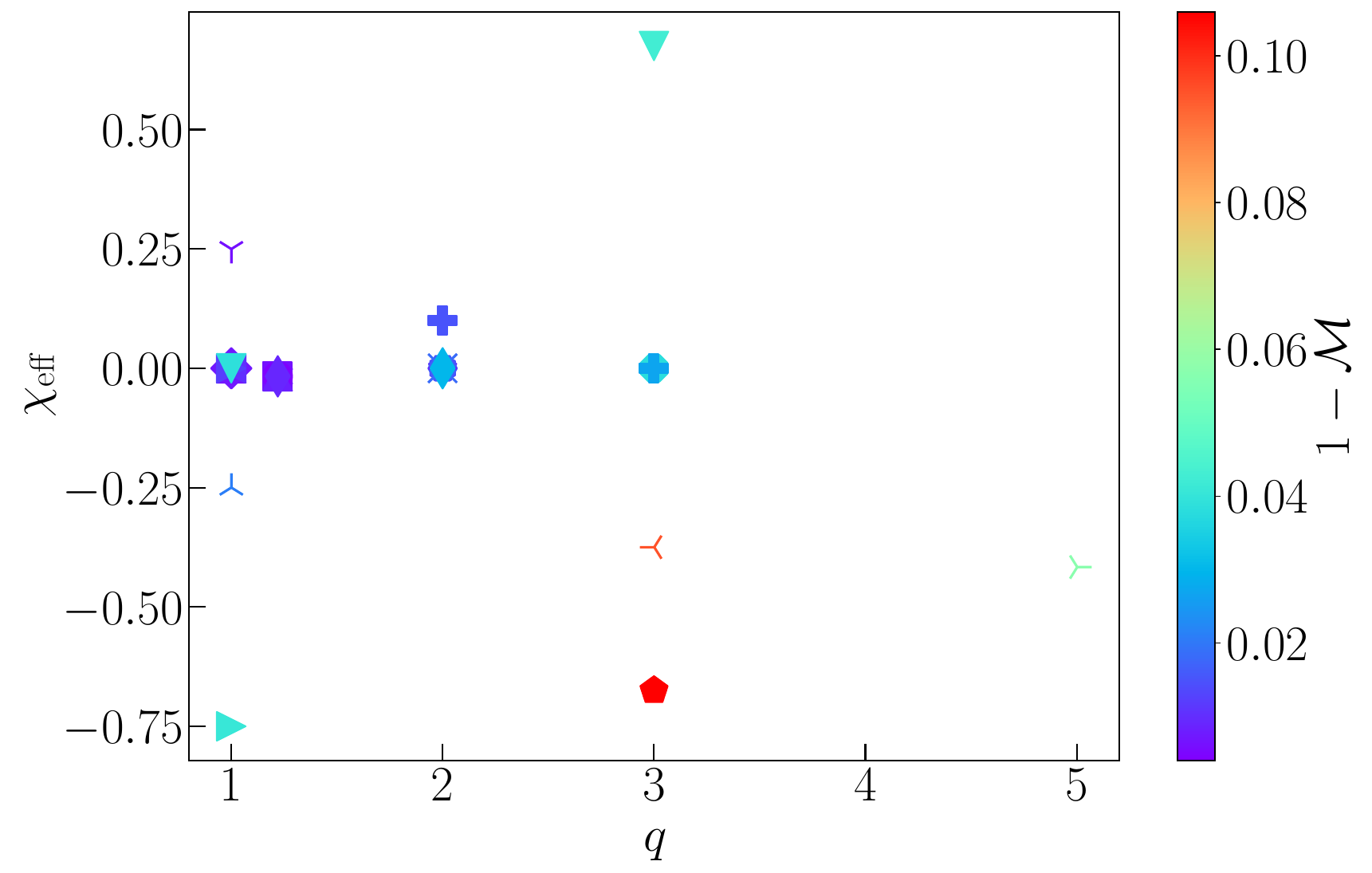}
   
        \caption{Similar to Fig.~\ref{fig:mismatch_NR_mass}, here we show (\textit{left}) mismatches 
        {{between anti-aligned spin NR simulations and \imresigma{} with $(2, \pm 2)$ modes included in both. The figure on the \textit{right} panel shows the mismatch between \imresigmahm{} and 32 eccentric NR simulations (see Table~\ref{table:sxs_id} for details) as a function of $\chi_{\rm{eff}}$ and $q$. Mismatch computation is performed using aLIGO power spectral density and with lower cutoff frequency 10Hz.}}}\label{fig:mismatch_anti_aligned}
\end{figure*}
\subsection{Validation of model in the quasi-circular limit}
In the previous sections, the construction of our full IMR model is described. As a first validation step, here we compare \inspiralesigma{} with existing quasi-circular spinning waveform models and quantify how accurately it is able to reproduce the true inspiral dynamics of spinning systems in the quasi-circular limit.   
Such binaries have been studied over the past decade and their dynamics and emitted gravitational waveforms have been modeled by carefully combining results from post-Newtonian theory and NR simulations. 
We subsequently validate \imresigma{} by comparing it with spinning NR simulations from SXS catalog in the quasi-circular limit.

In Fig.~\ref{fig:mismatch_qc}, we show matches between \inspiralesigma{} and \seobnrvfour{}~\citep{Bohe:2016gbl}, \imrphenom{}~\citep{Pratten:2020fqn}, and \seobnrvfive{}~\citep{Pompili:2023tna} in spinning quasi-circular limit for binaries with a fixed total mass $40M_{\odot}$ having dimensionless spins $(0.5, 0.5)$ using the dominant $\ell=|m|=2$ modes. We fix the lower cutoff frequency at $10$Hz for all cases and take the upper cutoff frequency of the match calculation to be the minimum frequency between the ISCO frequencies for Schwarzschild (i.e., non-spinning) and spinning cases~\citep{Husa:2015iqa}.\footnote{This choice corresponds to the inspiral to PMR transition frequency (\ftrans{}) for \imresigma{} (see Sec.~\ref{sec:mr}).} From the figure, we see that \inspiralesigmahm{} successfully reproduces the true dynamics of spinning systems in quasi-circular limit with match values $\mathcal{O}\gtrsim 99\%$ with both \texttt{EOB} and \texttt{Phenom} waveforms for mass-ratios up to $q=8$.

We next validate the full IMR model using quasi-circular NR waveforms. We choose a set of simulations with mass ratios between 1 and 5, and component spins (anti-) aligned with the orbital angular momentum from the SXS waveforms catalog~\citep{Boyle:2019kee}.
In  Fig.~\ref{fig:NR_comp_p1} we show \imresigma{} strain data $(h_{+})$ versus quasi-circular NR simulations for $32$ out of $42$ non-spinning and aligned-spin cases (see Table~\ref{table:sxs_id_qc_nr_4PN} of Appendix~\ref{sec:appendix_NR}). As before, we fix the source total mass to $40M_\odot$ and  use the dominant $\ell=|m|=2$ modes. The individual panels show the mass-ratio and component spins for each of the simulations. We notice that for both non-spinning and spinning systems, \imresigma{} reproduces NR simulations very closely.
Quantifying this further, we next compute matches between our model and the same set of NR simulations. We do so for a range of total masses for each simulation, and show the resulting mismatches in Fig.~\ref{fig:NR_qc_mismatches}. Dashed horizontal lines mark $1\%, 3\%$, and $ 10\%$ mismatches. We note that the agreement remains better than $3\%$ for most simulations, and better than $1\%$ for a majority of them. The largest mismatches $(3-7\%)$ are shown by a cluster of curves corresponding to binary configurations featuring anti-aligned component spins {\it{and}} unequal mass ratios. Other than for this corner of the binary black hole parameter space, we find \imresigma{} to be largely faithful to NR simulations in the quasi-circular limit. {\violet It is important to note while comparing \imresigma{} with quasi-circular NR simulations, we found that the best fit \imresigma{} waveform (that has the highest match with the NR simulation as given in Table~\ref{table:sxs_id_qc_nr_4PN}) has a very small $(\sim10^{-3})$ non-zero value of eccentricity. This can likely be fixed by including higher-order post-adiabatic initial conditions which we plan to explore in a future version of \imresigma{}.}

\section{Validation of model with eccentric Numerical Relativity Simulations}
\label{sec:validation_NR}

To demonstrate how accurately our \imresigma{} model reproduces the full dynamics of spinning eccentric binaries throughout inspiral, merger, and ringdown, we compare our model with all available eccentric NR simulations (both spinning and non-spinning) in the public SXS catalog~\citep{Boyle:2019kee}. Table~\ref{table:sxs_id} of Appendix~\ref{sec:appendix_NR} provides parameter values such as reference mass ratio $(q)$, reference eccentricity $(e_{\rm{ref}})$, $z$-component of reference dimensionless spins $(\chi_{1_z}, \chi_{2_z})$, and number of orbits $(N_{\rm{orb}})$ for these simulations as reported in their metadata.

When comparing with \imresigma{}, we separately consider the cases where we include the dominant $\ell = |m| =2$ modes only, as well as where we additionally include the subdominant modes: $(\ell, |m|)=(2,1),\,(3,3),\,(3,2),\,(4,4),\,(4,3)$ in the gravitational waveform.
In both scenarios we compute matches between \imresigma{} and reference NR waveforms to quantify their agreement. 
We start with defining a match function that explicitly depends on reference (initial) values of orbital eccentricity $(e_{\rm{ref}})$ and mean-anomaly $(l_{\rm{ref}})$ specified at a given reference (initial) frequency $(f_{\rm{ref}})$. We maximize this match to find optimal values of reference parameters $f_{\rm{ref}}$, $e_{\rm{ref}}$, and $l_{\rm{ref}}$. This maximization is performed in two stages. First a Monte-Carlo sampling is used to narrow down the relevant ranges of each of these parameters, and then we use the \texttt{Nelder-Mead} algorithm~\citep{NelderMead1965,2020SciPy-NMeth} to arrive at the optimal values within these ranges. These reference values are used to generate \imresigma{} waveforms to compute the final match values discussed below.

In Fig.~\ref{fig:NR_comp}, we visually compare our model with these spinning eccentric NR simulations. Each panel shows the mass ratio $(q)$, $z$-component of the dimensionless spins $(\chi_{1_z}, \chi_{2_z})$, and reference eccentricity $(e_{\rm{ref}})$ reported in the respective simulation's metadata for convenience of inspection.\footnote{For some simulations, the metadata does not report a concrete value of reference eccentricity, so we include the value of $e_{\rm{ref}}$ in the panel that produces optimal agreement between \imresigma{} and that simulation. We mark such simulations with an asterisk in the figures, with $e_{\rm{ref}}$ replaced by $e_{\rm{ref}}^{*}$.} We find that \imresigma{} reproduces most of these NR simulations faithfully.
Further, Figs.~\ref{fig:mismatch_NR_mass} and~\ref{fig:mismatch_anti_aligned} (left panel) together show mismatches between our model and all eccentric NR simulations listed in Table~\ref{table:sxs_id} of Appendix~\ref{sec:appendix_NR}. This is shown as a function of total mass, ranging from $20-120M_{\odot}$. We split these into two figures to discuss anti-aligned spinning equal and unequal mass-ratio cases separately.
Fig.~\ref{fig:mismatch_NR_mass} focuses on binaries with moderate mass-ratios and/or positively aligned binary spins (with the orbital angular momentum). It can be seen for most of such cases that the matches are well above 97\% across the range of total mass, with the majority cases reporting matches above 99\% as well. {\textit{This is a reassuring validation of our model in the region of BBH mass-spin space which most of the binary sources observed in the first three observing runs of LIGO-Virgo-KAGRA detectors occupy}}~\citep{LIGOScientific:2018mvr,LIGOScientific:2020ibl,LIGOScientific:2021djp}.

Next, Fig.~\ref{fig:mismatch_anti_aligned} (left panel) shows that mismatches with NR simulations that have one or both the spins anti-aligned with the orbital angular momentum vector (\texttt{SXS:BBH:0089}, \texttt{SXS:BBH:0105}, \texttt{SXS:BBH:0108}, \texttt{SXS:BBH:1136}, \texttt{SXS:BBH:1169}) can be $>3\%$ in the mass range $20-120 M_{\odot}$. Further, the mismatches are higher (can be as high as $10\%$) for the large mass ratio cases (\texttt{SXS:BBH:0105}, \texttt{SXS:BBH:0108}, \texttt{SXS:BBH:1169}) compared to others (\texttt{SXS:BBH:0089}, \texttt{SXS:BBH:1136}), as indicated by the right panel of the same figure which shows the highest mismatches to be for $q\gtrsim 3$ and $\chi_{\rm{eff}} < -0.3$. 
For simulations with larger initial eccentricity (such as \texttt{SXS:BBH:1136}), this is likely due to weakening of the assumption discussed in Sec.~\ref{sec:wf_model} that binary has circularised beyond the inspiral-merger transition frequency, which is smaller (and earlier) in case of anti-aligned spins as opposed to aligned spins. As a consequence, the attachment of eccentric inspiral with quasi-circular PMR is happening at a point where the binary hasn't sufficiently circularized yet. One remedy to this could be to use an eccentric PMR model instead of a quasi-circular one. This would accurately represent the dynamics of the eccentric signal throughout the merger and ringdown. As a consequence, the match will improve for these NR simulations.
For simulations with smaller initial eccentricity (such as \texttt{SXS:BBH:0105}) this could indicate the need to include higher-order spin information in \esigma{}. A remedy for this would be to include higher order spin-eccentricity dependent terms in the radiative sector and transition from inspiral to merger prescription closer to merger. We defer a detailed investigation to future work, and recommend the use of \imresigmahm{} to model binaries of moderately spinning black holes merging on moderately eccentric orbits.


\section{Importance of including eccentricity and higher-order waveform modes}
\label{sec:HM}
\begin{figure*}
    \includegraphics[width=0.32\linewidth, trim=10 0 0 0]{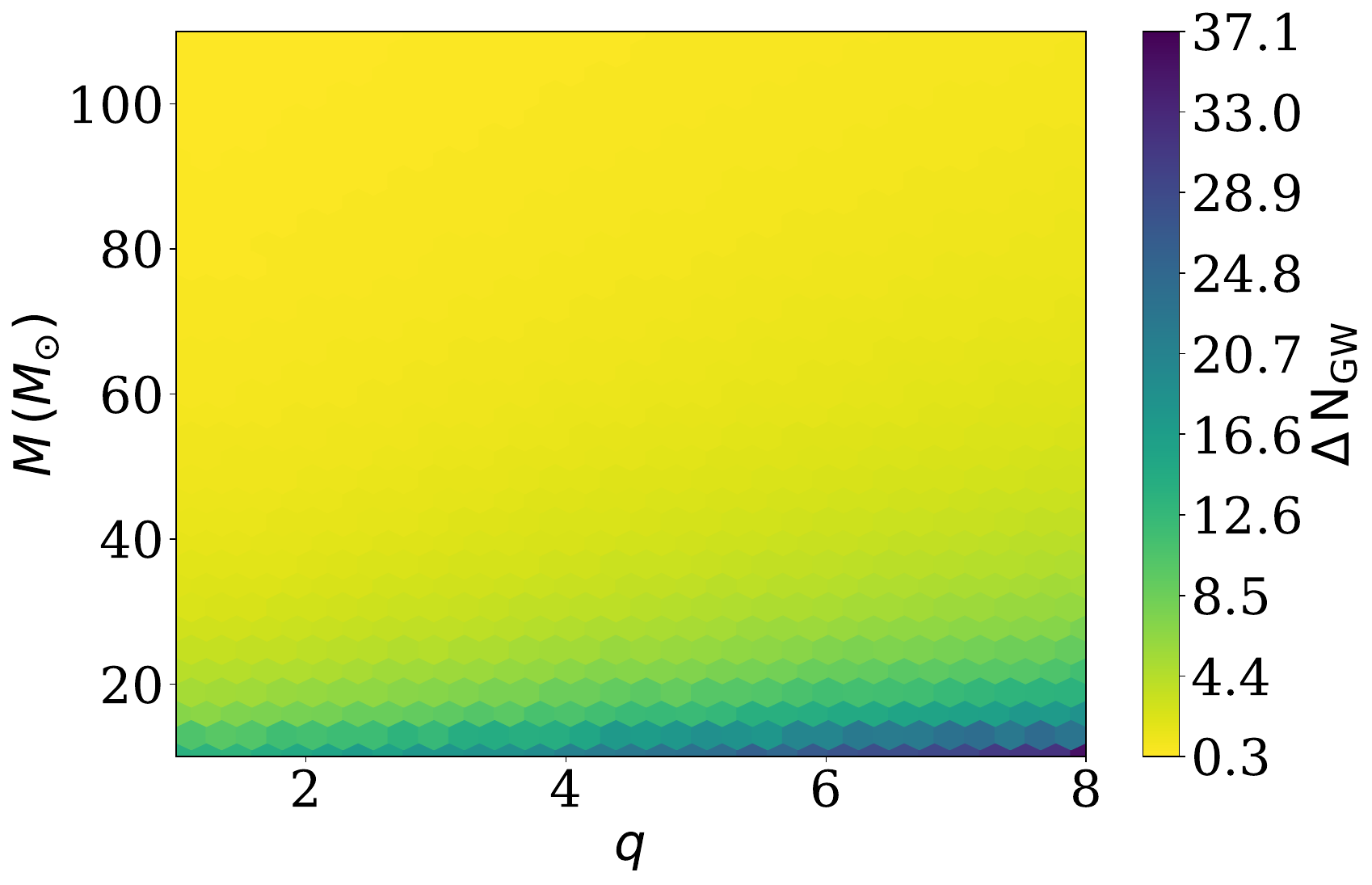}
    \includegraphics[width=0.32\linewidth]{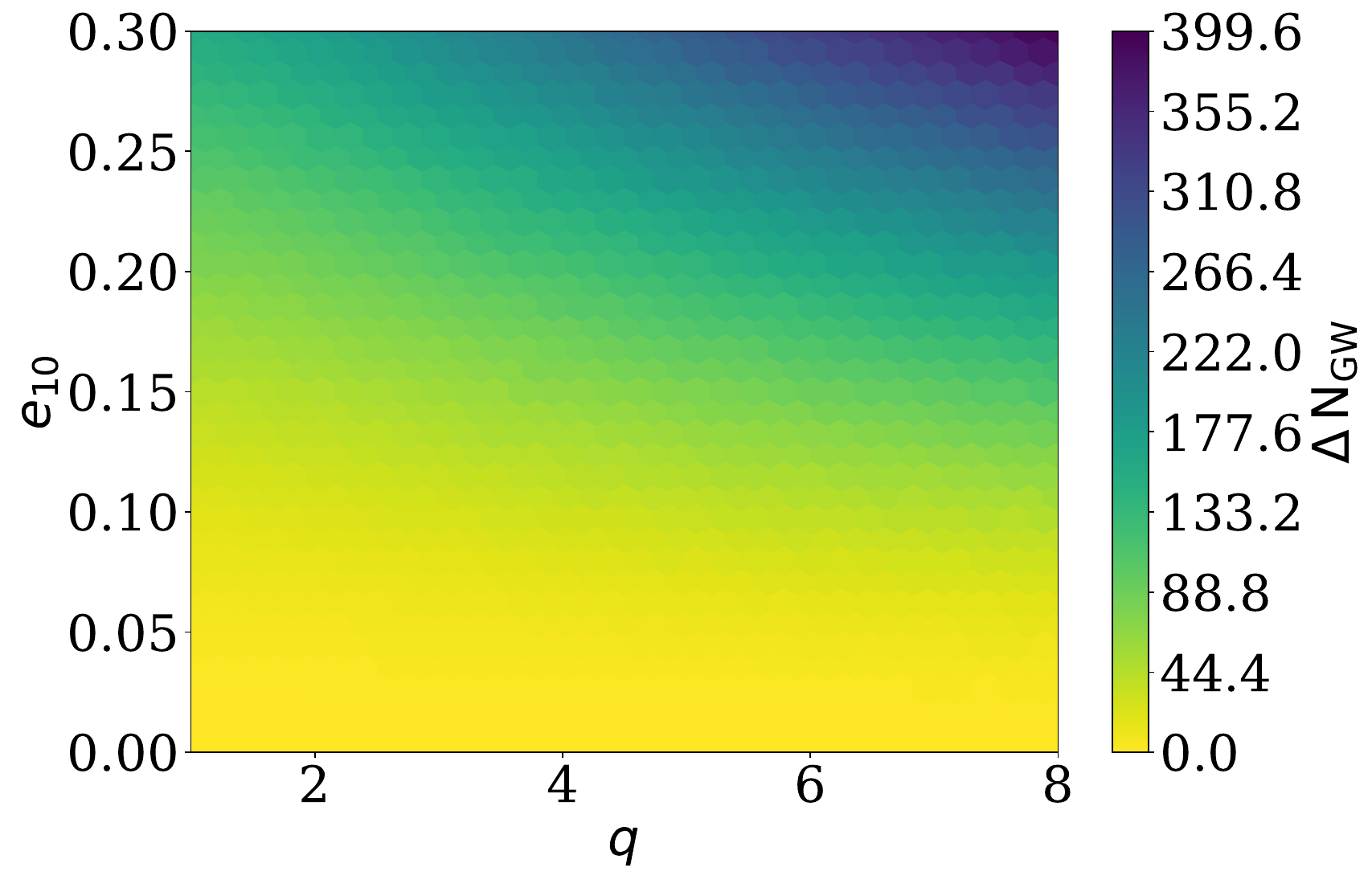}
    \includegraphics[width=0.32\linewidth]{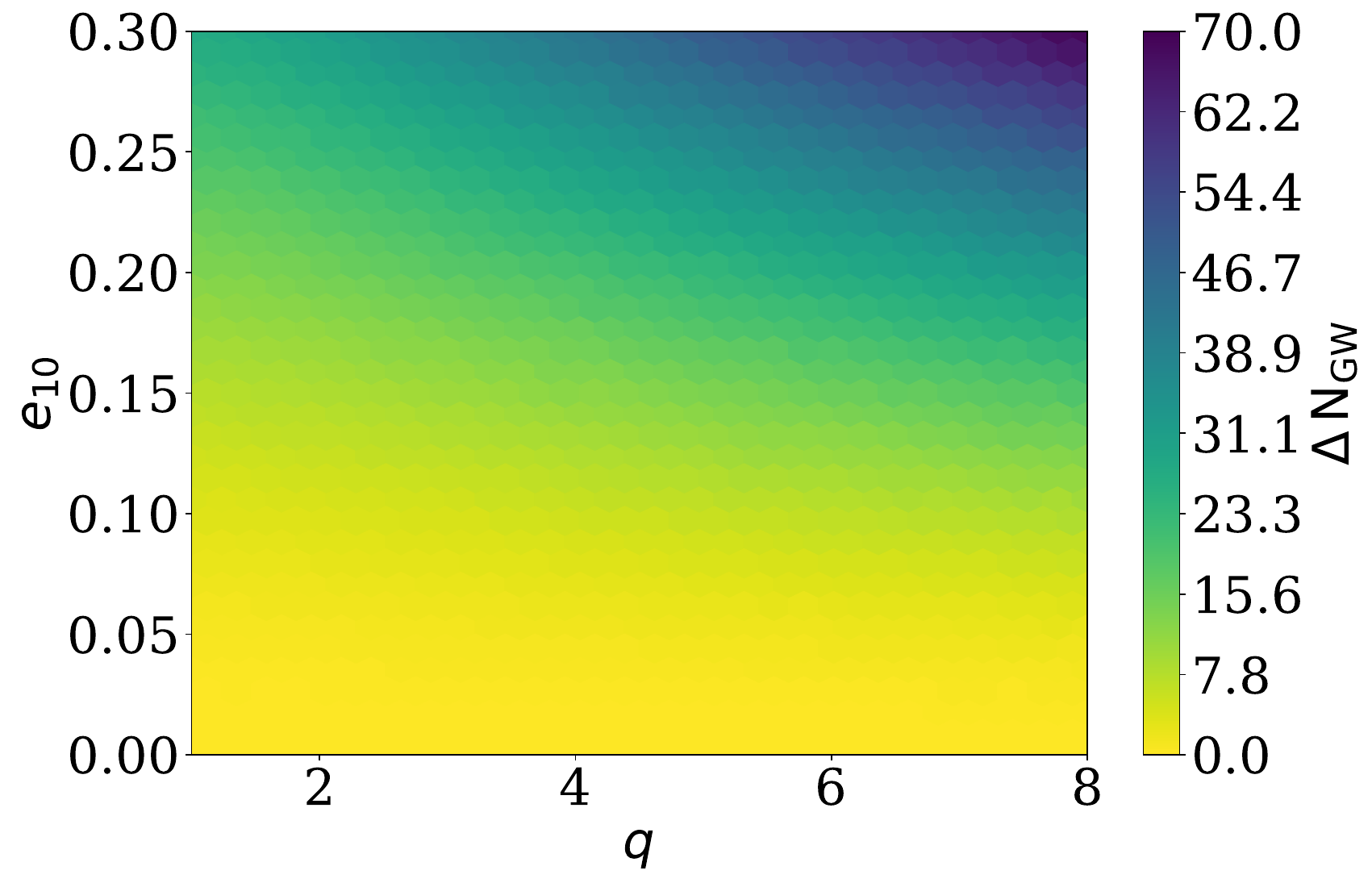}
    \caption{{{Effect of eccentricity on signal length is presented here. The difference in number of GW cycles $(\Delta \rm{N}_{\rm{GW}})$ as a function of (I) total mass and mass ratio (\textit{left}), (II) eccentricity and mass ratio (\textit{middle} and \textit{right}) between \inspiralesigma{} with (2,2) mode circular and eccentric version are shown. The left figure is for fixed eccentricity 0.05, while the middle and right figures are for fixed total masses $20M_{\odot}$ and $60M_{\odot}$, respectively. For all three cases, the dimensionless spins and low cutoff frequency is fixed to (0.5,0.5) and 10Hz, respectively.}}}
    \label{fig:Ngw_M_q_e}
\end{figure*}
\begin{figure*}
    \includegraphics[width=0.32\linewidth]{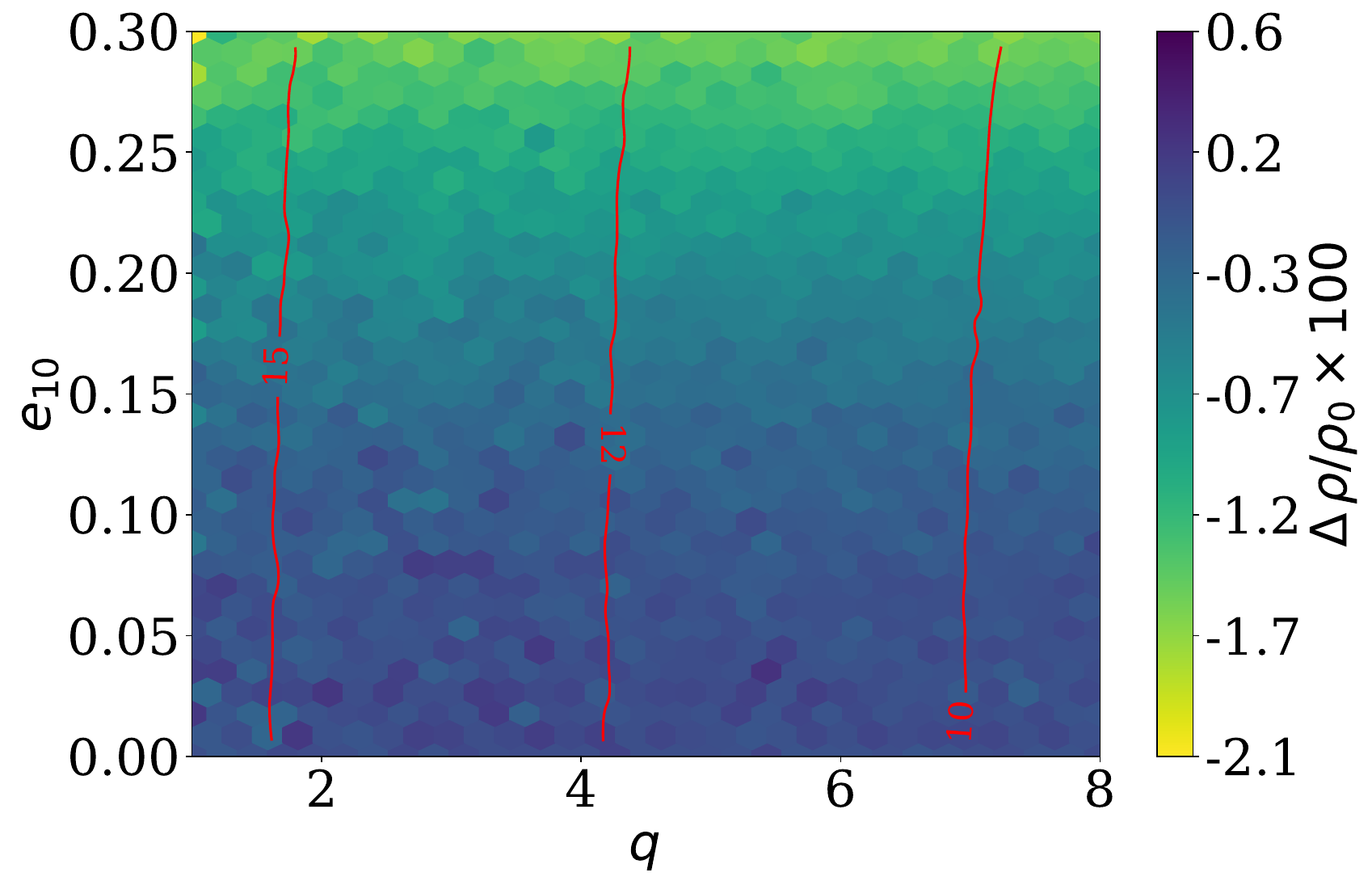}
    \includegraphics[width=0.32\linewidth]{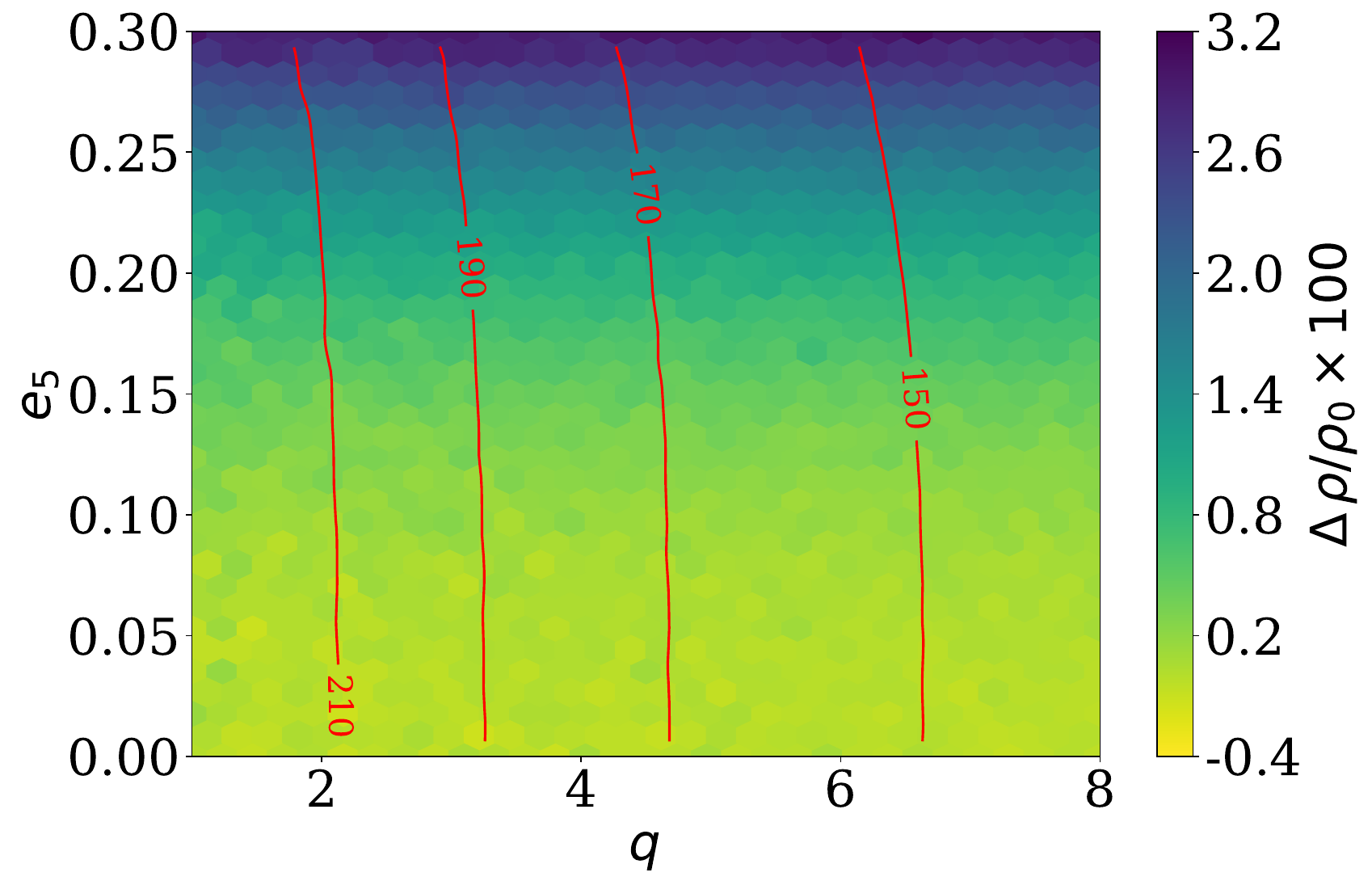}
    \includegraphics[width=0.32\linewidth]{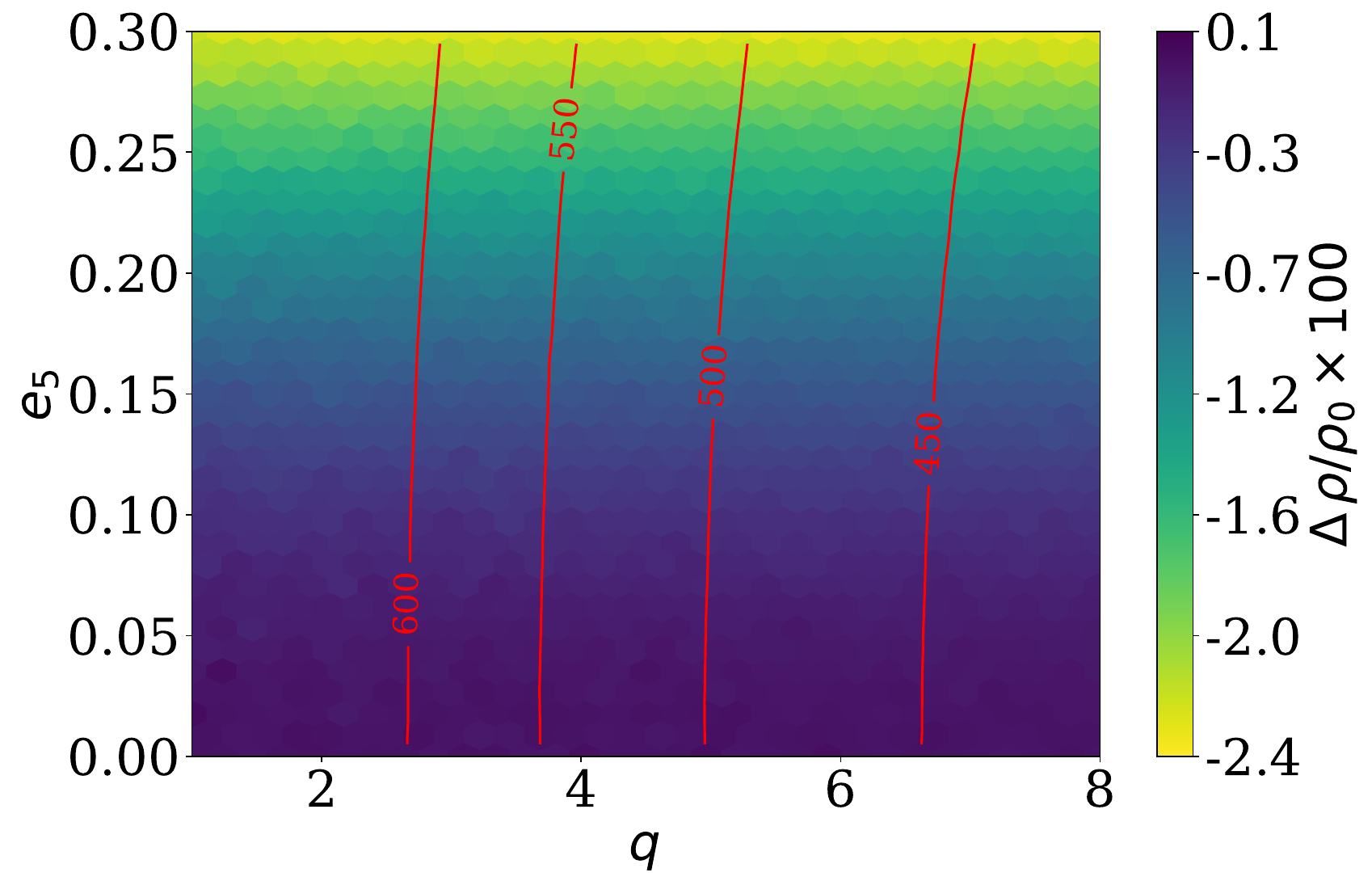}
    \includegraphics[width=0.32\linewidth]{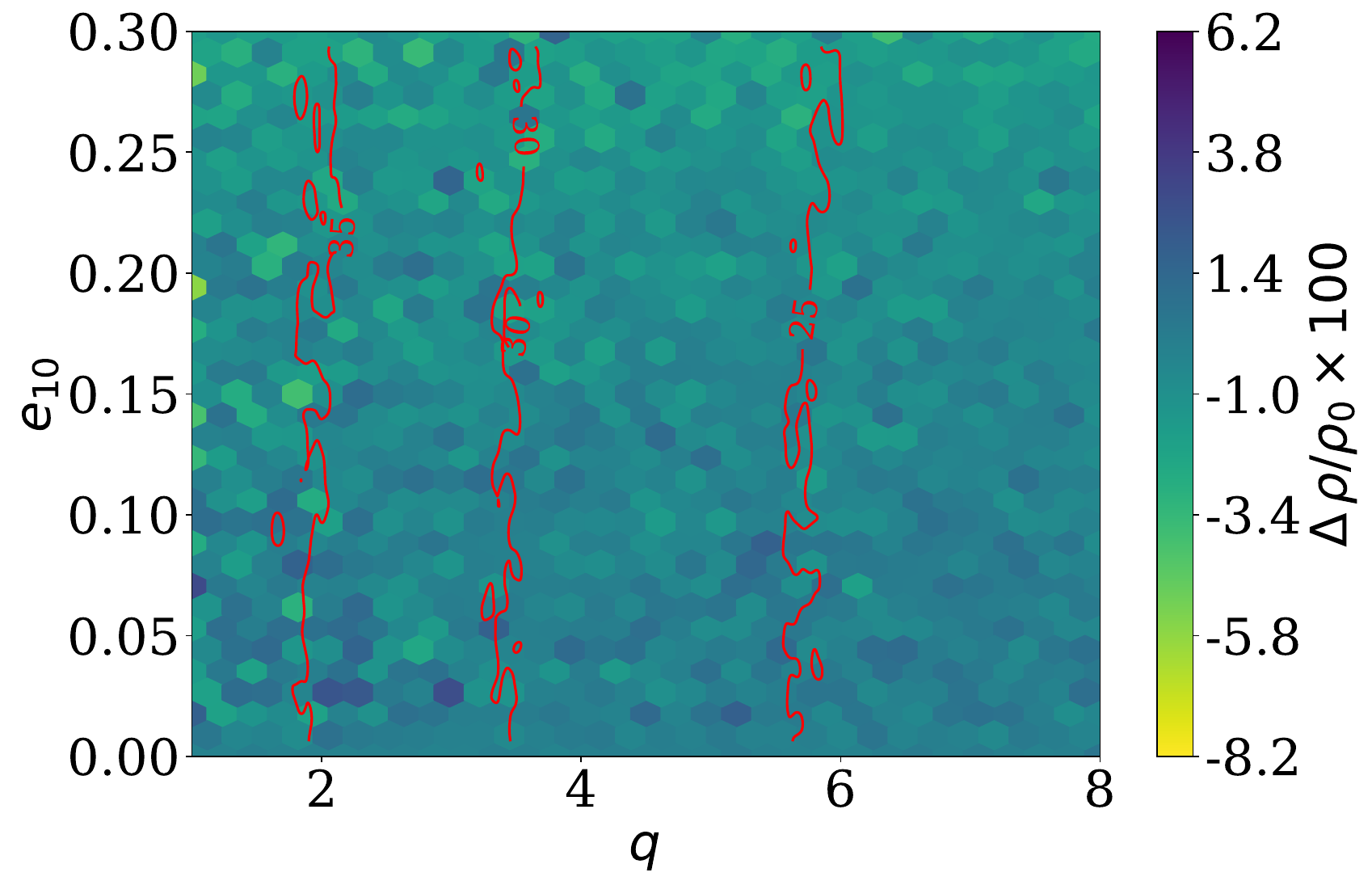}
    \includegraphics[width=0.32\linewidth]{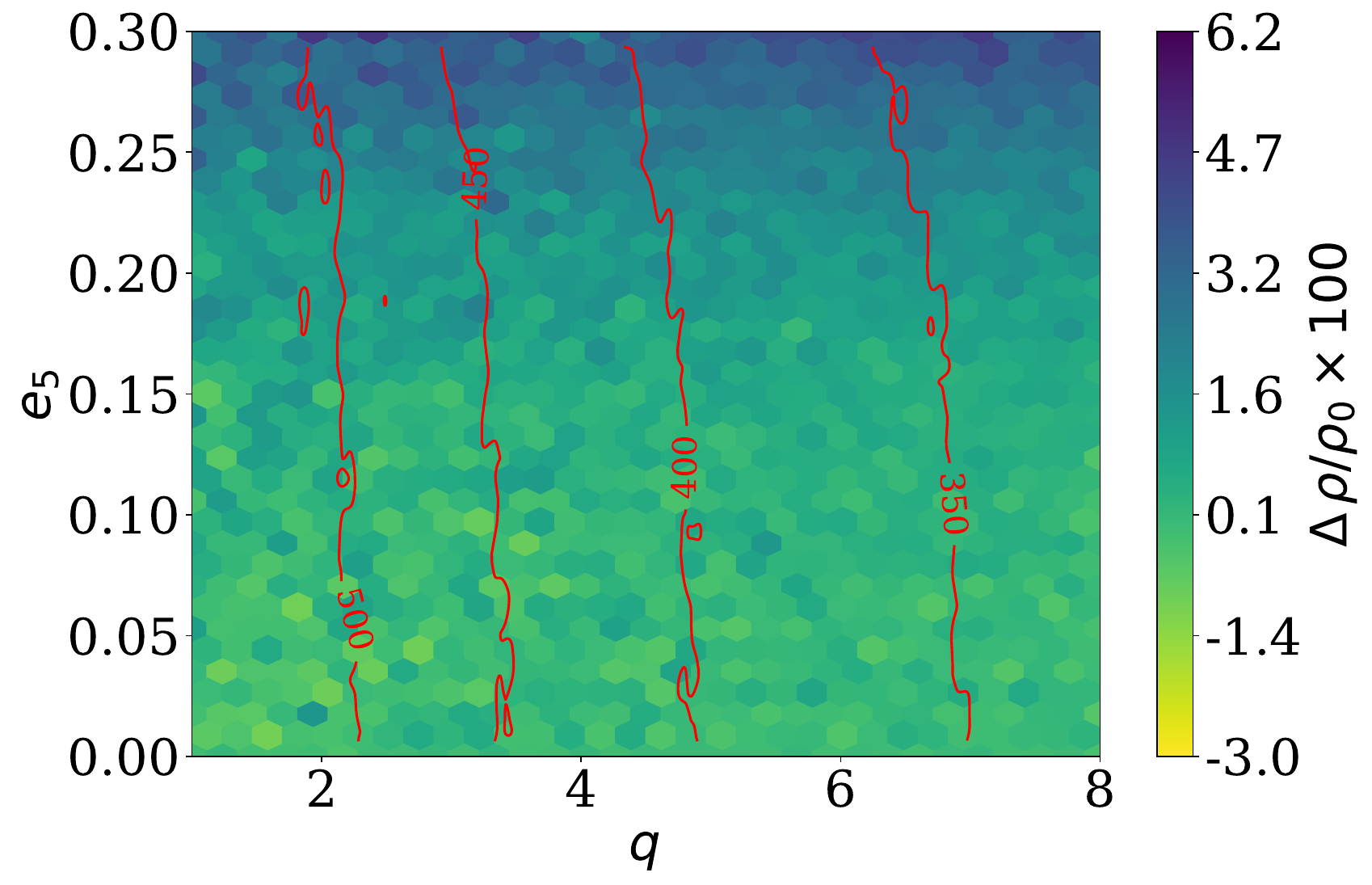}
    \includegraphics[width=0.32\linewidth]{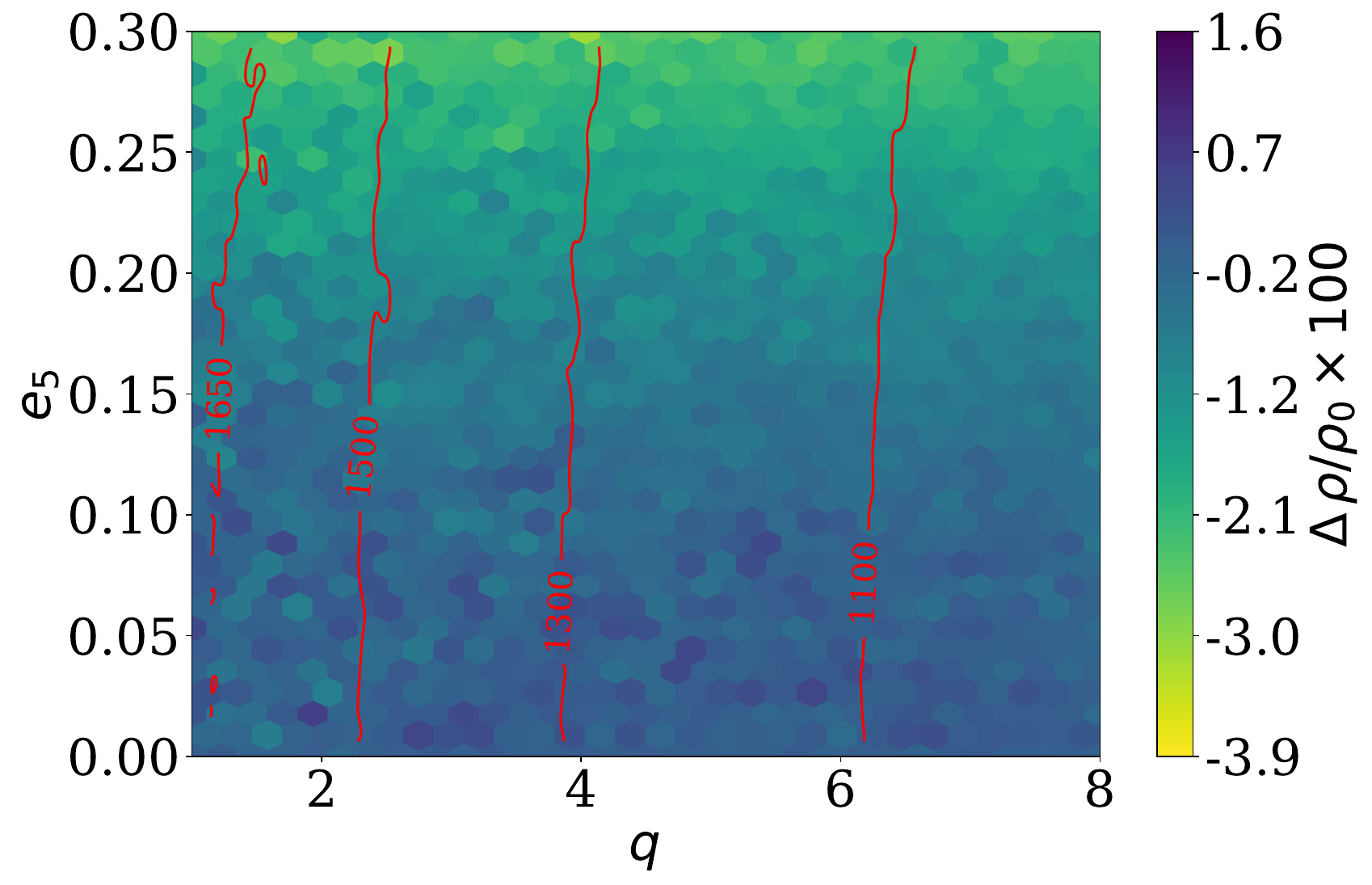}
    \caption{Relative difference in optimal SNR is shown as a function of eccentricity at $10$Hz $(e_{10})$ (for aLIGO) or at $5$Hz $(e_5)$ (for ET and CE) and $q$ between \inspiralesigma{} with (2,2) for circular $(\rho_0)$ and eccentric case for aLIGO (\textit{left}), ET (\textit{middle}), and CE (\textit{right}) detectors. The top and bottom row figures have fixed total mass $20M_{\odot}$ and $60M_{\odot}$, respectively. For both the cases, dimensionless spins are fixed to $(0.6, 0.6)$. The SNR computation is done using distance to source $500$Mpc at an inclination angle of $\pi/3$ radians. The lower cut off frequency for aLIGO and ET, CE are taken to be $10$Hz and $5$Hz, respectively. The red colored contours show the region of the SNR (the value is indicated on the contours) on the $e-q$ plane for eccentric signals.
    For higher mass binaries presence of orbital eccentricity changes the overall signal strength (optimal SNR) by up to $6-8\%$ for current and next-generation GW detectors. 
    }    
    \label{fig:SNR_e0}
\end{figure*}
\begin{figure*}
    \includegraphics[width=0.32\linewidth]{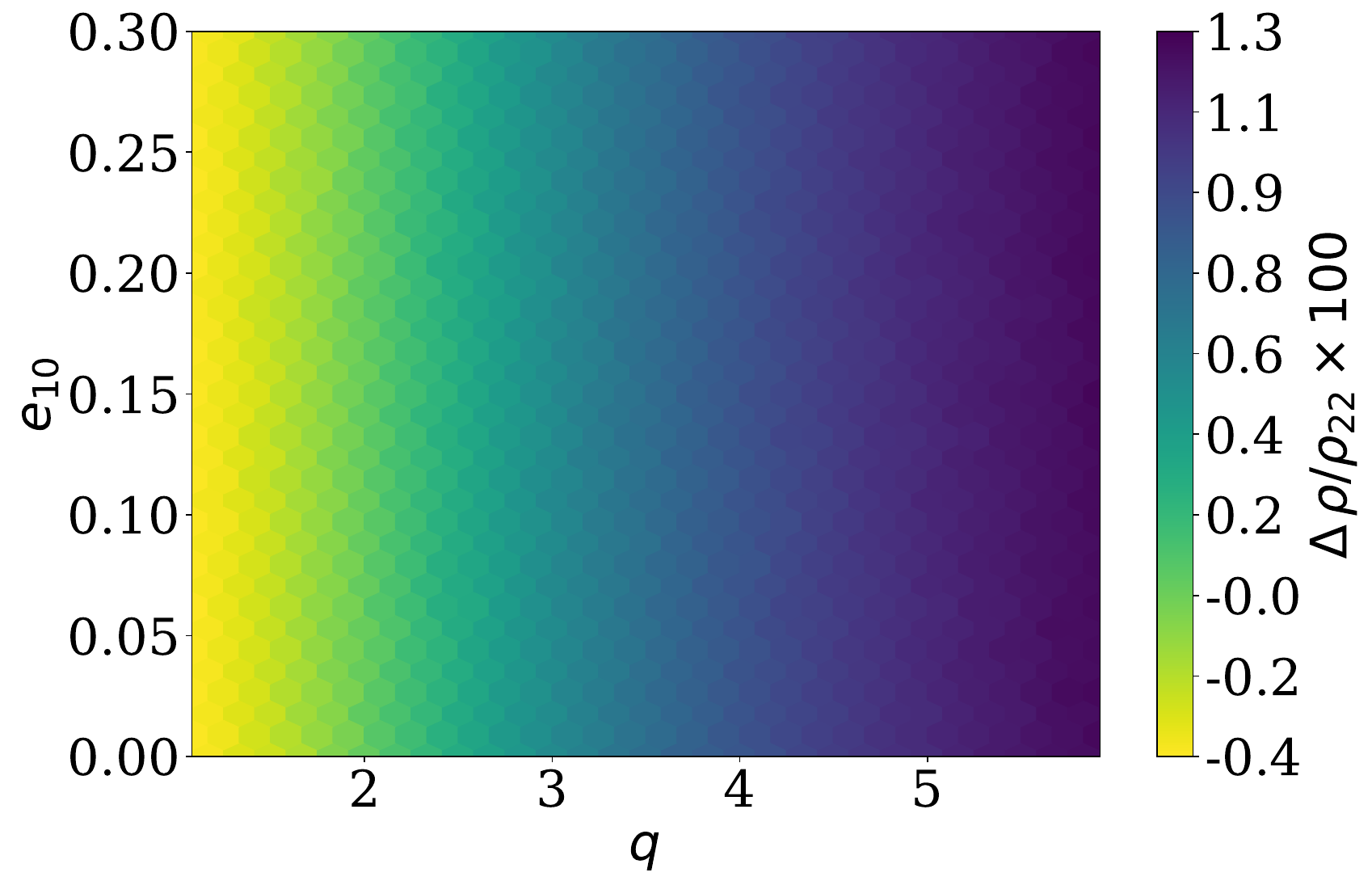}
    \includegraphics[width=0.32\linewidth]{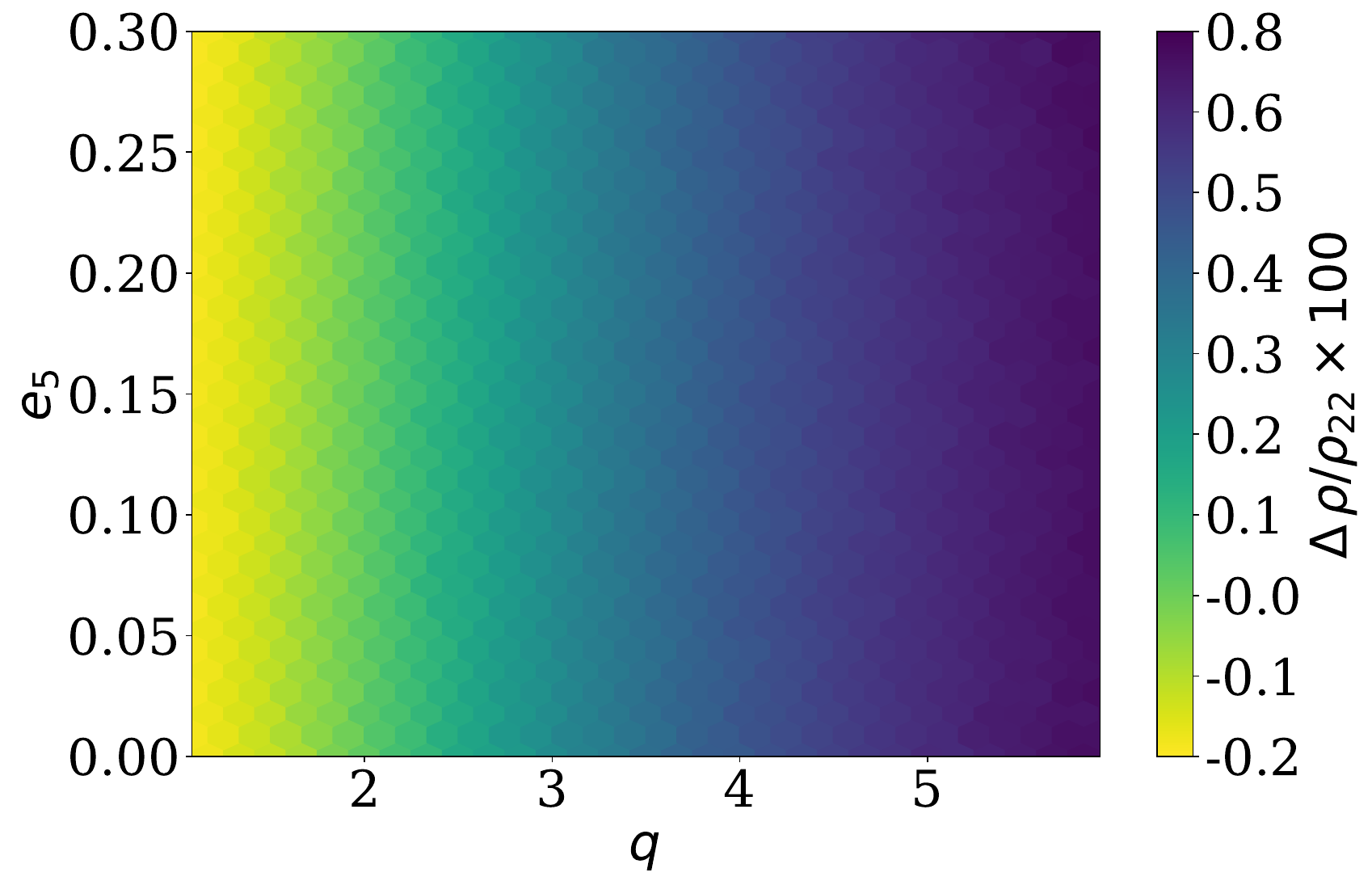}
    \includegraphics[width=0.32\linewidth]{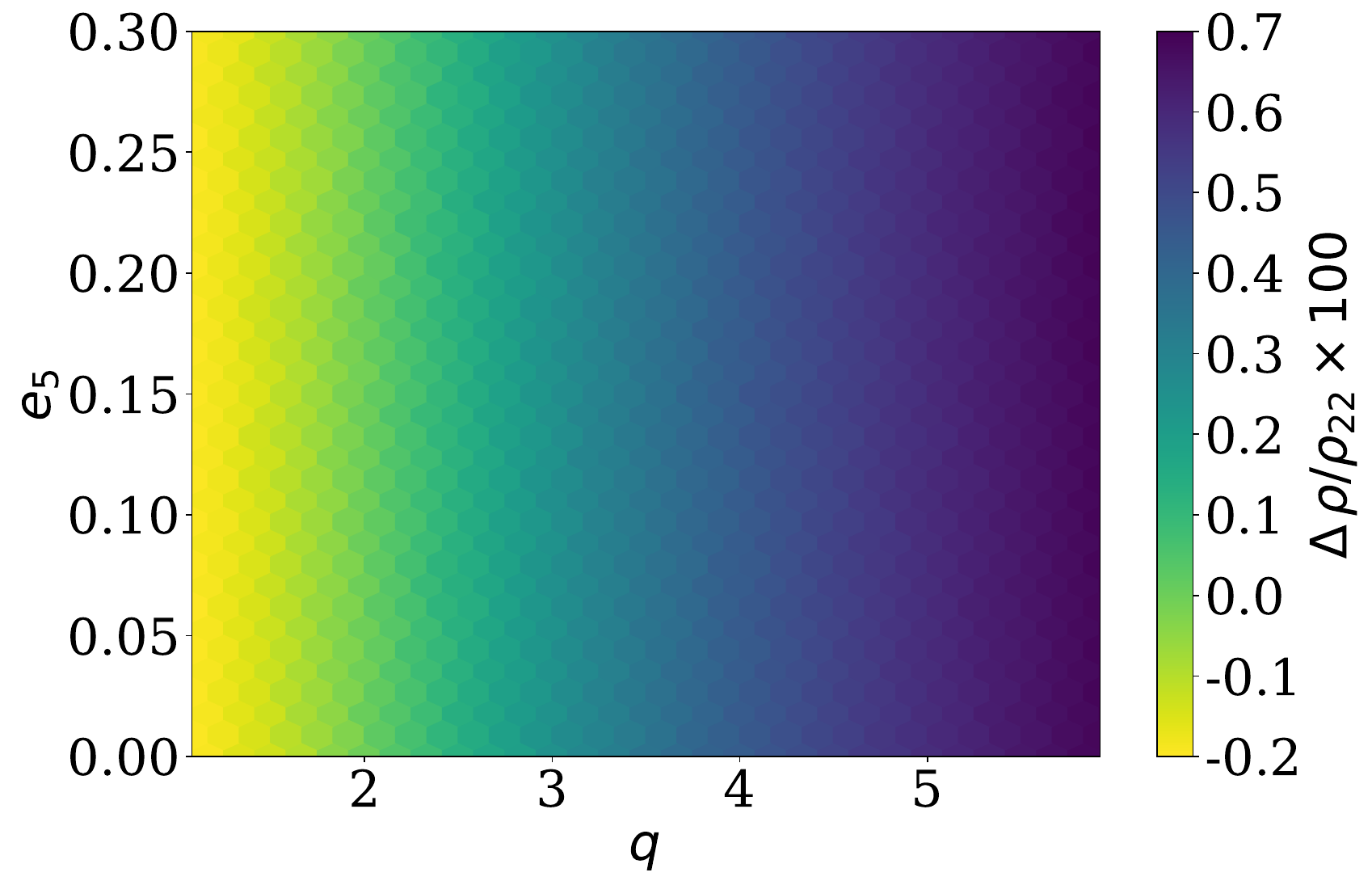}
    \includegraphics[width=0.32\linewidth]{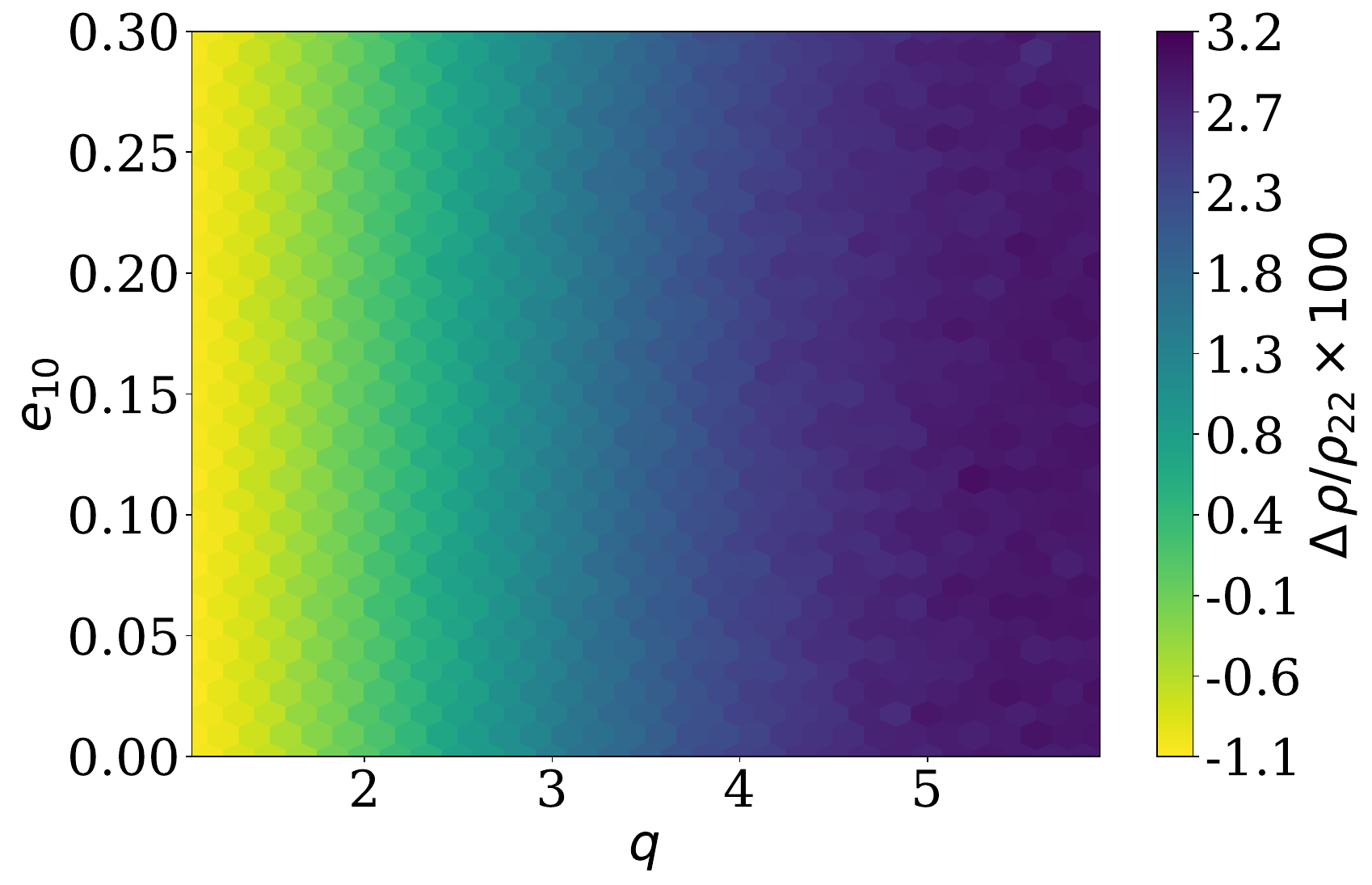}
    \includegraphics[width=0.32\linewidth]{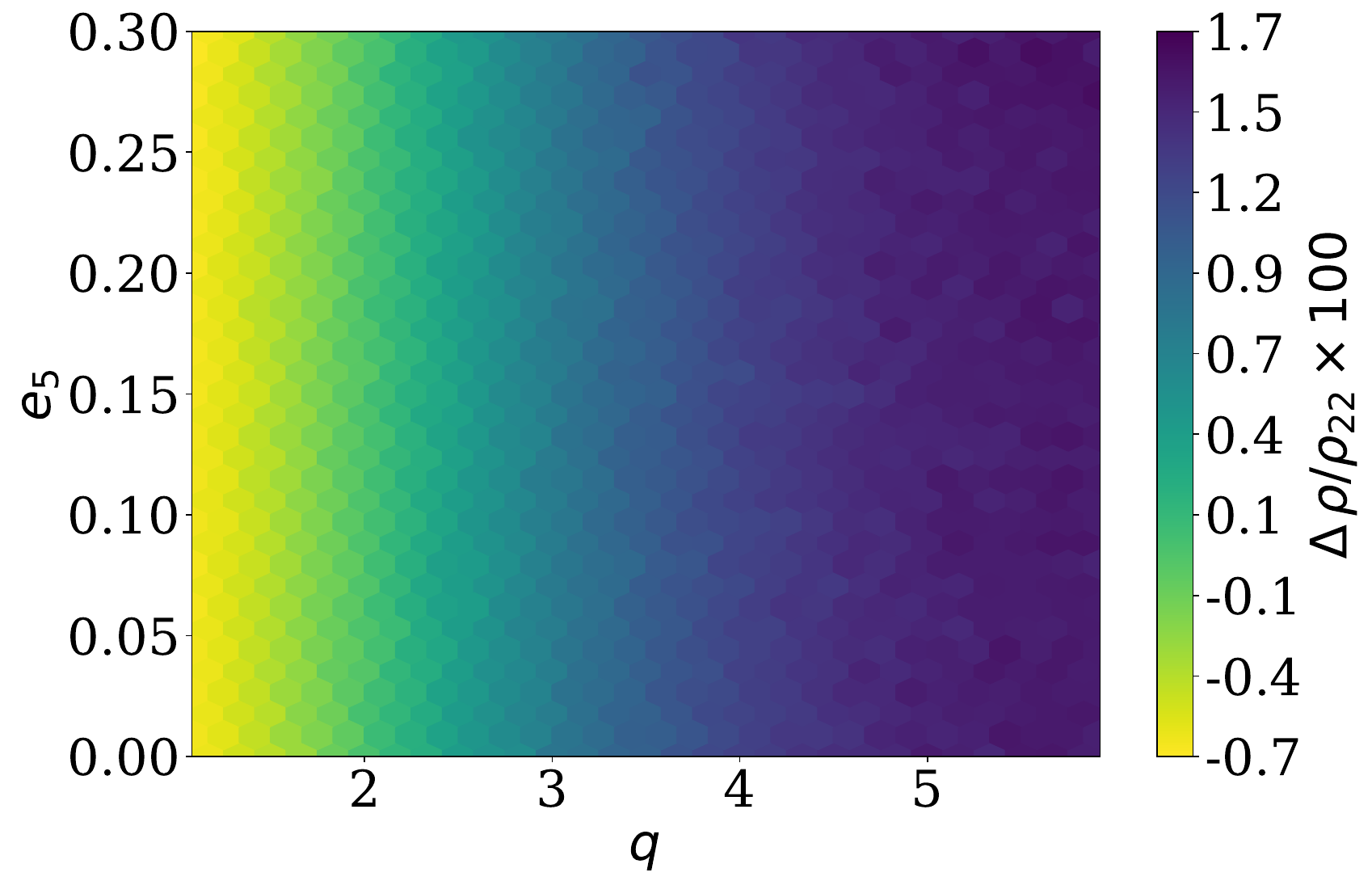}
    \includegraphics[width=0.32\linewidth]{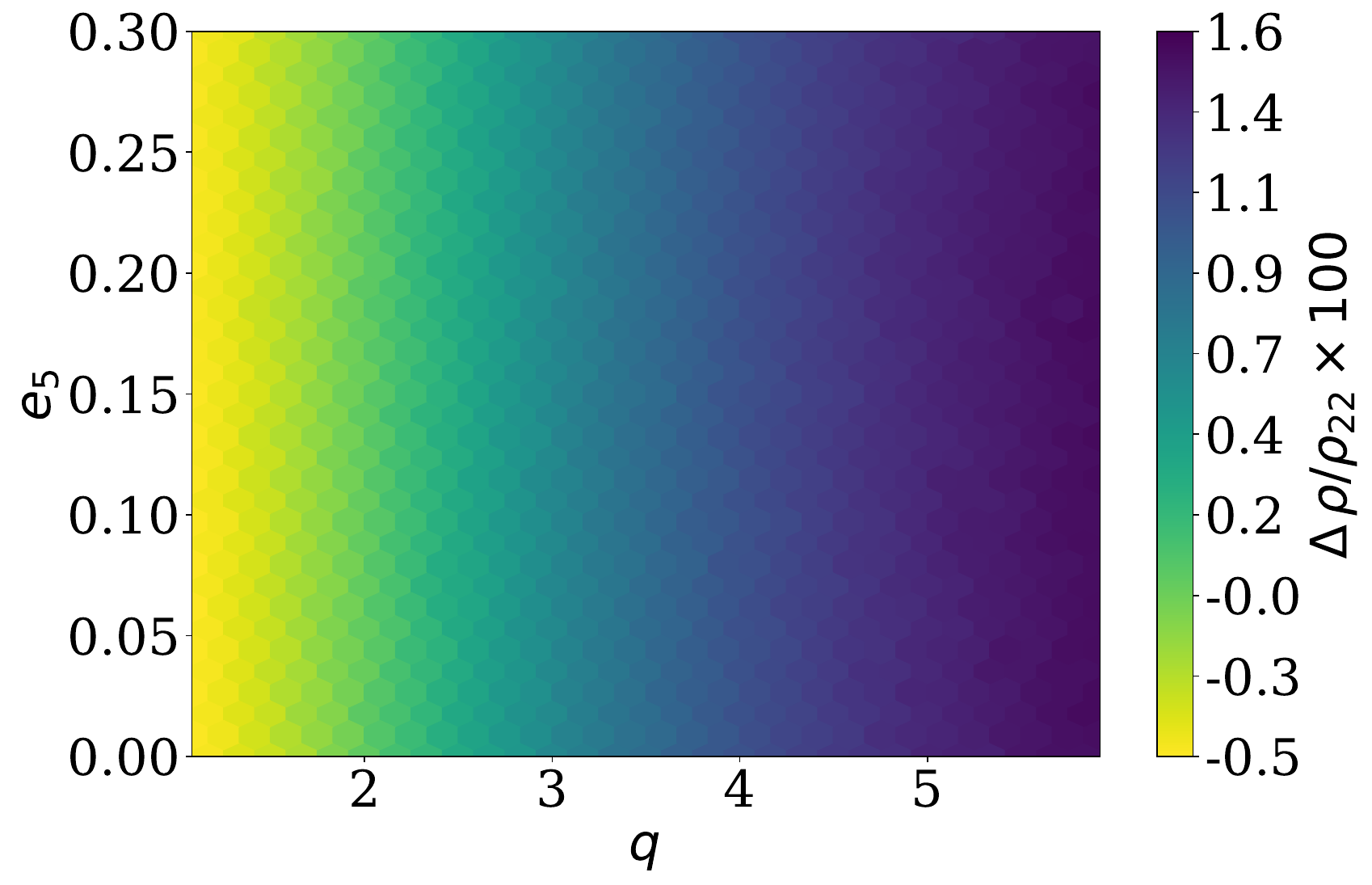}
    \caption{Effect of inclusion of higher-order modes is shown here. The relative SNR $(\Delta \rho/\rho_{22})$ for aLIGO (\textit{left}), ET (\textit{middle}), and CE (\textit{right}) detectors between \imresigmahm{}  and \imresigma{} with $(2, \pm 2)$ modes $(\rho_{22})$ is shown for binaries with fixed total mass $20M_{\odot}$ (top panel) and $60M_{\odot}$ (bottom panel). For all cases component (aligned) spins are fixed to $(0.6,0.6)$. The SNR computation is done using distance to source $500$Mpc and inclination angle $\pi/3$ radians. The lower cut off frequency for aLIGO is fixed to $10$Hz, while for ET and CE it is fixed to $5$Hz. 
    We find that the inclusion of high-order modes increases the overall signal strength (optimal SNR) by a few percent, but the same change is negligible for lower mass sources, indicating that higher-order modes are prominently excited primarily close to merger.
    }
    \label{fig:SNR_ET_CE}
\end{figure*}

Orbital eccentricity significantly changes both the length and shape of emitted GWs by imprinting additional modulations into them. As most GW searches and parameter estimation efforts focus on non-eccentric binary mergers, they are bound to introduce selection effects, possibly precluding the discovery of highly eccentric binaries.
In this section, we use the \esigma{} model to quantify the effect of orbital eccentricity on GW signals and the extent to which subdominant waveform harmonics contribute for eccentric binaries. We will study some aspects of this for current and third generation ground based GW observatories.

An important effect that immediately distinguishes an eccentric binary from a quasi-circular one is the increased rate of inspiral, leading to a shorter time to merger. {\violet We can quantify this by computing the number of cycles $\rm{N}_{\rm{GW}}$ present in the dominant GW mode from a given (orbit-averaged) frequency to merger, for both eccentric ($\rm{N}^{\rm{ecc}}_{\rm{GW}}$) and equivalent quasi-circular ($\rm{N}^{\rm{circ}}_{\rm{GW}}$) cases with the same black hole masses and spins.}
We compute the differences between them
\begin{equation}
    \Delta \rm{N}_{\rm{GW}} \equiv \rm{N}^{\rm{circ}}_{\rm{GW}} - \rm{N}^{\rm{ecc}}_{\rm{GW}},
\end{equation}
starting from a (orbit-averaged) GW frequency of $10$Hz. 
Positive values of $\Delta \rm{N}_{\rm{GW}}$ imply a shorter eccentric signal. We show $\Delta \rm{N}_{\rm{GW}}$ in Fig.~\ref{fig:Ngw_M_q_e} as a function of total mass and mass ratio $M-q$ (left panel), and eccentricity and mass ratio $e-q$ (middle and right panels). The component spins are fixed to moderate values of $(0.5, 0.5)$ in all panels, and orbital eccentricity is held constant at $0.05$ in the left panel. The middle panels correspond to a fixed binary mass of $20M_\odot$, while the right panel is for heavier binaries with mass $60M_\odot$. 
In the left panel showing $M-q$ plane, we see that there is a sharp increase in $\Delta \rm{N}_{\rm{GW}}$ with increasing mass ratio for small total masses, reaching as high as $(\sim 37.1)$ fewer GW cycles for $10M_\odot$ eccentric binaries with mass ratios $q \simeq 8$ even for relatively small value of eccentricity at $10$Hz $(e_{10})$ of $0.05$.
For masses larger than $20M_\odot$ the difference is relatively smaller $(\lesssim 4)$ irrespective of mass ratio. The mass dependence is also relatively weak indicating that binaries starting with $e_{10}=0.05$ with masses above $20M_\odot$ rapidly circularize in the band of ground-based detectors.  
In $e-q$ plane shown in the middle panel, we see clearly a stronger non-linear growth in $\Delta \rm{N}_{\rm{GW}}$ as we increase initial eccentricity or (to a lesser extent) the binary mass ratio. For these $20M_\odot$ binaries, $\Delta \rm{N}_{\rm{GW}}$ grows rapidly and reaches $\mathcal{O}(10^2)$ cycles when eccentricities is larger than $0.25$ for $q=2$ binaries, and when $e_{10}>0.15$ for $q=6$ systems. Compare this panel with the right panel to note the effect of increasing binary mass from $20$ to $60M_\odot$, a range which spans the masses of a bulk of sources observed by the current generation detectors so far~\citep{LIGOScientific:2018mvr,LIGOScientific:2020ibl,LIGOScientific:2021djp,Nitz:2019hdf}. At $60M_\odot$ we notice a smaller yet substantial change in the number of GW cycles from eccentric binaries compared to quasi-circular ones. This is consistent with there being much fewer signals cycles in frequency band for higher masses.
Note that the left and middle panels explore different parts of the parameter space, with only a horizontal strip of constant $M = 20M_\odot$ in the left panel being common with the $e_{10} = 0.05$ horizontal strip in the middle panel.

\begin{figure}
    \includegraphics[width=\linewidth]{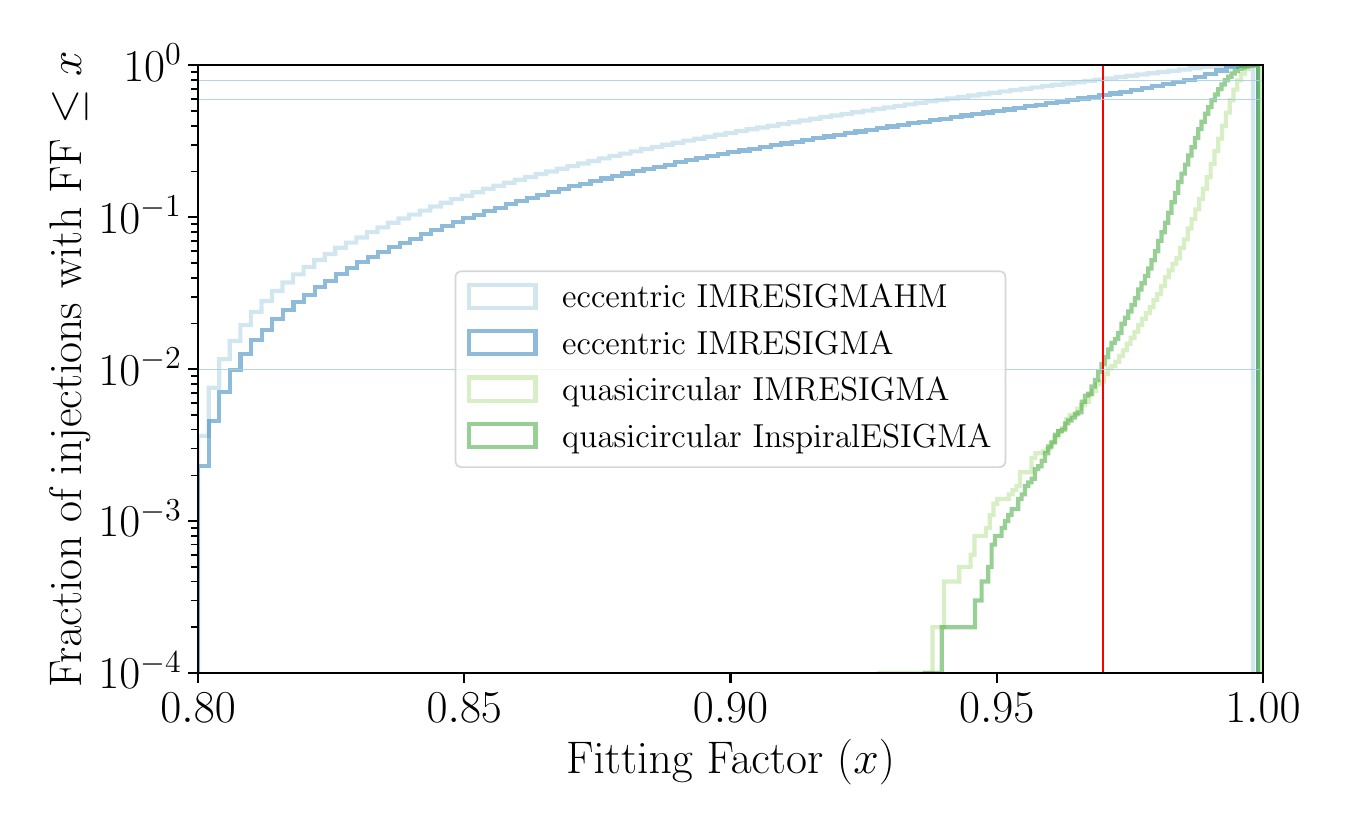}
    \caption{{Effect of eccentricity and higher-order modes are shown here on matched-filtering based GW searches. We show the distribution of fitting factors for different kinds of signal populations (as labelled in the legend), against a template bank of quasi-circular aligned-spin BBH waveform templates, similar to what is used in contemporary GW searches in LIGO-Virgo-KAGRA data.
    Comparing the blue and green curves shows the effectiveness of the same template bank in recovering eccentric vs quasi-circular BBH merger GW signals. This shows how currently used template banks will lose a sizable fraction of signal information from eccentric sources. The effect gets worse when sub-dominant harmonics are included in signal waveforms (comparing light and dark blue curves).
    }}
    \label{fig:FF_dist}
\end{figure}

\begin{figure*}
    \includegraphics[width=0.4\linewidth]{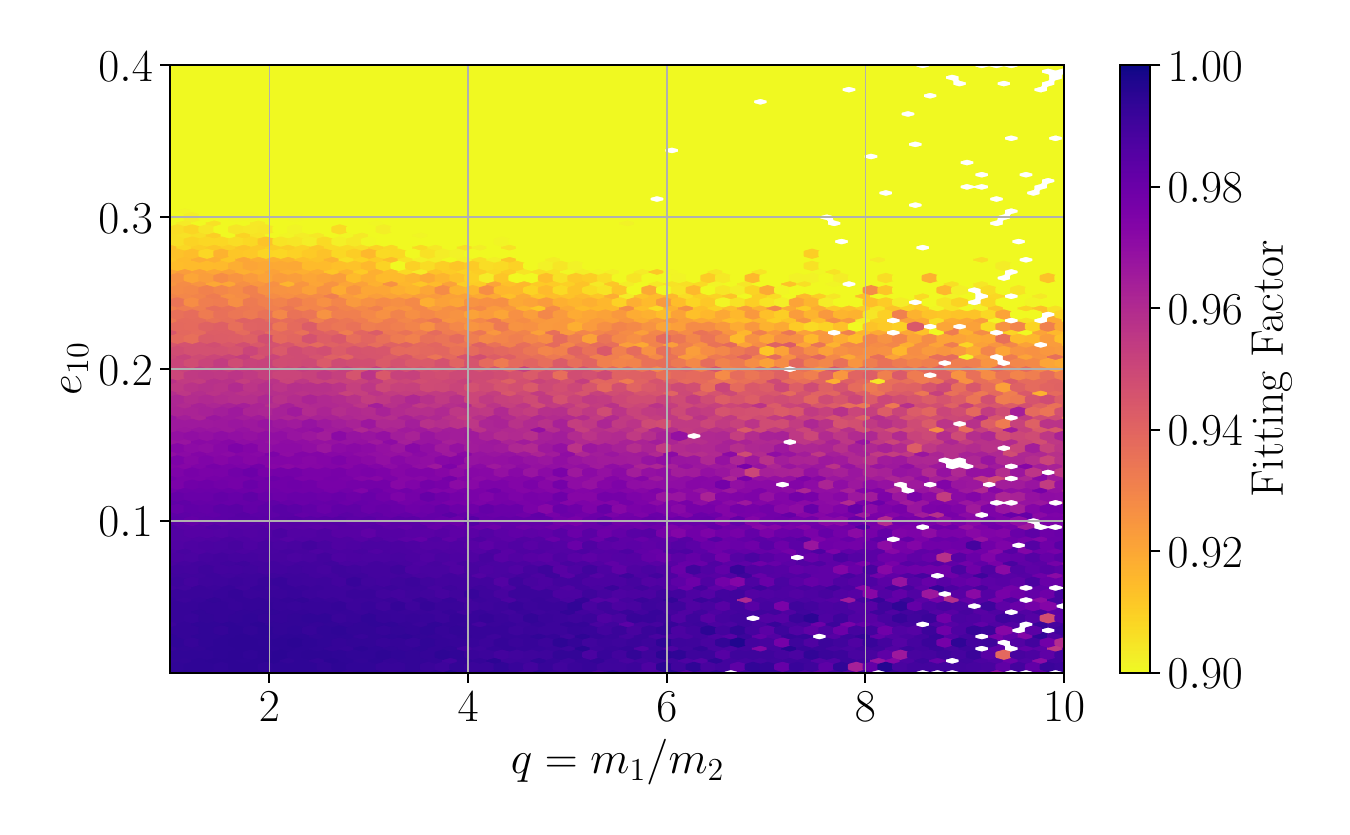}
    \includegraphics[width=0.4\linewidth]{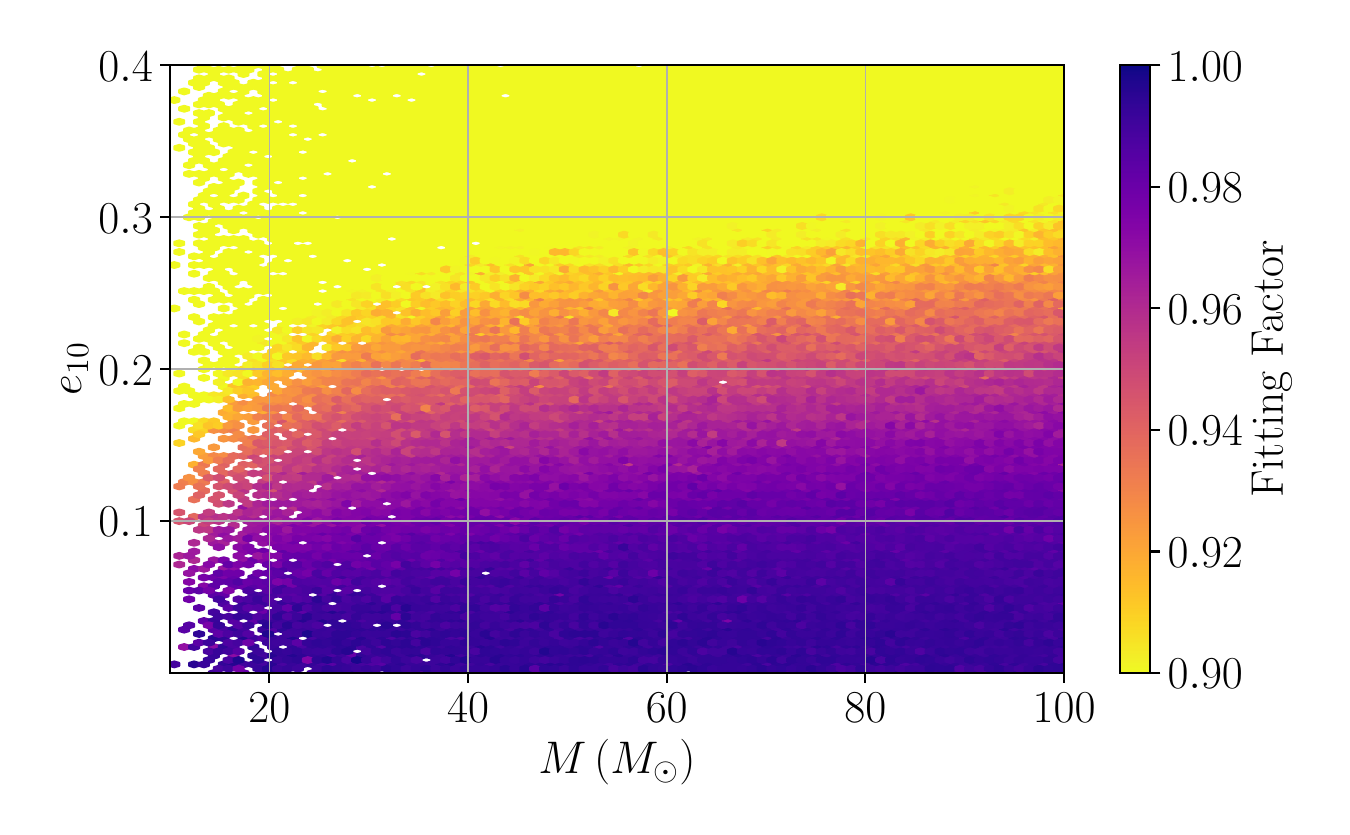}  \\
    \includegraphics[width=0.4\linewidth]{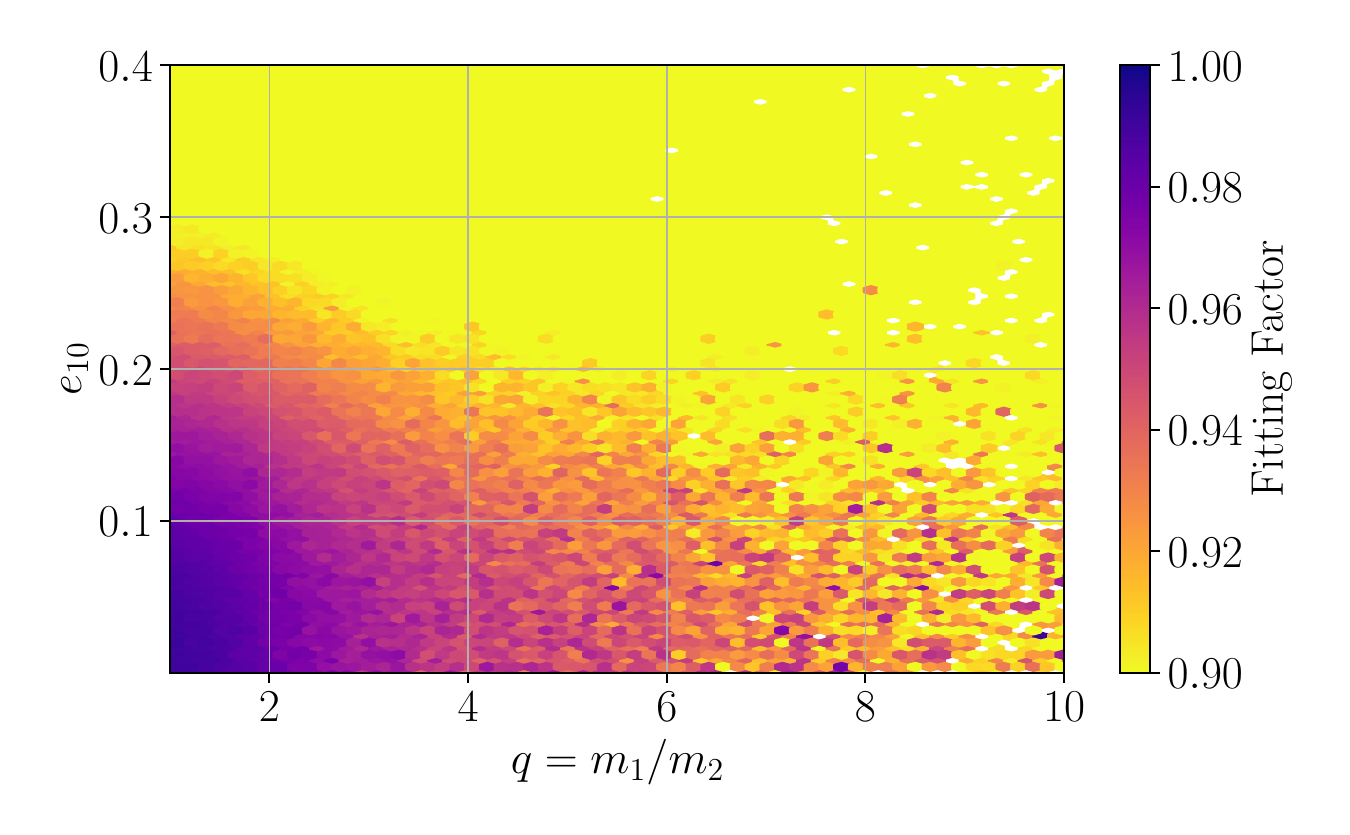}
    \includegraphics[width=0.4\linewidth]{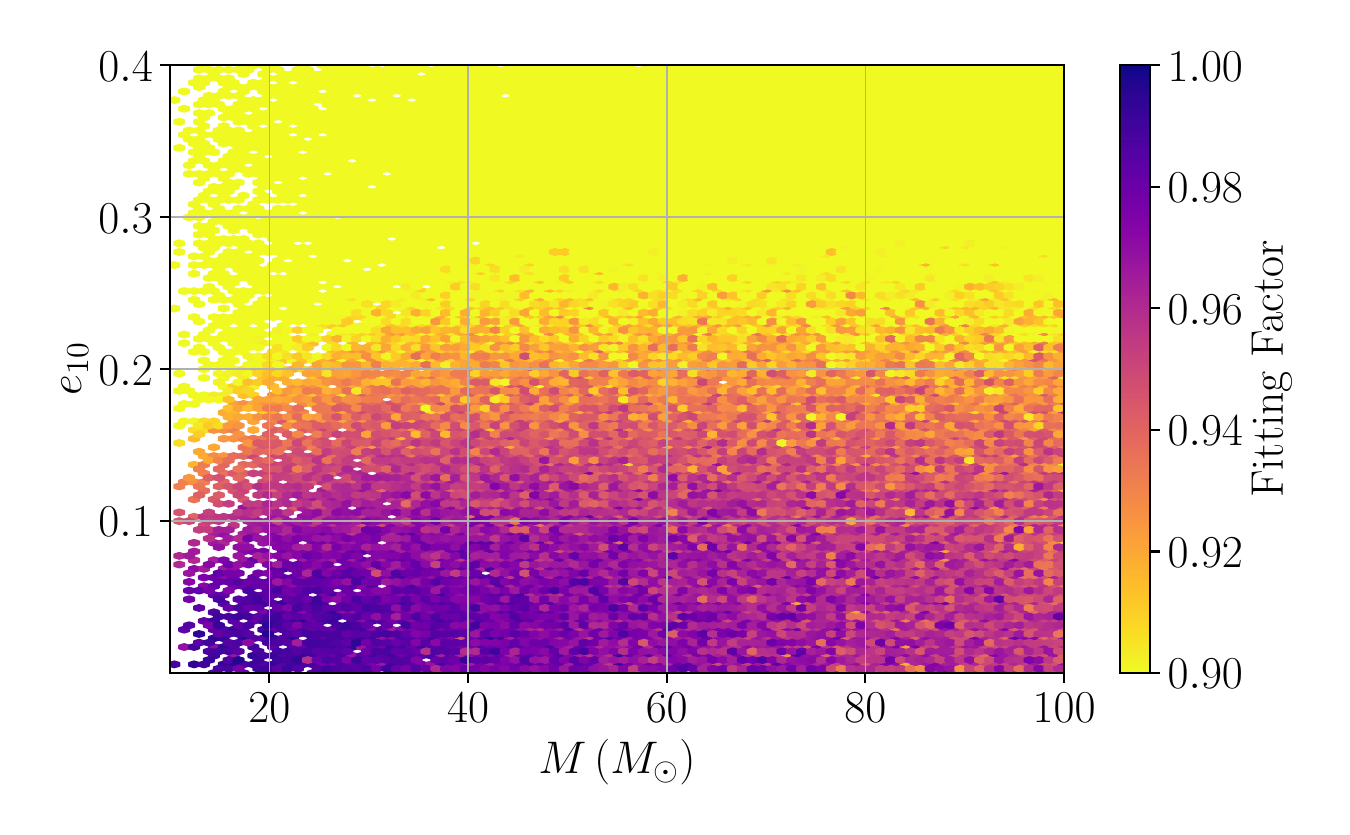}
    \caption{Effect of eccentricity and higher-order waveform modes are shown on the detectability of eccentric binary mergers by current LIGO-Virgo-KAGRA searches. We show fitting factors for eccentric dominant-mode signals in the top row, and for eccentric multi-modal signals in the bottom row. In each panel, the same is shown as a function of source eccentricity at $10$Hz, and either binary mass ratio (left column) or total mass (right column). Component masses are sampled uniformly between {\violet $5-50M_\odot$} and spins uniformly between $[-0.9, 0.9]$. 
    Even though fitting factor values go down to 0.7 or so (as seen in Fig.~\ref{fig:FF_dist}), we set the lower limit of these panels' colorbars to 0.9 in order to improve the visual contrast and highlight those parameter space regions where fitting factors remained above 0.97.
    We immediately note that for realistic signals, i.e. the bottom row, our current GW searches that employ only dominant-mode aligned-spin quasi-circular waveform templates, we will lose more than $10\%$ of signal information (optimal SNR) for sources with mass-ratio $q\gtrsim 4$, even for modest initial eccentricities. We will also lose more than 10\% of the signal information for the entire population of binary black holes that retain eccentricities $\gtrsim 0.2$ when they enter LIGO-Virgo-KAGRA band at 10Hz. These losses could amount to a $27\%$ loss in detection rate of high mass-ratio and/or moderate-to-high eccentric sources.}
    \label{fig:effectualness}
\end{figure*}

Next we look at the change in optimal SNR [as defined in Eq.~\eqref{eq:SNR_optimal}] for binaries with eccentricities compared to those without. This will help understand how much further will our current and future GW detectors be able to see eccentric binary sources compared to quasi-circular ones.
In Fig.~\ref{fig:SNR_e0} we show the relative change in optimal SNR for binaries as a function of initial eccentricity $e_{10}$. {\color{black}{The relative change in SNR is defined as $\frac{\Delta \rho}{\rho_{0}}$, where $\Delta \rho= \rho_{0}-\rho_{\rm{ecc}}$ and $\rho_{0}$, $\rho_{\rm{ecc}}$ are the optimal SNR for circular and eccentric signals, respectively.}} Each panel also shows contours (in red) of the actual optimal SNR values.
In the top row, the three panels correspond to aLIGO, ET and CE sensitivities and to $20M_\odot$ binaries. The bottom row is similarly arranged, but corresponds to more massive $60M_\odot$ sources.
The top left panel shows that eccentricity has a minimal effect in the overall loudness of signals coming from $20M_\odot$ BBHs as apparent to the current generation detectors {\color{black}{with a gain and loss in overall signal strength of up to $2.1\%$ and $0.6\%$ compared to quasi-circular sources, respectively}}. The panel below it shows that the same is true for more massive binaries as well {\color{black}{with gain and loss of up to $8.2\%$ and $6.2\%$, respectively.}}
{\color{black} The top middle panel shows that for the third generation detector Einstein Telescope we can gain up to $0.4\%$ and the loss can be $3.2\%$ in overall signal strength compared to quasi-circular sources of total mass $20M_{\odot}$. In the lower middle panel, we see a similar trend for more massive binaries as well, with the most eccentric sources recording $3\%$ larger SNRs for the most eccentric sources, corresponding to about $10\%$ larger detection volume}}. Note that these binaries have initial eccentricities defined at $5$Hz instead of $10$Hz, as will be for the next generation detectors, because they are projected to be sensitive to lower frequencies than current-generation instruments. Because of this, our sources in the middle and right columns will have lower $e_{10}$ by the time they enter the LIGO sensitive band than those in the left column.  Finally in the right top and bottom panels, we see that for Cosmic Explorer as well, eccentric sources will be up to $2-4\%$ louder than the quasi-circular ones. {\color{black}Although we mostly gain SNR, we can also lose SNR depending on the mass value and the sensitivity of the detector.}
From these results it is clear that the horizon range will be $10-20\%$ larger for eccentric sources than quasi-circular ones for next generation detectors, increasing the likelihood of our finding a large population of dynamical capture binaries.

{To assess the effect of the subdominant modes compared to the dominant (quadrupole) mode for eccentric binary sources, in Fig.~\ref{fig:SNR_ET_CE}, we show the difference in optimal SNR with and without the inclusion of subdominant modes in eccentric waveform templates.
We show this as a function of eccentricity $e$ and mass ratio $q$ in each panel, for LIGO (in the left column), ET (middle column), and CE (right column) detectors with total mass fixed to $20M_{\odot}$ (in the top row) and $60M_{\odot}$ (bottom row), dimensionless spins $(0.6,0.6)$ and source inclination angles of $\pi/3$ radians. 
As above, we keep the lower frequency cutoff $f_{\rm{low}}=10\rm{Hz}$ for LIGO and $5$Hz for CE and ET.
We immediately note that the overall signal strength (optimal SNR) is higher by up to $3\%$ for high total mass binaries of $60M_{\odot}$ when we include sub-dominant modes. The same remains below $1\%$ for $20M_{\odot}$ binaries. This indicates that similar to quasi-circular binaries, higher-order modes are strongly excited close to merger than during earlier inspiral cycles for eccentric binaries as well. While this does not seem to motivate the use of higher-order modes in templates very strongly, it mostly implies that the overall signal amplitude is minimally affected by them. {\violet Next we study the detectability of eccentric sources, which is a lot more sensitive to signal phasing than amplitude, we will find that indeed sub-dominant modes can lead to a non-trivial loss in recovered SNR for high mass-ratio systems.}}

We further our investigation by understanding the effectiveness of current GW searches on LIGO-Virgo-KAGRA data in finding eccentric binaries of spinning black holes. 
Matched-filtering searches that operated during the first three observing runs of the LIGO-Virgo-KAGRA detectors used waveforms for quasi-circular aligned-spin binaries as templates~\citep{LIGOScientific:2018mvr,LIGOScientific:2020ibl,LIGOScientific:2021djp,Nitz:2019hdf}. We therefore start with creating a standard search template bank for aligned-spin binaries~\citep{Brown:2012nn,Brown:2012qf} with individual black-hole masses between $5-50 M_{\odot}$ and $z$-component of the dimensionless spins in the range $[-0.9,0.9]$ using \texttt{PyCBC}~\citep{alex_nitz_2020_4134752}. We defer the reader to Ref.~\citep{Brown:2012qf} and references therein for details of the construction of template banks. Our bank was constructed for the design LIGO sensitivity with a design minimal match of $3\%$, and only includes the dominant $\ell = |m|=2$ modes. 
It was validated by taking $10,000$ quasi-circular signals in the same mass and spin range, using \imresigma{} (or \imresigmahm{}) to model the signals as well as filter templates. Each of these signals was filtered against all templates in the bank, recording the maximum SNR recovered for that signal over the entire bank. The ratio of that maximum recovered SNR with the optimal SNR for that signal is a quantity called the fitting factor, which measures {\it{the maximum fraction of the optimal SNR of a given signal that a given template bank is capable of recovering}}. 
A distribution of fitting factors for all our $10,000$ quasi-circular signals is shown by the green curves in Fig.~\ref{fig:FF_dist} with both signals and templates being modeled using \imresigma{} (light green) or \inspiralesigma{} (dark green). As expected more than $99\%$ of all quasi-circular signals are recovered with fitting factors above the design value of $97\%$, without higher-order modes being included in the signals. 
Next we take a set of $100,000$ eccentric signals with eccentricities $e_{10}$ distributed uniformly between $0-0.4$. For these sources, the fitting factors recovered for dominant-mode signals are shown in dark blue in Fig.~\ref{fig:FF_dist}. We immediately notice that the quasi-circular template bank recovers a much smaller fractions of optimal SNR for eccentric signals. We find that for only $40\%$ sources does the bank recover the intended $97\%$ of the optimal SNR. In fact for more than $10\%$ of the signals, at least $10\%$ of SNR is lost. 
The effect of sub-dominant mode is smaller than the overall effect of eccentricity, but still substantial. This is seen by focusing on the light blue curve and comparing it with the dark blue one in the same figure. In this case, for more than $20\%$ of the signals, at least $10\%$ of SNR is lost. Note that a $10\%$ loss in SNR results in approximately a $27\%$ reduction in overall detection volume (or observation rate, if sources are uniformly distributed in comoving volume).

Next we inspect these results in more detail in Fig.~\ref{fig:effectualness}. In the top row we show recovered fitting factors as a function of source eccentricity and mass ratio (total mass) in the left (right) panel. We find that up until eccentricities $e_{10} = 0.2$, if signals contained only the dominant $\ell=|m|=2$ modes our current aligned-spin quasi-circular template banks will recover more than $95\%$ of the optimal SNR for every signal. For larger eccentricities the loss in SNR increases rapidly. Real GW signals will of course include all possible harmonics, and that case is explored in the bottom row. The left and right panels are similar in presentation to the top panels. From the bottom row we see that for sources with mass ratios $q\gtrsim 4$, our current searches can easily miss out on $10\%$ of the optimal signal SNR for even small eccentricities. This implies a nearly $27\%$ reduction in overall rate of detection of high mass-ratio eccentric sources (with masses between $10-100M_\odot$) simply because of ignoring eccentricity in search template banks. We point out that this is an important result which strongly highlights the importance of including both orbital eccentricity effects and higher-order modes in GW models and (through them) in filter templates while searching for asymmetric ($q\gtrsim 4$) eccentric BBH signals.

\section{Conclusions and Future Outlook}
\label{sec:conclusion}

Orbital eccentricity is a unique signature of compact binaries formed in dense stellar environments via dynamical processes. Such binaries are expected to comprise an important sub-population of all signals that ground-based GW detectors are expected to observe. Contemporary GW searches (and signal analyses in general) predominantly rely on waveform models tailored to quasi-circular binaries, primarily because quasi-circular binaries are expected to remain a majority of all detected GW signals, and also because the dimensionality of their parameter space is more manageable. {\violet Theoretical IMR modeling of eccentric sources was largely ignored in literature in the pre-GW150914 detection era~\citep{LIGOScientific:2016aoc}, in favor of modeling non-eccentric sources with very high precision.}

{\violet However, since GW150914, more than 90 signals having been already seen during the first three observation runs of the LIGO-Virgo-KAGRA instruments, we have clearly entered a signal dominated era.} It is now timely to target astrophysically interesting sub-populations of compact binaries, such as eccentric BBHs. {\violet Many waveform models have also been under active development in the past few years catering to eccentric orbits ~\citep{Konigsdorffer:2006zt,Yunes:2009yz,Klein:2010ti,Mishra:2015bqa,Moore:2016qxz, Tanay:2016zog, Klein:2018ybm,Boetzel:2019nfw, Ebersold:2019kdc, Moore:2019xkm, Klein:2021jtd, Khalil:2021txt, Paul:2022xfy,Nagar:2022fep, Henry:2023tka, Albanesi:2023bgi,Albertini:2023aol, Nagar:2024dzj, Nagar:2024oyk, Klein:2013qda,Huerta:2014eca,Moore:2018kvz,Klein:2018ybm,Tanay:2019knc,Liu:2019jpg,Tiwari:2020hsu,Klein:2021jtd,Huerta:2016rwp,Hinderer:2017jcs,Huerta:2017kez,Hinder:2017sxy,Huerta:2017kez,Cao:2017ndf,Taracchini:2012ig,Chen:2020lzc,Chiaramello:2020ehz,Nagar:2018zoe, Nagar:2020pcj,Nagar:2021gss,Ramos-Buades:2021adz,Chattaraj:2022tay, Manna:2024ycx, Carullo:2023kvj, Carullo:2024smg,Islam:2021mha,Yun:2021jnh,Becker:2024xdi}.}
In this paper, we introduce \esigmahm{}, a time domain IMR waveform model for binaries of spinning black-holes on eccentric orbits, including the effect of higher-order modes.  This model builds on the \enigma{} framework that was the very first IMR waveform model for eccentric binaries~\citep{Huerta:2016rwp}.

{\violet\esigmahm{} is composed of two segments - an inspiral part attached to a PMR model.} The inspiral segment is built upon the results coming from the post-Newtonian (PN)~\citep{Blanchet:2013haa}, self-force approach and {{black hole perturbation theory (BHPT)}} developed in the last several decades. The gravitational wave strain is obtained from spherical harmonic modes following Eq.~\eqref{eq:GW_strain}. The instantaneous contribution to modes including the effect of spin and eccentricity are included up to 3.5PN order~\citep{Mishra:2015bqa, Paul:2022xfy, Henry:2023tka}. Additionally, the spinning and non-spinning hereditary contributions for quasi-circular orbits are included up to 3.5PN order following~\citep{Henry:2022dzx}. Recently reported~\citep{Blanchet:2023bwj, Blanchet:2023sbv} quasi-circular non-spinning 4PN piece in the $\ell=|m|=2$ modes is also included in our framework. The conservative and radiative dynamics of the system is governed by the evolution equations $\dot{\phi}$, $\dot{l}$, and $\dot{x}$, $\dot{e}$, respectively. Though the effect of spins and eccentricity are included up to 3PN in evolution equations, the effect of spins in quasi-circular orbits is updated up to 4PN in $\dot{x}$. Besides, the 4PN non-spinning quasi-circular piece is also included in $\dot{x}$ following~\citep{Blanchet:2023sbv,Blanchet:2023bwj}. The 3.5PN cubic-in-spin, 4PN non-spinning quasi-circular pieces in $\dot{x}$ are new and we list them in Appendix~\ref{sec:appendix}.
For the PMR part, we inherit the approximation of \enigma{} that the binary has circularized (i.e., it has radiated away its eccentricity via GW emission) by the time it has entered the sensitivity band of the ground-based detectors.
In Fig.~\ref{fig:orbital_evol} the time evolution of orbital angular velocity and the amplitude of the eccentric spinning (grey lines) and quasi-circular non-spinning (brown line), NR simulations are shown. It can be observed that at the transition from late inspiral to PMR (denoted by filled circles) most systems have decayed nearly all their eccentricity. {\violet Following this approximation, we use \nrsurdqfour{}~\citep{Varma:2019csw} - a state-of-the-art spinning quasi-circular surrogate model - to provide the PMR piece.\footnote{\violet EOB models of the SEOB and TEOB families attach a quasi-circular ringdown model at the peak of respective $(\ell, m)$ mode amplitudes. Their inspiral-merger piece is computed based on an eccentric dynamics. Smooth matching is achieved by applying next-to-quasi-circular corrections during the plunge. In \esigmahm{}, however, quasi-circular motion is assumed for a relatively larger portion of the signal, approximately starting from the Kerr or Schwarzschild ISCO.
}}

We quantify the impact of changes employed in \esigma{} over its predecessor \enigma{}~\citep{Chen:2020lzc}. This is done by computing fractional change in optimal SNR and difference in number of GW cycles between \enigma{} and \esigma{}, which are shown in Fig.~\ref{fig:SNR_Ngw}. 
We subsequently validate \imresigma{} in the quasi-circular limit. We first compare the model with existing \texttt{Phenom} and \texttt{EOB} models such as \imrphenom{}, \seobnrvfour{}, and \seobnrvfive{} by computing mismatches against them for a fixed total mass of $40 M_{\odot}$ and dimensionless spins (0.5, 0.5) in Fig.~\ref{fig:mismatch_qc}. We next compare \imresigma{} with publicly available non-eccentric NR simulations from the SXS catalog~\citep{SXS:catalog} (see Table~\ref{table:sxs_id_qc_nr_4PN} of Appendix~\ref{sec:appendix_NR}). We show a visual comparison in Fig.~\ref{fig:NR_comp_p1} and show mismatches against the same simulations as a function of binary total mass in Fig.~\ref{fig:NR_qc_mismatches}.
Next, we quantify the agreement between \imresigma{} with publicly available spinning {\it{eccentric}} NR simulations from the SXS catalog by comparing strain data $(h_{+})$ in Fig.~\ref{fig:NR_comp}. We also perform mismatch computation using only the dominant $\ell=|m|=2$ modes as well as with higher-order modes in Fig.~\ref{fig:mismatch_NR_mass} \& \textit{left} panel of Fig.~\ref{fig:mismatch_anti_aligned}. 
Table~\ref{table:sxs_id} of Appendix~\ref{sec:appendix_NR} outlines the SXS IDs, and the values of reference mass ratio, reference eccentricity, $z$-component of reference dimensionless spins, number of orbits, and the corresponding match values with our model for all the 32 eccentric NR simulations that have been used in the comparison.
Our validation studies indicate that \imresigma{} models moderately eccentric binaries well. It shows some disagreement with NR simulations for binaries with high mass-ratios {\it{and}} large anti-aligned spins. We therefore recommend the use of \imresigma{} to model the GW emission from moderately eccentric moderately spinning binary mergers, as well as for eccentric binaries with large positive-aligned spins.

We assess the impact of eccentricity by investigating its effect on signal length for current (aLIGO) and future (ET, CE) ground-based detectors, as measured in the number of GW cycles, in Fig.~\ref{fig:Ngw_M_q_e}.
We subsequently compute the change in overall signal strength due to the presence of orbital eccentricity in Fig.~\ref{fig:SNR_e0}.
We also assess the importance of inclusion of HMs in the waveform model by computing the change in overall optimal SNR due to the inclusion of higher-order modes for current and third-generation ground-based detectors in Fig.~\ref{fig:SNR_ET_CE}. 
Finally we investigate the impact of not including orbital eccentricity effects and higher-order modes in contemporary GW searches. We quantify the resulting SNR loss via fitting factors and show them in Figs.~\ref{fig:FF_dist} and~\ref{fig:effectualness}.

These investigations strongly indicate that orbital eccentricity should not be ignored in waveform models, and there is an urgent need to include its effects in GW searches and data analysis efforts. Our new waveform model \imresigma{} that includes effects such as spin, eccentricity, and higher-order modes is targeted as a major milestone along this direction. \imresigma{} can be employed for detection and analysis of the eccentric spinning GW signals. It can also help us understand eccentric mergers better and to constrain the eccentricity of BBH mergers detected by LIGO-Virgo-KAGRA detectors in the past as well as in the future. We show the mean evaluation cost of \imresigmahm{} as a function of total mass $M$ with lower cutoff frequency fixed to $10$Hz in Fig.~\ref{fig:cost} of Appendix~\ref{sec:appendix_cost}.

Having presented an upgraded IMR, spinning, eccentric waveform model with higher-order gravitational wave modes here, we plan to improve the same by comparison and calibration to a larger population of eccentric NR simulations in future work.
Additionally, \esigma{} currently allows for component spins that are parallel or anti-parallel to orbital angular momentum. We plan to include the effect of arbitrary orientation of spins about orbital angular momentum (spin precession) in future work. {\violet Besides, the \esigma{} framework aims to iteratively improve waveform models for the eccentric inspiral and plunge stages, with the ultimate goal of utilizing eccentric NR surrogates to accurately describe the plunge-merger-ringdown.}
\section{Acknowledgements}
We thank K. G. Arun and Guillaume Faye for discussions and insights into the modelling of spinning compact binaries. {We are also thankful to the SXS Collaboration for making  public a catalog of numerical relativity waveforms.} K.P. thanks members of the Gravitation and Cosmology group at the Department of Physics, Indian Institute of Technology (IIT) Madras, for useful discussions. C.K.M acknowledges the hospitality of Institut d'Astrophysique de Paris and Max Planck Institute for Gravitational Physics
(Albert Einstein Institute, Potsdam) during the final stages of writing of the paper. K.P. thanks the International Centre for Theoretical Sciences (ICTS) for the hospitality during the academic visits. 
A.M. and P.K. acknowledge support of the Department of Atomic Energy, Government of India, under project no. RTI4001 at the ICTS. P.K. also acknowledges support by the Ashok and Gita Vaish Early Career Faculty Fellowship at ICTS. C.K.M. acknowledges the support of SERB’s Core Research Grant No. CRG/2022/007959. Calculations in this paper were performed on the Sonic computing cluster at ICTS, and the Powehi workstation at IIT Madras. The authors are also grateful for computational resources provided by the LIGO Laboratory and supported by National Science Foundation Grants PHY-0757058 and PHY-0823459.
This document has LIGO preprint number LIGO-P2400383. We thank Antoni Ramos-Buades for useful comments and suggestions on our manuscript. K. S. acknowledges support from the Long Term Visiting Students Program at ICTS during the course of this work.

\appendix
\section{Spherical harmonic mode amplitudes}
\label{sec:appendix_hlm}
The explicit expression of the 3.5PN term of the instantaneous part of the $(2,2)$ mode is given in terms of general dynamical variables such as relative separation $r$, radial and orbital angular velocities $\dot{r}$, and $\dot{\phi}$ below, as,
\begin{align}
h^{22}_{\rm{3.5PN}} &= \frac{4 G M \nu}{c^4 R}\sqrt{\frac{\pi}{5}}e^{-2i\phi}H^{22}_{\rm{3.5PN}}
\end{align}
where,
\begin{widetext}
\begin{subequations}
\begin{align}
    H^{22}_{\rm{3.5PN}} &= \frac{G^2 M^2}{158760 c^7 r^4} \Bigg\{r^2 \Big(r \dot{\phi}\Big)^5 \Big(2940444 i \nu -1747608 i \nu ^2\Big)+r^2 \dot{r}^2 \Big(r \dot{\phi}\Big)^3 \Big(-10876680 i \nu -3279456 i \nu ^2\Big)+r^2
   \dot{r}^4 \Big(r \dot{\phi}\Big)\, \nonumber \\
   & \times \Big(-477504 i \nu -3520128 i \nu ^2\Big)+r^2 \dot{r}^5 \Big(362448 \nu +2064168 \nu ^2\Big)+r^2 \dot{r} \Big(r \dot{\phi}\Big)^4
   \Big(-5327910 \nu +2851968 \nu ^2\Big)\, \nonumber \\
   &+r^2 \dot{r}^3 \Big(r \dot{\phi}\Big)^2 \Big(-785412 \nu +4206756 \nu ^2\Big)+\bigg[r^2 \dot{r}^2 \Big(r \dot{\phi}\Big)^3 \delta 
   \Big(-665280+1942290 \nu -2994390 \nu ^2\Big)+r^2 \dot{r}^4 \Big(r \dot{\phi}\Big)\, \nonumber \\
   & \times \delta  \Big(463050-675045 \nu -2586780 \nu ^2\Big)+r^2 \dot{r}^3
   \Big(r \dot{\phi}\Big)^2 \delta  \Big(-1542240 i+3923640 i \nu +517860 i \nu ^2\Big)+r^2 \dot{r}^5 \delta \, \nonumber \\
   & \times \Big(-156870 i+715680 i \nu -758520 i \nu ^2\Big) +r^2
   \dot{r} \Big(r \dot{\phi}\Big)^4 \delta  \Big(-782460 i+1428210 i \nu +855540 i \nu ^2\Big)+r^2 \Big(r \dot{\phi}\Big)^5 \delta\, \nonumber \\
   & \times  \Big(-657720+2469285 \nu +207900 \nu
   ^2\Big)\bigg] \Big(\bm{\hat{\ell}}\cdot \bm{\chi}_{a}\Big) +\bigg[r^2 \dot{r}^4 \Big(r \dot{\phi}\Big) \Big(463050-2737035 \nu +1731240 \nu ^2-3533040 \nu ^3\Big)\, \nonumber \\
   & +r^2 \dot{r}^2
   \Big(r \dot{\phi}\Big)^3 \Big(-665280+2757510 \nu +3187170 \nu ^2 -120960 \nu ^3\Big) +r^2 \dot{r}^5 \Big(-156870 i+725760 i \nu -1433880 i \nu ^2 \, \nonumber \\
   &+211680 i \nu
   ^3\Big)+r^2 \dot{r}^3 \Big(r \dot{\phi}\Big)^2 \Big(-1542240 i+7585200 i \nu -5686380 i \nu ^2 +11511360 i \nu ^3\Big)+r^2 \dot{r} \Big(r \dot{\phi}\Big)^4 \Big(-782460
   i \, \nonumber \\
   &+5438790 i \nu -17554320 i \nu ^2+16177140 i \nu ^3\Big)+r^2 \Big(r \dot{\phi}\Big)^5 \Big(-657720+4184145 \nu -9662940 \nu ^2+6790140 \nu ^3\Big)\bigg] \, \nonumber \\
   & \times  \Big(\bm{\hat{\ell}}\cdot \bm{\chi}_{s}\Big) +G M \bigg[r \dot{r} \Big(r \dot{\phi}\Big)^2 \Big(49728360 \nu -11712024 \nu ^2\Big)+r \dot{r}^3 \Big(4810464 \nu -1274976 \nu ^2\Big)+r
   \Big(r \dot{\phi}\Big)^3 \Big(13949520 i \nu \, \nonumber \\
   &-2288088 i \nu ^2\Big)+r \dot{r}^2 \Big(r \dot{\phi}\Big) \Big(-20524176 i \nu +3674016 i \nu ^2\Big) +\bigg[r \Big(r \dot{\phi}\Big)^3
   \delta  \Big(384300+3891405 \nu -746340 \nu ^2\Big) \, \nonumber \\
   &+r \dot{r}^3 \delta  \Big(-1582560 i+6296640 i \nu +1347360 i \nu ^2\Big)+r \dot{r} \Big(r \dot{\phi}\Big)^2
   \delta \Big(1499400 i-5015640 i \nu -2611560 i \nu ^2\Big) \, \nonumber \\
   & +r \dot{r}^2 \Big(r \dot{\phi}\Big) \delta  \Big(-587160+8890140 \nu +4601100 \nu ^2\Big)\bigg]
   \Big(\bm{\hat{\ell}}\cdot \bm{\chi}_{a}\Big)+\bigg[r \Big(r \dot{\phi}\Big)^3 \Big(384300+8644545 \nu +7772100 \nu ^2 \, \nonumber \\
   &-7546140 \nu ^3\Big) +r \dot{r}^3 \Big(-1582560 i+7691040 i \nu -7203840 i \nu
   ^2 +1285200 i \nu ^3\Big)+r \dot{r} \Big(r \dot{\phi}\Big)^2 \Big(1499400 i-4064760 i \nu \, \nonumber \\
   &+24863160 i \nu ^2 -18698400 i \nu ^3\Big)+r \dot{r}^2 \Big(r \dot{\phi}\Big)
   \Big(-587160+5944680 \nu -10484040 \nu ^2+4062240 \nu ^3\Big)\bigg] \Big(\bm{\hat{\ell}}\cdot \bm{\chi}_{s}\Big)\bigg] \, \nonumber \\
   &+G^2 M^2 \bigg[\dot{r} \Big(5779200 \nu -93240 \nu
   ^2\Big)+\Big(r \dot{\phi}\Big) \Big(5816048 i \nu -2415840 i \nu ^2\Big) +\bigg[\Big(r \dot{\phi}\Big) \delta  \Big(289380+4509120 \nu \, \nonumber \\
   &-765660 \nu ^2\Big)+\dot{r} \delta 
   \Big(-6727560 i +5714100 i \nu -1145760 i \nu ^2\Big)\bigg] \Big(\bm{\hat{\ell}}\cdot \bm{\chi}_{a}\Big)+\bigg[\dot{r} \bigg[-238140 i \lambda_{1} +238140 i \lambda_{2} \, \nonumber \\
   &+\big(714420 i
   \kappa_{1}-714420 i \kappa_{2}+714420 i \lambda_{1} -714420 i \lambda_{2}\big) \nu +\delta  \bigg[-238140 i \lambda_{1} -238140 i \lambda_{2}+\big(714420 i \kappa_{1} \, \nonumber \\
   &+714420 i \kappa_{2}+238140 i \lambda_{1}+238140 i \lambda_{2} \big) \nu \bigg]\bigg]+\Big(r \dot{\phi}\Big) \bigg[-238140 \lambda_{1}+238140 \lambda_{2}  +\big(714420 \kappa_{1}-714420 \kappa_{2} \, \nonumber \\
   &+714420 \lambda_{1}-714420 \lambda_{2}\big) \nu +\delta  \bigg[-238140
   \lambda_{1}-238140 \lambda_{2} +\big(714420 \kappa_{1}+714420 \kappa_{2} +238140 \lambda_{1}+238140 \lambda_{2}\big) \nu \bigg]\bigg]\bigg] \, \nonumber \\
   & \times \Big(\bm{\hat{\ell}}\cdot \bm{\chi}_{a}\Big)^3+\bigg[\dot{r} \Big(-6727560 i+12494580 i \nu -5984160 i \nu ^2 +6417600 i \nu ^3\Big)+\Big(r \dot{\phi}\Big)  \Big(289380-1871520 \nu \, \nonumber \\
   & -9697800 \nu ^2+4001760 \nu
   ^3\Big)+\bigg[\Big(r \dot{\phi}\Big) \bigg[-714420 \lambda_{1}-714420 \lambda_{2} +\big(1111320 \kappa_{1}+1111320 \kappa_{2} +2143260 \lambda_{1} \, \nonumber \\
   &+2143260 \lambda_{2}\big) \nu +(-1587600-793800 \kappa_{1}-793800 \kappa_{2}) \nu ^2+\delta  \bigg[-714420 \lambda_{1} +714420 \lambda_{2}+ \big(1111320 \kappa_{1} \, \nonumber \\
   & -1111320 \kappa_{2}+714420 \lambda_{1}-714420 \lambda_{2}) \nu \bigg]\bigg]+\dot{r} \bigg[-714420 i \lambda_{1}-714420 i \lambda_{2}+\big(1508220 i \kappa_{1} +1508220 i \kappa_{2} \, \nonumber \\
   &+2143260 i \lambda_{1} +2143260 i \lambda_{2}\big) \nu +\big(-3175200
   i-1587600 i \kappa_{1}-1587600 i \kappa_{2}\big) \nu ^2+\delta  \bigg[-714420 i \lambda_{1} \, \nonumber \\
   &+714420 i \lambda_{2} +\big(1508220 i \kappa_{1} -1508220
   i \kappa_{2}+714420 i \lambda_{1}-714420 i \lambda_{2}\big) \nu \bigg]\bigg]\bigg] \Big(\bm{\hat{\ell}}\cdot \bm{\chi}_{a}\Big)^2\bigg] \Big(\bm{\hat{\ell}}\cdot \bm{\chi}_{s}\Big)+\bigg[\dot{r}\, \nonumber \\
   & \times
   \bigg[-714420 i \lambda_{1}+714420 i \lambda_{2} +\big(873180 i \kappa_{1} -873180 i \kappa_{2}+2143260 i \lambda_{1}-2143260 i
   \lambda_{2}\big) \nu +\big(-3175200 i \kappa_{1} \, \nonumber \\
   &+3175200 i \kappa_{2}\big) \nu ^2+\delta  \bigg[-714420 i \lambda_{1}-714420 i \lambda_{2} + \big(873180 i
   \kappa_{1}+873180 i \kappa_{2}+714420 i \lambda_{1}+714420 i \lambda_{2}\big) \nu \bigg]\bigg] \, \nonumber \\
   &+\Big(r \dot{\phi}\Big) \bigg[-714420 \lambda_{1}+714420 \lambda_{2}+(79380 \kappa_{1}-79380 \kappa_{2} +2143260 \lambda_{1}-2143260 \lambda_{2}) \nu +(-1587600 \kappa_{1} \, \nonumber \\
   &+1587600 \kappa_{2}) \nu ^2+\delta  \bigg[-714420 \lambda_{1}-714420 \lambda_{2}+ \big(79380 \kappa_{1}+79380 \kappa_{2} +714420
   \lambda_{1}+714420 \lambda_{2}\big) \nu \bigg]\bigg]\bigg] \Big(\bm{\hat{\ell}}\cdot \bm{\chi}_{a}\Big) \, \nonumber \\
   & \times  \Big(\bm{\hat{\ell}}\cdot \bm{\chi}_{s}\Big)^2+\bigg[\Big(r \dot{\phi}\Big) \bigg[-238140 \lambda_{1}-238140
   \lambda_{2}+\big(-317520 \kappa_{1}-317520 \kappa_{2} +714420 \lambda_{1}+714420 \lambda_{2}\big) \nu \, \nonumber \\
   &+\big(1587600-793800 \kappa_{1}-793800 \kappa_{2}\big) \nu ^2+\delta  \bigg[-238140 \lambda_{1}+238140 \lambda_{2}+\big(-317520 \kappa_{1} +317520 \kappa_{2} +238140
   \lambda_{1} \, \nonumber \\
   &-238140 \lambda_{2}\big) \nu \bigg]\bigg]+\dot{r} \bigg[-238140 i \lambda_{1}-238140 i \lambda_{2}+\big(79380 i \kappa_{1}+79380 i
   \kappa_{2}+714420 i \lambda_{1}+714420 i \lambda_{2}\big) \nu \, \nonumber \\
   &+\big(3175200 i-1587600 i \kappa_{1}-1587600 i \kappa_{2}\big) \nu ^2+\delta 
   \bigg[-238140 i \lambda_{1}+238140 i \lambda_{2}+\big(79380 i \kappa_{1}-79380 i \kappa_{2} \, \nonumber \\
   &+238140 i \lambda_{1} -238140 i \lambda_{2}\big)
   \nu \bigg]\bigg]\bigg] \Big(\bm{\hat{\ell}}\cdot \bm{\chi}_{s}\Big)^3\bigg]\Bigg\}\,,
\end{align}
\end{subequations}
\end{widetext}
where, $\nu=m_1 m_2/M^2$, $\delta=(m_1 - m_2)/M$, $\bm{\chi}_{s}=\frac{1}{2}(\bm{\chi}_1 + \bm{\chi}_2)$, $\bm{\chi}_{a}=\frac{1}{2}(\bm{\chi}_1 - \bm{\chi}_2)$. Here, $m_1$, $m_2$ are the component masses of the binary and $M=m_1 +m_2$ is the total mass of the source, $R$ is the distance to the source, $G$ and $c$ are the universal gravitational constant and speed of light respectively. Also, $\bm{\hat{\ell}}\,,\bm{\chi}_{1}\,,\bm{\chi}_{2}\,\,$ denote the unit vector along orbital angular momentum vector and dimensionless spin vectors, respectively. Finally, the constants $\kappa_A$ and $\lambda_A$ represent the spin-induced quadrupolar and octupolar deformations of body $A$.

\section{Derived Quantities in Orbital Dynamics}
\label{sec:appendix}
    The derived pieces in the evolution equation of $\dot{x}$ are given here. The spinning quasi-circular pieces up to 4PN have been given in Ref.~\citep{Cho:2022syn} (except cubic-in-spin; SSS), however, we list them in addition to newly derived non-spinning and spinning pieces for convenience of the reader. The explicit expressions for relevant pieces of $M\dot{x}$ are given as, 
    \begin{widetext}
\begin{subequations}
    \begin{align}
    \dot{x}^{\rm{SO}}_{\rm{3.5PN}} &= \frac{1}{18144}\Bigg[-3 \delta  \Big(1042600-2388207 \nu +700133 \nu ^2\Big) \Big(\bm{\hat{\ell}}\cdot\bm{\chi}_{a}\Big)+\Big(-3127800+11028703 \nu -7176813 \nu ^2 \nonumber \\
    &+908796 \nu ^3\Big) \Big(\bm{\hat{\ell}}\cdot\bm{\chi}_{s}\Big)\Bigg]\,, \\
   \dot{x}^{\rm{SS}}_{\rm{3.5PN}} &= \frac{1}{8} \pi  \Bigg\{2 \Big(\bm{\hat{\ell}}\cdot\bm{\chi}_{a}\Big) \Big(\bm{\hat{\ell}}\cdot\bm{\chi}_{s}\Big) \bigg[\delta +80 \kappa_1 (1+\delta -2 \nu
   )-80 \kappa_2 (1-\delta -2 \nu )\bigg]+\Big(\bm{\hat{\ell}}\cdot\bm{\chi}_{a}\Big)^2 \bigg[1+80 \kappa_1 (1+\delta -2 \nu ) \nonumber \\
   &-320
   \nu +80 \kappa_2 (1-\delta -2 \nu )\bigg]+\Big(\bm{\hat{\ell}}\cdot\bm{\chi}_{s}\Big)^2 \bigg[1+80 \kappa_1 (1+\delta -2 \nu
   )+316 \nu +80 \kappa_2 (1-\delta -2 \nu )\bigg]\Bigg\}\,, \\
   \dot{x}^{\rm{SSS}}_{\rm{3.5PN}} &= \frac{1}{24} \Bigg\{\bigg[-508 \kappa_1+508 \kappa_2-264 \lambda_1+264 \lambda_2+(2826
   \kappa_1-2826 \kappa_2+792 \lambda_1-792 \lambda_2) \nu +\delta  \bigg[-15-508
   \kappa_1 \nonumber \\
   &-508 \kappa_2-264 \lambda_1-264 \lambda_2+(2036+1810 \kappa_1+1810
   \kappa_2+264 \lambda_1+264 \lambda_2) \nu \bigg]\bigg] \Big(\bm{\hat{\ell}}\cdot\bm{\chi}_{a}\Big)^3+\bigg[-45-1524
   \kappa_1 \nonumber \\
   &-1524 \kappa_2-792 \lambda_1-792 \lambda_2+(2165+6622 \kappa_1+6622
   \kappa_2+2376 \lambda_1+2376 \lambda_2) \nu +(-2992-1496 \kappa_1 \nonumber \\
   &-1496
   \kappa_2) \nu ^2+\delta  \bigg[-1524 \kappa_1+1524 \kappa_2-792 \lambda_1+792
   \lambda_2+(3574 \kappa_1-3574 \kappa_2+792 \lambda_1-792 \lambda_2) \nu
   \bigg]\bigg] \Big(\bm{\hat{\ell}}\cdot\bm{\chi}_{a}\Big)^2 \nonumber \\
   & \times \Big(\bm{\hat{\ell}}\cdot\bm{\chi}_{s}\Big)+\bigg[-1524 \kappa_1+1524 \kappa_2-792
   \lambda_1+792 \lambda_2+(4766 \kappa_1-4766 \kappa_2+2376 \lambda_1-2376
   \lambda_2) \nu +(-2992 \kappa_1 \nonumber \\
   &+2992 \kappa_2) \nu ^2+\delta  \bigg[-45-1524
   \kappa_1-1524 \kappa_2-792 \lambda_1-792 \lambda_2+(-1958+1718
   \kappa_1+1718 \kappa_2+792 \lambda_1 \nonumber \\
   &+792 \lambda_2) \nu \bigg]\bigg] \Big(\bm{\hat{\ell}}\cdot\bm{\chi}_{a}\Big)
   \Big(\bm{\hat{\ell}}\cdot\bm{\chi}_{s}\Big)^2+\bigg[-15-508 \kappa_1-508 \kappa_2-264 \lambda_1-264
   \lambda_2+(-1967+970 \kappa_1+970 \kappa_2 \, \nonumber \\
   &+792 \lambda_1+792 \lambda_2) \nu
   +(2956-1496 \kappa_1-1496 \kappa_2) \nu ^2+\delta  \bigg[-508 \kappa_1+508
   \kappa_2-264 \lambda_1+264 \lambda_2+(-46 \kappa_1+46 \kappa_2 \nonumber \\
   &+264
   \lambda_1-264 \lambda_2) \nu \bigg]\bigg]\Big(\bm{\hat{\ell}}\cdot\bm{\chi}_{s}\Big)^3\Bigg\}\,, \\
        \dot{x}^{\rm{NS}}_{\rm{4PN}} &= \Bigg[\frac{917 \nu ^4}{1152}-\frac{1909807 \nu ^3}{62208}+\frac{504221849 \nu ^2}{326592}+\pi ^2
   \Big(-\frac{22099 \nu ^2}{384}-\frac{1472377 \nu }{16128}-\frac{361}{126}\Big)+\nu 
   \bigg[-\frac{58250 \gamma _E}{441} \,\nonumber \\
   &-\frac{29125 \log
   (x)}{441}-\frac{317589100793}{391184640}+\frac{47385 \log (3)}{392}-\frac{850042 \log
   (2)}{2205}\bigg]+\frac{124741 \gamma _E}{4410}+\frac{124741 \log
   (x)}{8820}\, \nonumber \\
   &+\frac{3959271176713}{25427001600}-\frac{47385 \log (3)}{1568}+\frac{127751 \log (2)}{1470}\Bigg]\,, \\
   \dot{x}^{\rm{SO}}_{\rm{4PN}} &= \frac{\pi}{2016}  \bigg[\delta  (-307708+660165 \nu ) \Big(\bm{\hat{\ell}}\cdot\bm{\chi}_{a}\Big)+\Big(-307708+1035821 \nu -411636 \nu
   ^2\Big) \Big(\bm{\hat{\ell}}\cdot\bm{\chi}_{s}\Big)\bigg]\,, \\
   \dot{x}^{\rm{SS}}_{\rm{4PN}} &= \frac{1}{72576}\Bigg\{\bigg[41400957+5338168 \kappa_1+5338168 \kappa_2+(-217740271-19802588
   \kappa_1-19802588 \kappa_2) \nu \nonumber \\
   &+(172641789+16943508 \kappa_1+16943508 \kappa_2)
   \nu ^2+(-6318144-3159072 \kappa_1-3159072 \kappa_2) \nu ^3 \nonumber \\
   &+\delta  \bigg[5338168
   \kappa_1-5338168 \kappa_2+(-9126252 \kappa_1+9126252 \kappa_2) \nu +(4589172
   \kappa_1-4589172 \kappa_2) \nu ^2\bigg]\bigg] \Big(\bm{\hat{\ell}}\cdot\bm{\chi}_{a}\Big)^2 \nonumber \\
   &+\bigg[10676336
   \kappa_1-10676336 \kappa_2+(-39605176 \kappa_1+39605176 \kappa_2) \nu +(33887016
   \kappa_1-33887016 \kappa_2) \nu ^2 \nonumber \\
   &+(-6318144 \kappa_1+6318144 \kappa_2) \nu
   ^3+\delta  \bigg[82801914+10676336 \kappa_1+10676336 \kappa_2+(-162578766 \nonumber \\
   &-18252504
   \kappa_1-18252504 \kappa_2) \nu +(57322314+9178344 \kappa_1+9178344 \kappa_2)
   \nu ^2\bigg]\bigg] \Big(\bm{\hat{\ell}}\cdot\bm{\chi}_{a}\Big) \Big(\bm{\hat{\ell}}\cdot\bm{\chi}_{s}\Big) \nonumber \\
   &+\bigg[41400957+5338168 \kappa_1+5338168
   \kappa_2+(-110442323-19802588 \kappa_1-19802588 \kappa_2) \nu \nonumber \\
   &+(80036553+16943508
   \kappa_1+16943508 \kappa_2) \nu ^2+(-11752020-3159072 \kappa_1-3159072 \kappa_2)
   \nu ^3 \nonumber \\
   &+\delta  \bigg[5338168 \kappa_1-5338168 \kappa_2+(-9126252 \kappa_1+9126252
   \kappa_2) \nu +(4589172 \kappa_1-4589172 \kappa_2) \nu ^2\bigg]\bigg] \Big(\bm{\hat{\ell}}\cdot\bm{\chi}_{s}\Big)^2\Bigg\}\,, 
   \end{align}
   \end{subequations}
   \end{widetext}
\begin{widetext}
\section{Model computational cost}\label{sec:appendix_cost}

An important consideration in choosing a waveform model for follow-up studies of GW events is the evaluation cost of the model. Models have been found to be expensive enough in the past that parameter estimation analyses with them can take months of wallclock time~\citep{LIGOScientific:2016wkq}. Therefore, we quantify here the computational cost of using the \esigmahm{} waveform model. 

We sample $10,000$ binary black hole source configurations and measure the evaluation time of \imresigmahm{} for each of them, starting from an initial GW frequency of $10$Hz {\violet at a sampling rate of $4096$Hz}. We choose $1000$ binary configurations for $10$ distinct values of total mass between $10-100 M_\odot$. We sample mass-ratios uniformly between $[1, 6]$, component spins (both independently) between $[-0.9, 0.9]$, initial eccentricity between $[0, 0.3]$, mean anomaly between $[0, 2\pi)$ and inclination angle between $[0, \pi]$. The resulting distributions of evaluation times, along with their median values, for each binary total mass are shown in Fig.~\ref{fig:cost}. We report the evaluation times with \imresigmahm{}, and also with \imresigma{} using only the dominant $\ell=m=2$ modes. 

\begin{figure}[h!]
    \centering
    \includegraphics[width=0.5\columnwidth]{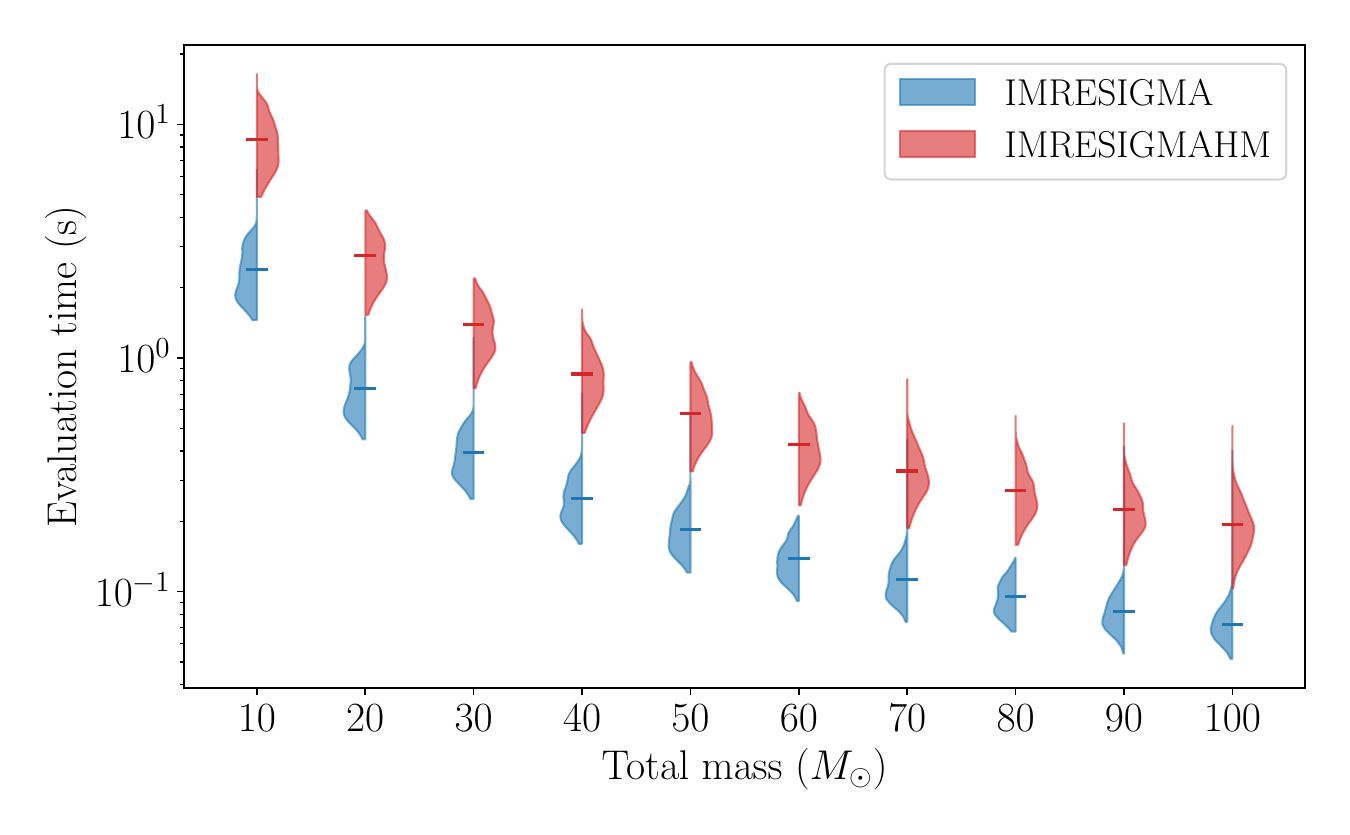}
    \caption{Evaluation cost of \imresigmahm{} and \imresigma{} {\violet sampled at $4096$Hz} generated from an initial GW frequency of $10$Hz is shown as a function of binary total mass. For each value of binary total mass, we sample evaluation times over a uniform distribution of mass-ratio, component spins, eccentricity, mean anomaly and inclination angle. The corresponding median evaluation times are also indicated by the respective markers. The study was performed  on an AMD EPYC 7352 processor operating at 2.3 GHz.}
    \label{fig:cost}
\end{figure}

For sources with total mass above $30M_\odot$ the median evaluation time of \imresigmahm{} (\imresigma{}) remains below $1$ $(0.3)$ second, and decreases significantly for higher masses. For the lowest mass of $10M_\odot$ the evaluation time can reach up to $\sim10$ $(\sim2)$ seconds. {\violet The inspiral mode generation dominates the computational cost, taking roughly twice the time taken for the ODE integration of the inspiral orbital dynamics for \imresigma{}, and becoming about 10 times that for \imresigmahm{}. The generation and attachment of the plunge-merger-ringdown piece generated via \nrsurdqfour{} takes 0.5-7 times the inspiral ODE integration depending on the initial parameters and the mode-content of the waveform, but the inspiral mode generation always dominates the cost. We expect better parallelization of mode generation to further mitigate the overall computational cost of the model. Additionally, we also plan to develop surrogate models for \imresigma{} to address this point.} It is to be noted that these reported time measurements are with minimal optimizations, and we endeavor to improve upon them in our implementation.
\end{widetext}
\section{NR comparison tables}\label{sec:appendix_NR}
\vskip -5pt
Details of the quasi-circular and eccentric NR simulations from SXS catalog~\citep{SXS:catalog} used to compare with \imresigma{} are given in the tables below.
\begin{widetext}
\begin{center}
\begin{table}[h]
\caption{Details of all the quasi-circular NR simulations used in the paper (publicly available in SXS catalog~\citep{SXS:catalog}) such as mass ratio $(q)$, reference eccentricity $(e_{\rm{ref}})$, $z$-component of dimensionless spins ($\chi_{1_z}$, $\chi_{2_z}$), number of orbits $(N_{\rm{orbits}})$, and the corresponding match ($\mathcal{M}$) with \imresigma{} with dominant $\ell = |m| = 2$ modes only. {\violet The ordering of the NR simulations is done first by increasing mass ratio and then by increasing the value of the dimensionless spin parameter corresponding to the primary object.}} \label{table:sxs_id_qc_nr_4PN} 
\begin{tabular}{|c|c|c|c|c|c|c|c|c|c|}
\hline
\hline
Count & Simulation ID & $q$ & $e_{\rm{ref}}$ & $\chi_{1_z}$ & $\chi_{2_z}$ & $N_{\rm{orbits}}$  & $\mathcal{M}$ \\
\hline
1 & SXS:BBH:2091 & 1 & $<10^{-4}$ & -0.6 & 0.6 & 22.6 & 99.3 \\
\hline
2 & SXS:BBH:0222 & 1 & $<10^{-4}$ & -0.3 & $<10^{-6}$ & 23.6 & 98.5 \\
\hline
3 & SXS:BBH:0389 & 1 & $<10^{-4}$ & $<10^{-4}$ & $<10^{-4}$ & 18.6 & 99.3 \\
\hline
4 & SXS:BBH:2097 & 1 & $<10^{-4}$ & 0.3 & $<10^{-4}$ & 23 & 99.6 \\
\hline
5 & SXS:BBH:0304 & 1 & $<10^{-4}$ & 0.5 & -0.5 & 29 & 99.3 \\
\hline
6 & SXS:BBH:0394 & 1 & $<10^{-4}$ & 0.6 & 0.4 & 20.3 & 99.3 \\
\hline
7 & SXS:BBH:0019 & 1.5 & $<10^{-4}$ & -0.5 & 0.5 & 20.4 & 98.1 \\
\hline
8 & SXS:BBH:0007 & 1.5 & $<10^{-3}$ & $<10^{-7}$ & $<10^{-6}$ & 29.1 & 98.9 \\
\hline
9 & SXS:BBH:0593 & 1.5 & $<10^{-4}$ & $<10^{-4}$ & $<10^{-4}$ & 18.8 & 99.1 \\
\hline
10 & SXS:BBH:0025 & 1.5 & $<10^{-4}$ & 0.5 & -0.5 & 22.4 & 99.5 \\
\hline
11 & SXS:BBH:0009 & 1.5 & $<10^{-4}$ & 0.5 & $<10^{-6}$ & 17.1 & 99.6 \\
\hline
12 & SXS:BBH:0013 & 1.5 & $<10^{-3}$ & 0.5 & $<10^{-6}$ & 23.8 & 99.7 \\
\hline
13 & SXS:BBH:2111 & 2 & $<10^{-4}$ & -0.6 & 0.6 & 22.6 & 96.1 \\
\hline
14 & SXS:BBH:1222 & 2 & $<10^{-4}$ & $<10^{-4}$ & $<10^{-3}$ & 28.8 & \textcolor{violet}{98.0} \\
\hline
15 & SXS:BBH:2125 & 2 & $<10^{-4}$ & 0.3 & 0.3 & 23 & 99.7 \\
\hline
16 & SXS:BBH:0255 & 2 & $<10^{-4}$ & 0.6 & $<10^{-4}$ & 23.3 & 99.6 \\
\hline
17 & SXS:BBH:0513 & 2 & $<10^{-3}$ & 0.6 & -0.4 & 20.4 & 99.6 \\
\hline
18 & SXS:BBH:1512 & 2.4 & $<10^{-3}$ & 0.2 & $<10^{-3}$ & 23.7 & 99.1 \\
\hline
19 & SXS:BBH:1453 & 2.4 & $<10^{-4}$ & 0.8 & -0.8 & 21.1 & 99.7 \\
\hline
20 & SXS:BBH:0191 & 2.5 & $<10^{-3}$ & $<10^{-4}$ & $<10^{-4}$ & 22.5 & 97.3 \\
\hline
21 & SXS:BBH:0259 & 2.5 & $<10^{-4}$ & $<10^{-7}$ & $<10^{-6}$ & 28.6 & \textcolor{violet}{97.0} \\
\hline
22 & SXS:BBH:1462 & 2.6 & $<10^{-3}$ & -0.8 & 0.5 & 17.4 & 93.6 \\
\hline
23 & SXS:BBH:0046 & 3 & $<10^{-3}$ & -0.5 & -0.5 & 14.4 & 92.9 \\
\hline
24 & SXS:BBH:2144 & 3 & $<10^{-4}$ & -0.3 & 0.3 & 22.2 & 94.8 \\
\hline
25 & SXS:BBH:1183 & 3 & 0.007 & $<10^{-3}$ & $<10^{-4}$ & 15.2 & 97.5 \\
\hline
26 & SXS:BBH:0280 & 3 & $<10^{-4}$ & 0.3 & 0.8 & 23.6 & 99.6 \\
\hline
27 & SXS:BBH:0031 & 3 & $<10^{-4}$ & 0.5 & $<10^{-6}$ & 21.9 & 99.8 \\
\hline
28 & SXS:BBH:0047 & 3 & $<10^{-3}$ & 0.5 & 0.5 & 22.7 & 99.3 \\
\hline
29 & SXS:BBH:2162 & 3 & $<10^{-3}$ & 0.6 & 0.4 & 27 & 99.8 \\
\hline
30 & SXS:BBH:0288 & 3 & $<10^{-3}$ & 0.6 & -0.4 & 23.5 & 99.8 \\
\hline
31 & SXS:BBH:0294 & 3.5 & $<10^{-4}$ & $<10^{-6}$ & $<10^{-6}$ & 28 & \textcolor{violet}{95.0} \\
\hline
32 & SXS:BBH:1489 & 3.5 & $<10^{-3}$ & 0.3 & -0.2 & 21.2 & 98.5 \\
\hline
33 & SXS:BBH:1906 & 4 & $<10^{-3}$ & $<10^{-4}$ & $<10^{-4}$ & 20.4 & 95.7 \\
\hline
34 & SXS:BBH:2036 & 4 & $<10^{-3}$ & $<10^{-4}$ & -0.4 & 20.1 & 94.8 \\
\hline
35 & SXS:BBH:1942 & 4 & $<10^{-4}$ & 0.4 & -0.8 & 21.6 & 99.5 \\
\hline
36 & SXS:BBH:1937 & 4 & $<10^{-3}$ & 0.4 & $<10^{-4}$ & 22.2 & 99.1 \\
\hline
37 & SXS:BBH:1938 & 4 & $<10^{-3}$ & 0.4 & 0.8 & 22.8 & 99.7 \\
\hline
38 & SXS:BBH:0295 & 4.5 & $<10^{-4}$ & $<10^{-6}$ & $<10^{-6}$ & 27.8 & 93.9 \\
\hline
39 & SXS:BBH:0190 & 4.5 & $<10^{-4}$ & $<10^{-4}$ & $<10^{-4}$ & 20.1 & 95.5 \\
\hline
40 & SXS:BBH:0056 & 5 & $<10^{-3}$ & $<10^{-6}$ & $<10^{-6}$ & 28.8 & 93.5 \\
\hline
41 & SXS:BBH:0110 & 5 & $<10^{-3}$ & 0.5 & $<10^{-6}$ & 24.2 & 99.6 \\
\hline
42 & SXS:BBH:1463 & 5 & $<10^{-3}$ & 0.6 & 0.2 & 24.4 & 99.3 \\
\hline
\hline
\end{tabular}
\end{table}
\end{center}
\begin{center}
\begin{table}[h]
\caption{Details of all the eccentric NR simulations used in the paper (publicly available in SXS catalog~\citep{SXS:catalog}) such as mass ratio $(q)$, reference eccentricity $(e_{\rm{ref}})$, $z$-component of the dimensionless spins ($\chi_{1_z}$, $\chi_{2_z}$), number of orbits $(N_{\rm{orbits}})$, and the corresponding match ($\mathcal{M}$) with \imresigmahm{}. {\violet The ordering of the NR simulations is done first by increasing mass ratio and then by increasing the value of the dimensionless spin parameter corresponding to the primary object.}} \label{table:sxs_id}
\begin{tabular}{|c|c|c|c|c|c|c|c|}
\hline
\hline
Count & Simulation ID & $q$ & $e_{\rm{ref}}$ & $\chi_{1_z}$ & $\chi_{2_z}$ & $N_{\rm{orbits}}$ & $\mathcal{M}$ \\ 
\hline
1 & SXS:BBH:1136 & 1     & 0.124 & -0.75       & -0.75       & 9.5  & 95.9 \\ \hline
2 & SXS:BBH:0089 & 1     & 0.06  & -0.5        & $< 10^{-6}$ & 31.1 & 97.9 \\ \hline
3  & SXS:BBH:1360 & 1     & 0.364 & $< 10^{-6}$ & $< 10^{-6}$ & 13.1 & {\violet 99.0} \\ \hline
4  & SXS:BBH:1356 & 1     & 0.198 & $< 10^{-5}$ & $< 10^{-5}$ & 22.3 & 99.1 \\ \hline
5  & SXS:BBH:1361 & 1     & 0.333 & $< 10^{-5}$ & $< 10^{-5}$ & 13.0 & 98.8 \\ \hline
6  & SXS:BBH:1363 & 1     & 0.24  & $< 10^{-5}$ & $< 10^{-5}$ & 12.2 & 96.1 \\ \hline
7  & SXS:BBH:1355 & 1     & 0.067 & $< 10^{-4}$ & $< 10^{-4}$ & 13.9 & 99.5 \\ \hline
8  & SXS:BBH:1357 & 1     & 0.221 & $< 10^{-4}$ & $< 10^{-4}$ & 14.8 & {\violet 99.0} \\ \hline
9  & SXS:BBH:1358 & 1     & 0.219 & $< 10^{-4}$ & $< 10^{-4}$ & 14.1 & 99.3 \\ \hline
10  & SXS:BBH:1359 & 1     & 0.218 & $< 10^{-4}$ & $< 10^{-4}$ & 13.8 & 99.3 \\ \hline
11  & SXS:BBH:0083 & 1     & 0.025 & 0.5         & $< 10^{-6}$ & 32.4 & 99.4 \\ \hline
12 & SXS:BBH:0320 & 1.22  & 0.023 & 0.33        & -0.44       & 13.5 & 99.4 \\ \hline
13 & SXS:BBH:0321 & 1.22  & 0.061 & 0.33        & -0.44       & 15.0 & 99.6 \\ \hline
14 & SXS:BBH:0322 & 1.22  & 0.107 & 0.33        & -0.44       & 15.0 & 99.2 \\ \hline
15 & SXS:BBH:0323 & 1.22  & 0.194 & 0.33        & -0.44       & 14.6 & 99.1 \\ \hline
16 & SXS:BBH:0309 & 1.221 & 0.029 & 0.33        & -0.44       & 15.8 & 98.9 \\ \hline
17 & SXS:BBH:0246 & 2     & 0.05  & $< 10^{-6}$ & 0.3         & 22.9 & 98.5 \\ \hline
18 & SXS:BBH:1364 & 2     & 0.079 & $< 10^{-4}$ & $< 10^{-3}$ & 16.1 & 98.4 \\ \hline
19 & SXS:BBH:1365 & 2     & 0.114 & $< 10^{-4}$ & $< 10^{-4}$ & 16.1 & 99.1 \\ \hline
20 & SXS:BBH:1366 & 2     & 0.215 & $< 10^{-4}$ & $< 10^{-3}$ & 15.6 & 98.6 \\ \hline
21 & SXS:BBH:1367 & 2     & 0.213 & $< 10^{-4}$ & $< 10^{-4}$ & 15.3 & 98.4 \\ \hline
22 & SXS:BBH:1369 & 2     & 0.24  & $< 10^{-4}$ & $< 10^{-4}$ & 13.9 & 98.2 \\ \hline
23 & SXS:BBH:1370 & 2     & 0.25  & $< 10^{-4}$ & $< 10^{-4}$ & 13.2 & {\violet 97.0} \\ \hline
24 & SXS:BBH:1368 & 2     & 0.212 & $< 10^{-3}$ & $< 10^{-3}$ & 15.0 & 99.1 \\ \hline
25 & SXS:BBH:1169 & 3     & 0.044 & -0.7        & -0.6        & 22.1 & 89.4 \\ \hline
26 & SXS:BBH:1371 & 3     & 0.109 & $< 10^{-4}$ & $< 10^{-4}$ & 18.2 & 98.5 \\ \hline
27 & SXS:BBH:1372 & 3     & 0.214 & $< 10^{-4}$ & $< 10^{-5}$ & 17.7 & 97.7 \\ \hline
28 & SXS:BBH:1373 & 3     & 0.209 & $< 10^{-4}$ & $< 10^{-5}$ & 17.3 & 96.2 \\ \hline
29 & SXS:BBH:1374 & 3     & 0.24  & $< 10^{-4}$ & $< 10^{-4}$ & 15.6 & 97.3 \\ \hline
30 & SXS:BBH:1149 & 3     & 0.046 & 0.7         & 0.6         & 24.1 & 95.7 \\ \hline
31 & SXS:BBH:0105 & 3.001 & 0.022 & -0.5        & $< 10^{-6}$ & 29.7 & 90.5 \\ \hline
32 & SXS:BBH:0108 & 5     & 0.034 & -0.5        & $< 10^{-6}$ & 20.7 & 94.3 \\ \hline
\end{tabular}
\end{table}
\end{center}
\end{widetext}

\clearpage
\bibliographystyle{apsrev4-1}
\bibliography{master}
\end{document}